\documentclass[a4paper,nobind]{ociamthesis}
\usepackage[style=numeric-comp, sorting=none, backend=biber, doi=true, isbn=false
, giveninits=false % just print the initial of the author's first name.
, maxbibnames=100
, backref=true
]{biblatex}
\AtEveryBibitem{\clearfield{number}%\clearfield{url} % Below: code ensures url is printed only if there is no eprint.
\clearfield{issn}\clearfield{day}\clearfield{month}\clearfield{journaltitle}\clearfield{shortjournal}\clearfield{volume}\clearfield{date}\clearfield{pages}} 
\AtEveryBibitem{\clearfield{urldate}\clearfield{urlyear}
\clearfield{pubstate}\clearfield{urlmonth}\clearfield{eprintclass} \clearfield{note} % FUTURE restore this on Zotero re-coupling
\clearfield{language}}
\renewbibmacro{in:}{}
\DeclareNameAlias{default}{given-family}
\newcommand*{\bibtitle}{References}
\newcommand{\inspirebox}[1]{[\href{#1}{\textsc{Inspire}}]} % For appendies

\DeclareSourcemap{
  \maps[datatype=bibtex]{
    \map[overwrite]{
      \step[fieldsource=eprint, final]
      \step[fieldset=url, null]
    }  
  }
}

\DefineBibliographyStrings{english}{
  backrefpage={},
  backrefpages={}
}

\usepackage{xpatch}
\DeclareFieldFormat{backrefparens}{\addperiod\scriptsize{#1}}
\xpatchbibmacro{pageref}{parens}{backrefparens}{}{}

\usepackage{iftex}
\ifLuaTeX %
\usepackage{fontspec} %
\else
\fi
\usepackage{verbatim} %

\usepackage{subcaption}
\usepackage{standalone}
\usepackage{pgfplots}
\pgfplotsset{compat=1.16}
\usepackage{derivative}
\DeclareDifferential{\Dd}{\mathrm{D}}[style-notation=single,
style-notation-*=mixed]

\usepackage{xparse}

\usepackage{amssymb}
\usepackage{amsmath}

\usepackage{float}
\usepackage{xcolor}
\usepackage{cancel}
\usepackage{bm}
\usepackage{mathtools}
\usepackage{mdframed}
\usepackage{hyperref}
\usepackage{cleveref}
\usepackage{ytableau}

\usepackage{amstext} %
\usepackage{nicematrix}
\usepackage{xspace} %

\usepackage{slashed}

\newcommand{\mm}[1]{{\mathfrak{m}_{#1}}}
\newcommand{\Mtm}{{\tilde{\mathfrak{M}}}}
\newcommand{\dotwo}{\tfrac{d}{2}}

\def\det{\mathop{\rm det}\nolimits}

\def\tr{\mathop{\rm tr}\nolimits}

\def\Re{\mathop{\rm Re}\nolimits}

\def\Tr{\mathop{\rm Tr}\nolimits}
\def\bbra{{\langle\kern-2.5pt\langle}}
\def\kket{{\rangle\kern-2.5pt\rangle}}
\def\Bbra{{\Big\langle\kern-3.5pt\Big\langle}}
\def\Kket{{\Big\rangle\kern-3.5pt\Big\rangle}}

\newcommand\swapifbranches[3]{#1{#3}{#2}}
\makeatletter
\MHInternalSyntaxOn
\patchcmd{\DeclarePairedDelimiter}{\@ifstar}{\swapifbranches\@ifstar}{}{}
\MHInternalSyntaxOff
\makeatother

\DeclarePairedDelimiter\abs{\lvert}{\rvert}

\DeclarePairedDelimiterX\braket[2]{\langle}{\rangle}{#1\,\delimsize\vert\,\mathopen{}#2}
\DeclarePairedDelimiter\expval{\langle}{\rangle}

\newcommand   \half{\frac 1 2}
\newcommand   \lptl{\raise .8ex\hbox{$^\leftarrow$} \hspace{-9pt} \partial}
\newcommand   \lrptl{\raise .8ex\hbox{$^\leftrightarrow$} \hspace{-9pt} \partial}

\renewcommand \L  {\Lambda}

\newcommand   \R  {\mathbb{R}}
\newcommand   \Z  {\mathbb{Z}}
\newcommand   \N  {\mathbb{N}}
\newcommand   \Cplx  {\mathbb{C}}
\newcommand   \Ibb {\mathbb{I}}

\newcommand   \SO    {\mathrm{SO}}
\newcommand   \gO    {\mathrm{O}}
\newcommand   \gU    {\mathrm{U}}

\newcommand   \Sp    {\mathrm{Sp}}

\renewcommand \Re    {\mathop{\mathrm{Re}}}

\DeclareMathOperator{\vol}{vol}

\newcommand   \cD {\mathcal{D}}

\newcommand   \cF {\mathcal{F}}

\newcommand   \cL {\mathcal{L}}
\newcommand   \cM {\mathcal{M}}
\newcommand   \cN {\mathcal{N}}
\newcommand   \cO {\mathcal{O}}

\newcommand   \cZ {\mathcal{Z}}

\newcommand{\be}{
  \begin{equation}
  \begin{aligned}
}
\newcommand{\ee}{
  \end{aligned}
  \end{equation}
}

\makeatletter
\DeclareRobustCommand\bea{\@ifnextchar[{\@@bea}{\@bea}}
\def\@@bea[#1]#2\eea{\begin{subequations}\begin{align}#2\end{align}\label{#1}\end{subequations}}
\def\@bea#1\eea{\begin{subequations}\begin{align}#1\end{align}\end{subequations}}
\makeatother

\newcommand\Pochhammer[2]{{\left(#1\right)_{#2}}}

\usepackage[compat=1.1.0]{tikz-feynman}
\tikzfeynmanset{warn luatex=false}

\usepackage{array}   %
\newcolumntype{L}{>{$}l<{$}} %
\makeatletter
\edef\savedcodes{\catcode`\noexpand\_=\the\catcode`\_}
\edef\@tempa{\csname opt@newtxmath.sty\endcsname}
\def\@tempb{{subscriptcorrection}}
\expandafter\expandafter\expandafter
  \in@\expandafter\@tempb\expandafter{\@tempa}
\ifin@
  \catcode`\_=12 %
\fi
\makeatother
\usepackage{tensor}
\savedcodes
\newcommand{\od}[2]{\frac{\mathrm{d}#1}{\mathrm{d}#1}}
\newcommand{\verticalcenter}[1]{\vcenter{\hbox{#1}}}

\newcommand\hp{h_p}
\newcommand\hw{h_w}
\newcommand{\hmelonic}{\ensuremath{h_{\text{melonic}}}\xspace}
\newcommand{\hprismatic}{\ensuremath{h_{\text{prismatic}}}\xspace}
\newcommand{\lammelonic}{\ensuremath{\lambda_{\text{melonic}}}\xspace}
\newcommand{\hlammelonic}{\ensuremath{h\lambda_{\text{melonic}}}\xspace}
\newcommand{\hlamprismatic}{\ensuremath{h\lambda_{\text{prismatic}\xspace}}}

\newcommand{\id}{\mathbb{I}}
\newcommand{\spinid}{\mathbb{I}_s}
\newcommand{\thalf}{\tfrac{1}{2}}
\DeclareMathOperator{\Str}{\mathrm{Str}}
{%
}%

\newcommand{\Gglobal}{\mathcal{G}}
\newcommand{\rhoext}{\rho^\prime}
\newcommand{\Ft}{\tilde{F}}
\newcommand{\FttextOrPDF}{\texorpdfstring{$\Ft$}{F~}\xspace}
\newcommand{\FtextOrPDF}{\texorpdfstring{$F$}{F}\xspace}
\newcommand{\epsPDF}{\texorpdfstring{$\epsilon$}{epsilon}\xspace}
\newcommand{\NPDF}{\texorpdfstring{$N$}{N}\xspace}

\newcommand{\twoPt}{D}

\usepackage{empheq} %
\newcommand*\widefbox[1]{\fbox{\hspace{2em}#1\hspace{2em}}}

\usepackage{silence} % silence warnings
\usepackage{appendix}
\usepackage{chngcntr}
\usepackage{etoolbox}
\usepackage{listings}
\AtBeginEnvironment{subappendices}{%
\chapter*{Appendices}
\addcontentsline{toc}{chapter}{Appendices}
\counterwithin{figure}{section}
\counterwithin{table}{section}
}

\AtEndEnvironment{subappendices}{%
\counterwithout{figure}{section}
\counterwithin{figure}{chapter}
\counterwithout{table}{section}
}

\title{Quantum field theories with many fields}
\author{Ludo Fraser-Taliente}
\college{Wadham College}

\degree{Doctor of Philosophy}
\degreedate{Hilary 2025}

\begin{document}

%%%%% CHOOSE YOUR LINE SPACING HERE
% This is the official option.  Use it for your submission copy and library copy:
%\setlength{\textbaselineskip}{22pt plus2pt}
% This is closer spacing (about 1.5-spaced) that you might prefer for your personal copies:
\setlength{\textbaselineskip}{18pt plus2pt minus1pt}
% DOUBLE SPACING YES?NO

% You can set the spacing here for the roman-numbered pages (acknowledgements, table of contents, etc.)
\setlength{\frontmatterbaselineskip}{18pt plus1pt minus1pt}

% Leave this line alone; it gets things started for the real document.
\setlength{\baselineskip}{\textbaselineskip}

%%%%% CHOOSE YOUR SECTION NUMBERING DEPTH HERE
% You have two choices.  First, how far down are sections numbered?  (Below that, they're named but
% don't get numbers.)  Second, what level of section appears in the table of contents?  These don't have
% to match: you can have numbered sections that don't show up in the ToC, or unnumbered sections that
% do.  Throughout, 0 = chapter; 1 = section; 2 = subsection; 3 = subsubsection, 4 = paragraph...

% The level that gets a number:
\setcounter{secnumdepth}{2}
% The level that shows up in the ToC:
\setcounter{tocdepth}{2}

%%%%% ABSTRACT SEPARATE
% This is used to create the separate, one-page abstract that you are required to hand into the Exam
% Schools.  You can comment it out to generate a PDF for printing or whatnot.
\begin{abstractseparate}
	The large-$N$ quantum field theories provide a window into the regime of strongly-coupled physics. Our principal object of study in this thesis is the large-$N$ family of melonic QFTs, which contain the Sachdev-Ye-Kitaev-like models, tensor models, and vector models.
We begin with a review of this limit of a large number of degrees of freedom (large-$N$) as an approach to the solution of QFTs.
Two toy models are used to clarify this approach: a zero-dimensional field theory and the flow of a generalized free field theory.
Both models are solvable, and so we can explicitly demonstrate: using the former, the simplifications at large $N$; using the latter, the tools used to study scale-dependence of physics -- the renormalization group.
We develop $\Ft$-extremization, a simple method of solution for an arbitrary large-$N$ melonic QFT in its strongly-coupled limit. 
The infrared conformal field theories show remarkable simplicity, in that they are entirely solved by the requirement that they have as many degrees of freedom as possible, up to a simple constraint arising from the interaction between the fields.
We measure the number of degrees of freedom of the conformal infrared theory via $\Ft$, the universal part of the free energy.
We then present the example of the quartic Yukawa model in continuous dimension.
This model is considered as a tensor field theory, and solved for its conformal limit; we then illustrate its multiplicity of fixed points and their stability, as well as its operator spectrum, matching the data between the large-$N$ and dimensional expansions. 
These features reflect general characteristics of melonic conformal field theories: their existence, stability, and spectral characteristics.
We conclude with future directions of exploration for the melonic theories.
 % Create an abstract.tex file in the 'text' folder for your abstract.
\end{abstractseparate}

% JEM: Pages are roman numbered from here, though page numbers are invisible until ToC.  This is in
% keeping with most typesetting conventions.
\begin{romanpages}

% JEM: By default, this template uses the traditional Oxford "Belt Crest". Un-comment the following
% line to use the newer, "Blue Square" logo:
% \renewcommand{\crest}{{\includegraphics[width=4.2cm, height=4.2cm]{figures/newlogo.pdf}}}

% Title page is created here
\maketitle

%%%%% DEDICATION -- If you'd like one, un-comment the following.
%\begin{dedication}
%This thesis is dedicated to\\
%someone\\
%for some special reason\\
%\end{dedication}

%%%%% ACKNOWLEDGEMENTS -- Nothing to do here except comment out if you don't want it.
%\begin{acknowledgements}
 	%\input{text/acknowledgements}
%\end{acknowledgements}

%%%%% ABSTRACT -- Nothing to do here except comment out if you don't want it.
\begin{abstract}
	The large-$N$ quantum field theories provide a window into the regime of strongly-coupled physics. Our principal object of study in this thesis is the large-$N$ family of melonic QFTs, which contain the Sachdev-Ye-Kitaev-like models, tensor models, and vector models.
We begin with a review of this limit of a large number of degrees of freedom (large-$N$) as an approach to the solution of QFTs.
Two toy models are used to clarify this approach: a zero-dimensional field theory and the flow of a generalized free field theory.
Both models are solvable, and so we can explicitly demonstrate: using the former, the simplifications at large $N$; using the latter, the tools used to study scale-dependence of physics -- the renormalization group.
We develop $\Ft$-extremization, a simple method of solution for an arbitrary large-$N$ melonic QFT in its strongly-coupled limit. 
The infrared conformal field theories show remarkable simplicity, in that they are entirely solved by the requirement that they have as many degrees of freedom as possible, up to a simple constraint arising from the interaction between the fields.
We measure the number of degrees of freedom of the conformal infrared theory via $\Ft$, the universal part of the free energy.
We then present the example of the quartic Yukawa model in continuous dimension.
This model is considered as a tensor field theory, and solved for its conformal limit; we then illustrate its multiplicity of fixed points and their stability, as well as its operator spectrum, matching the data between the large-$N$ and dimensional expansions. 
These features reflect general characteristics of melonic conformal field theories: their existence, stability, and spectral characteristics.
We conclude with future directions of exploration for the melonic theories.

\end{abstract}

%%%%% MINI TABLES
% This lays the groundwork for per-chapter, mini tables of contents.  Comment the following line
% (and remove \minitoc from the chapter files) if you don't want this.  Un-comment either of the
% next two lines if you want a per-chapter list of figures or tables.
\dominitoc % include a mini table of contents
%\dominilof  % include a mini list of figures
%\dominilot  % include a mini list of tables

% This aligns the bottom of the text of each page.  It generally makes things look better.
\flushbottom

% This is where the whole-document ToC appears:
\tableofcontents

%\listoffigures
%	\mtcaddchapter
% \mtcaddchapter is needed when adding a non-chapter (but chapter-like) entity to avoid confusing minitoc

% Uncomment to generate a list of tables:
%\listoftables
%	\mtcaddchapter

%%%%% LIST OF ABBREVIATIONS
% This example includes a list of abbreviations.  Look at text/abbreviations.tex to see how that file is
% formatted.  The template can handle any kind of list though, so this might be a good place for a
% glossary, etc.
%\include{text/abbreviations}

% The Roman pages, like the Roman Empire, must come to its inevitable close.
\end{romanpages}

%%%%% CHAPTERS
% Add or remove any chapters you'd like here, by file name (excluding '.tex'):
\flushbottom
%\chapter{Introduction}
\chapter{Introduction: QFT and RG}

The space of quantum field theories is vast. 
The space of quantum field theories that match basic observations of the universe -- its particle content, locality, Lorentz invariance, and conservation of probability -- is tamed, but still enormous.
We know one additional fact: that whatever the ultimate theory that will replace QFT is, it appears at an energy scale as yet beyond our reach.
Adding this alone, we can derive all known particle physics.
The standard model is pinned down exactly, up to a few parameters that must be experimentally measured.

This approach is agnostic to the true theory of reality. 
We do not expect quantum field theory to describe the world at arbitrarily high energies -- rather, it is a description that should work to describe any quantum theory in the limit of field-like degrees of freedom. 
In this thesis, then, we take the following perspective on QFT:
\begin{quote}
    \begin{itemize}
        \item Lorentzian QFT is an effective theory of our quantum reality.
        \item A Lorentzian QFT is a Wick-rotated Euclidean QFT.
        \item A Euclidean QFT is statistical mechanics carefully taken to the continuum limit.
        \item Statistical mechanics is a weighted counting of configurations.
    \end{itemize}
\end{quote}

Unfortunately, solving a quantum field theory -- actually understanding its dynamics -- is hard.
One area where we can make progress is with quantum field theories that contain a large number of fields.
To this end, we seek here to understand the so-called \textit{melonic quantum field theories}, which provide a rare window into strongly-coupled physics; a realm which is typically analytically intractable.

We shall explain our general perspective on QFT and the flows between them further in the remainder of this introduction, paying attention to the role of the free energy, a quantity which is central to this work.
In \cref{chap:largeNQFT}, we will use this perspective to understand the flow of QFTs, via a pair of solvable toy models: a zero-dimensional field theory, and the flow of a generalized free field theory.
In doing so, we will introduce the limit of a large number of degrees of freedom (large-$N$) as an approach to the solution of QFTs and explicitly demonstrate the large-$N$ simplifications in the solvable $0$d model.
We will then review the large-$N$ family of melonic QFTs, which contain the critical vector models, tensor models, and Sachdev-Ye-Kitaev-like models.

In \cref{chap:fextr} we present a general method for solving an arbitrary large-$N$ melonic QFT in its strongly coupled low-energy limit, named $\Ft$-extremization -- work based on \cite{Fraser-Taliente:2024hzv}.
As we will show, this approach is identical to a known supersymmetric solution technique, and demonstrates a remarkable simplicity of the low energy limit of both the large-$N$ and supersymmetric quantum field theories: essentially, we demand that the conformal infrared theory have as many degrees of freedom as possible, up to a simple constraint that arises from the interaction between the fields.
We measure the number of degrees of freedom of the conformal infrared theory via $\Ft$, which is the universal part of the free energy.

In \cref{chap:3dyuk}, based on \cite{Fraser-Taliente:2024rql}, we demonstrate general aspects of the family of melonic conformal field theories, using the specific example of the quartic Yukawa model in continuous dimension. We study its tensor field theory realisation, and solve for the conformal limit.
This model is then used to illustrate the stability, unitarity, and spectral characteristics of the melonic conformal field theories.
We conclude in \cref{chap:conclusion} with possible future directions of exploration for the melonic theories.

\section{Quantum field theory as statistical mechanics}

\newcommand{\NumDofs}{N_\phi}
Consider a collection of $\NumDofs$ degrees of freedom $\{\phi_x\}$: that is, real numbers labelled by $x$ which are free to take any value in $\mathbb{R}$.
We postulate that this assembly explores every possible configuration.
This is an extremely large, continuous space, of size $\R^{\NumDofs}$, so we need to constrain it somehow.

Each $\phi_x$ takes every possible value, but different configurations should contribute differently.
To implement this, to every configuration we assign an action
\begin{equation}
S[\phi] = S[\{\phi_x\}] \in \mathbb{R},
\end{equation}
such that each configuration is taken with a probability $\sim e^{-S[\phi]}/Z$. The normalizing constant $Z$, called the \textit{partition function}, is fixed by demanding that the probabilities sum to $1$. It (and its logarithm, the free energy) are of great importance for the melonic QFTs.

\subsection{The partition function, free energy, and observables}\label{sec:introZandF}
This partition function
\begin{equation}
Z = \sum_{\phi \text{ configurations}} e^{-S[\phi]},
\end{equation}
is just a pure number that depends only on $S$. If each $\phi_x$ is a continuous real-valued parameter, as it will be in full QFT, this formal sum can be implemented as $\NumDofs$ integrals,
\begin{equation}
\sum_{\phi \text{ configurations}} e^{-S[\phi]}\quad \mapsto \quad \int \Dd{\phi} \, e^{-S[\phi]} \equiv \prod_x \left(\int_{-\infty}^\infty \frac{\odif{\phi_x}}{\sqrt{2\pi}}\right) e^{-S[\phi]},
\end{equation}
where the conventional factor of $\sqrt{2\pi}$ is there to cancel otherwise omnipresent factors.

It is perhaps enlightening to note that $Z$ can be thought of as counting the number of \enquote*{active} configurations, where \enquote*{active} may be defined by having a small action: for configurations with $S\ll 1$, $e^{-S} \simeq 1$; for configurations with $S \gg 1$, $e^{-S} \ll 1$. If every single configuration were active (i.e. if $S[\phi]=0$), we would find a horribly divergent quantity
\begin{equation}\label{eq:horriblyDivergent}
Z \sim (\text{size of }\R)^{\NumDofs}.
\end{equation}
Fortunately, we know that typically, most configurations of the universe are unimportant.
One simple way of constraining this which allows us to actually do the integrals is to enforce that each $\phi$ have lowest action when it lies in the range $(-1/a,1/a)$ -- the physical consequence of which is that $\phi$ will \enquote*{usually} be in this range.
For simplicity, we model this with a Gaussian.
Ignoring any higher order terms for the moment, we find
\begin{equation}
Z = \int \Dd{\phi}\, e^{-S[\phi]} = \prod_x \left(\int_{-\infty}^\infty \frac{\odif{\phi_x}}{\sqrt{2\pi}} \, e^{-\half a^2 \phi_x^2}\right) = a^{-\NumDofs},
\end{equation}
to which we return shortly. 
First, we must understand how to probe the system to understand what it is doing.

\subsubsection{Observables and macrostate free energy minimization}

We define the observables as the expectation values of the operators, which are simply functions $\cO(\phi) = \cO(\{\phi_x\})$ of the $\phi_x$s.
Expectation values are weighted by the probabilities,
\begin{equation}
\expval{\cO} \equiv \sum_{\phi \, \mathrm{configurations}} \cO(\phi) P(\phi), \quad P(\phi) \equiv \begin{cases}\frac{e^{-S[\phi]}}{Z} & \text {discrete}\\
\frac{\Dd{\phi}\,e^{-S[\phi]}}{Z} & \text{continuous}\end{cases}.
\end{equation}
This means that each configuration is then given a probability $P(\phi)$; in a mathematical sense, therefore, a choice of action is simply a choice of measure on the configuration space, which can be discrete or continuous.
As we might hope, $\expval{1}=1$. 

The way we will attempt to understand calculations of these expectation values is the following. 
Let us say that there are $\Omega$ very similar configurations $\phi^{\mathrm{min},i}$, called \textit{microstates}, each of which has the smallest possible action value, $S[\phi^{\mathrm{min},i}] \equiv S_\mathrm{min}$. 
Further, imagine that $\cO$ is a sufficiently blurry observable that it cannot tell the difference between the $\Omega$ configurations: $\cO(\phi^\text{min,i}) = \cO(\phi^\text{min,1}) \, \forall i$ -- so $\cO$ depends only on the \textit{macrostate}, not the microstate. 
If all other configurations take an extremely large action value, then
\begin{equation}
\expval{\cO} = \frac{\Omega\, e^{-S_\mathrm{min}} \cO(\phi^\mathrm{min,1}) + \cdots}{\Omega\, e^{-S_\mathrm{min}} + \cdots} \simeq \cO(\phi^\mathrm{min,1})
\end{equation}
and so in this limit, we can approximately calculate $\expval{\cO}$.
What if there are several competing sets of configurations (macrostates) with small values of $S$? Well, if each macrostate $m$ has $\Omega_m$ microstates, then
\begin{equation}
\expval{\cO} = \frac{\sum_m \Omega_m e^{-S[\phi^{m,1}]} \cO(\phi^{m,1}) + \cdots}{\sum_m \Omega_m e^{-S[\phi^{m,1}]} + \cdots}.
\end{equation}
So, approximately $\expval{\cO} \simeq \cO(\phi^{m,1})$ for the macrostate $m$ with the lowest value of $-\log(\Omega_m)+S[\phi^{m,1}]$.
Therefore, it is not $S[\phi]$ that needs to be minimized; it is the \textit{macrostate free energy}
\begin{equation}\label{eq:macrostateFE}
F_m \equiv-\log(\Omega_m)+S[\phi^{m,1}]
\end{equation}
that we minimize over the space of macrostates $\{m\}$. 
The actual expected value of the blurry observable $\cO$ comes from the macrostates that balance a maximization of $\Omega$ with a minimization of the action $S$: that is, the macrostate with lowest $F_m$.

To explicitly disambiguate: in standard statistical mechanics at finite temperature (SM), the action is equal to $U/T$, or rather, the energy divided by the temperature $T$.
Recalling Boltzmann's formula, which in natural units defines the entropy $S_\text{Entropy} = \log \Omega$, we arrive at the usual definition of free energy of a macrostate from SM,
\begin{equation}
\frac{F_\mathrm{SM}}{T} \equiv \frac{U}{T} - S_\text{Entropy},
\end{equation}
such that $\Omega_m \, e^{-S_m} = e^{-F_\mathrm{SM}/T}$. Therefore, in SM the dominant macrostate minimizes the conventional SM free energy $F_\text{SM}$ -- unsurprisingly, a balance between energy and entropy. 

\subsubsection{Free energy of the system}

A related but separate concept is \textit{free energy of the entire system}:
\begin{equation}
F\equiv -\log(Z) = \NumDofs \log a.
\end{equation} 
Evidently, the relation to \eqref{eq:macrostateFE} is that if one free energy $F_m$ is much lower than that of all others, then $F \simeq F_m$. 
The logarithm means that this free energy $F$ is extensive, in the sense that if we have two decoupled systems, $Z=Z_1 Z_2$, then the free energies of each system just add: $F=F_1 +F_2$.
Therefore, $F$ scales with the number of degrees of freedom $\NumDofs$ -- this is a general fact about the logarithm of the partition function.
We can relate this to the interpretation of $Z$ as counting \enquote*{active} configurations (from around \eqref{eq:horriblyDivergent} above): generically, the number of configurations is exponential in the number of degrees of freedom; thus $\log Z$ is proportional to the number of degrees of freedom.

We are allowed to shift the action by arbitrary constants $S' = S + c$. This modifies the partition function while leaving all correlation functions invariant, since $e^{-S'}/Z[S'] = e^{-S}/Z[S]$. Thus, there is an inherent ambiguity in the definition of the free energy -- an ambiguity which also exists in the full QFT case.

\subsubsection{Free energy in QFT}

In this thesis we will be considering QFTs on a Euclidean sphere $S^d$, in which case the free energy is still the logarithm of the partition function $F\equiv -\log Z_{S^d}$.
This is a definition of what we mean by the \textit{sphere free energy}\footnote{$F_{S^d}$ should not be confused with the free energy of a Euclidean QFT at finite temperature $T$, $F_{S^1_{1/T} \times \cM^{d-1}} /T \equiv -\log Z_{S^1_{1/T} \times \cM^{d-1}} \equiv (\vol \cM^{d-1}/T) f'$, with $f'$ the \textit{free energy density}.}.

For continuum QFT the space of possible configurations is continuous.
We may then be able to approximately evaluate the partition function as a sum over the competing saddle points by the standard method of steepest descent.
These saddle points, defined as stationary points of the action, are precisely the solutions to the classical equations of motion.
We generally work in natural units, but briefly restoring $\hbar$, we have that
\begin{equation}\label{eq:saddlePointApprox}
Z = \int_\text{configs.} \Dd{\phi} \, e^{-\frac{S}{\hbar}}\simeq \sum_{\text{saddles}} e^{-\frac{S\rvert_\text{saddle}}{\hbar}} \left(\text{1-loop determinant}\right)(1+ \cdots)
\end{equation}
where the $\cdots$ indicate higher order corrections in whatever dimensionless parameter is used to make the action large. 
\begin{enumerate}[noitemsep]
\item If the action is large compared to $\hbar$, then we can approximate the partition function as just $\sum_{\text{saddles}}e^{-S/\hbar}$ -- of course, this is just the classical limit.
\item If $S \gtrsim \hbar$ we must keep the higher orders; these are quantum corrections.
\item If $S \ll \hbar$, the approximation \eqref{eq:saddlePointApprox} is invalid. 
This is what we mean when we say that the system is strongly coupled; a different approach is required to solve the theory.
\end{enumerate}
If the expansion is valid, we see that the idea of a macrostate in the discrete case is approximately the same as that of a saddle point in the continuum.
What this should make clear is that the fundamental object is the weighted sum over paths (Feynman's sum over histories), and classical physics emerges as a particular limit\footnote{In this sense, the notion of \enquote*{quantizing} a classical system is poorly defined -- all we can do is find a quantum system which reduces to a classical system in a particular limit.
There may be multiple different limits of classical systems -- this is another way of phrasing the notion of duality \cite{Polchinski:2014mva}: that two different classical systems, which when \enquote{quantized} turn out to describe the same quantum system. 
Typically, when one is at strong coupling the other is at weak coupling, and vice versa. 
This is how two classical descriptions can be compatible when one might be in a different number of dimensions than the other, as in AdS/CFT.}.
In this thesis, we will be taking the large-$N$ limit of QFTs: we rewrite our field theories in terms of a new set of degrees of freedom which, in the limit of many fields, behave classically -- in the sense that the expansion \eqref{eq:saddlePointApprox} is under control.

As before, observables are the expected values
\begin{equation}
    \expval{\cO} \equiv \int_\phi  \cO(\phi) P(\phi), \quad P(\phi) \equiv \frac{\Dd{\phi} \, e^{-S[\phi]}}{Z}.
\end{equation}
We stress that $P(\phi)$ is a measure on configuration space: that is, $P$ is a map from $\phi$ configurations to probabilities (or probability amplitudes, in the Lorentzian case), satisfying $\int P =1$.

Finally: in QFT we might expect the free energy to be minimized as well; however, the continuum QFT case is more subtle, and the naive minimization procedure is not quite correct. 
In Lorentzian signature, where the partition function is $\int \Dd{\phi}\, e^{iS}$, it becomes clear that all extrema will provide saddle points that can contribute significantly to the path integral, not just the minima.
In Euclidean signature the situation is less clear, but the saddle points of the path integral that are not minima generically play a role (which we will explicitly demonstrate in \cref{sec:FthmsFmax}).
We discuss the relation between Lorentzian and Euclidean theories further in \cref{sec:WickRotation}. 

\subsubsection{Counting degrees of freedom in QFT}

In the context of a $d$-dimensional QFT (QFT$_d$) the naive $F=-\log Z$ will not count the number of \textit{effective} degrees of freedom in the QFT. 
This is due to many extra degrees of freedom associated with the high-energy details of the theory that do not change physics at low energy scales -- we do not want to count these extra degrees of freedom.
From the perspective of the low-energy theory, these provide ambiguous (regulator-dependent) additional terms.
Happily, the requirement of a continuum limit combined with Lorentz invariance heavily constrains the possible ambiguities.

We shall see that we can still use the non-ambiguous -- or \textit{universal} -- parts of the free energy to quantify the number of degrees of freedom of a conformal field theory. However, it is worth stressing that what to use for a quantum field theory in generic $d$ remains unknown at present.
That universal part is the part which actually changes when the physics changes, rather than being sensitive to irrelevant details like the ground state energy, that do not affect the physics of a QFT.
In particular, for a $d$-dimensional conformal field theory (CFT$_d$) the universal part of $F$ is called $\Ft$, and in $F$ it appears multiplying the volume of $(d+1)$-dimensional hyperbolic space. We shall return to this in \cref{sec:FthmsFmax}.
As mentioned above, this measure of the number of degrees of freedom is what will be extremized when solving our large-$N$ field theories.

There is one, final, extremely important point to make.
The parametrization of the basis of microstates is not absolute.
We can select any basis we want, $\phi'\equiv\phi'(\phi)$, and it may just so happen that one way of performing the calculation is more useful than another.
In physics, this is often phrased as duality: in one duality frame, a calculation is significantly easier; the string dualities are the canonical example of this \cite{Witten:1995ex}. 
To take a drastic example, consider holography: we re-express the degrees of freedom of a strongly-coupled quantum field theory in terms of the degrees of freedom of a weakly coupled quantum gravitational theory in one dimension higher \cite{Polchinski:2014mva}. 
In the latter, we are permitted to expand the partition function around certain saddles -- an invalid approximation in the former.
In \cref{chap:fextr} we will rewrite the partition function from being computed in terms of the fields to being computed in terms of the two-point functions. 
When we have a large number of fields, this is precisely a reparametrization that makes a simple saddle point expansion \eqref{eq:saddlePointApprox} possible for arbitrary coupling, controlled by powers of $1/N$.
Doing so will drastically simplify the process of computing the free energy of the melonic QFTs.

To get to full QFT we must now impose Lorentz invariance on the theory, and take the continuum limit. Let us do so.

\section{Constraining the action} \label{sec:constrainingAction}

With $\NumDofs$ oscillators, the space of possible actions is incredibly complicated: the set of all possible functions of $\NumDofs$ variables.
This is natural, however -- it reflects the ability of the QFT framework to describe arbitrary systems.

We do not know what our underlying reality is, but let us attempt to describe what we can see. 
We can now input some observational data -- the following three observations about the real world:
\begin{enumerate}
    \item $d$-dimensional space exists, with fieldlike degrees of freedom that take values at each point.

    To reproduce this, we make the $x$s label points on some Euclidean $d$-dimensional lattice:
    the $\phi_x$s become fields. We can only measure up to some finite precision in reality, so regardless of the underlying nature of reality, we can model it as a lattice for the moment.
    We shall discuss why we work in Euclidean space shortly, in \cref{sec:WickRotation}, but for the moment we comment that it is a mechanical necessity to make our integrals converge.
    \item Locality: objects that are separated do not directly interact with each other.

To ensure this, we demand that our action depend only locally on the configuration. 
This is highly constraining.
So, we are allowed terms like $\sum_x \phi_x^4$ or $\sum_x \sum_{\hat{n}} (\phi_x - \phi_{x+\hat{n}})^2$ in our action, but not terms like $\sum_x \sum_y {\phi_x \phi_y}$, which couple together $\phi$s that are arbitrarily far apart.
    \item Lorentz invariance.

This means that we cannot single out individual directions in any way. Hence, terms like $\sum_x(\phi_x - \phi_{x+\hat{e}})^2$, for some fixed lattice vector $\hat{e}$, are forbidden. %
\end{enumerate}

To complete the story, we now take a particular continuum limit:
\begin{equation}\label{eq:continuumLimit}
\text{take the limit $\NumDofs \to \infty$, \textit{for fixed observables of the QFT}}.
\end{equation}
Our italicised comment corresponds to the renormalization conditions, which we will explain ultimately in \cref{sec:reparametrizingAction}.
Ignoring the many crucial subtleties such as the mathematical construction of the path integral, the presence of fermions\footnote{For example, we require Grassmann-valued fields, rather than $\mathbb{R}$-valued fields, to describe fermions.}, or gauge symmetries, we are done.
After analytically continuing in time to Lorentzian signature, %
from this framework, it is possible to describe all of quantum field theory and the standard model of particle physics \cite{wittenPhysicsGeometry1987}.

It is important to note that neither QFT nor any lattice construction of it are expected to be fundamental: QFT is simply a general way of describing field-like degrees of freedom, which are what we observe in reality. %
Specifically, Weinberg's \enquote{folk theorem} holds that any quantum theory which at low energy and large distances looks Lorentz invariant and local (technically, satisfying \textit{cluster decomposition}) must be a QFT \cite{Weinberg:1996kw}.
However, for certain QFTs it is possible to take the continuum limit, which then grants access to true (non-gravitational) fundamental theories: the conformal field theories.

It might seem that even with the requirement of locality we have far too much freedom: %
if any interaction at all can be in the fundamental definition $S[\phi]$ of the theory, how can our theory be predictive? 
Worse, not only are we allowed, but we are forced to include all possible terms in the action compatible with the symmetries of the desired model.
This is a manifestation of Gell-Mann's totalitarian principle:
\begin{equation}\label{eq:GellMann}
    \text{everything not forbidden is compulsory.} %
\end{equation}
Even with the requirement of locality, we ought to need to make an infinite number of measurements to fully constrain the function $S[\phi]$: one for every possible term in the action. But if we need to make an infinite number of measurements before we can predict anything, we will never be able to predict anything at all!

Fortunately, physics is possible. The reason for this is the \textit{renormalization group} (RG).
The RG is a very general organizing principle in physics: it states that the best description of nature changes as we look at different scales.
In particular, it tells us that for any given physical theory, \textit{if} it is only used at length scales much larger than the microscopic scale at which it is defined, we need only consider a small number of interactions. %
Concretely, these are the terms in the action that have relevant or marginal scaling dimension, which we explain in \cref{sec:relevanceIrrelevance}.

First, now that we have introduced spacetime, we briefly digress to discuss as promised the connection between physics in Euclidean and Lorentzian signature.

\subsection{Interlude: why do we work in Euclidean space?} \label{sec:WickRotation}

The three observations of \cref{sec:constrainingAction} were implicitly in $3+1$ dimensional Lorentzian space, since we discussed the real world.
However, it proves to be necessary to work in Euclidean space in the context of our large-$N$ QFTs: first and foremost, to make the integrals converge, but also to access our desired quantity $\Ft$. We shall therefore remain there, and not discuss Lorentzian physics.

Thanks to the Osterwalder-Schrader theorem this is entirely permissible even if we want to understand Lorentzian physics.
What this theorem proves is that a well-defined Lorentzian QFT is the analytic continuation of a Euclidean theory to Lorentzian signature by Wick rotation \cite{Kravchuk:2021kwe} \cite{Glimm:1987ylb,Kravchuk:2020scc}.
This is necessary because formally the Lorentzian path integral does not exist -- but the Euclidean path integral does. %
Indeed, the Feynman $i\epsilon$-prescription, used to make Lorentzian theories well-defined, \textit{is} the infinitesimal remainder of a Wick rotation, $\theta\propto \epsilon$.
Hence, we can relate $0+d$ Euclidean QFTs to $1+(d-1)$ dimensional Lorentzian QFTs: the correlators of Lorentzian QFTs are obtained by analytic continuation of the Euclidean correlators\footnote{
This has an interesting consequence: one might hope that in a Euclidean QFT, the $e^{-S}$ will suppress any wild field configurations.
This is not so.
In fact, the wild configurations are crucial: in the Lorentzian quantum theories that are \textit{defined} by analytic continuation in time of sensible Euclidean QFTs, the non-commutativity of a degree of freedom and its conjugate momentum (such as $[x_i,p_j] = i\hbar \delta_{ij}$ for quantum harmonic oscillators) follows directly from the non-differentiability of some of the paths that must be included in the path integral \cite{skinnerAQFT}.
}. 
There is then a one-to-one correspondence between sensible Lorentzian theories (as defined with the $i\epsilon$ prescription) and Euclidean theories, implemented by Wick rotation. Indeed, the unitary (probability-preserving) Lorentzian theories are precisely the Euclidean QFTs that are reflection positive.

Purely Euclidean QFTs are also worthy of study in their own right, as there exist $0+1,0+2$, and $0+3$-dimensional systems which they describe well: statistical mechanical theories. 
One additional datum, should one still be sceptical: there exist physical Lorentzian critical points which have critical exponents matching those of the corresponding physical Euclidean critical points.
In this way Wick rotation has been directly observed in nature \cite{Simmons-Duffin:2016gjk}!

\subsection{Relevance and irrelevance}\label{sec:relevanceIrrelevance}

Wilson taught us that the importance of a particular interaction term in a QFT action depends on the scale at which we probe the theory; all but some finite number of interaction terms in the action become irrelevant (in both a literal and technical sense) as we make measurements at lower and lower energy scales.
They can therefore be ignored.

This is very easy to understand for local Lagrangian field theories at a simple level. Let us say that we define a theory at some extremely high (UV) energy scale $\Lambda$. For example, $\Lambda$ could be the string scale or a lattice spacing. 
The UV theory is under no obligation to look like a local QFT, and indeed in the string theory case will have non-locality on the scale of the strings. 
If we look at this theory at lower and lower energies, the generality of the QFT framework means that we ought to be able to approximate it by a continuum quantum field theory with a fixed cutoff $\Lambda$.
The dimensionless action of that QFT must look like
\begin{equation}
S[\{\phi_i\}]=\int \odif[d]{x} \, \sum_i c^i \Lambda^{d-\Delta_i} \cO_i,
\end{equation}
a sum of local scalar operators $\cO_i[\{\phi_i\}]$ integrated over the spacetime, with dimensionful bare coupling constants $g_b^i \equiv c^i \Lambda^{d-\Delta_i}$,
where all operators allowed by the symmetries of the microscopic definition should be there.
Since we are trying to approximate some unknown theory with a QFT, there should be an infinite number of operators -- this is identical to how the Taylor expansion approximation of an arbitrary function will almost always have an infinite number of terms.
More concretely: the unknown theory is under no obligation to have an action with a finite expansion in powers of $\phi$, particularly since $\phi$ is most likely not the fundamental degree of freedom.

Crucially, $\Lambda$ is the characteristic scale of every term in this action and the way it appears is entirely fixed by dimensional analysis.
Here, $\Delta_i$ is the UV scaling dimension of the operator $\cO_i$, and $c^i$ must be naturally order-one dimensionless constants.
By scaling dimension, we mean the following\footnote{We could make this precise by considering a small perturbation around some conformal field theory with scaling dimensions $\{\Delta_i\}$, but for our purposes this heuristic picture suffices.}: when we calculate $\int \odif[d]{x} \, \cO_i$ for a configuration of the fields $C$ which is of energy scale $\mu$, we find that $\cO_i|_C$ is approximately $\mu^{\Delta_i}$ in a region of volume $\mu^{-d}$, and zero elsewhere:
\begin{equation}
\int \odif[d]{x} \, \cO_i|_C \simeq \mu^{-d} \mu^{\Delta_i} o_i
\end{equation}
for $o_i$ an order one number.

This action has precisely the problem of \eqref{eq:GellMann}: to predict anything we must measure the values of all of the $c^i$s.
Nonetheless, we now attempt to measure any given observable, where that observable is at the scale $\mu \ll \Lambda$.
We therefore only probe configurations $\{C\}$ around that scale. %
Then the action is approximately
\begin{equation}\label{eq:Sapproximation}
S|_C = \sum_i c^i \left(\frac{\Lambda}{\mu}\right)^{d-\Delta_i} o_i  \simeq \sum_i^{\Delta_i \le d} c^i \left(\frac{\Lambda}{\mu}\right)^{d-\Delta_i} o_i = \int \odif[d]{x} \, \sum_i^{\Delta_i \le d} c^i \Lambda^{d-\Delta_i} \cO_i|_C
\end{equation}
since we can drop all of the operators for which $d-\Delta_i$ is negative, as they contribute negligibly! We define
\begin{enumerate}[noitemsep]
\item the operators with $\Delta_i > d$ to be \textit{irrelevant};
\item the operators with $\Delta_i = d$ to be \textit{marginal};
\item the operators with $\Delta_i < d$ to be \textit{relevant}.
\end{enumerate}
Hence, if we are only making measurements at scale $\mu \ll \Lambda$, we are free to set all of the bare couplings $g_b^i=0$ in the action for all of the irrelevant operators, safe in the knowledge that their actual values do not matter up to corrections of order $\mu/\Lambda \ll 1$.

For a typical physical system, there are only a few relevant or marginal operators (i.e. with $\Delta \le d$) that we do need to measure, so we are done.
We conclude that for an \textit{effective field theory} (EFT) intended for use at the scale $\mu \ll \Lambda$, we need only make a finite number of measurements, to fix the values of the $c^i$s for the relevant/marginal operators. All of the irrelevant $c^i$s can be set to zero, without measuring them, and then the theory is predictive.
This is what we are actually doing in practice when considering the so-called \enquote*{renormalizable} theories.
If we try to use an EFT at a scale that is too close to $\Lambda$, the values of all of the other $c^i$s start to become important. Thus, the theory can no longer predict anything: we would need to make an infinite number of measurements to fix all of the $c^i$s.

This is \textit{universality}: regardless of the details of the high energy (UV) microscopic definition, in the low-energy infrared (IR), physics is well described by a small set of numbers $\{c^i\, | \, \Delta_i \le d\}$. 
The precise set of numbers (operators) is fixed by what the degrees of freedom are, the dimension $d$, and the symmetries of the system.

Let us return to the standard model. Since we as yet have not measured the scale $\Lambda$ in our reality, we know that it is large. %
We know the particle content of the universe, and so its field content.
Hence, we need only consider the operators of dimension $\Delta \le 4$ that are Lorentz invariant\footnote{The only hint of irrelevant ($\Delta_i^{\text{free}}>d$) operators that need to be added to the standard model considered as an effective field theory are the neutrino masses -- as well as gravity, of course.}.
We simply construct a linear combination of these operators: this \textit{is} the action of the standard model!

Once we know the particle content of the universe, the form of the standard model action is entirely determined: as we commented at the beginning of this thesis, it is the most general \textit{effective} action compatible with observations.
$\sim 27$ continuous dimensionless parameters must then be measured \cite{Duff:2014mva}, and we can then predict any standard model process. 
Though that is a few more than we would like, it is nonetheless a triumph of reductionism.

\section{Why does physics work at all?}

The discussion above provides an answer to the fundamental question: why does physics work at all? Why do we not need to solve quantum gravity in order to understand a game of billiards? %

If we are performing an experiment at an arbitrary scale, and we want to predict the result, we do need the full quantum-gravitational ultraviolet completion. 
But we are not performing experiments at arbitrary scales. We are stuck down at the bottom of the IR-UV scale, only briefly able to peek at even TeV scales via the work of our experiment colleagues.

Now, in classical physics, we are familiar with the fact that the effective parameters change 
\begin{itemize}
\item at different scales;
\item and due to the presence of interactions.
\end{itemize}
For the former, we know that at smaller scales, the Reynolds number changes. 
This is the reason that flows around bacteria in a microscope appear very viscous, and why for bees, air is an extremely viscous medium. %
For the latter, take a sphere of known \enquote{real} mass $m_0$ and volume $V$, and submerge it in a perfect fluid of density $\rho$ (perhaps a ping-pong ball underwater).
If we attempt to measure its mass, perhaps by applying known forces to it, we instead measure an \enquote{effective} mass
\begin{equation}\label{eq:sphereEffectiveMass}
m=m_0 + \half \rho V;
\end{equation}
the sphere has changed its mass due to its interactions with the continuum\footnote{Specifically, if a sphere moves through a fluid at velocity $v$, the fluid must flow to get out of the way. 
If the total momentum of the equilibrium configuration is calculated, we find precisely $mv$, for $m$ as defined in \cref{eq:sphereEffectiveMass} }\cite[\S 10.1]{Coleman:2018mew}!
This is a mass renormalization which is similar to that which appears in QFT; though in that case, we cannot remove our spheres (particles, like the electron) from the fluid (the interacting vacuum) and therefore the \enquote{effective} mass $m$ is the physical mass, and $m_0$ is an unobservable parameter.

It is also possible for the best description of nature to change entirely at different scales. This is well known in the case of QCD, where strong coupling causes a complete transmutation of the degrees of freedom: in the UV, we have weakly-coupled quarks and gluons; in the deep IR, that theory of quarks and gluons becomes strongly coupled.
However, we can still calculate observables perturbatively if using the low-energy effective field theory of a different set of degrees of freedom: the pions. %

In principle, abstract reductionism would suffice. In practice, it cannot be implemented because of the exponentially increasing amount of computation that would be required to, say, simulate a game of billiards using:
\begin{enumerate}[(a),noitemsep]
\item electromagnetism at the atomic level; \label{item:electromagnetism}
\item the effective theory of electrons and nuclei;
\item the full Standard Model;
\item string theory.
\end{enumerate}
More is different \cite{andersonMoreDifferent1972}, but only because our computers will never be big enough (by orders \textit{of} orders of magnitude).
We can use the laws of classical mechanics instead, trading the $\sim 10^{23}$ degrees of freedom of \ref{item:electromagnetism} for 22 (plus of course two cues).

The phenomenon of universality tells us that this must be what happens.
Regardless of the details of the UV description (the arbitrarily complicated $S[\{\phi\}]$), we end up with some theory governed by only a few small parameters; in the case of electrodynamics and gravity, just the fine structure constant and the electron/nuclei masses.

This is both a blessing and a curse. To build a bridge, we do not need to have studied quantum chromodynamics -- good! The RG tells us that physics deliberately washes out the details of our UV theory -- less good. This makes it hard to find out what the true theory of our reality is.
Universality is to blame for the fact that the way to test what happens at the Planck scale is to build a Planck-scale collider (or use the one given to us by the early universe).

QFT is the best approximation of whatever that theory is (and no more than that).
We can and will understand reality one day. But whatever that reality is, it must contain quantum fields at low scales.

\subsubsection{Flow in theory space}

At any given scale, the best theory to use is the one with the right degrees of freedom and the right choice of parameters, such that the system is near a configuration that we can understand and calculate with.

The reason that the best description changes is the following. 
For an experiment at energy scale $E_\text{ex}$ we could always attempt to use a theory defined at $E_R \gg E_\text{ex}$. 
This works, if we have unlimited computing ability. 
However, the theory at scale $E_R$ will contain too much extra detail: for example, the presence of massive particles $M \sim E_R$ which we cannot excite, and so should not consider.
Additionally, if working perturbatively we would discover large factors $\log(E_R/E_\text{ex})$ at each order of our perturbative series, meaning that the series does not converge. 
The right thing to do is to remove those heavy degrees of freedom from our QFT, and modify the coupling constants to account for this: this defines an \textit{effective field theory} at scale $E_\mathrm{ex}$. %
Changing our scale then smoothly changes the values of the parameters and the number of effective degrees of freedom -- smoothly, as the disappearing degrees of freedom just become so much heavier than the scale of interest, so as to never be excited \cite{Polchinski:1983gv,Rosten:2010vm}.

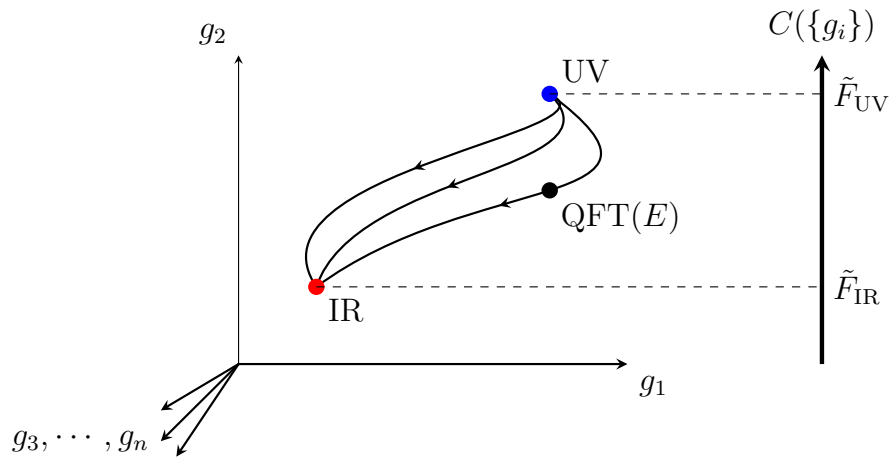
\begin{figure}
\begin{center}
    \begin{tikzpicture}[>=stealth]
	\draw[->, thick] (0,0) -- (5,0) node[below right] {$g_1$};
	\draw[->] (0,0) -- (0,4) node[above left] {$g_2$};
	\draw[->, thick] (0,0) -- (-1,-1) node[left] {$g_3,\cdots, g_n$};
	\begin{scope}[thick,decoration={ markings, mark=at position 0.5 with {\arrow{>}}} ] 
	\draw[postaction={decorate}]  (4,3.5) node[above right] {UV} .. controls (5,2.5) and (1.5,2.5) .. (1,1) node[below right] {IR};
	\draw[postaction={decorate}] (4,3.5) .. controls (5,3) and (0,2.5) .. (1,1);
	\draw[postaction={decorate}] (4,3.5) .. controls (6,2) and (3,2.5) .. (1,1);
\end{scope}
	\draw[->, thick] (0,0) -- (-1,-0.6);
	\draw[->, thick] (0,0) -- (-0.8,-1.2);
	\fill[blue] (4,3.5) circle (3pt); %
	\fill[red] (1,1) circle (3pt); %
	\fill[black] (4,2.25) circle (3pt) node[below right] {QFT($E$)}; %
	
	\draw[->, ultra thick] (7.5,0) -- (7.5,4) node[above] {$C(\{g_i\})$};
	
	\draw[dashed] (4,3.5) -- (7.5,3.5) node[right] {$\tilde{F}_\mathrm{UV}$};
	\draw[dashed] (1,1) -- (7.5,1) node[right] {$\tilde{F}_\mathrm{IR}$};
	
	\end{tikzpicture}
\end{center}
\caption{Flow in the infinite-dimensional theory space. A QFT has some UV definition (which may not be a CFT: perhaps a lattice or stringy UV completion); then as we make measurements at lower and lower energy scales $E$ the best description of reality changes, so we have a QFT($E$); the number of effective degrees of freedom of that QFT, encoded by $C$, should decrease.
At asymptotically low scales (in the IR) we find a fixed point of the flow, which may be empty -- or it may be a \textit{conformal field theory}, which has a number of degrees of freedom measured by $\Ft$. In some scenarios, we also find a CFT as we approach the UV.}\label{fig:UVIRflow}
\end{figure}

The way the change happens is made precise by the \textit{renormalization group flow}, and it is sketched in \cref{fig:UVIRflow} as a flow in the space of renormalized coupling constants $\{g_i\}$. 
We have some high energy definition as $E\to \infty$; then as $E$ decreases\footnote{Naturally, we cannot reverse this flow: we cannot use the standard model to make predictions at the quantum gravity scale, because we have dropped information (the irrelevant $c^i$s).
Thus, the renormalization group is a semigroup: we forget UV details on the way down the energy scale.}, we move in parameter space, until in the IR limit $E \to 0$, we expect a fixed point of the flow: these fixed points are usually the conformal field theories, as we will discuss in \cref{sec:CFTdef}. %

In this thesis we will be exploring precisely this flow in the space of theories -- where it is possible to make sense of it.
As was described in \cref{sec:introZandF}, we can attempt to put a height function $C(\{g^i\})$ on the RG flow such that the UV theory always has more degrees of freedom than the IR does: we do not have such a function for a generic $d$-dimensional QFT, but we have a candidate for its values at the endpoints of the flow (the CFTs): $\Ft$ -- see \cref{fig:UVIRflow}.

At the moment, we have only written down the partition function: we have not solved it. 
It turns out that physics simplifies in the limit of many degrees of freedom, thanks to self-averaging among those degrees of freedom; hence, we now turn to the upper reaches of \cref{fig:UVIRflow}, with $C \sim \Ft \propto N \to \infty$.
In those reaches, we find flows from the UV to the IR that are understandable -- these are the melonic QFTs; the fixed points of those flows are the melonic CFTs.
Solving QFTs is hard, but the large-$N$ limit allows us to see aspects of QFT that are otherwise hidden.

\chapter{Large-\NPDF QFT and CFT} \label{chap:largeNQFT}

In this chapter, we use two solvable toy models to illustrate points about renormalization in the large-$N$ limit, each of which will be useful to understand the melonic theories. In \cref{sec:0dQFT}, based on \cite{Fraser-Taliente20260dQFT}, we study
\begin{enumerate}[(I)]
    \item the zero-dimensional $\phi^4$ QFT. 
    It can be solved exactly for all values of $N$ and the coupling, and serves a twofold purpose: first as a very simple forum for renormalization, and second as a demonstration of the properties of the large-$N$ expansion. 
    However, because it has no notion of spacetime there is no notion of scaling and so no natural flow in theory space.
\end{enumerate}
We therefore move on to $d>0$.
In \cref{sec:CFTdef} we introduce our toolkit of renormalization flow and define the theories that arise at the fixed points of this flow: the conformal field theories (CFTs).
In \cref{sec:solvingCFTs} we discuss the large-$N$ expansion and the $d=4-\epsilon$ expansion in the context of the Wilson-Fisher fixed point.
Then, in \cref{sec:GFFflow}, to illustrate the flow between CFTs, we introduce our second model of
\begin{enumerate}[(I),start=2]
    \item the flow of a generalized free field to another generalized free field.
    This provides a clear example of full-fledged renormalization in a case where we can understand the theory exactly, and it does have a notion of scaling. 
    Further, we can use it to illustrate the continuous-$d$ solution of CFTs, and \textit{precisely} model the RG behaviour of the melonic CFTs that will appear in \cref{chap:fextr}.
\end{enumerate}

\section{Toy model I: zero-dimensional QFT}\label{sec:0dQFT}

A critical observation about QFTs and renormalization is that we never measure the bare parameters that define the action.
We can only measure the observables, which are correlation functions.
Attempting to calculate with those naive bare parameters leads to strong dependence on the regulator that we use to take the continuum limit.
That is fine, as we should never have used them: if we want to use a QFT to predict reality, it is essential to parametrize the action in terms of the results of its correlation functions, such that we can appropriately take the continuum limit with observables fixed \eqref{eq:continuumLimit}. 

We can make this concrete in the case of zero-dimensional QFT.
Take spacetime to be a single point on which we place $N$ degrees of freedom, so $N_\phi=N$. We can think of this as the static ultra-local limit of a single point in real QFT, where nearby points decouple entirely \cite{Argyres:2001sa}.
The real value of this model comes from its exact solvability.

Specifically, we consider the $d=0$ equivalent of $\gO(N)$-symmetric $\phi^4$ theory,
\begin{equation} \label{eq:0dphi4action}
S[\phi_i] \equiv \half \phi_i Z_\phi \phi_i + \frac{1}{8} \frac{\lambda_b}{N} Z_\phi^2 (\phi_i \phi_i)^2.
\end{equation}
Here: the factor of $N$ is chosen to ensure a nice parametrization for large values of $N$; the factor of $8$ is a standard Feynman-diagrammatic automorphism factor which also makes the expressions neat; and we require $\lambda_b>0$ for convergence of the integrals, which ensures stability of the theory.

$Z_\phi$ is the zero-dimensional version of the field renormalization, as there is no kinetic term in zero dimensions.
Naively, we see that as the bare coupling $\lambda_b$ increases $\phi_i$ is localized closer and closer to 0 because the action becomes extremely large away from $\phi_i=0$.
This is only the naive picture: in fact, we will see that \textit{it does not make sense to consider variation of $\lambda_b$ for fixed $Z_\phi$}: in order to be able to compare two theories for different values of $\lambda_b$ we must fix some observable to be the same in each theory.
This \textit{is} renormalization, and the same will apply in the QFT case.

We shall chiefly use this model to introduce renormalization and the large-$N$ limit, and at the end of this section we will compare our results to the higher-dimensional case.
We do not need any regulator here, but the need for renormalization, in the sense of a correct parametrization of the action, will be evident.

\subsubsection{Calculating with unrenormalized quantities}

We can calculate the partition function of this theory \eqref{eq:0dphi4action} by changing to hyperspherical coordinates $r^2 \equiv \phi_i \phi_i$:
\begin{equation}\label{eq:0dZintegral}
Z = \int \Dd{\phi_i} \, e^{-S[\phi_i]} = \frac{\vol S^{N-1}}{(2\pi)^{N/2}} \int \odif{r} \, r^{N-1} \exp\left( -\half Z_\phi r^2 - \frac{1}{8}\frac{\lambda_b Z_\phi^2}{N} r^4\right).
\end{equation}%
Assuming $Z_\phi>0$ and $\lambda_b>0$, substituting $t=\sqrt{\tfrac{\lambda_b}{N}} Z_\phi \tfrac{r^2}{2}$ reveals this to be an integral representation of the parabolic cylinder function $U(a,z)$ \cite[\href{https://dlmf.nist.gov/12.5.1}{(12.5.1)}]{NIST:DLMF}\footnote{
    The function $U(a,z)=$\lstinline|ParabolicCylinderD[-a-1/2, z]| can be rewritten using a Tricomi confluent hypergeometric function $U(a,b,z)$, or \lstinline|HypergeometricU[a,b,z]|, as \cite[\href{https://dlmf.nist.gov/12.7.14}{(12.7.2-14)}]{NIST:DLMF}
\begin{equation}\label{eq:TricomiEqCylinderEqHermite}
U(a,\thalf,\tfrac{z}{2}) = 2^a e^{z/4} U(2a-\thalf, \sqrt{z})=2^{2a} H_{-2a}(\sqrt{\tfrac{z}{2}}) \text{ for } z>0,
\end{equation}
in which case $Z = (2\lambda_b Z_\phi^2/N)^{-\frac{N}{4}} U\left(\tfrac{N}{4},\tfrac{1}{2},\tfrac{N}{2\lambda_b}\right)$ for $Z_\phi>0$ -- which matches the results of \cite{Argyres:2001sa,Benedetti:2022twd}. 
Although strictly valid only for $-2a\in \mathbb{Z}_0$, \eqref{eq:TricomiEqCylinderEqHermite} also generalizes the Hermite polynomials $H_n(z)$ to arbitrary real $n$ \cite[\href{https://dlmf.nist.gov/12.7.2}{(12.7.2)}]{NIST:DLMF}. %
For arbitrary $\lambda_b$ and $Z_\phi$, \eqref{eq:0dZintegral} converges for $\Re\lambda_b Z_\phi^2 >0$. 
Then the $U$ in \eqref{eq:0dphi4PFevaluated} is modified to $U(\tfrac{N-1}{2}, \sqrt{N/\lambda_b} \, \mathrm{sgn}\, \Re \sqrt{\lambda_b} Z_\phi )$. %
}
\begin{equation}\label{eq:0dphi4PFevaluated}
Z = e^{\frac{N}{4\lambda_b}} (\lambda_b Z_\phi^2/N)^{-\frac{N}{4}} U\left(\tfrac{N-1}{2},\sqrt{\tfrac{N}{\lambda_b}}\right).
\end{equation}

Then the propagator, or two-point function, is easy to find by differentiating with respect to $Z_\phi$ at constant $\lambda_b Z_\phi^2$:
\begin{equation}
\sum_i \expval{\phi_i\phi_i}=-2\left(\odv{\log Z}{Z_\phi}\right)_{\lambda_b Z_\phi^2} = \frac{N}{Z_\phi} \frac{ U\left(\tfrac{N+1}{2},\sqrt{\tfrac{N}{\lambda_b}}\right)}{\sqrt{\tfrac{\lambda_b}{N}} U\left(\tfrac{N-1}{2},\sqrt{\tfrac{N}{\lambda_b}}\right)}.
\end{equation}
Assuming $\gO(N)$ symmetry, we must have $\expval{\phi_i\phi_j}= G^{(2)} \delta_{ij}$, and so
\begin{equation}\label{eq:0dPropUnreno}
\expval{\phi_i\phi_j}=\frac{1}{Z_\phi} \frac{ U\left(\tfrac{N+1}{2},\sqrt{\tfrac{N}{\lambda_b}}\right)}{\sqrt{\tfrac{\lambda_b}{N}} U\left(\tfrac{N-1}{2},\sqrt{\tfrac{N}{\lambda_b}}\right)} \delta_{ij} \equiv \frac{P_N}{Z_\phi}\, \delta_{ij},
\end{equation}
which defines a convenient function $P_N(\lambda_b)$. We can go further and also calculate the four-point function. Differentiating with respect to $\lambda_b$, we see that
\begin{equation}\label{eq:phii4full}
\expval{(\phi_i \phi_i)^2} = -\frac{8 N}{Z_\phi^2} \odv{\log Z}{\lambda_b}= \frac{2N^2}{\lambda_b Z_\phi^2}\left(1-P_N\right).
\end{equation}
Once again, $\gO(N)$ symmetry tells us that the uncontracted version of this correlator should be proportional to $\delta_{(ij}\delta_{kl)}\equiv\frac{1}{3}(\delta_{ij}\delta_{kl} +\delta_{ik}\delta_{jl} +\delta_{il}\delta_{jk})$. This contracts to $N(N+2)/3$ when summed with $\delta_{ij} \delta_{kl}$, so we conclude that the full four-point function correlator is
\begin{equation}
\expval{\phi_i \phi_j\phi_k \phi_l} = \frac{3}{Z_\phi^2}\frac{2 N}{N+2}\frac{1-P_N}{\lambda_b} \cdot \delta_{(ij}\delta_{kl)}.
\end{equation}
To get the connected component, we must subtract off $3P_N^2Z_\phi^{-2} \delta_{(ij}\delta_{kl)}$:
\begin{equation}\label{eq:0dPhi4cUnreno}
G^{(4)}_c\equiv \frac{\expval{\phi_i \phi_j\phi_k \phi_l}_c}{\delta_{(ij}\delta_{kl)}} = \frac{3}{Z_\phi^2}\left(\frac{2N}{N+2}\frac{1-P_N}{\lambda_b} - P_N^2\right).
\end{equation}
The full and connected six-point functions are thus
\begin{align}\label{eq:phii6full}
G^{(6)} &\equiv \frac{\expval{\phi_i \phi_j \phi_k \phi_l \phi_m\phi_n}}{\delta_{(i j} \delta_{kl} \delta_{mn)}} = \frac{1}{Z_\phi^3\lambda_b^2}\frac{15 N^2\left(2P_N (\lambda_b (N+2)+2 N)-4 N\right)}{(N+2) (N+4)}\\
G^{(6)}_c &\equiv G^{(6)}-15G^{(2)}G^{(4)} +30G^{(2)}G^{(2)}G^{(2)}.\label{eq:phii6c}
\end{align}
The form of \eqref{eq:0dphi4PFevaluated} has clearly forced the simple $Z_\phi$ dependence of these correlators: indeed, any correlator containing $2n$ $\phi_i$s can only depend on $Z_\phi$ via a factor of $Z_\phi^{-n}$. 
Specifically, we can calculate %
\begin{equation}
\expval{(\phi_i \phi_i)^n} =\frac{1}{Z_\phi^n} 
\left(\frac{N}{\lambda_b}\right)^{n/2}  \frac{2^n\Gamma \left(n+\tfrac{N}{2}\right)}{\Gamma \left(\tfrac{N}{2}\right) } \frac{U\left(n+\tfrac{N-1}{2},\sqrt{\tfrac{N}{\lambda_b}}\right)}{U\left(\tfrac{N-1}{2},\sqrt{\tfrac{N}{\lambda_b}}\right)}.
\end{equation}
Since $\delta_{(i_1 i_2} \cdots \delta_{i_{2n-1}i_{2n})}$, when fully contracted, gives the ratio of Pochhammer symbols $(\tfrac{N}{2})_n/(\tfrac{1}{2})_n$, the uncontracted full correlator must be %
\begin{equation}\label{eq:uncontractedFullCorrelatorUnreno}
G^{(2n)} \equiv \frac{\expval{\phi_{i_1} \phi_{i_2} \cdots \phi_{i_{2n-1}} \phi_{i_{2n}}}}{\delta_{(i_1 i_2} \cdots \delta_{i_{2n-1}i_{2n})}} 
=
\frac{1}{Z_\phi^n} 
\left(\frac{N}{\lambda_b}\right)^{n/2}  \underbrace{\frac{2^n\Gamma \left(n+\tfrac{1}{2}\right)}{\Gamma \left(\tfrac{1}{2}\right) }}_{(2n-1)!!} \frac{U\left(n+\tfrac{N-1}{2},\sqrt{\tfrac{N}{\lambda_b}}\right)}{U\left(\tfrac{N-1}{2},\sqrt{\tfrac{N}{\lambda_b}}\right)}.
\end{equation}
Unfortunately, there is no such nice expression for the general connected correlator $G^{(2n)}_c$. %
However, notably, we can use hypergeometric identities to reduce all correlators $Z_\phi^n \expval{\phi^{2n}}$ to fractions containing only $N$, $\lambda_b$, and $P_N$, such as \eqref{eq:phii4full} and \eqref{eq:phii6full}.
Further, observe that once we have stripped off the index structure, our results \eqref{eq:uncontractedFullCorrelatorUnreno} and therefore also $G_c^{(2n)}$ are analytic in the originally discrete parameter $N$.
That this is possible is due to the combined efforts of hyperspherical coordinates and the continuation to non-integer $N$ of the Pochhammer symbols\footnote{
More mathematically, the fact that the computations still make sense reflects the Deligne-categorical definition of $\gO(N)$ symmetry for non-integer $N$ \cite{Binder:2019zqc}.}. 

\begin{figure}
\centering
\includegraphics[width=0.8\textwidth]{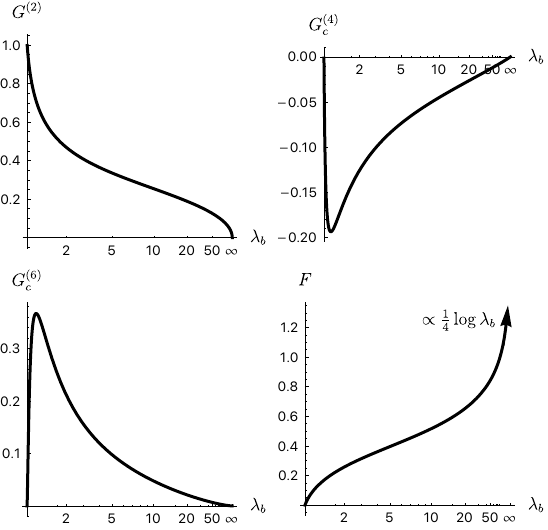}
\caption{The propagator \eqref{eq:0dPropUnreno}, connected four-point \eqref{eq:0dPhi4cUnreno} and six-point \eqref{eq:phii6c} functions, and the free energy $F=-\log \eqref{eq:0dphi4PFevaluated}$ for the \textit{unrenormalized} $N=1$ $\phi^4$ model ($Z_\phi =1$ fixed). All correlators vanish as $\lambda_b\to\infty$, and the free energy diverges (i.e. $Z\to 0$).}
\label{fig:0dphi4unrenormalized}
\end{figure}

If we take $Z_\phi=1$, we can see in \cref{fig:0dphi4unrenormalized} that we have no non-trivial strong-coupling limit: at infinite bare coupling the two-point, connected four-point, and connected six-point couplings all vanish, and the free energy $F=-\log Z$ diverges.
Worse, as we crank up the coupling we find non-monotonic behaviour of the higher-point functions $\expval{\phi^{4,6}}_c$.
The critical observation now is that the fundamental action has no reason to have $Z_\phi=1$: indeed we can set $Z_\phi$ to whatever we want -- and we should set it to something useful. 
The same logic applies in full QFT.

\subsubsection{Renormalization of the two and four-point functions}

Evidently, varying $Z_\phi$ only changes the normalization of correlators -- the functional forms are otherwise unchanged -- so it is a \textit{redundant parameter}. 
$\lambda_b$ by contrast is not redundant. 
If our aim is to compare the results from theories for different values of $\lambda_b$, we should therefore apply two \textit{renormalization conditions} which will fix $Z_\phi$ and $\lambda_b$ respectively.
\begin{enumerate}
\item Our first renormalization condition is to remove the redundant parameter $Z_\phi$ by fixing the propagator $G^{(2)}=1$:
\begin{subequations}\label{eq:0drenoconds}
\begin{equation}\label{eq:0dPhi4PropRenoCondition}
\expval{\phi_i \phi_j} = \delta_{ij} \quad \implies \quad Z_\phi = P_N.
\end{equation}
Even though we are in zero dimensions, and there is no notion of scale, this is precisely analogous to the usual renormalization condition for a scalar field in QFT, where we unit-normalise the propagator\footnote{This is straightforward to motivate from the QFT perspective, as in QFT scattering experiments we only detect single outgoing states -- we have no notion of the amplitude of the quantum field, and therefore it makes sense to unit-normalize the propagator.}. In both cases, the field normalization does not affect the physics, and we simply demand that the fundamental excitation have a particular amplitude.

\item Secondly, we can never actually observe $\lambda_b$: all we can actually observe is the four-point correlator. Again, this is just as in full-fledged QFT.
So, let us re-parametrize our action in terms of the observed four-point coupling instead: defining $\lambda_b \equiv Z_\lambda(\lambda) \lambda$, our second renormalization condition is therefore\footnote{The factor of $3=4!/8$ comes from the normalization of $\lambda_b/8$ in \eqref{eq:0dphi4action}.}
\begin{equation}\label{eq:lambdaDef}
-\frac{3\lambda}{N} \, \delta_{(ij}\delta_{kl)} \equiv \expval{\phi_i \phi_j\phi_k \phi_l}_c,
\end{equation}
\end{subequations}
where $\lambda$ is the renormalized coupling.
Again, the factor of $N$ matches that of the bare action, and has been chosen to ensure that $\lambda$ will be $O(1)$ in the large-$N$ limit. Hence,
\begin{equation}
 \lambda \equiv N+ \frac{2N^2}{N+2} \frac{P_N^{-1}-P_N^{-2}}{\lambda_b}.
\end{equation}
This implicitly defines $Z_\lambda(\lambda)=\lambda_b(\lambda)/\lambda$.
As a check, we find a perturbative expansion $\lambda=\lambda_b+O(\lambda_b^2)$.
\end{enumerate}
\begin{figure}
\centering
\includegraphics[width=0.7\textwidth]{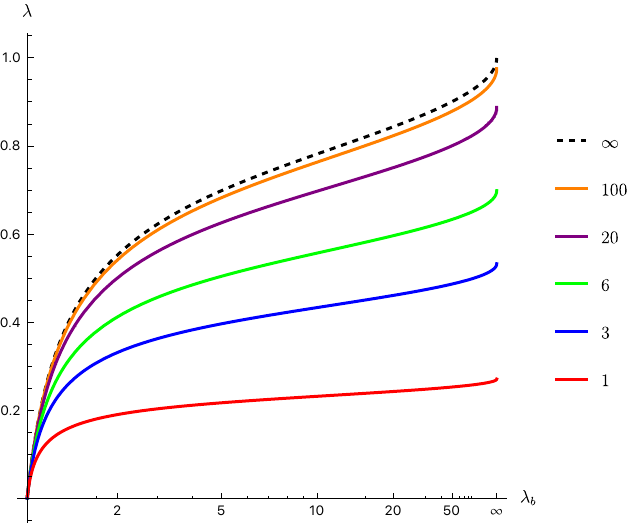}
\caption{The renormalized four-point coupling against the bare coupling $\lambda_b$. That is, here we set $Z_\phi(\lambda)$ and $\lambda_b(\lambda)$ such that $\expval{\phi\phi} =1$ and $\expval{\phi^4}_c =- \frac{3}{N} \lambda  \delta_{(ij}\delta_{kl)}$. We note that for increasing $N$, the four-point function converges very well to the $N=\infty$ result \eqref{eq:lamvsLam0nInfty}.}
\label{fig:0dphi4LambdaAgainstLam0}
\end{figure}
Both of these renormalization conditions are identical to those of the standard approach to perturbative QFT in $\phi^4$ theory, where we unit-normalize the propagator and set the connected four-point function to be minus the coupling constant.

For strong bare coupling, the limit of the renormalized correlators is
\begin{equation}
\lim_{\lambda_b \to \infty} G^{(2n)} = (2n-1)!! \frac{\left[\Pochhammer{\tfrac{N+2}{4}}{\thalf}\right]^n}{\Pochhammer{\frac{N+2}{4}}{\tfrac{n}{2}}},
\end{equation}
so the renormalization condition $G^{(4)}_c \equiv -3 \lambda/N$ gives that $\lambda$ has a finite limit,
\begin{equation}
\lambda_\text{max} \equiv \lim_{\lambda_b \to \infty} \lambda = N\left(1- \frac{\Gamma(\tfrac{N+4}{4})^2}{\Gamma(\tfrac{N+2}{4})\Gamma(\tfrac{N+6}{4})}\right), %
\end{equation}
despite how $Z_\phi \to 0$.
Even better, $\lambda(\lambda_b)$ lies between $0$ and $\lambda_\text{max} \le 1$ for all values of $N$; this is demonstrated in \cref{fig:0dphi4LambdaAgainstLam0}. 
This justifies our convenient normalization of $\lambda_b/N$ in the action \eqref{eq:0dphi4action}, as it means that $\lambda$ remains $O(1)$ for all values of $N$. %
With this done, all quantities depend only on $\lambda$, because $\lambda_b(\lambda)$ and $Z_\phi(\lambda)=P_N(\lambda_b(\lambda))$ are determined by the renormalization conditions \eqref{eq:0drenoconds}.
For example, the renormalized partition function is
\begin{equation}
Z(\lambda) = \int \Dd{\phi} \, e^{-\half \phi_i Z_\phi \phi_i - \frac{\lambda_b}{8 N} Z_\phi^2 (\phi_i \phi_i)^2} = 
e^{\frac{N}{4\lambda_b}} \frac{U\left(\tfrac{N-1}{2},\sqrt{\tfrac{N}{\lambda_b}}\right)^{\frac{N}{2}+1}}{U\left(\tfrac{N+1}{2},\sqrt{\tfrac{N}{\lambda_b}}\right)^{\frac{N}{2}}},
\end{equation}
where we give its integral form for explicitness.

There are no further parameters in our model, so we require no further renormalization conditions. 
We stress that in this context, we \textit{choose} to renormalize based on wanting to compare theories for different values of the coupling, with a fixed two-point function.
Effectively, the question we are asking after renormalization is \enquote{given that our fundamental excitation $\expval{\phi_i\phi_j}$ has a fixed unit amplitude, and the four-point interaction has a given strength $\lambda$, what do the higher-point correlation functions look like?}
The question we ask in higher-dimensional QFT is identical.

\subsubsection{A scaling symmetry -- of sorts}

We note from \cref{fig:0dphi4LambdaAgainstLam0} that as $\lambda_b$ runs to infinity in the renormalized theory the observed coupling $\lambda$ becomes constant.
Our zero-dimensional QFT could therefore be considered as an extremely simple example of a non-trivial field theory with a scaling symmetry, in that if $\lambda_b$ is extremely large, increasing the bare coupling further does nothing -- we are at an approximate \enquote*{fixed point}\footnote{This is not precise, as usually the relevant scaling symmetry is a scaling symmetry of spacetime, which causes flow of the coupling; however, here, spacetime consists of a single point, so there is nothing to scale.
Instead, we just change the coupling. If we had not renormalized, in the different limit $\lambda_b \to \infty|_{Z_\phi=1}$ we find only a trivial theory with all correlators zero, as shown in \cref{fig:0dphi4unrenormalized}.
}.

Thus, this is a precursor of the conformal field theories which we will discuss in the remainder of this chapter.
Further, we will find behaviour analogous to $\lambda(\lambda_b)$ and $F(\lambda)$ in the melonic models (i.e. a finite renormalized coupling and free energy as the bare coupling goes to infinity), so it is important to point out that this behaviour is generic for fixed points.

\subsubsection{Observables in the renormalized $0$d $\gO(N)$ model: the free energy and six-point function}

After renormalizing, we have a theory described by one parameter $\lambda$ such that $G^{(2)}=1$. All other predictions of our model are now fixed -- most trivially, the connected four-point correlator is $G^{(4)}_c = -\tfrac{3}{N} \lambda$, by definition of $\lambda$.
Since we will be focusing on the renormalized free energy $F(\lambda)=-\log Z(\lambda)$ in QFT heavily in \cref{chap:fextr}, we study it next.

\begin{figure}[h]
    \centering
    \begin{subfigure}[t]{0.49\textwidth}
        \centering
        \includegraphics[width=\textwidth]{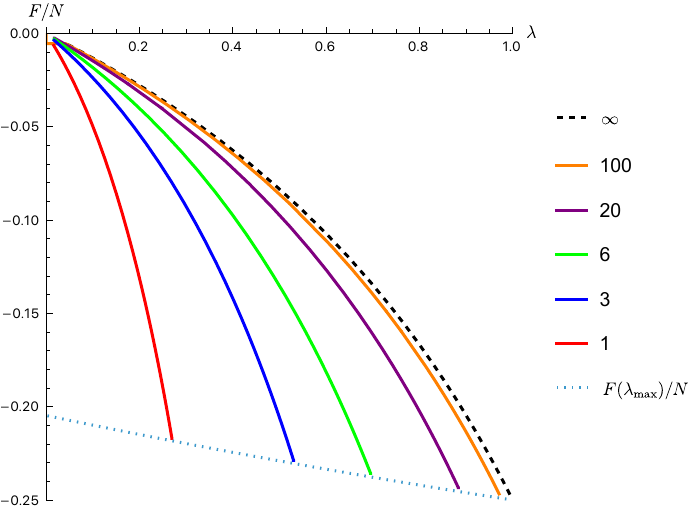}
        \caption{Specific free energy $F(\lambda)/N$.}
        \label{fig:0dphi4_free_energy}
    \end{subfigure}
    \hfill
    \begin{subfigure}[t]{0.49\textwidth}
        \centering
        \includegraphics[width=\textwidth]{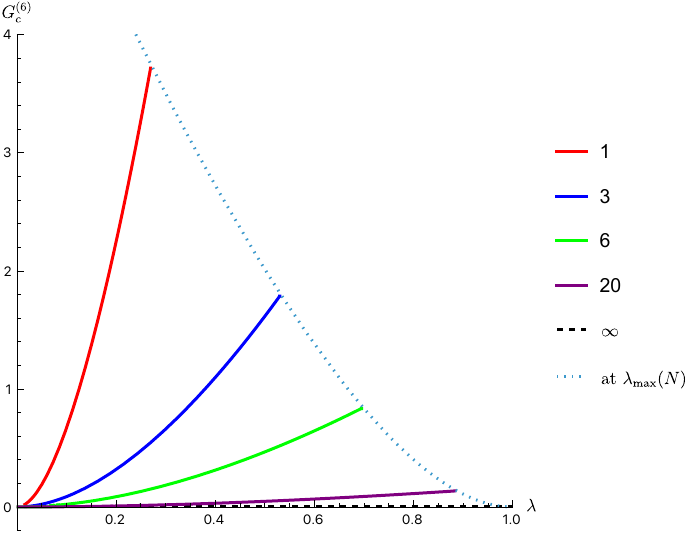}
        \caption{Connected six-point function $G^{(6)}_c(\lambda)$.}
        \label{fig:0dphi4_phi6}
    \end{subfigure}
    \caption{Observables of the renormalized $0$d $\phi^4$ model are fixed after we have set $Z_\phi(\lambda)$ and $\lambda_b(\lambda)$ such that $G^{(2)}=1$ and $G^{(4)}_c=-\frac{3\lambda}{N}$. In each case, we show the $N\to\infty$ limits (zero in the six-point function case), as well as the strong coupling limit at $\lambda=\lambda_{\text{max}}(N)$ (the latter makes it clear why each line suddenly ends).}
    \label{fig:0dphi4_free_energy_and_phi6}
\end{figure}
Thanks to our choice of normalization of the measure, $F$ begins at zero, as $Z=1^N$.
As we increase $\lambda$ from zero, both $F$ and Z decrease, which now intuitively makes sense: by adding a potential, we are constraining the field and so reducing the number of configurations it can take. Hence, $F(\lambda)< F_{\text{free}} = F(0)=0$. %
Crucially, the strong coupling limit of this renormalized free energy ($\lambda \to \lambda_{\text{max}}$, $\lambda_b \to \infty$) is finite. We will see this finite free energy at strong coupling precisely in the case of the melonic models, which will be made most explicit for the long-range models of \cref{sec:LRmodels}. We expect the free energy to be proportional to the number of oscillators, and so we plot not $F$ but the specific free energy $F/N$ in \cref{fig:0dphi4_free_energy}.

Using \cite[\href{https://dlmf.nist.gov/12.10.v}{(12.10(v))}]{NIST:DLMF} $P_\infty\equiv \lim_{N\to\infty} P_N = \frac{2}{1 + \sqrt{1+2\lambda_b}}$, the large-$N$ $\lambda$ is exactly
\begin{equation}\label{eq:lamvsLam0nInfty}
\lambda_{\infty}(\lambda_b) = \lim_{N \to \infty} \lambda = 1-\frac{1}{\sqrt{1+2\lambda_b}}.
\end{equation}
Then the large-$N$ specific free energy takes a simple (finite) form
\begin{equation}\label{eq:0dphi4LargeNF}
\lim_{N\to\infty}\frac{F(\lambda_\infty)}{N} = -\frac{\lambda_\infty}{4} \frac{1}{2-\lambda_\infty},
\end{equation} %
which is valid for $0\le \lambda_\infty \le \lambda_{\text{max}}|_{N=\infty} = 1$, and is shown in \cref{fig:0dphi4_free_energy}.

We might also wonder what happens to other observables, such as the connected six-point correlator: we see from \cref{fig:0dphi4_phi6} that it also has a finite strong-coupling limit for finite $N$.
Note that it vanishes as $G_c^{(6)} \sim 75 N^{-2}$ for large $N$; this is a single case of a general large-$N$ result: 

\subsection{The large-\NPDF limit is a mean field theory} \label{sec:MFTintro}

Consider the full uncontracted correlator $\expval{\phi^{2n}}$ in \eqref{eq:uncontractedFullCorrelatorUnreno} after renormalization,
\begin{equation}
G^{(2n)} =(2n-1)!! \, U\left(n+\tfrac{N-1}{2},\sqrt{\tfrac{N}{\lambda_b}}\right) \frac{U\left(\tfrac{N-1}{2},\sqrt{\tfrac{N}{\lambda_b}}\right)^{n-1}}{ U\left(\tfrac{N+1}{2},\sqrt{\tfrac{N}{\lambda_b}}\right)^n}.
\end{equation}
Taking the large-$N$ limit, we plug in 
\begin{equation}
Z_\phi = P_N = \frac{2}{1+\sqrt{1+2\lambda_b}}\left(1- \frac{1}{N}\frac{\lambda_b}{1+2\lambda_b} + O(1/N^2)\right)
\end{equation}
and \eqref{eq:lamvsLam0nInfty} to satisfy the two renormalization conditions, and so find
\begin{align}\label{eq:0dphi4Gncalc}
    G^{(2n)} &= (2n-1)!!\, \left[1-\frac{1}{N} \frac{n(n-1)}{2} \left(1-\frac{1}{\sqrt{1+2 \lambda_b}}\right) +O(1/N^2)\right]\\
    &= (2n-1)!!\, \left[1-\binom{n}{2}\frac{\lambda_\infty}{N} +O(1/N^2)\right].
\end{align}
Manifestly, $G^{(2n)}$  does not depend on $\lambda_\infty$ to leading order in $N$, and the first subleading order is extremely simple, even though it is non-perturbative in $\lambda_\infty$.
Additionally, the prefactor $(2n-1)!! \equiv 2^n \Pochhammer{\thalf}{n}$ is recognizable as the number of ways to pairwise Wick contract $2n$ free fields. 
Together, these strongly hint that the large-$N$ limit is also simple.

If we consider instead the connected correlators, we can find 
\begin{equation}\label{eq:0dconnectedsNexp}
G_c^{(2)}=G^{(2)}=1, \quad G^{(4)}_c = - \frac{3\lambda_{\infty}}{N}, \quad G^{(6)}_c = \frac{15 \lambda_\infty^2}{N^2}(6-\lambda_\infty) + O(1/N^3),
\end{equation}
where of course the first two are exact due to the renormalization conditions.
This pattern %
continues: all the connected $2n$-point correlators vanish $\propto N^{1-n}$ in the large-$N$ limit, for all values of $0<\lambda_\infty <\lambda_\text{max}$. 
This is easily proven by considering Feynman diagrams: it is evident that in this theory connected contributions to a $2n$-point correlator are obtained by splitting a line in a $(2n-2)$-point correlator diagram -- doing so then reduces its contribution by a factor of $N$.
\eqref{eq:0dphi4LargeNF} shows that $F\propto N$, so the $0$-point correlators (vacuum diagrams) scale $\propto N$, and we are done.

 So, let us sum up. 
 After renormalization, our theory is a function only of $0<\lambda<\lambda_{\text{max}}\le 1$, and:
\begin{enumerate}[noitemsep]
    \item $\expval{\phi_i \phi_j} = \delta_{ij}$.
    \item $\expval{\phi_i \phi_j \phi_k \phi_l}_c = -\frac{3\lambda}{N}\delta_{(ij} \delta_{kl)} \propto N^{-1}$.
    \item $\expval{\phi_i \phi_j \phi_k \phi_l \phi_m \phi_n}_c = G_c^{(6)}(\lambda) \, \delta_{(ij} \delta_{kl} \delta_{mn)}$, with $\lim_{N\to\infty} G_c^{(6)}(\lambda) \propto N^{-2}$.
    \item $\expval{\phi_{i_1}\cdots \phi_{i_{2n}}}_c = G_c^{(2n)}(\lambda) \, \delta_{(i_1 i_2} \cdots \delta_{i_{2n-1}i_{2n})}$, with $\lim_{N\to\infty}  G_c^{(2n)}(\lambda) \propto N^{1-n}$.
    \item $\expval{\phi^{2n+1}} = 0$. 
\end{enumerate}
Thus, any correlation function that we choose is dominated by the disconnected contributions. For example, though we have been discussing the connected four-point function, the full large-$N$ four-point correlator appears to be that of a free theory, up to the $1/N$ corrections that define $\lambda$:
\begin{equation}\label{eq:4ptfull}
\expval{\phi_i \phi_j \phi_k \phi_l} = \delta_{(ij} \delta_{kl)}
-\frac{3\lambda}{N}\delta_{(ij} \delta_{kl)}.
\end{equation}
We describe this behaviour as being that of a \textit{mean field theory}. %
This is a general expectation for theories with many fields: that we can solve to leading order in $N$ in a straightforward way.
If the action is invariant under some symmetry group $G$ of size order $N$, then for $G$-invariant quantities we expect a self-averaging to occur, leading to very small (order $1/N$) fluctuations about the expectation value -- hence we say \textit{mean field}. 
The demonstration of this for a general large-$N$ theory is the same as shown above: in correlators of single-trace operators, connected $n$-point functions for single-trace operators must come from breaking lines in lower-point functions, and therefore have fewer factors of $N$.
Large-$N$ factorization will be the crucial component which allows the solution of the melonic CFTs by $\Ft$-extremization\footnote{Note that for tensor models of rank $r \ge 3$, only restricted families of observables satisfy factorization in the large-$N$ limit -- but the family of melonic observables is one that does \cite{Gurau:2025evo}.}.

\subsection{Comparison to higher dimensions}

A higher-dimensional version of $\phi^4$ theory on a discretized lattice is implemented by putting a copy of \eqref{eq:0dphi4action} on each lattice site; then the fields $\phi_i(x)$ at each site can be coupled by adding terms of the form $\sim \phi_i(x) \phi_i(y)$ to the action.
The arbitrary-$d$ action can then be written
\begin{equation}
S = \sum_{x,y} \frac{Z_\phi}{2} \phi_i(x) C^{-1}(x,y) \phi_i(y) + \sum_x \frac{1}{8} \frac{\lambda_b}{N} Z_\phi^2 (\phi_i(x) \phi_i(x))^2
\end{equation}
for some function\footnote{
The form of $C^{-1}$ is constrained by requiring that this action be local.
After taking the continuum limit, one valid choice is $C^{-1}(x,y)=(-\partial_x^2+m_0^2) \delta^{(d)}(x-y)$, which gives the canonical $\phi^4$ theory.
Other choices are possible, but they are more exotic \cite{Fraser-Taliente:2025udk}.} $C^{-1}(x,y)$; this is the inverse of the $\phi_i$ propagator in the free ($\lambda_b=0$) theory.

Now, in the original zero-dimensional theory, we had to manually change the value of $\lambda_b$.
For actual QFTs, %
recall from \eqref{eq:Sapproximation} that basic dimensional analysis tells us that the coupling constant associated with each term in the action effectively receives a prefactor of $(\Lambda/\mu)^{d-\Delta_i}$. 
Hence, performing experiments at different energy scales is effectively equivalent to varying the bare couplings! %
Manifestly, relevant couplings effectively become stronger for lower $\mu$.
We shall make it explicit with our second toy model, in \cref{sec:GFFflow}.

In these QFTs, then, we may find that as we consider the theory on larger and larger length scales, eventually the theory stops changing -- this is the point marked by IR in \cref{fig:UVIRflow}. 
Thus, the renormalization group singles out a special set of quantum field theories, which are the \textit{fixed points}. 
Under standard assumptions, one of three things happens \cite{Rychkov:2016iqz}:
\begin{enumerate}[(a)]
\item The IR theory has a mass gap, and is trivial (empty).
\item The IR theory is one of massless particles.
\item The IR theory is scale invariant, and has a continuous spectrum.
\end{enumerate}
The case that we shall focus on is the last. 
These QFTs have the special property of scale invariance: the result of any experiment does not depend on its scale.
Usually, this symmetry is enhanced to conformal symmetry, and so these are CFTs. 
The CFTs have no notion of a particle or a scattering matrix, and so the only observables are correlation functions.
The CFTs are therefore signposts in the space of QFTs. %

Indeed, any QFT which has a UV completion is a deformation by some relevant operator of that UV completion. 
If that UV completion is field-like, then it must be scale-invariant (and so conformal), since a UV completion is necessarily a fixed point. 
Therefore, the CFTs are the only QFTs with true continuum definitions.
In other QFTs, we always require either a cutoff of some kind, or a CFT as the UV definition. 
In the latter case, the QFT is the flow away from the CFT, as was sketched in \cref{fig:UVIRflow}. 
For example, this can be done for the QFT of QCD. In this sense QCD was the first UV-complete theory of nature, as QCD \textit{by itself} makes sense for arbitrarily high energy scales due to its asymptotic freedom: for $N_c=3$ and $N_f=6$, QCD becomes a free CFT of quarks and gluons in the UV \cite{Politzer:2005kc}.

We want to understand QFT.
One fruitful approach to studying fluid systems is to study the sources and sinks; with that in mind, we now focus on the sources and sinks of RG flow, the conformal field theories\footnote{
Of course, CFTs do not just turn up in the definitions of QFTs.
They keep popping up in physics, whether in the study of critical phenomena in statistical mechanics, as the dual description of a quantum gravity system in holography, or on the worldsheet in string theory: each appearance compounds the importance of understanding them.
As the physicist once wrote \cite[p. 33]{Einstein1949-AutobiographicalNotes},
\begin{quote}
    A theory is the more impressive the greater the simplicity of its premises is, the more different kinds of things it relates, and the more extended is its area of applicability.
\end{quote}
}.

\section{Conformal field theory}\label{sec:CFTdef}

At the fixed points, typically the Lorentz symmetry $\SO(d)$ is enhanced to the conformal group, $\SO(d+1,1)$, which is highly constraining: hence, we find conformal field theories. That is, typically scale invariance, which by definition exists at the scaling fixed points, is enhanced to conformal symmetry\footnote{There exist various proofs of this, though they do not necessarily apply in arbitrary fractional dimension due to the automatic non-unitarity of such theories which we will discuss in \cref{sec:continuousD}. Some cases of this enhancement not occurring are known, but they are non-unitary \cite{Dorigoni:2009ra, Rychkov:2016iqz}.
} \cite{Polchinski:1987dy,Papadopoulos:2024uvi,Nakayama:2024jwq}.

For practical purposes, this means that the results of correlation functions in flat space must be covariant not only under Poincaré symmetry and scaling, but also under inversions (though technically inversions need not be a symmetry for chiral CFTs \cite{Osborn:2019cftlectures}):
\begin{equation}\label{eq:inversions}
x^\mu \to x'^{\mu} =\frac{x^\mu}{x^2}.
\end{equation}
By considering an inversion, followed by translation $x \to x' = x+b$, followed by another inversion, we generate the special conformal transformations (SCTs)
\begin{equation}
	x^\mu \to x' = \frac{x^\mu + b^\mu x^2}{1+2 b \cdot x + b^2 x^2}.
\end{equation}
Translations correspond to moving the origin; since inversions swap the origin and the point at infinity, these SCTs correspond to moving the point at infinity. 
The SCTs give the extra $d$ symmetries that combine with Poincaré group $+$ scaling to give the global conformal group $\SO(d+1,1)$ of flat space, which has dimension $\frac{1}{2}(d+1)(d+2)$. 
We ignore here the infinite enlargement of the global conformal group to the Virasoro group that occurs in $d=2$.

\subsection{Weyl transformations and conformal transformations}\label{sec:WeylConf}

In the context of QFT, conformal transformations are interpreted as a particular subgroup of Weyl transformations, with an extra compensating diffeomorphism that then restores the metric to its previous form. %

\subsubsection{Weyl transformations}

Weyl transformations are physical modifications of the geometry (i.e. the metric) and (primary) fields of the form
\begin{equation}\label{eq:WeylT}
\text{Weyl: } g_{\mu\nu}(x) \mapsto \Omega(x)^2 g_{\mu\nu}(x), \quad \phi(x) \mapsto \Omega(x)^{-\Delta_\phi} \phi(x),
\end{equation}
which are manifestly local, and generically change the curvature of space; of course, most theories do not have their correlators transforming in a simple way under Weyl transformations. 
The (global conformal) primary fields/operators are the operators that are local enough that they do not feel the effect of the \textit{variation} of the scale factor, and so transform as \eqref{eq:WeylT}; they are exactly the operators that cannot be written as the derivative of another operator (i.e. descendant operators). 

\subsubsection{Conformal transformations}

For an arbitrary metric, the group of conformal transformations are the Weyl transformations for which there exists a coordinate transformation $x^\mu \mapsto x'^\mu(x)$ that makes the Weyl-transformed metric $\Omega(x)^2g_{\mu\nu}(x)$ diffeomorphic to the original metric $g_{\mu\nu}(x)$. 
That is, there exists an $x^\mu \mapsto x'^\mu(x)$ such that the metric in the new $x'$ coordinates,
\begin{subequations} \label{eq:WeyledGisDiffeometric}
\begin{equation}
g'_{\mu\nu}(x') = \pdv{x^\rho}{x'^\mu} \pdv{x^\sigma}{x'^\nu} g_{\rho \sigma}(x(x'))
\end{equation}
is the same as the rescaled metric:
\begin{equation}
    g'_{\mu\nu}(x'(x)) = \Omega(x)^2 g_{\mu\nu}(x).
\end{equation}
\end{subequations}
Hence the actual geometry does not change.
Naturally, under this diffeomorphism the fields transform as usual: 
\begin{equation}
\phi'(x') = \phi(x(x')). %
\end{equation}
In flat space $g_{\mu\nu}= \delta_{\mu\nu}$, a calculation of which $\Omega(x)$s have an associated $x^\mu \mapsto x^{\prime\mu}$ that satisfies this property\footnote{They are the group of transformations with a Jacobian that at each point is proportional to an element of $\SO(d)$; hence they look locally like a rotation and scale transformation \cite{Rychkov:2016iqz}.} identifies exactly the \textbf{global conformal group} $\SO(d+1,1)$ described above.

Flipping this argument around, we see that the flat space conformal transformations $x^\mu \to x'^\mu(x)$ are precisely those that leave the metric of flat space invariant up to some particular conformal factor $\Omega$.
For example, for inversions \eqref{eq:inversions} of flat space $g_{\mu\nu} = \delta_{\mu\nu}$, we find $g'_{\mu\nu}(x'(x)) = (x^2)^{-2} \delta_{\mu\nu}$, so $\Omega=(x^2)^{-1}$ satisfies \eqref{eq:WeyledGisDiffeometric}.
Crucially, the physical metric does not change, since $g'$ and $g$ are diffeomorphic; however all the fields do transform. 
This in turn makes line elements in the new coordinates the same as the old ones up to a local scale factor $\odif{x'^2} = \Omega(x)^2 \odif{x^2}$. 
We therefore say that a conformal field theory is invariant under transformations that preserve angles but not necessarily lengths. 

Following the Weyl transformation with the compensating diffeomorphism, we see that conformal transformations are
\begin{equation}\label{eq:conformalMap}
    \text{Conformal: } g_{\mu\nu}(x) \mapsto {\Omega}(x)^2 g_{\mu\nu}(x)=g'_{\mu\nu}(x'(x)), \quad \phi(x) \mapsto \Omega(x)^{-\Delta_\phi} \phi(x),
\end{equation}
for the ${\Omega}$s that satisfy \eqref{eq:WeyledGisDiffeometric} \cite{Farnsworth:2017tbz}. 
Thus, a conformal transformation involves both a diffeomorphism which transforms the metric by a conformal factor and a particular Weyl transformation that removes that factor; if the measure of a QFT is invariant under this conformal transformation, we say that it is conformally symmetric.

\subsection{Conformal invariance and Weyl invariance}

\subsubsection{Conformal invariance}

We take the case of the theory of a scalar field $\phi$ and vector field $A_\mu$. If under the conformal transformation encoded by $x \mapsto x'$ and the associated $\Omega(x)$ the metric transforms as \eqref{eq:WeyledGisDiffeometric}, then under the replacements
\begin{subequations}
\begin{align} %
\phi(x) &\mapsto \tilde{\phi}(x') = \Omega(x)^{-\Delta_\phi} \phi(x)\, ,
\\
A_\mu(x) &\mapsto \tilde{A}_\mu(x') =\Omega(x)^{-\Delta_A} \frac{\partial x^\nu}{\partial{x'}^\mu}A_\nu(x)\, ,
\\
g_{\mu\nu}(x) &\mapsto \tilde{g}_{\mu\nu}(x') = g'_{\mu\nu}(x') \stackrel{\eqref{eq:WeyledGisDiffeometric}}{=} \Omega(x)^2 g_{\mu\nu}(x), %
\end{align}
\end{subequations}
the (unit-normalized) probability measure does not change
\begin{equation}\label{eq:Pinvariant}
    P_{g}(\phi,A) = \Dd{\phi} \Dd{A} \, e^{-S[\phi,A,g]}/Z = \Dd{\tilde\phi}\Dd{\tilde{A}} \, e^{-S[\tilde\phi,\tilde{A},\tilde{g}]}/Z' = P_{\tilde{g}}(\tilde{\phi},\tilde{A}),
\end{equation}
then we say the theory is conformally invariant. 

This has the following consequences. We have that the correlation functions of $\tilde{\phi}$, $\tilde{A}$ are the same as those of $\phi$ and $A$. 
\begin{equation} \label{eq:sameCorrelators}
       \expval*{\phi(x)\phi(y)}= \expval*{\tilde{\phi}(x) \tilde{\phi}(y)}
\end{equation}
This is because $\tilde{\phi}$ can be thought of as the image of $\phi$ under a spatially non-uniform RG transformation that leaves the measure invariant; for this reason the $\sim$ is often dropped \cite{Rychkov:2016iqz}. 
We can now find constraints from \eqref{eq:Pinvariant}.
For example, take the two-point function evaluated on two different pairs of points on flat space, related to each other by the map $x \mapsto x'$: 
\begin{equation}
    \begin{aligned}\label{eq:twoPointFuncConstraint}
&\expval*{\phi(x')\phi(y')} \stackrel{\eqref{eq:sameCorrelators}}{=} \expval*{\tilde{\phi}(x') \tilde{\phi}(y')}  = \int P_{\tilde{g}}(\tilde{\phi}) \, \tilde{\phi}(x') \tilde{\phi}(y')\\
&\stackrel{\eqref{eq:Pinvariant}}{=} (\Omega(x)\Omega(y))^{-\Delta_\phi} \int P_g(\phi) \, \phi(x) \phi(y) = (\Omega(x)\Omega(y))^{-\Delta_\phi}  \expval{\phi(x) \phi(y)}.
    \end{aligned}
\end{equation}
Should this identity hold only for $\Omega(x) =1$, then this precisely describes the usual Poincaré invariance. 
If this identity holds for arbitrary $x \mapsto x'$ in $\SO(d+1,1)$, the functional form of the correlators in the theory is heavily constrained.

\subsubsection{Weyl invariance}

If this identity of measures \eqref{eq:Pinvariant} holds for arbitrary Weyl transformations \eqref{eq:WeylT} away from a general background metric $g_{\mu\nu}$, then we say that the theory is Weyl invariant/covariant at the quantum level. 
This applies even if there is a Weyl \enquote*{anomaly}, as Weyl Ward identities still hold, and Weyl invariance can still be regarded as a good symmetry; the \enquote*{anomaly} just encodes the response of the partition function to the geometry \cite{Farnsworth:2017tbz}.
Then correlation functions of primary operators in the two different geometries are related by
\begin{equation}
\text{Weyl invariance: }\expval{\phi(x) \phi(y) \cdots}_{\Omega^2 g} = \Omega^{-\Delta_\phi(x)} \Omega^{-\Delta_\phi(y)}\expval{\phi(x) \phi(y) \cdots}_{ g}.
\end{equation}
We note that Weyl transformations\footnote{These are not strictly a symmetry under the usual definition, as a background field -- the metric -- transforms non-trivially.} are an infinite-dimensional group, larger even than the $d=2$ conformal group, the Virasoro group -- which roughly corresponds to the restricted set of holomorphic Weyl transformations $\Box \Omega =0$. 

These definitions make it manifest that Weyl invariance of a QFT implies conformal invariance, but the converse is not necessarily true \cite[\S 2.3]{Fraser-Taliente:2026gdh}.
However, a CFT can generally be coupled to a metric $g_{\mu\nu}$ in such a way that the conformal symmetry descends from Weyl invariance -- although there are exceptions in certain non-unitary theories and not all metric couplings are Weyl invariant \cite{Farnsworth:2021ycg} (for example, for certain even integer values of $d$s, the higher-derivative free CFTs cannot be made Weyl covariant due to the presence of \enquote{obstruction tensors} \cite{Stergiou:2022qqj,Fraser-Taliente:2026gdh}).
Thus, we typically can find the following pattern of enhancement of invariance for unitary Poincaré-invariant field theories:
\begin{equation}
    \text{Flat space scaling} \quad \to \quad \text{Flat space conformal} \quad \to \quad \text{Weyl}.
\end{equation}
The full conditions for these uplifts are not known at present. 
However, to sum up:
\begin{itemize}
\item Flat space conformal and scale invariance (\eqref{eq:conformalMap} and the same \eqref{eq:conformalMap} but with $\Omega$ constant, respectively) relate correlators of theories defined on the same geometry (flat space);
\item Weyl invariance generically relates correlators of theories defined on different geometries.
\end{itemize}
In \cref{chap:fextr} we will consider CFTs on Euclidean spheres with the round metric
\begin{equation}
g_{\mu\nu}^{S^d} = \Omega(x)^2 \delta_{\mu\nu}, \quad \Omega(x) \equiv \frac{2R}{1+x^2},
\end{equation}
which manifestly is obtained from flat space $g_{\mu\nu}=\delta_{\mu\nu}$ by a Weyl transformation \eqref{eq:WeylT} encoded by $\Omega$. Hence, if the CFT is Weyl invariant the correlators on the two spaces are very simply related to each other.

If we have a CFT in flat space, Weyl invariance of the theory means that we also can compute its correlators in any space that differs from flat space by a Weyl map \eqref{eq:WeylT}. 

Critically, for our purposes, the Euclidean sphere can be Weyl transformed to flat space: this is explicit when using the coordinate system of the stereographic projection, as then the metric is proportional to the flat one.
So, when calculating the conformal data of a CFT it does not matter whether we do so on the plane or on the sphere -- in \cref{chap:fextr} we will be doing the latter.

\subsection{The constraints of conformal symmetry}

The simplest observables of a CFT are very highly constrained. For one, two, and three-point functions of \textit{primary operators} (as above, operators that cannot be written as a derivative of another operator), there are very few functions that transform covariantly, and so:
\begin{enumerate}
    \item One-point functions are forced to be zero.
    \item Two-point functions are forced to take a simple form that depends only on the normalization $Z_\Phi$ of the fields, their Lorentz representation, and one more piece of data: their scaling dimension $\Delta_\phi$. For example, for scalar fields $\phi$, the conformal invariance encoded by \eqref{eq:twoPointFuncConstraint} requires that the flat-space propagator take the form
\begin{equation}
\expval{\phi(x) \phi(y)} = \frac{\cZ_\phi}{\abs{x-y}^{2\Delta_\phi}}.
\end{equation}
We say that the two-point function of any primary field $\Phi$, up to normalization, depends only on the \textbf{conformal representation} of $\Phi$, which is a combination of its scaling dimension $\Delta_\Phi$, and its Lorentz representation $\rho_\Phi$. This is written compactly as
\begin{equation}
(-\Delta_\Phi, \rho_\Phi).
\end{equation}
These are almost standard group-theoretic Dynkin labels of $\SO(d+1,1)$: see \cite[eqs. (2.1-5)]{Kravchuk:2017dzd} for the precise isomorphism \cite[\S 3.3.1]{Karateev:2018oml}. 
Since there is one non-compact direction in $\SO(d+1,1)$, there is one entry that is not forced to be (half-)integer: the scaling dimension\footnote{In Lorentzian theories, there is a second non-compact direction in $\SO(2,d)$. Hence, the spin $J$ can be continuously varied to give the light-ray operators, which are non-local for non-integer $J$ \cite{Kravchuk:2018htv}. }.
We typically choose to unit-normalize all of the primaries, so $\cZ_\Phi=1$.
There may also be some global symmetry group $\Gglobal$ in the CFT. 
In this case, the operator's data also includes further quantum numbers, which encode its representation $R_\Phi$ of $\Gglobal$; we can package this into a representation $\rhoext_\Phi=\rho_\Phi \times R_\Phi$ of $\SO(d) \times \Gglobal$. 

\item Three-point functions are constrained to depend on a finite number of numbers $\{C^{(n)}_{ijk}\}$, called the \textbf{OPE coefficients}
\begin{equation}
\expval{\phi_i(x)\phi_j(y) \phi_k(z)} = \sum_n C_{ijk}^{(n)} G^{(n)}(x,y,z),
\end{equation}
where the range of $n$ is typically small -- for example, there is only one conformal structure in the case of three scalar fields \cite{Kravchuk:2016qvl}. The $G^{(n)}$s are simple universal functions that depend only on the conformal representations of the fields $\phi_{i,j,k}$, and for unitary theories, $C_{ijk}$ should be real.
\end{enumerate}
Correlators of descendant fields, which are obtained as derivatives of primary fields, can naturally be obtained from those of primaries.

\subsection{Defining a CFT: the spectrum and coefficients}

For higher-point functions, the situation is more complicated. 
There is one final tool which is of great importance: inside any correlation function, the product of two operators can be expanded via the \textbf{operator product expansion}
\begin{equation}
\expval{\phi_i (x) \phi_j (y) \cdots} = \sum_{\cO \text{ primary}} \sum_n C_{ij\cO}^{(n)}\, H^{\cO,(n)}_{ij}(x-y, \partial_y)\, \expval{\cO(y) \cdots}, %
\end{equation}
where we assume that there are no other fields in the way: we require $\abs{x-y}< L$, where $L$ is the distance to the nearest other field in the correlator.
Once again, the functions $H^{\cO,(n)}_{ij}(x-y)$ are universal, and depend only on the conformal representations of the three fields. This is extremely powerful, as by performing successive OPEs we can express any correlator of local fields that we want as a function only of the spectrum $\{(-\Delta_i, \rho_i)\}$ and the OPE coefficients $\{C^{(n)}_{ijk}\}$.

That is, to define (at least the local part of) a conformal field theory, it suffices to have a spectrum -- a list of conformal representations -- and then the OPE coefficients.
For $d>2$, a full definition may require data for extended operators too -- but this is less clear (see section 6.2 of \cite{Belin:2016yll}), and if we only care about correlators of local operators, $\{(-\Delta_i,\rho_i), C_{ijk}\}$ suffices.
This defines what we mean by \textit{conformal data}. 
In the remainder of this thesis, when we talk about solving CFTs, we mean computing these scaling dimensions.

If any of the scaling dimensions or OPE coefficients are complex, then the theory, when Wick rotated to be Lorentzian, is not unitary: but it still may define a \textit{non-unitary CFT}.
In this thesis, when we refer to unitary CFTs, we mean that the CFT is unitary (probability-conserving)  when continued to Lorentzian signature. Back in Euclidean signature, this means that the theory satisfies \textit{reflection positivity}, in that the norms of all states are positive \cite{Rychkov:2016iqz}.
More fatally, the conformal data may not be consistent, in that it may not satisfy crossing symmetry: such data therefore cannot define a CFT\footnote{It may define an approximate CFT, which has been studied in context of AdS/CFT \cite{Belin:2023efa,Jafferis:2025vyp}.}.

\subsection{Justification: continuous dimension}\label{sec:continuousD}

It will prove useful to consider the dimension of a conformal field theory to be continuous.
Continuous $d$ allows us to connect \textit{a priori} different unitary field theories in integer dimensions by a line of non-unitary field theories.
As we will see in \cref{sec:solvingCFTs}, the $4$d free boson, the $3$d critical vector model, and the $2$d $\mathcal{M}_{3,4}$ minimal model are three unitary CFTs within a continuous line of Wilson-Fisher CFTs that exist for arbitrary $d$.
This fact will permit us to analytically approach the strongly-coupled 3d CFT which is otherwise intractable: we just analytically continue in $d$.

The most rigorous construction\footnote{We might also attempt to justify non-integer dimension by appealing to field theories defined on fractal lattices, which naturally have non-integer (Hausdorff and spectral) dimensions \cite{Eyink:1989dv, carmonaCriticalPropertiesIsing1998, Delporte:2019tof}. Though such theories can be scale-invariant, 
they will not have the $\SO(d+1,1)$ symmetry which we require, and therefore have no reason to be fully conformal.} of fractional-dimensional CFTs with full $\SO(d+1,1)$ symmetry (whatever that means) is via the \textit{conformal bootstrap}. 
This is an approach which constrains the conformal data by imposing that the OPE be consistent when done in a different order (i.e. crossing-symmetric).
The required functions, called conformal blocks, exist for arbitrary $d\ge 2$, are always functions of only two variables, and are constructed as eigenfunctions of the conformal Casimir operator -- which is analytic in $d$. The notion of crossing-symmetric conformal data can therefore be extended to non-integer dimensions, and so it is possible to bootstrap for non-integer $d$ \cite{El-Showk:2013nia} (see also \cite{Cappelli:2018vir,Henriksson:2022gpa,Bonanno:2022ztf,Henriksson:2025kws,Golden:2014oqa}).
Additionally, some fractional-$d$ CFT data can be calculated in the FRG approximation \cite{Codello:2014yfa}, and the data from the various analytic and numerical sources seem to agree \cite{El-Showk:2013nia}.
More surprisingly, fractional-dimensional CFTs arise as an auxiliary structure in non-conformal D$p$-brane holography; the holographic non-conformal correlators can be found by integrating correlators from an auxiliary CFT defined in fractional dimension $d= p+1+\eta$ over the extra fractional dimensions, $\expval{\cO} = \int \odif[\eta]{x}\, \expval{\cO}_{d=p+1+\eta}$, where $\eta=(3-p)^2/(5-p)$ \cite{Bobev:2025idz}.
These appearances suggest that these models may actually exist, though the definition of a field theory in non-integer dimension beyond perturbation theory remains unknown.%

\subsubsection{Automatic non-unitarity in fractional dimensions}

These theories in non-integer dimension are automatically non-unitary: \cite{Hogervorst:2015akt} found that the so-called evanescent operators, which decouple (and have zero norm) in integer dimension, have complex scaling dimensions -- albeit only for operators of dimension $\Delta \gg 1$, which normally we do not deal with.
This non-unitarity is inherent to the theory and cannot be decoupled: generically the scaling operators in non-integer $d$ are a combination of physical and evanescent operators \cite{Hogervorst:2015akt}. %

We note that there does appear to be something special about the non-unitary CFTs that are unitary when continued to integer $d$, as opposed to other CFTs.
For example, they are seemingly unitary at low orders, with violation only due to the aforementioned high-dimension evanescent operators. 
This is not understood at present. 
In any case, we shall always consider the theories of interest in this thesis as defined in continuous dimension; we will remain interested in all of the CFTs that arise, despite any non-unitarity.

\subsubsection{Practical aspects of DREG}

As a practical justification of the use of continuous dimension, it allows a straightforward regularization of power law divergences: only the logarithms then need to be tamed.
Essentially, whenever integrating, we always implicitly analytically continue to low enough dimension that all power law divergences vanish: this is termed dimensional regularization (DREG).
DREG makes it evident that the need to regularize infinities is often a distraction from the actual work of renormalization (something which we shall discuss again in \cref{sec:reparametrizingAction}), as most of the infinities disappear \enquote{automatically}.
In practice, this means that analytic continuation in $d$ corresponds to an implicit selection of local Lorentz-invariant counterterms for power-divergent observables, which exactly cancel the seemingly automatically removed divergences.

More concretely, in an EFT approach to QFT, we can consider a QFT with a finite cutoff $\Lambda$ in an arbitrary dimension $d$.
Then, given the standard prescription for how to integrate over $d$-dimensional space for $d \notin \mathbb{N}$, physicists \textit{notice} that, in perturbation theory, observables are typically\footnote{Though not always: for example, in the critical vector model's large-$N$ expansion, at higher orders the bare interaction term is marginal in any $d$, and so a further analytic regulator beyond DREG, called DREG$+\delta$, is required \cite{zinn-justin_quantum_2002,Gracey:2018ame,Fraser-Taliente:2025udk}.
However, after this further regularization is correctly handled, the same conclusions apply: we can analytically continue in $d$.
We must be careful, though, because the analytic continuation in $d$ may lead to non-unitary (and possibly complex) theories that may not correspond to the physical system which we want to study.
This is clear for CFTs, both in fractional dimension, as described above, and in integer dimension (e.g. $d=5$ \cite{Giombi:2019upv}); it has also been noticed in non-relativistic QM -- see \cite[around (32)]{Phillips:1997xu}.
}
meromorphic functions of $d$, having poles but not branch cuts.
Then, in perturbation theory, we observe that in a (theory-dependent) open set in the complex $d$-plane, we can send $\Lambda\to\infty$ and obtain a finite value for our observables. 
These integrals, and the corresponding observables, can then be analytically continued to the full $d$-plane.
Thus, analytic continuation in $d$ can be used as its own form of UV cutoff: and we call it dimensional regularization.

This can be understood using the following simple example. 
In arbitrary $d$, the standard one-loop momentum diagram contributing to $\expval{\phi^4}$, with an arbitrary cutoff $\Lambda$, 
\begin{equation}
B_\Lambda(p, m)= \int_{\abs{k} < \Lambda} \frac{1}{k^2 + m^2} \frac{1}{(p-k)^2+m^2}
\end{equation}
is evaluated as the following in the $m \to 0$ limit:
\begin{align}
B_\Lambda(p,0) &\stackrel{2<d<4}{=} b(d) p^{d-4} -a(d) \Lambda^{d-4} + O(\Lambda^{d-6} p^2) \\%(18.49) of ZJ
b(d) &= -\frac{\pi  \Gamma \left(\frac{d}{2}\right)^2}{\sin \left(\frac{\pi  d}{2}\right) \Gamma (d-1)} N_d, \quad N_d = \frac{\text{Area}(S^{d-1})}{(2\pi)^d},
\end{align}
where $b(d)$ is universal, but $a(d)$ depends on the details of the cutoff procedure \cite{zinn-justin_quantum_2002}.
Hence, for $2< \Re d<4$ (where there are no IR divergences as $m\to 0$), we can then take $\Lambda \to \infty$, and obtain
\begin{equation}
B_\infty(p,0) = b(d) p^{d-4}
\end{equation}
which has a $1/\epsilon$ pole in $d=4-\epsilon$.
By considering the full expression, we see that really, this is a remnant of $B_{\Lambda\to\infty}(p,0)|_{d=4} \propto \log(p/\Lambda)$. 
However, if we forget about $\Lambda$, we can reparametrize in terms of the physical observables, working in generic $d$. 
Having done so, we can safely take the limit $\epsilon \to 0$ (analytically continuing to $d=4$), with the same effect as if we had worked for $\Lambda$ finite in $d=4$ exactly, reparametrized with an \textit{explicit} $\Lambda$-dependent counterterm, and then taken $\Lambda \to \infty$. %

We do not need to consider the snail-type contribution to the $\phi$ propagator (which leads to a power-law divergence which would renormalize the mass), since in DREG it vanishes in the $m\to 0$ limit: this is because scaleless integrals vanish, $\int_k 1/k^2 =0$. %
However, if we were working at finite $m,\Lambda$ we would have needed to introduce a mass counterterm to make the mass finite. 
By using DREG, we can self-consistently ignore this requirement -- this is the sense in which DREG corresponds to an \textit{implicit} selection of Lorentz-invariant counterterms for power-divergent observables.
Not needing to keep track of an explicit cutoff, and the accompanying  power-law divergences which lead to extra counterterms, is extremely helpful when performing higher-loop computations\footnote{Additionally, unlike a hard cutoff, DREG respects (non-chiral) gauge invariance -- and any regulator which respects the symmetries of the problem is always better.
This was one of the original motivations for DREG given by 't Hooft and Veltman \cite{tHooft:1972tcz}.
See \cite[\S 4]{Collins:1984xc} for more rigorous commentary on the DREG procedure. 
}.

\section{Solving CFTs: Wilson-Fisher \epsPDF and large-\NPDF} \label{sec:solvingCFTs}

Above, it was suggested that it suffices to give a spectrum of conformal representations and a list of OPE coefficients to define a CFT.
This is true.
However, solving non-trivial quantum field theories is hard. 
Practically speaking, we only know how to do one functional integral: the generalized Gaussian,
\begin{subequations}\label{eq:GaussianIntegral}
\begin{equation}
    \begin{aligned}
\int \Dd{\phi}\,  e^{-\half \phi_i [C^{-1}]_{ij} \phi_j + J_i \phi_i} &=  \prod_i \int_{-\infty}^\infty \frac{\odif{\phi_i}}{\sqrt{2\pi}} e^{-\half \phi^T C^{-1} \phi + J^T \phi} =  (\det C^{-1})^{-\half}  e^{+\half J^T C J}\\
&= \exp\left(-\half \log \det C^{-1} + \half J^T C J\right)
    \end{aligned}
\end{equation}
and its real Grassmann cousin
\begin{equation}\begin{aligned}
\int \Dd{\psi}\, e^{+\half \psi_i [C^{-1}]_{ij} \psi_j + \eta_i \psi_i}&= \int \odif{\psi_{n}} \cdots \odif{\psi_1} e^{+\half \psi^T C^{-1} \psi + \eta^T \psi} = (\det C^{-1})^{\half} e^{+\half \eta^T C \eta}\\
& = \exp\left(+\half\log \det C^{-1} + \half \eta^T C \eta\right),
\end{aligned} \end{equation}
\end{subequations}
in the conventions of \cite{Zinn-Justin:1991ksq}. %
If we consider these integrals as written in DeWitt notation, these are exactly infinite-dimensional functional integrals.
That is, if we upgrade $\phi_i \to \phi(x)$, $[C^{-1}]_{ij} \to C^{-1}(x,y)$, and $\sum_i \to \int_x$, these expressions describe the functional Gaussian integrals\footnote{We have been careful to define the bosonic measure $\Dd{\phi}$ with the factors of $(\sqrt{2\pi})^\infty$.}.

This means that we are only really capable of solving for small perturbations around free theories. The solvable non-trivial CFTs are the following\footnote{For completeness, we mention here the rational CFTs, which exist only in $d=2$, where they rely on the existence of the infinite-dimensional symmetry algebra.} %
\begin{enumerate}
	\item Free (Gaussian) field theories;
	\item SUSY CFTs;
	\item Large-N CFTs.
\end{enumerate}
The main result of this thesis is that one of our best tools for solving strongly-coupled SUSY CFTs applies identically in the large-$N$ case: this will be the topic of \cref{chap:fextr}.
Indeed, we will see that the only functional integral needed in this procedure is the only one we can do: the Gaussian.

\subsection{Measuring RG flow: conventions}\label{sec:measuringRG}

We will be interested in the fixed points of the RG flow of QFTs, which are typically conformal field theories.
Recall that CFTs in $d$ dimensions are defined by their conformal data, which consists of a set of conformal representations $\{(-\Delta_i, \rho_i)\}$, and three-point coefficients $C_{ijk}^{(n)}$. These are observable, in the sense that they can be measured (at least for gauge-invariant operators).
Free bosons and fermions are trivially conformal field theories, and have the following scaling dimensions:
\begin{equation}
    \Delta_\phi^{\mathrm{free}} = \frac{d-2}{2}, \quad \Delta_\psi^{\mathrm{free}}= \frac{d-1}{2}.
\end{equation}
In this thesis, we will be finding interacting IR fixed points in the large-$N$ limit by adding a UV perturbation to a free CFT of bosons and/or fermions. 
In the IR, we will find that the operator which in the free theory had scaling dimension $\Delta_\phi^{\text{free}}$ now has a scaling dimension $\Delta_\phi$. Therefore, in general, we define the anomalous dimension of an operator $\cO$ to be $\gamma_\cO$, defined by $\gamma_\cO = \Delta_\cO - \Delta_\cO^{\text{free}}$.

In \cref{chap:3dyuk,sec:SDEanalysis,sec:bilinearCalculationOverview}, we will be calculating these dimensions using a non-perturbative approach, working directly in the IR; however, these dimensions are usually found via a perturbative renormalization group analysis, which applies throughout the flow.
This analysis will be done for a melonic QFT in \cref{sec:betas}, and hence we now introduce the required quantities: the beta functions.

\subsection{Reparametrizing the action}\label{sec:reparametrizingAction}

To find the flow in theory space and calculate scaling dimensions at fixed points, we follow the usual process of rewriting the bare Lagrangian, which we define as
\begin{equation}\begin{aligned}
\cL = \cL_{\mathrm{kinetic}}[\phi_0] + \sum_i g^i_b \cO_i^{\mathrm{bare}},
\end{aligned}\end{equation}
where the $g^i_b$s are dimensionful: from \cref{sec:constrainingAction} we recall that the natural scale of these $g^i_b$s is set by the UV definition's $\Lambda$.
Then we replace all bare couplings with dimensionless renormalized versions, and all fields with renormalized versions,
\begin{equation}
g^i_b = Z_{g^i} g^i \mu^{d-\Delta_i^{\text{free}}},\quad 
 (\phi_i)_{\text{bare}} = \sqrt{Z_{\phi_i}} \phi_i,
\end{equation}
where the $Z_j(\{g^i\})$s are as-yet-undetermined functions.
The dimensions are made up by factors of the renormalization group scale parameter $\mu$; we define $\Delta_i^{\text{free}}$ to be the scaling dimension in the free theory. 
For example, for the operators $\cO_i^{\mathrm{bare}} = (\phi_0)^n =Z_{\phi}^{n/2} \phi^n$ in scalar field theory, $\Delta_i^{\text{free}} = n \frac{d-2}{2}$. 

\subsubsection{Renormalizing}

Just as in $d=0$, we then impose a renormalization scheme. The only difference is that being able to manipulate the action to do the reparametrization requires regularization of the theory.

That is, we take some QFT with a regulator, and apply some renormalization conditions, which are precisely analogous to the renormalization conditions in the $0$d case: 
\begin{enumerate}[noitemsep]
\item At the scale $\mu$, the fields must have a unit-normalized propagator,
\item We tune the function $Z_{g^{(4)}}(\{g^i\})$ such that the observed $4$-point function at the scale $\mu$ is equal to the renormalized couplings, $\expval{\phi^4}|_\mu = - g^{(4)}$\footnote{Other renormalization schemes lead to the same conclusions. For practical purposes, in \cref{sec:betas} we use $\overline{\mathrm{MS}}$ scheme. Taking for example $\phi^4$ in DREG around $d=4-\epsilon$, this means that we tune $Z_\phi$ and $Z_g$ such that $\expval{\phi^2}_{p^2=\mu^2} = 1/p^2 +O(\epsilon^0)$ and $\expval{\phi^4}_{\sqrt{s}\simeq \mu} = -g+O(\epsilon^0)$. That is, once we have ensured that the correlators are finite, we do not bother also subtracting the $O(\epsilon^0)$ terms. This suffices to remove the strong regulator dependence. \label{footnote:MSBAR}}. 
\end{enumerate} 
Similar conditions are applied to any other interaction terms.
Reparametrizing in this way reduces the sensitivity of observables to the regularization -- which is the point, as when this has been done we can remove the regulator. 
That this reparametrization is possible is under the assumption that we are using an effective theory at some $\mu \ll \Lambda$; in that case Wilson told us that the physics should indeed not depend on the high-energy details of the cutoff procedure, so we should be able to factor it out in this way.
In particular, different regulators correspond to slightly different modifications of the UV physics, all of which yield the same IR physics. In the case of UV-complete theories, we can take the scale of all of these possible regulators $\Lambda \to \infty$, and therefore obtain a unique UV CFT.

One should not be confused by the fact that these two types of condition seem slightly different: in both cases, we are simply nailing down the value of a correlator at a particular scale. 
In the field case, precisely because $Z_\phi$ is a redundant deformation that only changes the normalization of correlators, not the physics, we must effectively mod out by it.
This can be done by tuning it such that the propagator is unit-normalized -- identical to \eqref{eq:0dPhi4PropRenoCondition} in the $0$d QFT.  %

In theory, we would have to do this for all of the operators that could possibly be in the action, including the infinite number of irrelevant ones.  However, thanks to the arguments of \cref{sec:constrainingAction}, we are able to set all the irrelevant operators' couplings to zero, as their values should not affect physics at our scale $\mu \ll \Lambda$.

Once we have renormalized, we can send any regularizing parameter, such as the cutoff $\Lambda$, off to infinity, \textit{for fixed $g^i$} -- and from now on we need not worry about infinities. The $g^i$s are what we meant by \enquote{fixed observables} in \eqref{eq:continuumLimit}. Thus, the famous infinities of quantum field theory are and always were an artifact of using bad parametrizations of the continuum limit: that is, expanding in $g_b$ instead of $g$ \cite{Neumaier:2015ren}.
As Weinberg comments in QFT I \cite[Section~10.3]{Weinberg:1995mt},
\begin{quote}
    The renormalization of masses[, couplings], and fields has nothing directly to do with the presence of infinities and would be necessary even in a theory in which all momentum space integrals were convergent.
\end{quote}
We have seen this already in the classical case of the effective mass \eqref{eq:sphereEffectiveMass} of a submerged sphere, and mentioned it in the context of DREG, but we shall make it explicit for QFT when we consider the flow between generalized free fields in \cref{sec:GFFflow}.

\subsubsection{RG functions: \texorpdfstring{$\beta$ and $\gamma$}{beta and gamma} describe scale dependence}

These renormalized couplings necessarily depend on the renormalization scale $\mu$. We describe this via the beta functions $\beta[g^i]$ of each of the couplings
\begin{equation}\begin{aligned}
\beta[g^i] &= \odv{g^i}{\log \mu},
\end{aligned}\end{equation}
which, once we've tuned the $Z_{g^i}(\{g^j\})$s to satisfy the renormalization conditions, are calculated by using the fact that $g^i_b =Z_{g^i} g^i \mu^{d-\Delta_i^{\text{free}}}$ must be independent of the scale $\mu$. 
Remembering that the renormalization conditions mean that the coupling constants have physical meaning (in that the $g^i$s \textit{are} the values of certain correlators at the scale $\mu$) we see that these beta functions encode how the dynamics of a theory change as it is probed at different energy scales: the renormalization group flow!
As promised, the choice of a renormalization scheme yields the couplings and beta functions that make sense of the flow depicted in \cref{fig:UVIRflow}.

In usual QFT examples, once we have sent the cutoff to infinity, this RG flow should be local in the space of coupling constants. 
That is, these differential equations should not depend on the numerical value of the scale $\mu$; only the current position in theory space, encoded by $\{g^i\}$ should matter -- so $\beta[g^i](\{g^i\})$ only.
This follows immediately from dimensional analysis: if the cutoff has been removed there are no longer any scales except $\mu$ -- so since $\beta[g^i]$ is dimensionless, $\mu$ cannot appear. 
The physical fact encoded by this is that only the ratio of the renormalization scale $\mu$ to scales in correlators should matter.  %
The fixed points are found by solving $\beta[g^i](\{g^i_\star\})=0$. 

The anomalous ($\gamma_{\phi_i}$) dimensions of the fields are defined by $\odv{}{\log\mu} \phi_i(\mu) = -\gamma_{\phi_i} \phi_i(\mu)$. Since the bare fields are independent of $\mu$, $\odv{}{\mu} \phi_0=0$ gives that
\begin{equation}\begin{aligned}
\gamma_{\phi_i} &= \odv{\log\sqrt{Z_{\phi_i}}}{\log\mu}, \quad \Delta_{\phi_i} = \Delta_{\phi_i}^{\mathrm{free}} + \gamma_{\phi_i},
\end{aligned}\end{equation}
where $\Delta_{\phi_i}$ are the full scaling dimensions. %
The $\phi_i$ conformal dimension at the fixed point is simply $\Delta_{\phi_i}|_{\{g_*\}}$ -- this is a physical observable at a fixed point, but $\gamma_{\phi_i}$ is scheme-dependent away from fixed points. 
At a fixed point, $\beta[g^i](\{g^i_\star\})=0$, and we can then define the \textit{stability matrix} 
\begin{equation}\begin{aligned}\label{eq:stabMat}
S_{ij} \equiv \odv{\beta[g^i]}{g_j} \Big\rvert_{\{g_*\}},
\end{aligned}\end{equation}
the eigenvalues of which are the dimensions $\Delta_i - d$ of the conformal operators, which accordingly may be linear combinations of the bare operators in the action.
Thus, positive eigenvalues of $S_{ij}$ signify an irrelevant operator in the IR, while negative eigenvalues signify a relevant operator. Complex eigenvalues indicate the fixed point is non-unitary. We shall apply this to a QFT in \cref{sec:stabmats}.

In \cref{sec:GFFflow} we demonstrate RG flow in the case of an exactly solvable QFT.
However, it will first be useful to consider how QFTs are solved in practice: all of the difficulty is contained in calculating the correlators.%

\subsection{Various ways of solving conformal field theories} \label{sec:VariousWaysOfSolvingCFTs}

In general, conformal field theories are strongly coupled -- perturbation theory around some integrable theory, such as the free theory, does not suffice, as the coupling is not small. 
There is a way out: in cases where no such parameter seems to exist, we \textit{invent} a perturbation parameter and perturb around some solvable field theory that exists in the limit of that parameter being taken to zero. In decreasing order of commonness of usage, three such perturbation parameters are 
\begin{enumerate}[noitemsep]
    \item The Wilson-Fisher $\epsilon$ expansion.
    \item The large-$N$, $1/N$ expansion.
    \item The Zamolodchikov $1/m$ expansion in the minimal models \cite{Zamolodchikov:1987rg,Poghossian:2013fda}.
\end{enumerate}
Of course, this is not a complete list -- see e.g. \cite{Segev:1991fq,Gracey:2015uaa,Giombi:2024zrt,Behan:2025ydd}.
Each of these corresponds to analytic continuation in a parameter that one should reasonably expect to be a fixed integer, namely: the dimension $d=4-\epsilon$; the number of fields\footnote{Note that the analytic continuation of the $d=2,3$ critical $\gO(N)$ CFTs can be completely justified by considering the IR of a loop model, for which $N$ is naturally a continuous parameter \cite[footnote 3]{Gorbenko:2020xya} \cite{Liu:2012ca}.} $\phi_{i=1,\cdots, N}$; the exponent of the polynomial in the potential $V \propto \phi^{2m}$.
As each of these now-continuous parameters $\epsilon,1/N,1/m$ are taken to zero, there exists a CFT which is solvable.
Hence, in these cases, we can solve for the interacting CFT by perturbing about the solvable CFT\footnote{As an aside: the absence of something solvable to perturb about is the root of the difficulties of M-theory, as the quantum supermembrane has no such free parameters, and so cannot be attacked in this way \cite[43:06]{wittenCommentsStringDynamics1995}.}
In this thesis, we discuss only the first two, mentioning the third \cite{Ribault:2018jdv} only for completeness. 

\subsubsection{The critical $\gO(N)$ vector model}

The canonical example used for discussion of the $\epsilon$ and $1/N$ expansions is the $\phi^4$ theory with $N$ fields. 
This is precisely the model of \cref{sec:0dQFT} but in higher dimensions.
Including all operators that are relevant in $d<4$, this is
\begin{equation}
S= \int \odif[d]{x} \half Z_\phi \phi_i(-\partial^2+Z_{m^2} m^2) \phi_i + \frac{Z_g g \mu^{4-d} Z_\phi^2}{8}  (\phi_i \phi_i)^2.
\end{equation}
In the following, we will consider the solution of this theory for generic values of $d\le 4$ and $N$. 
For $d<4$, the interaction $g$ is relevant, in that it causes a flow away from the free theory. 
The mass coupling $m^2$ is also relevant, so it needs to be tuned to a particular value.
If we perform the flow for the tuned value of $m^2$, in the IR we will reach a CFT, the critical $\gO(N)$ CFT$_d$, which describes the $\gO(N)$ symmetric second-order phase transitions in $d$ dimensions.
\begin{enumerate}[(a)]
\item In $d=2$, the critical $N=1$ CFT is exactly the $\mathcal{M}_{3,4}$ minimal model, which is exactly solvable: $\Delta_{\phi_i} = 1/8$ \cite{Antunes:2022vtb,Antunes:2024mfb}.
\item In $d=3$, there exists a strongly coupled conformal field theory, which is a Ginzburg-Landau description of the effective theory of the order parameters of various physical systems near their critical points. For example: 
\begin{enumerate}[(i),noitemsep]
    \item For $N=1$, the critical Ising model and the liquid-gas transition in water and carbon dioxide. %
    \item For $N=2$, ferromagnetic phase transitions.
    \item For $N=3$, isotropic magnets.
\end{enumerate}
This is nicely reviewed in \cite{Henriksson:2022rnm}.
\item For $d=4$, there is no interacting CFT. The only fixed point (i.e. continuum limit) is at  $g=0$; that is, when the renormalized coupling is zero, and we have a free CFT \cite{Poland:2022zhe}.
\item For $d>4$, the critical vector CFTs are non-unitary \cite{Fei:2014yja,Giombi:2019upv}.
\end{enumerate}
Given our discussion of continuous-$d$ CFTs in \cref{sec:continuousD}, we might now ask -- what about the dimensions in between?

\subsubsection{QFTs in $3.99$ dimensions with $99.99$ fields}
\begin{figure}
	\centering
\begin{subfigure}[t]{0.45\textwidth}
    \centering
 	\includegraphics[width=0.8\textwidth]{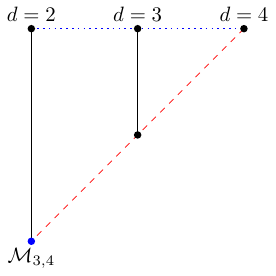}
    \caption{The $4-\epsilon$ expansion for $N=1$.}
    \label{fig:epsilonExpansion}
\end{subfigure}
\hfill
\begin{subfigure}[t]{0.45\textwidth}
    \centering
    \includegraphics[width=\textwidth]{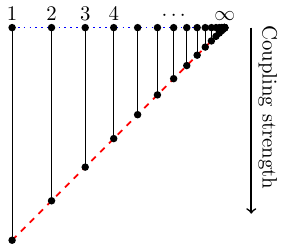}
    \caption{The large-$N$ expansion in $d=3$.}
    \label{fig:largeNExpansion}
\end{subfigure}
   \caption{Schematic $4-\epsilon$ expansion (with $N=1$) and large-$N$ expansion for the critical $\gO(N)$ $\phi^4$ CFT. The vertical axis represents the renormalized coupling \cite{Antunes:2022vtb,Antunes:2024mfb}.
   The top line is the UV free field theory; the diagonal line is the IR Wilson-Fisher CFT -- RG flow proceeds downwards.
   The IR coupling strength becomes weak as we take $d \to 4$ or $N\to\infty$: we can expand about each to find perturbative solutions.
   We find unitary CFTs for integer $(d,N)$, indicated with black dots; the black lines are unitary QFTs.
Non-unitary QFT/CFTs exist in any $d,N$. 
In $4$d the IR and UV CFTs are the same, and for $N=1$ in $2$d (the blue dot) the CFT is the minimal model $\mathcal{M}_{3,4}$.
   }
   \label{fig:phi4epsilonExpansion}
\end{figure}

\textbf{$3.99$ dimensions}. 
No exact solution exists in $d=3$, where physics is strongly coupled. However, we can solve for the beta function of the theory in $d=4-\epsilon$ dimensions, 
\begin{equation}\label{eq:phi4beta}
    \beta_g = \odv{g}{\log \mu} = -\epsilon g(1- g/g_{\mathrm{\star}}) + O(g^3).
\end{equation}
We find two fixed points: the first is the free field theory ($g=0$), a UV fixed point; the second is an IR Wilson-Fisher fixed point with renormalized $g_{\star} \propto \epsilon +O(\epsilon^2)$ \cite{Wilson:1971dc}.

We then find the values of observables, like anomalous dimensions, as power series in $\epsilon$ that describe the generically non-unitary CFTs in $d=4-\epsilon$.
Assuming analyticity in $d$, we can then calculate for $d=3$ by using quantities in $d=4-\epsilon$. That is, we expand in a perturbative series (often Padé approximated), and then set $\epsilon=1$ -- and hope that this matches the actual quantity! 
We draw this schematically in \cref{fig:epsilonExpansion}\footnote{These comments apply to all of the $A$-series of minimal models $M_{m+1,m+2}$, which have Ginzburg-Landau description as a single field with action \cite{zamolodchikovConformalSymmetryMulticritical1986,Katsevich:2024jgq} %
\begin{equation}\label{eq:minimalGL}
	S_{m+1,m+2} = \int \odif[d]{x}\, \left(\half Z_\phi \phi(-\partial^2)\phi + \frac{Z_g g \mu^{d+(2 - d)m}}{(2m)!} (Z_\phi \phi^2)^m \right),
\end{equation}
where of course counterterms are generically required for all the operators $\phi^{2n < 2m}$, but they can be ignored when working in DREG. Since $N=1$ here, we modify the normalization of $g$.
\eqref{eq:minimalGL} has only the free CFT for $d=d_c\equiv 2m/(m-1)$ (though said CFT is non-unitary for $m>3$, as $d_c \notin \Z$).
In $d=2$, the IR CFT is $M_{m+1,m+2}$.
This makes contact with the $1/m$ expansion: for $m \gg 1$ these CFTs are weakly coupled in $d=2$. Thus, the 2d CFT is reachable by a perturbative expansion around the critical dimension: $d=d_c-\epsilon$, with the 2d CFT at $\epsilon \sim 2/m \ll 1$.}. %
\vspace*{0.2cm}

\noindent\textbf{$99.99$ fields}. Another approach which permits exact results in generic dimension is the $1/N$ expansion.
For the reasons discussed in \cref{sec:MFTintro}, we expect the large-$N$ limit to be mean field-like, with a renormalized four-point coupling $g=\lambda/N$ at the fixed point, with $\lambda\sim O(1)$. Thus, we can solve perturbatively in the small parameter $1/N$: this is sketched in \cref{fig:largeNExpansion} for $d=3$.
We expect the IR CFT$_d$ in the strict large-$N$ limit to be almost the same as the free CFT$_d$, with one difference: the scaling dimension of the mass operator $\phi_i \phi_i$ changes from $d-2$ to $2$ as is expected on general grounds for a double-trace deformation \cite{Diaz:2007an,Goykhman:2019kcj}.

$1/N$ becomes a quasi-continuous parameter for $N$ large, which could motivate analytic continuation in $N$ as well. 
For $\gO(N)$-symmetric models, we can certainly do so, as in the $\epsilon$ expansion we obtain conformal data that are analytic in $N$.
Though this might seem purely formal, a mathematical framework justifying non-integer values of $N$ was presented in \cite{Binder:2019zqc} (see also \cite{Jepsen:2020czw} for more). %

In one case, we know the system described by these CFTs. 
For continuous $-2 \le N \le 2$, the $d=2$ $\gO(N)$ model describes the fixed point behaviour of self-avoiding loops: in this system, some conformal data can even be obtained exactly \cite{Gorbenko:2020xya,Zan:2026oyb}, and they interpolate between the known $N=1,2$ cases. 
For similar reasons to $d$, we expect these QFTs for non-integer $N$ to be non-unitary \cite{Gorbenko:2020xya,Zan:2026oyb}, but we can nonetheless consider the UV and IR lines of \cref{fig:largeNExpansion} to be CFTs for all $N$.

\subsubsection{Solving in continuous $d$ and $N$}

\begin{figure}
	\centering
    \includegraphics[width=\textwidth]{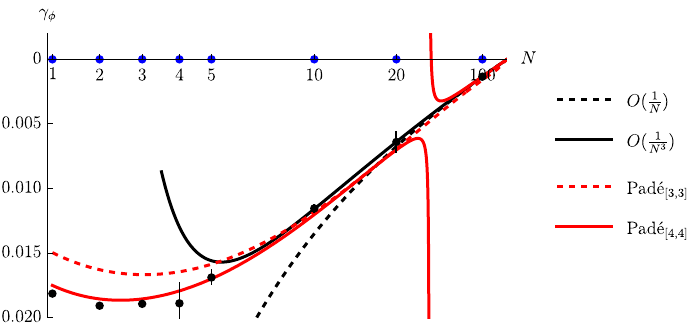}
   \caption{A practical implementation of the $4-\epsilon$ expansion for $\epsilon=1$, showing the calculated anomalous dimension, which is a good proxy for the renormalized coupling strength. As before, the blue dots are the free CFTs, and the black dots indicate the numerical IR data \cite{Henriksson:2022rnm} from the conformal bootstrap (for $N=5,10,100$, from the derivative expansion).
   The red curves were obtained by calculating a standard Padé$_{[a,b]}$ approximant of the $\gO(\epsilon^8)$ results of \cite{Henriksson:2022rnm} for each value of $N$, and then taking $\epsilon=1$.  We also show the $d=3$ $O(1/N)$ and $O(1/N^3)$ results in black, which show remarkable accuracy down to $N \simeq 5$.}
   \label{fig:phi4largeNexpansion3d}
\end{figure}

How well do these analytic approaches work?
In $d=3$, we can compare the results of the two approaches to the conformal bootstrap, which obtains approximate numerical values on the $\phi_i$ anomalous dimension, $\gamma_\phi$: the results are shown in \cref{fig:phi4largeNexpansion3d}. The $\epsilon$ expansion proves to be excellent, but the large-$N$ expansion also proves to be remarkably accurate down to $N\simeq 5$. 
This anomalous dimension is scheme-independent, unlike the coupling, and so is a good candidate to measure the size of the IR fixed point's deviation from the free UV theory, since $\gamma_\phi \equiv \Delta_\phi - \Delta_\phi^{\text{free}}$. 

\begin{figure}
    \centering
\includegraphics[width=\textwidth]{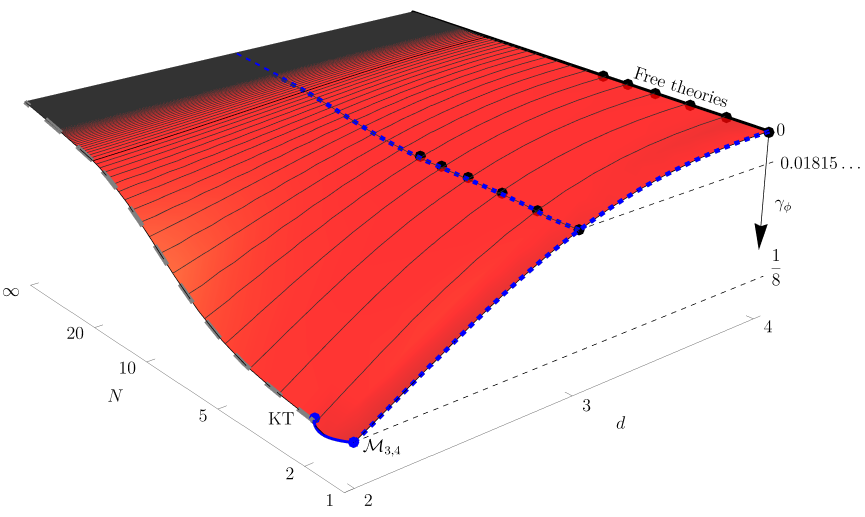}
  \caption{A schematic of the $(d,N)$ surface of the critical $\gO(N)$ model, where, as before, the depth below zero gives the IR anomalous $\gamma_{\phi_i} = \Delta_{\phi_i} - \Delta_{\phi_i}^{\text{free}}$ which measures the size of the deviation from the free CFT. 
  The parallel lines mark integer $N$; the dots (shown only for $N\le 6$) indicate the unitary physical theories that also have $d \in \mathbb{Z}$.
  In $d=2$, at strongest coupling (largest $\gamma_{\phi_i}$), we have the blue dots at $N=1,2$ indicating $\cM_{3,4}$ and the Kosterlitz-Thouless CFTs respectively; there exists a (solid blue) line of non-unitary CFTs interpolating between them, where we have some exact data \cite{Gorbenko:2020xya}.
  In $d=2$, such CFTs only exist for $-2 \le N \le 2$ (the fixed points are complex otherwise), so we indicated with a grey dashed line that $d>2$ is required there. The two blue dashed lines correspond to the slices that approximately reproduce \ref{fig:epsilonExpansion} and \ref{fig:largeNExpansion} respectively; they intersect at the $d=3$, $N=1$ CFT.
  }
   \label{fig:phi4epsilonAndN}
\end{figure}

It is then tempting to combine \cref{fig:epsilonExpansion} for arbitrary $N$ and \cref{fig:largeNExpansion} for arbitrary $d$ to obtain the anomalous dimension for any value of $(d,N)$. 
A schematic of the resulting surface is given in \cref{fig:phi4epsilonAndN}, and in the caption we remark on notable features of the landscape obtained.
The overall shape is correct, but it is not precise, due to the difficulty of calculating $\gamma_\phi$ in the bulk region, away from the perturbative boundaries. 

This figure \ref{fig:phi4epsilonAndN} then serves to summarise the two main solution techniques of this work: large $N$ and $\epsilon$ expansion.
\begin{itemize}
    \item In \cref{chap:fextr}, we will focus on approaching a more general family of CFTs from the large-$N$ side (the rear left): we will be able to solve for $O(1)$ anomalous dimensions exactly for all values of $d$. We shall return to the critical vector models in \cref{sec:vectorModels}, and show how they fit into the family of melonic CFTs.

    \item In \cref{chap:3dyuk}, we will take a particular QFT which has its free and interacting fixed points collide at $d=3$; then, we will solve it in the $d=3-\epsilon$ expansion (approaching from the equivalent of the rear right), finding IR couplings that scale $\sim \sqrt{\epsilon}$ or $\sim \epsilon$.
    We then also solve the QFT in the melonic large-$N$ expansion; the two approaches give access to different data, but we will be able to match some results in the overlapping region -- this corresponds to data in the back corner of \cref{fig:phi4epsilonAndN}.
\end{itemize}

Finally, in both of these cases we are perturbing with a parameter that is no longer small: $\epsilon \simeq 1$ and $N \simeq 5$; the large-$N$ expansion has even been known to work remarkably well for $N\sim 1$ \cite{Klebanov:2011td,Chester:2022wur}. 
As Witten commented, we should not be surprised by this: \enquote{it simply illustrates the important rule that approximations that are correct qualitatively tend to be successful also quantitatively, although sometimes for reasons that are understood only in hindsight} \cite{Witten:1979pi}.

\section{Toy model II: flow of a generalized free field}\label{sec:GFFflow}

Having described the toolkit of RG, we apply it in the context of a simple exactly solvable model, the flow of a generalized free field (GFF). 
This toy model proves to be extremely similar to the melonic theories: the melonic QFTs are also a family of QFTs that in the large-$N$ limit interpolate between generalized free CFTs in the UV and the IR.
The crucial difference is that the IR scaling dimensions in the melonic theories are dynamically determined by the $\Ft$-maximization procedure of \cref{sec:fundamentalClaim} rather than being fixed.
Additionally, an almost identical flow occurs in the cases of the large-$N$ double-trace deformation \cite{Hartman:2006dy} and of a fermion confined to a $2$d wire interacting with a $d$-dimensional bulk photon \cite{Fraser-Taliente:2024lea}.

\subsection{Construction of a running GFF}

Consider a quadratic QFT with two kinetic terms for the same field, each with a different power of the momentum dimension.
Not only is this solvable, it can also be renormalized, in the sense of reparametrization in terms of observables. 
Despite having no divergences, it nonetheless displays many of the RG features of a fully interacting QFT: a beta function, fixed points, anomalous dimensions, and the problem of large logarithms.

We define this Euclidean QFT by
\begin{equation}
S = \int_p \half \phi_0(p) ((p^2)^\alpha + g_b (p^2)^{\alpha-\delta}) \phi_0(-p).
\end{equation} %
This action is that of a generalized free field, in the sense that it is quadratic, but does not have the canonical propagator\footnote{Hence, this action is non-local; as we shall see in \cref{sec:GFFsAndFt}, it is bilocal.}. We can solve it in any $d$, and we end up with a propagator of the form 
\begin{equation}
\expval{\phi_0(p) \phi_0(q)} = \delta^{(d)}(p+q) \frac{1}{p^{2\alpha} +g_b p^{2(\alpha-\delta)}}
\end{equation}
for some dimensionful bare coupling constant $g_b>0$, which we assume to be fixed by the microscopic definition of the theory. %
This is then a generalized free field theory, as all correlators of this field $\phi$ can be determined by Wick contractions. 

Assuming without loss of generality that $\delta>0$ (not necessarily small), we see that for fixed $g_b$: 
\begin{itemize}
\item for large $p$, the first term dominates, and we obtain a field which looks like a conformal field of dimension $\Delta_{\mathrm{UV}} \equiv \frac{d}{2}-\alpha$, since $\expval{\phi_0(p) \phi_0(-p)} \sim 1/p^{2\alpha}$;
\item for small $p$, the converse occurs, and so $\expval{\phi_0(p) \phi_0(-p)} \sim 1/p^{2(\alpha-\delta)}$; that is, we obtain a new generalized free field of dimension $\Delta_\mathrm{IR} \equiv \Delta_\mathrm{UV}+\delta$. 
\end{itemize}
This matches up with general expectations. 
We have conformal fixed points in the scaling limits, and in between we find a generic QFT with a functional behaviour that is not determined by symmetries!
This all seems very reasonable: but because we know the answer exactly, we have not needed to perform any renormalization.

\subsection{Renormalization of GFFs}

Let us now make this more precise, and renormalize this theory as we would if we didn't know the full correct answer. Due to the simplicity of the theory, we do not need to regulate it. First, we exchange the bare quantities for renormalized ones:
\begin{itemize}
    \item The bare field, which has engineering dimension $\dotwo -\alpha$, is exchanged for the renormalized field, $\phi_0 = \sqrt{Z_\phi} \phi$.
    \item The bare coupling $g_b$ is exchanged for the (dimensionless) renormalized coupling $g$, by defining
\begin{equation}
g_b = Z_g g \mu^{2\delta}
\end{equation}
for some renormalization scale $\mu$, which we will choose shortly.
\end{itemize}
The new propagator is then
\begin{equation}
    \expval{\phi(p) \phi(q)} = \delta^{(d)}(p+q) \frac{1}{Z_\phi}\frac{1}{p^{2\alpha} + Z_g g \mu^{2\delta} p^{2(\alpha-\delta)}}.
\end{equation}
We now impose our renormalization conditions. 
\begin{enumerate}
\item We require that we observe a massless field of scaling dimension $\alpha$ at scale $\mu$; hence we demand that at scale $\mu$, the $\phi$ propagator is precisely that of such a field,
\begin{equation}
\expval{\phi(p) \phi(q)}|_{p^2 = \mu^2} = \delta^{(d)}(p+q)\frac{1}{p^{2\alpha}}|_{p^2=\mu^2},
\end{equation}
and so we can determine immediately that
\begin{equation}\label{eq:Zphi2val}
Z_\phi = \frac{1}{1+Z_g g}.
\end{equation}
\item Our second renormalization condition is that the $\phi(-\partial^2)^{\alpha-\delta}\phi$ term in the action is not renormalized.
That is, $g$ is such that $Z_g Z_\phi=1$; \eqref{eq:Zphi2val} then immediately gives that $Z_\phi = 1-g$. 
In this simplified model, this is an arbitrary renormalization condition that serves only to make $g$ finite. 
In a more complicated model, it might be justified by appealing to some other sector of the theory that also contains contributions from $g$; or, for generic $\delta$, from the fact that non-local terms in the action are not corrected by RG flow \cite{Fraser-Taliente:2024lea,Behan:2017emf}.
This $g$ will have a beta function almost identical to the standard one-loop RG beta function of $\phi^4$ theory, but with the helpful property of being exact. 
\end{enumerate}
There are no further renormalization conditions, as there are no more observables: all of the connected higher-point functions are zero, since this is a Gaussian theory.

\begin{figure}
\begin{subfigure}[t]{0.45\textwidth}
    \centering
    \includegraphics[width=\textwidth]{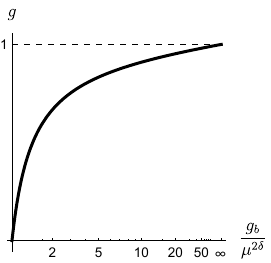}
    \caption{The renormalized coupling plotted against the bare coupling.}
    \label{fig:GFFbareVsReno}
\end{subfigure}
\hfill
\begin{subfigure}[t]{0.45\textwidth}
    \centering
    \includegraphics[width=\textwidth]{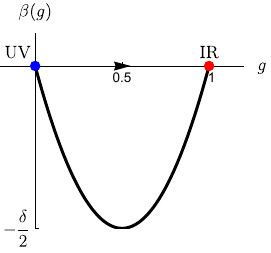}
    \caption{The beta function; the arrow indicates the way the coupling moves as $\mu$ decreases.} %
    \label{fig:combinedExpansions}
\end{subfigure}
\caption{The bare and renormalized couplings of the flow between GFFs, and the flow of $g$ described by the beta function.}
\end{figure}

We have now determined the bare coupling $g_b$ in terms of our renormalized $g$
\begin{equation}\label{eq:GFFcouplings}
g_b = \frac{g \mu^{2\delta}}{1-g}, \quad g = \frac{g_b}{g_b+\mu^{2\delta}},
\end{equation}
which we show in \cref{fig:GFFbareVsReno}.
This relationship is precisely analogous to the $\lambda(\lambda_b)$ from \cref{fig:0dphi4LambdaAgainstLam0} in the $0$d $\phi^4$ case: the only well-defined $g$s are those between $0$ and $1$, as only those values can arise from a $g_b \ge 0$.
We shall see a very similar plot in the case of the long-range tensor models, in \cref{fig:LRcouplings}.
There is only one critical difference: the presence of $\mu$, which means that even if the fundamental parameter $g_b$ is fixed, $g_b/\mu^{2\delta}$ varies.
It is this factor that leads to the RG flow of QFTs.

\subsection{The beta function and anomalous dimension of a GFF}
We can now calculate the beta function from the scale independence of $g_b$:
\begin{equation}\label{eq:GFFbeta}
	\beta_g \equiv \odv{g}{\log \mu} = -2\delta g(1-g),
\end{equation}
which given $\delta >0$ clearly encodes a UV fixed point at $g=0$ and an IR fixed point at $g=1$. 
We can even integrate this equation again, which unsurprisingly yields that $g(\mu)= (1+ C\mu^{2\delta})^{-1}$: precisely \eqref{eq:GFFcouplings}. 

\eqref{eq:GFFbeta} is very similar to \eqref{eq:phi4beta}, the one-loop beta function of $g\phi^4$ theory in $d=4-\epsilon$ (and, more generally, the leading beta function in conformal perturbation theory \cite{Klebanov:2011gs}). The parallel is even closer in the large-$N$ limit, where \eqref{eq:phi4beta} becomes exact, with no cubic corrections \cite{zinn-justin_quantum_2002}.
They both encode a reparametrization of the bare coupling: in each case, there is some limit as $g_b \to \infty$, which we reparametrize to express as a function of a finite renormalized coupling.
For $\phi^4$ in $d=4-\epsilon, \epsilon \ll 1$, we have $g_\star \propto \epsilon$, and so the IR fixed point turns out to be at weak renormalized coupling, and therefore under control as a perturbative series in $\epsilon$.

We can find the field anomalous dimension
\begin{equation}
\gamma_\phi \equiv \odv{\log \sqrt{Z_\phi}}{\log \mu} = g \delta,
\end{equation}
which is $0$ in the UV ($g=0$) and $\delta$ in the IR ($g=1$), showing how the scaling dimension begins as $\Delta_\mathrm{UV}$ and ends as $\Delta_\mathrm{IR}$\footnote{If we had chosen another renormalization condition for $g$, such as $Z_g=1$, we find different RG functions but identical conclusions: $Z_\phi^{-1}=1+g$, so $\beta_g =-2\delta g$, which has fixed points at $0$ and $\infty$. Then $\gamma_\phi=\frac{g \delta}{1+g}$.
We want the renormalized coupling to always be finite, and so choose this condition. 
As usual, the value of the anomalous dimension away from fixed points is scheme-dependent.
}.

The most enlightening quantities to see here are the action and the two-point function after the application of the renormalization conditions, namely
\begin{equation}\label{eq:toyModelFullSolution}
\begin{aligned}
S&=\int_p \half \phi(p) p^{2\alpha}\left((1-g)+g (\mu^2/p^2)^\delta \right) \phi(-p)\\
\expval{\phi(p)\phi(q)} &= \delta^{(d)}(p+q) \frac{1}{p^{2\alpha}}\frac{1}{(1-g)+g (\mu^2/p^2)^\delta},
\end{aligned}
\end{equation}
which manifestly interpolate between the two unit-normalized conformal fields as $g$ is varied from $0$ to $1$.
We could now use the Gaussian integral \eqref{eq:GaussianIntegral} to compute the free energy of these conformal GFFs on the sphere, but we postpone that to \cref{sec:GFFsAndFt}.
In a generic QFT, whenever we renormalize, we are trying to find a nice parametrization of the QFT analogous to \eqref{eq:toyModelFullSolution}; the difficulty is that we are often forced to work perturbatively in the analogues of $g$ and $\delta$. 

\subsubsection{Why we must renormalize at the right scale}

Since in more realistic situations such an exact solution as \eqref{eq:toyModelFullSolution} is not accessible, we only have the perturbative data for $\beta_g$ and $\gamma_\phi$. 
We can now illustrate why it is dangerous to use a renormalized theory away from the scale at which it was renormalized.
Take $\delta \ll 1$ to be a small parameter such that the UV is near the IR -- analogous to $\epsilon$ or $1/N$ from \cref{sec:solvingCFTs}.
Then when calculating the two-point function, we find a series of corrections to the bare propagator
\begin{equation}
\expval{\phi(p)\phi(q)} = \delta^{(d)}(p+q) \frac{1}{p^{2\alpha}}\sum_{n,m} a_{n,m} (\delta\log(\mu/p))^n g^m.
\end{equation}
If we tried to use this perturbation series in $g$ and $\delta$ (not knowing that it could be resummed exactly) for $\mu/p \gg 1$ or $\mu/p \ll 1$, then the $\log(\mu/p)$s will cause the series coefficients to be extremely large, and the series will not get close\footnote{We do not say that the series converges, as generically we expect the perturbation series in QFT to be asymptotic -- to have zero radius of convergence -- due to an argument of Dyson \cite{Dyson:1952tj,skinnerAQFT} (reviewed in \cite[\S 2.2]{Strocchi:2013awa}). %
Nonetheless, it does get close to the exact answer before diverging, and is amenable to resummation techniques such as Pad\'e approximants and Borel transforms \cite{Zinn-Justin:1999opn,marinoIntroductionResurgenceQuantum}.
The same applies to the $\epsilon$ expansion \cite{Rychkov:2015naa, Brezin:1976vw}. 
The vector large-$N$ expansion has a reduced growth of the number of diagrams at fixed order in $1/N$ (polynomial \cite{Alanne:2019vuk,Balduf:2024njk} or exponential), and hence can have a finite radius of convergence \textit{if} no renormalons are present \cite[\S 4]{Serone:2024uwz}.
} to the exact answer \eqref{eq:toyModelFullSolution}.
For this reason, in QFT we should always renormalize with a $\mu$ that is around the energy scale of interest for our calculations.

\section{Diagrammatic approaches to vector and matrix models} \label{sec:diagrammaticApproaches}

\begin{figure}[h]
    \centering
    \begin{subfigure}[t]{0.4\textwidth}
        \centering
        \includegraphics[width=\textwidth]{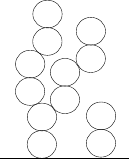}
        \caption{Cactus diagrams in the vector model: here we show a contribution to the two-point function.}
        \label{fig:cactusDiagrams}
    \end{subfigure}
    \hfill
    \begin{subfigure}[t]{0.49\textwidth}
        \centering
        \includegraphics[width=\textwidth]{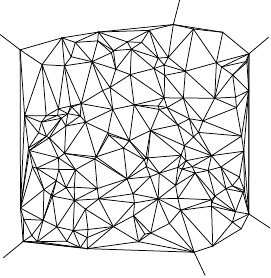}
        \caption{A planar diagram in a matrix model, contributing to the connected six-point function.}
        \label{fig:planarDiagram}
    \end{subfigure}
    \caption{Leading-order diagrams in large-$N$ vector and matrix models.}
    \label{fig:leadingOrderDiagrams}
\end{figure}

We now consider the diagrammatic techniques used to solve for the fixed points in practice. 
In the case of the large-$N$ QFT$_d$
\begin{equation}
S = \int_x \half Z_\phi \phi_i C^{-1}\phi_i + \frac{Z_g g Z_\phi^2 \mu^{4-d}}{8} (\phi_i \phi_i)^2,
\end{equation}
we are halfway there already. The combinatorics of this model are necessarily identical to the zero-dimensional case of \cref{sec:0dQFT}, and so follow through automatically.
The difficulty is then in actually evaluating the relevant Feynman diagrams, since the free propagator now is nontrivial: $C^{-1}=-\partial^2$ means that $\expval{\phi(p)\phi(-p)}_{\text{free}}=C(p)/Z_\phi =1/(Z_\phi p^2)$. Happily, the leading-$N$ diagrams are extremely simple.
When computing diagrams, we observe that when evaluated for arbitrary $N$, the leading-$N$ contributions to the connected $(2n)$-point function are the so-called \textit{cactus diagrams}, drawn in \cref{fig:cactusDiagrams} which scale $\propto N^{1-n}(Z_g g N)^\text{number of loops}$.
This is very easy to understand: adding a single loop to the cactus contributes a $\delta_{ii}=N$ and a factor of the coupling constant. Since the vacuum diagrams scale as $N^1$, and higher-point diagrams are obtained by breaking loops open, the $(2n)$-point function with no loops must scale as $N^{1-n}$.
\newcommand{\snail}{%
  \begin{tikzpicture}[baseline=-0.5ex, scale=0.6]
    \begin{feynman}
      \vertex (a) at (0,0);
      \vertex (b) at (1,0);
      \diagram* {
        (a) -- [scalar] (b),
        (a) -- [half left, looseness=1.5, scalar] (a)
      };
    \end{feynman}
  \end{tikzpicture}%
}

Motivated by our $0$d example, we might define the \textit{optimal scaling} for our dimensionless coupling constant $\lambda \equiv g N$ ($Z_\lambda \equiv Z_g|_{g=\lambda/N}$).
Then, we would observe a nice large-$N$ limit, where \textit{only} the cactus diagrams contribute to leading order, at every loop order! 
These cactus diagrams manifestly form a summable geometric series, as they are constructed by repeatedly multiplying a simple one-loop diagram.
As we will see in \cref{sec:vectorModels}, we can then solve the two-point function Schwinger-Dyson equations to find the infrared anomalous dimension, which is $O(1/N)$, as expected from the large-$N$ discussion of this model in \cref{sec:solvingCFTs} \cite{zinn-justin_quantum_2002}.

We have done this to lead into an identical diagrammatic analysis which can be performed for the melonic theories. 
The advantage of the melonic CFTs that we will find is that they generically have an $O(1)$ anomalous dimension in the scaling limit: they therefore grant a rare glimpse of a strongly-coupled infrared that is \textit{far} from the ultraviolet free theory, $\Delta-\Delta^\text{free} \gg 1/N$.

\subsection{Matrix models}
 
Matrix models are the natural next step after vector models. 
We take the dynamical field to have one extra index, $\phi_{ab}$, and require the action to be symmetric under $\gO(N)^2$. 
Continuing with the $\phi^4$ theme, one such action is
\begin{equation}
S = \int_x \frac{1}{2} \phi_{ab}C^{-1} \phi_{ab} + \frac{g_\mathrm{dt}^b}{8} (\phi_{ab}\phi_{ab})^2 + \frac{g_\mathrm{sq}^b}{4} (\phi_{ab}\phi_{bc}\phi_{cd}\phi_{da}),
\end{equation}
where $g_\mathrm{dt}$ is the coupling for the double-trace term, and $g_\mathrm{sq}$ is the coupling for the square (quartic) interaction term. 
If we scale the terms here suitably, we also find a large-$N$ limit that is a sum over a restricted class of diagrams at leading order; however, it is much more complicated \cite{Coleman:1985rnk}. 
The result is that only planar diagrams contribute to leading order in $N$ -- the diagrams that can be drawn on a genus-zero surface. 
These can be arbitrarily complicated (see \cref{fig:planarDiagram}) and are certainly not summable.
Though of great importance in the context of quantum gravity and AdS/CFT, we treat these models no further here, and simply comment that they are harder to obtain results for than the melonic theories.

\subsection{What's next?}

These large-$N$ vector $\phi_i$ and matrix $\phi_{ij}$ scalar field theories, invariant under $\gO(N)$ and $\gO(N)^2$ respectively, are well known. The natural continuation is to consider higher rank fundamental fields:
\begin{equation}
\phi_{i_1 i_2 \cdots i_r}(x).
\end{equation}
That is, we take standard scalar field theories in $\mathbb{R}^d$ but enforce that the fields transform in some representation of the symmetry group $\gO(N)^{r>2}$.
Naturally, these models contain limits that are, for example, $N^{r-1}$ vector models, or $N^{r-2}$ matrix models \cite{Ferrari:2017jgw,Flodgren:2024utm}; however, there exists a novel limit dominated by the graphs which are planar in every pair of indices-- these are the melonic graphs.
Just as in the vector case, these graphs are resummable, allowing us to solve the theory. 
However, there exists an entire family of QFTs which also have this property: the \textit{melonic QFTs}. 
We now turn to them.

\section{Melonic QFTs}

The melonic QFTs are a family of large-$N$ QFTs that have a resummable diagrammatic expansion. Recently discovered, they have a structure that is simpler than that of the matrix models, and lead to exactly solvable large-$N$ CFTs \cite{Benedetti:2020seh}.
It is useful to distinguish three principal types of melonic QFTs, all of which occur in the strict large-$N$ limit:
\begin{itemize} 
    \item the Sachdev-Ye-Kitaev (SYK) model and its generalizations \cite{Gross:2016kjj};
    \item the critical vector models, such as the $\gO(N)$ $\phi^4$ model \cite{zinn-justin_quantum_2002}; 
    \item the tensor models $\phi_{a_1\cdots a_r}$ with $\gO(M)^{\times r}$ symmetry for rank $r\ge 3$ and $M^r=N$ \cite{Witten:2016iux,Giombi:2017dtl}.
\end{itemize}
The solvability of each of these models arises from the exact resummation of their Feynman-diagrammatic expansions at leading order in $N$. Thus, all the known melonic CFTs can be and have been solved individually, whether in the SYK-like \cite{Maldacena:2016hyu,Turiaci:2017zwd,Murugan:2017eto,Liu:2018jhs, Rosenhaus:2018dtp,Bulycheva:2017ilt,Yoon:2017nig,Bulycheva:2017uqj,Gu:2019jub,Marcus:2018tsr,Fu:2016vas,Berkooz:2017efq,Chang:2021fmd,Chang:2023gow,Biggs:2023mfn}, tensor \cite{Witten:2016iux, Choudhury:2017tax, Klebanov:2018fzb, Gurau:2019qag, Gubser:2018yec,Fraser-Taliente:2024rql, Prakash:2017hwq, Klebanov:2019jup,Benedetti:2019rja,Chang:2018sve}, vector \cite{zinn-justin_quantum_2002,Chang:2021wbx}, or other \cite{Benedetti:2020iku} cases.  The resummability occurs for a slightly different (albeit related) reason in each case: for SYK-like theories, a disorder average over the coupling; for the vector and tensor models, the tuned combinatorics. However, for our purposes, the particular mechanism used is irrelevant -- to leading order in $N$ (and leading order only, \cite{Gurau:2019qag,Choudhury:2017tax}) the solvable CFT found in the IR is not sensitive to those details.

\subsection{Why are melonic theories solvable?}

These theories lie, perhaps surprisingly, between the vector models and the matrix models of \cref{sec:diagrammaticApproaches}. Recall that the diagrammatic expansion of the vector models is dominated by contributions from the cactus/snail diagrams in the large-$N$ limit, which are completely summable via a geometric series; they are in a sense too simple to be interesting, leading to \textit{ultralocal} dynamics. The diagrammatic expansion of the matrix models is dominated by the subset of Feynman diagrams that can be drawn on the plane; they are \textit{not} so directly summable, and analytic progress is more difficult. %

The graphs that dominate in the melonic limit of tensor models, and indeed after the disorder average of the SYK model, are the \textit{melonic} graphs, which are a simpler subset of the planar graphs.

In the tensor models, for any diagram, we can construct the \textit{(Gurau) degree}. This is conceptually straightforward to calculate. We take each of the $\tbinom{r}{2}$ pairs of indices, and ignore the rest. This transforms the graph into a ribbon graph, which just as in the matrix model case has an associated non-orientable genus.
The sum of these genera for each pair defines twice the Gurau degree \cite{Gurau:2010ba}. 

\begin{figure}
    \centering
    \includegraphics[width=0.5\textwidth]{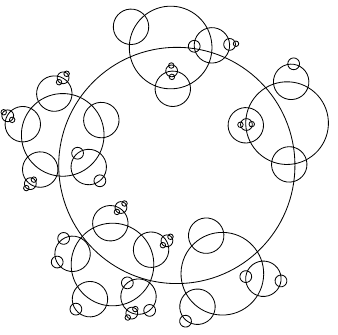}
    \caption{All graphs in the melonic limit are constructed from the iterated melon; here we show a high-order vacuum diagram, which contributes to the free energy at leading order in $N$, in an arbitrary quartic melonic theory.}
    \label{fig:quarticMelon}
\end{figure}

When we scale the coupling constants appropriately, we find that a consistent large-$N$ limit exists, where only the graphs of lowest Gurau degree contribute: the melonic graphs.
These graphs are simple enough to be summable, but complex enough to be interesting, and in that sense lie between the vector and matrix models -- in richness, but not in rank ($1 < 3 \nless 2$). Figure \ref{fig:quarticMelon} displays a complicated-looking melonic contribution to the free energy of a quartic theory, which nonetheless can be resummed.

Unlike the case of the vector model, we find non-trivial dynamics at order $N^0$, such as a non-zero anomalous dimension for operators; and unlike the matrix models, it is straightforward to make exact statements. 

\section{Melonic field theories and their IR limits} \label{sec:MelonicFTs}

\begin{figure}
    \centering
    \includegraphics[width=0.8\textwidth]{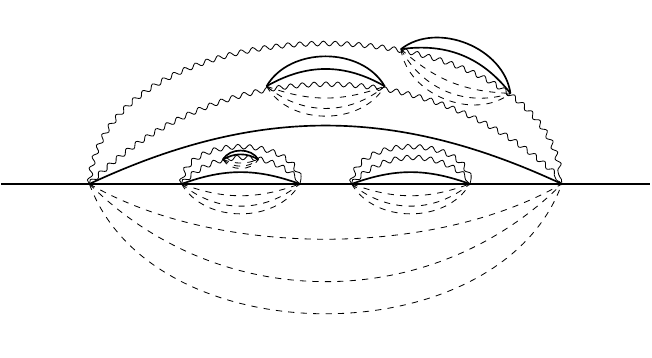}
    \caption{Iterated insertions on a melonic contribution to a two-point function of a theory containing three fields, with $q_i = \{3,2,2\}$. It is only diagrams of this kind that contribute to the leading order of a melonic theory.}
    \label{fig:recursiveMelons}
\end{figure}

The melonic QFTs are found in the large-$N$ limit of QFTs with interaction terms of schematic form
\begin{equation}
\begin{aligned}
g_m \prod_{\text{fields }\phi} \phi^{q^m_{\phi}}, \quad q^m_{\phi} \in \mathbb{N}_0,
\end{aligned}
\end{equation} 
where we have suppressed any index structure or constants. These theories are then melonic if they possess a Feynman-diagrammatic expansion of their two-point functions $\expval{\Phi(x) \Phi(y)}$ that, at leading order in the large-$N$ limit, is dominated by diagrams of melonic form \cite{Benedetti:2023mli}
\begin{equation}\begin{aligned}
\Pi_{\Phi} \supset  \Phi \vcenter{\hbox{
\begin{tikzpicture}[anchor=base,baseline]
  \pgfmathsetmacro{\L}{4}
  \coordinate (L) at (-\L,0);
  \coordinate (R) at (\L,0);
  
  \draw[black, decorate, decoration={snake, amplitude=.4mm, segment length=2mm}]  (L) to[bend left=90,looseness=1.25] (R);
  \draw[black, decorate, decoration={snake, amplitude=.4mm, segment length=2mm}]  (L) to[bend left=75,looseness=1.20] (R);
  \draw[loosely dotted, thick] (0, {0.65*\L}) to (0, {0.55*\L});
  \node at (0.4, {(0.58*\L)}) {$q^m_{\phi_i}$};
  \draw[black, decorate, decoration={snake, amplitude=.4mm, segment length=2mm}]  (L) to[bend left=60] (R);
  
  \draw[black, thick] (L) to[bend left=50] (R);
  \draw[black, thick] (L) to[bend left=40, looseness=1] (R);
  \draw[loosely dotted, thick] (0, {0.32*\L}) to (0, {0.22*\L});
  \node at (0.8, {(0.25*\L)}) {$(q^m_{\Phi} -1)$};
  \draw[black, thick] (L) to[bend left=25, looseness=0.7] (R);

  \draw[black, thick] (-\L-1.5, 0) to (L);
  \draw[loosely dotted, thick] (0, {0.09*\L}) to (0, {-0.09*\L});
  \draw[black, thick] (R) to (\L+1.5,0);
  
  \draw[black, dashed] (L) to[bend right=40, looseness=1] (R);
  \draw[loosely dotted, thick] (0, {-0.23*\L}) to (0, {-0.33*\L});
  \node at (0.4, {-(0.29*\L)}) {$q^m_{\phi_k}$};
  \draw[black, dashed] (L) to[bend right=25, looseness=0.7] (R);
  \draw[black, dashed] (L) to[bend right=70] (R);
  \draw[black, dashed]  (L) to[bend right=90,looseness=1.25] (R);
\end{tikzpicture}
}} \Phi. 
\end{aligned}\end{equation} 
We refer to this diagram as the melon diagram, for obvious reasons.
The full self-energy $\Pi_{\Phi}$ of the field $\Phi$ is then given by the melon diagram, and all possible nested re-insertions of the melon.
That is, each $\phi_j$ propagator within the melon can have a recursive insertion of a melon containing $q^m_{\phi_k}$ propagators of $\phi_k$ for each $\phi_k\neq j$, and $(q^m_{\phi_j}-1)$ propagators of $\phi_j$, as illustrated in \cref{fig:recursiveMelons}.
The sum of all of these \textit{melonic diagrams} gives a geometric series, which is resummable. 
Mathematically, in the case of the tensor models, these dominant diagrams are those of leading \textit{Gurau degree} \cite{Gurau:2019qag,Benedetti:2023mli}. %

The melonic diagrams are a summable subset of the planar diagrams, and therefore also lie between the vector models and matrix models in terms of complexity: above the vector models, because they do not have the ultralocal dynamics of the vector models; but below the matrix models, since the melonic diagrams can in fact be resummed. Therefore, the data that defines a melonic field theory is simply the collection of integers $q^m_\phi$, which determine which melons contribute to the resummation of each $\phi$ propagator.
The result of this chapter is that once a list of fields is given, the conformal IR solution is completely specified by this set of integers $q^m_{\phi}$, up to a discrete choice of vacua.

The canonical theories in the melonic class are those containing only a single field:
\begin{itemize}
    \item The $\psi^q$ SYK model is defined by $q_{\psi} = q$ \cite{Kitaev:2017awl,Maldacena:2016hyu}.
    \item The $\phi^q$ tensor model is defined by $q_{\phi}=q$ \cite{Witten:2016iux,Giombi:2017dtl,Carrozza:2018psc,Benedetti:2019rja,Harribey:2021xgh}.
\end{itemize}
The \textit{multi-field} melonic models, i.e. those with multiple distinct fundamental fields, display much richer IR structure, because the melonic constraints \eqref{eq:melonicConstraints} no longer completely solve the theory:
\begin{itemize}
    \item The generalized SYK model is defined \cite{Gross:2016kjj} by $q_{\psi_i} = q_i$ (this is the origin of our choice of notation).
    \item The large-$n$ $\gO(n)$ $\phi^4$ critical vector model, rewritten with an auxiliary field $\sigma=\phi_I\phi_I$, is defined by $q_{\sigma}=1$ and $q_{\phi} =2$. The same holds for the large-$n$ $\gU(n)$ Gross-Neveu model. We elaborate on this in \cref{sec:vectorModels}.
    \item We can also consider multi-field tensor models, as in \cite{Giombi:2018qgp,Prakash:2022gvb,Fraser-Taliente:2024rql}, with fields $\phi^i$. For a rank-$r$ model, each field $\phi_{a_1 \cdots a_r}$ transforms in $r$ copies of $\gO(M)$ (i.e. $\gO(M)^{\times r}$, with $N=M^r$) -- or indeed $\Sp(M)^{\times r}$ \cite{Gurau:2022dbx}.
    We consider one such model, the quartic Yukawa model, in \cref{chap:3dyuk}.
    \item The supersymmetric SYKs or tensor models can be considered in two different ways:
    \begin{itemize}
        \item They can be formulated directly in superspace, using superfields $\Phi$. In that case, they correspond to single-field melonic models. Taking a superpotential of the form $W\sim \Phi^q$, we have $q_{\Phi}=q$ \cite{Chang:2018sve, Popov:2019nja, Lettera:2020uay}.
        \item They can also be formulated in components, $\Phi= \phi + \theta \psi - \theta^2 X$, in which case we obtain the simplest example of a multi-interaction melonic theory. A superpotential $W\sim \Phi^q$ (see (3.3) of \cite{Chang:2018sve}) decomposes into the two melonic interactions
    \begin{equation}\begin{aligned}
       g_1: \, \{q^1_\phi = q-2, \, q^1_\psi = 2, q^1_F =0\}, \quad g_2: \,
        \{q^2_\phi = q-1, \, q^2_\psi = 0, q^2_F =1\}.
    \end{aligned}\end{equation} 
    Component form permits the consideration of supersymmetry-breaking dynamics; we expand upon this in \cref{sec:SUSYmelons}.
    \end{itemize}
\end{itemize}
There are also further melonic mechanisms which do not fit so neatly into these categories: one such is the Amit-Roginsky model. This is the $d$-dimensional theory of $N$ scalar fields in an irrep of $\SO(3)$, with a cubic coupling that takes the form of a Wigner 3-$j$ symbol \cite{Amit:1979ev,Benedetti:2020iku,Nador:2023inw}.
However, now that we have specified that melonic dominance occurs due to these melonic mechanisms, we need not discuss them any further: see the various reviews \cite{Rosenhaus:2018dtp,Gurau:2019qag,Benedetti:2020seh} for details of their implementation.

The long-range models are only briefly considered as an aside in \cref{sec:LRmodels}, because they must have the same scaling dimension as these short-range models by construction \cite{Gross:2017vhb,Benedetti:2019rja,Benedetti:2020rrq,Benedetti:2024mqx,Benedetti:2021wzt,Shen:2023srk}. 
Constructing them therefore requires already knowing the relevant IR scaling dimensions.
Additionally, in \cref{chap:fextr} we will not be investigating the renormalization group (RG) flow between these CFTs, and the associated QFTs \cite{Berges:2023rqa, Benedetti:2019rja,Berges:2024ydj,Fraser-Taliente:2024rql}, as our non-perturbative calculation of $\Ft$ requires working strictly in the conformal limit such that the two-point function is constrained; we delay such considerations to \cref{chap:3dyuk}. %

We note that there exist solvable models of melonic theories in zero dimensions \cite{Carrozza:2024gnh,Gurau:2011xp,Bonzom:2011zz,Carrozza:2015adg,Gurau:2019qag,Carrozza:2021qos}, analogous to our model of \cref{sec:0dQFT}.
However, we are principally interested in infrared or ultraviolet limits of QFTs, and so we do not consider zero-dimensional models any further.

\subsection{The \texorpdfstring{$\phi^4$}{phi\^4} tensor model}

We now begin by considering the standard single-field real bosonic tensor model \cite{Giombi:2017dtl,Benedetti:2019rja}, to set up for the more complicated tensor models that we will consider in \cref{chap:3dyuk}.
This model has some large number of real scalar fields $\phi_I$, $I=1,\ldots,\cN$, and, without interaction terms, has an $O(\cN)$ symmetry group.
The choice of interaction can break this down to a smaller subgroup $\gO(N)^r$, with $N^r = \cN$, where we call $r$ the rank. The well-studied vector and matrix models correspond to $r=1$ and $r=2$ respectively \cite{Moshe:2003xn}. We will focus on the case $r=3$, but for a recent summary of rank $r>3$ tensors, see \cite{Jepsen:2023pzm}. 

Taking $r=3$, we are naturally led towards organizing our $\cN=N^3$ fields into a tensorial field $\phi_{abc}$, $a,b,c=1,\ldots,N$, as then terms in the Lagrangian allowed by the symmetry are tensor invariants of these fields, such as $\sum_{abc} \phi_{abc}\phi_{abc}$. Note that each index\footnote{In the $\gO(N)$ model, we do not distinguish raised and lowered indices.} here transforms in the fundamental of a different $\gO(N)$ global symmetry, such that the action of the symmetry group $\gO(N)^3$ is
\begin{equation}\begin{aligned}
\phi_{abc} \mapsto O_{a a'} P_{b b'} Q_{c c'} \phi_{a'b'c'},%
\end{aligned}\end{equation}
and we require that the Lagrangian is invariant under this global transformation of all the fields. In order to obtain a conformal field theory in $d\le 3$, we restrict to renormalizable interactions. The melonic limit is then taken by choosing an optimal scaling in $N$ for the coupling constants, and then taking $N \to \infty$ in the graphical expansion.

Then, the only diagrams that contribute are those of Gurau degree zero. 
We recall that the Gurau degree of any diagram is simply (half of) the sum of the genera of each of the ribbon graphs obtained by forgetting one of the indices. 
Given that the genus is nonnegative, we might amusingly describe the leading diagrams as hyper-planar, in the sense of being \enquote*{even more planar than planar} \cite{Witten:2016iux} -- and we call them melonic.

We therefore permit in the interaction potential of tensorial $\phi^4$ theory in $d<4$ terms like the \textit{tetrahedron interaction}
\begin{equation}\begin{aligned}
V(\phi) \supset \frac{g_t}{4! N^{3/2}} \phi_{abc} \phi_{ade} \phi_{fbe} \phi_{fdc} \equiv \frac{g_t}{4! N^{3/2}} \quad 
\begin{tikzpicture}[baseline={([yshift=-.5ex]current bounding box.center)}]
\draw[color=red] (2,0) -- (2,2) (0,2) -- (0,0);
\draw[color=green] (2,2) -- (0,2) (0,0) -- (2,0);
\draw[color=blue] (0,0)--(2,2) (0,2)--(2,0);
\fill[color=black] (0,0) circle (0.05) (2,0) circle (0.05) (2,2) circle (0.05) (0,2) circle (0.05);
\end{tikzpicture},
\end{aligned}\end{equation}
where a (red, green, blue) line denotes contraction of the (first, second, third) index of $\phi_{abc}$;
the $4!$ is the usual automorphism factor associated with the interaction $\frac{1}{4!} \phi^4$ (although some authors modify this to pre-empt the melonic limit);
and the $N^{3/2}$ gives the optimal scaling for this interaction to ensure that this melonic interaction dominates in the large-$N$ limit. 
This term is manifestly invariant under the $\gO(N)^3$ symmetry, and so we can put it in the Lagrangian -- but as we will see in \cref{chap:fextr}, there are a multiplicity of other possible invariants, which we neglect in the melonic limit.

With only one field in the Lagrangian (and therefore only one field in the melonic-dominant Feynman diagrams) the scaling dimension of $\phi$ in the conformal large-$N$ limit is completely determined by dimensional analysis.
For the full propagator $G(p)=\expval{\phi(p)\phi(-p)}$, in the melonic limit the IR Schwinger-Dyson equation is simply
\begin{equation}
G(p)^{-1} \propto \int_{k,l}G(p+k+l)G(k)G(l),
\end{equation}
which immediately gives that $d-2\Delta_\phi = 2d-3(d-2\Delta_\phi)$, and so $\Delta_\phi = d/4$ exactly \cite{Giombi:2017dtl}.
When more fields are introduced we have the possibility of more large-$N$ fixed points and scaling dimensions that are non-linear in $d$: in particular, the scaling dimensions are far from their free-field values.
We analyse in general all such melonic CFTs in \cref{chap:fextr}, and analyse in detail a specific CFT of this kind in \cref{chap:3dyuk}.

\chapter{\FttextOrPDF-extremization in melonic CFTs}\label{chap:fextr}

\section{Introduction}

In this chapter, we show that the conformal data of the large-$N$ melonic CFTs are determined by constrained extremization of $\Ft$, the universal part of the sphere free energy of a collection of generalized free fields.

The constraints arise directly from the interaction terms, and are linear in the conformal scaling dimensions of the fields. Put another way, we demonstrate that the melonic CFTs are precisely the conformal mean field theories with constrained extremal $\Ft$. 
Notably, this procedure turns out to be identical to the $F$ and $a$-maximization principles used to determine the $R$-charges and scaling dimensions of SCFTs with four supercharges \mccorrect{consisting only of chiral superfields} \cite{Giombi:2014xxa, Pufu:2016zxm}.

We show that for any melonic QFT$_d$, as defined in \cref{sec:MelonicFTs}, regardless of the individual complexities of the model in question, the IR CFT$_d$ is determined by a universal principle: extremization of $\Ft$, where $\Ft$ is defined for a mean field theory with the same field content as the QFT but arbitrary conformal scaling dimensions. 
This reflects our expectation that in the large-$N$ limit, factorization means that the correlators of the fundamental field to leading order in $N$ should be essentially Gaussian, i.e. mean field. 
Since $\Ft$ is thought to count the effective number of degrees of freedom of a CFT$_d$ \cite{Giombi:2014xxa,Fei:2015oha,Giombi:2024zrt} (being a candidate weak $C$-function in the sense of Zamolodchikov \cite{Zamolodchikov:1986c,Cardy:1988cwa}), this has an appealing simplicity.

In outline, the $\Ft$-extremization procedure is as follows. In arbitrary dimension $d$, we define a UV theory of order $\sim N$ free fields $\{\phi\}$ in arbitrary Lorentz and global symmetry representations. 
We perturb by a particular relevant interaction and follow the renormalization group flow to the deep IR, where we reach the melonic CFT of interest. There, the conformal symmetry means that the fields have some conformal scaling dimensions $\Delta_\phi$.
Thanks to the simplifications of the melonic limit, we can compute the universal part of the sphere free energy as a function of the unknown $\Delta_\phi$s,
\begin{equation}\begin{aligned}\label{eq:melonicFtSum}
\Ft \equiv \sum_{\text{fields } \phi} \Ft_{\phi}(\Delta_{\phi}).
\end{aligned}\end{equation}
\mccorrect{This can be interpreted as $\Ft$ for a trial mean field theory that has the same field content as the original theory, but each field $\phi$ has an arbitrary \textit{trial} dimension $\Delta_\phi$. 
Then, the \textit{actual} IR scaling dimensions of the true interacting theory are precisely those that extremize this function $\Ft(\{\Delta_\phi\})$, subject only to a particular melonic constraint. 
Additionally, the value of $\Ft_{\mathrm{CFT}}$ for this CFT to leading order in $N$ is indeed precisely the value of} \eqref{eq:melonicFtSum} at the extremum. 
This procedure always leads to a finite (albeit possibly zero) number of conformal vacua in the IR for each $d$.

Explicitly, for a perturbing melonic interaction of schematic form\footnote{We have suppressed the large-$N$ structure that ensures the melonic resummability.}
\begin{equation}\begin{aligned}
    S_{\mathrm{int}} \supset \sum_m g_m\int \odif[d]{x} \prod_{\text{fields }\phi} \phi^{q_{\phi}^m},
\end{aligned}\end{equation}
the IR \textit{melonic constraints} are
\begin{equation}\begin{aligned} \label{eq:melonicConstraints}
\forall \, m: \quad [\odif[d]{x}] +\sum_{\text{fields } \phi} q^m_{\phi} \Delta_{\phi} = 0, \quad [\odif[d]{x}]=-d.
\end{aligned}\end{equation} 
However, we must allow for the possibility that any given $g_m$ runs to zero in the IR, in which case the associated constraint is not applied.

Remarkably, this formulation makes manifest that any apparent dependence on the details of the global symmetry representations, or any particular complicated form for the interaction (especially for non-scalar Lorentz representations) is washed out by the conformal melonic limit. The only pieces of data that matter are:
\begin{enumerate}
    \item the monomial form of the relevant melonic interactions (given by $\{q^m_\phi \in \N_0\}$);
    \item and the dimensions of the global symmetry representations of the fields.
\end{enumerate}
Therefore, the melonic CFTs have the same scaling dimensions as the trial conformal mean field theories with the given field content that extremize $\Ft$, subject to the constraints \eqref{eq:melonicConstraints}. 
\mccorrect{
In the large-$N$ limit, any non-Gaussian behaviour in the correlation functions of the fundamental field of the CFT is subleading in $N$; thus the correlators of the fundamental fields at separated points are described by the conformal mean field theory to leading order in $N$. 
Of course, the other operators in the CFT are not captured by this simple mean field theory}; in particular, as we will see in \cref{sec:bilinears}, the bilinears have $O(1)$ anomalous dimension.

Because the universal part of the sphere free energy ($\Ft$) interpolates between the Weyl anomaly in even $d$ and the free energy in odd $d$, this procedure is identical to the mechanisms of $a$- and $F$-extremization in superconformal field theories (SCFTs) with four supercharges, without the large-$N$ limit\footnote{If the SCFTs are unitary, it can be shown that the extrema are actually maxima \cite{Pufu:2016zxm}.} \cite{Giombi:2014xxa}: we need only swap out the $\phi$s for chiral superfields $\Phi$, and space for superspace.
The constraints \eqref{eq:melonicConstraints} are then exactly the SUSY-preserving constraints in the IR. A melonic version of such an SCFT can be solved using either the supersymmetry or the melonicity; indeed, the two procedures are identical in the SUSY-preserving sector. 
However, the melonic procedure extends to SUSY-breaking vacua.

The $\Ft$-extremization procedure is proved using the two-particle-irreducible (2PI) effective action \cite{Benedetti:2018goh}, or, equivalently, using the Schwinger-Dyson equations. 
In this chapter, we give both proofs. 
The effective action approach is more enlightening, and we focus on it: in particular, $\Ft(\{\Delta_\phi\})$ \textit{is} the 2PI effective action evaluated with the conformal ansatz on the sphere. 
At its extremum, this function coincides by construction with the universal part of the CFT sphere free energy, which is the usual quantity $\Ft_\mathrm{CFT}=\sin(\pi d/2)\log Z_{S^d}$ \cite{Giombi:2014xxa,Fei:2015oha}. %

The proof reduces to two components. 
\begin{enumerate}
\item The first is straightforward: since the full quantum solution functionally extremizes the 2PI effective action, the non-perturbative conformal scaling dimensions must also extremize $\Ft$.
\item Second, we demonstrate that in the conformal limit the running coupling constants become Lagrange multipliers implementing the constraints \eqref{eq:melonicConstraints}.
There are no further contributions, so the effective action becomes simply the sum of the sphere free energies of generalized free fields: the constraint naturally vanishes on its solution.
\end{enumerate}
This gives a concrete understanding of why the $\Ft$-extremization procedure works. 
Crucially, however, the proof that the interaction leads to exactly linear constraints on the IR scaling dimensions rests on the particular properties of the melonic limit; except for the supersymmetric melonic case mentioned above, this structure will not persist to subleading orders in $N$, and the subleading structure explicitly depends on which of the various melonic mechanisms is used.

\subsection{Chapter summary}

We defined the melonic field theories in \cref{sec:MelonicFTs}, and they fit neatly into the perspective of $\Ft$-extremization.
For completeness, we therefore briefly discuss them.
In \cref{sec:FthmsFmax} we review the role of the universal part of the sphere free energy in QFT, generalized free fields, and the various maximization procedures appearing in supersymmetric quantum field theories.
We give a careful explanation of our claim of $\Ft$-extremization in \cref{sec:fundamentalClaim}, and illustrate it with a simple example.
Our claim is proved in \cref{sec:FtMaxFrom2PI}, using the 2PI effective action (\cref{app:diagrammaticProof} contains an alternative Feynman-diagrammatic proof using the Schwinger-Dyson equations).
The long-range melonic models follow from the $\Ft$ analysis immediately, and are discussed in \cref{sec:LRmodels}.
In the remainder of the chapter, we explore the IR structure of these melonic CFTs as a function of $d$.
Much of this structure can be understood by comparison with the more familiar critical vector models, which we examine from an $\Ft$-extremization point of view in \cref{sec:vectorModels}; in \cref{sec:melonicModels}, we consider as examples three melonic CFTs, one of which is an SCFT, and outline the similarities. 
Finally, in \cref{sec:MelonicFmaxLimitations}, \mccorrect{we discuss the limitations of $\Ft$-extremization: does the Lagrangian QFT that we write down actually flow to the melonic CFTs we identify?}
We postpone our conclusions to \cref{sec:FextrOutlook}.

\newcommand{\fr}{\mathfrak{r}}
\newcommand{\nrows}[1]{r(#1)}

\section{Review: the \FtextOrPDF-theorems and \FttextOrPDF-maximization} \label{sec:FthmsFmax}

\subsection{The free energy in QFT and \texorpdfstring{$C$}{C}-functions} %

In the Wilsonian renormalization group, each RG step is composed of a Kadanoff blocking followed by a rescaling \cite{Wilson:1973jj}.
The Kadanoff blocking clearly decreases the number of degrees of freedom towards the IR; however, the rescaling step reintroduces degrees of freedom, complicating this interpretation.
Nonetheless, we still need to capture our intuition that the number of effective degrees of freedom of a QFT decreases under RG flow.
That is, as we sketched in \cref{fig:UVIRflow}, we want to define a $C$-function (in the sense of Cardy \cite{Cardy:1988cwa}) such that $\odv{C(E)}{E} > 0$.
Such functions are not clear in continuous dimension, but they are known in some integer dimensions \cite{Zamolodchikov:1986c,Komargodski:2011vj,Liu:2012eea,Pufu:2016zxm,Shore:2016xor}, so a slightly weaker condition is to require a \textit{weak $C$-function}, defined only at the ends of the flow (where generically we expect a CFT), such that $C_{UV} > C_{IR}$.

As reviewed in \cref{sec:introZandF}, in statistical mechanics at finite temperature, the free energy can be thought of as counting the number of degrees of freedom.
Unfortunately, in Euclidean CFTs, $F=-\log Z$ is not a suitable weak $C$-function, as it is mired in subtleties, possessing a volume divergence $\int \odif[d]{x}$, cutoff dependence, scheme dependence, and gauge dependence. 
All of these obscure the universal information hiding within it. $\Ft$, a modified version of the \textit{sphere} free energy, defined below, evades these issues -- thus providing a candidate weak $C$-function.

\subsection{\FttextOrPDF and the generalized \FttextOrPDF-theorem}

For a given CFT in continuous dimension $d$, we define 
\begin{equation}\begin{aligned}\label{eq:Ftdef}
\Ft \equiv -\sin(\pi d/2) F = \sin(\pi d /2) \log Z_{S^d},
\end{aligned}\end{equation}
where $Z_{S^d}$ is the partition function evaluated on the sphere $S^d$. The computation on the sphere regulates the volume divergence. The computation in continuous dimension removes the cutoff dependence of $F=-\log Z_{S^d}$, except when approaching even dimensions $d=2\mathbb{N}-\epsilon$, where the Weyl anomaly gives a $\sim a/\epsilon$ divergence in $F$.
Generically, $F$ has a series of power law divergences proportional to the cutoff, so this procedure involves analytically continuing in $d$ to dimension low enough that all power-law divergences in $F$ vanish. %
We notice that we can make this regularized $F$ finite in all $d$ by dividing through by the analytically continued volume of Euclidean hyperbolic space, $\vol \mathbb{H}^{d+1}$\mccorrect{, a computation which we discuss in} \cref{sec:volHd}. 
We ensure that $\Ft$ is dimensionless by actually multiplying by $\half \vol S^{d+1}/\vol \mathbb{H}^{d+1}$: this yields the overall factor of $-\sin(\pi d/2)$ in \eqref{eq:Ftdef} \cite{Giombi:2014xxa}.%

$\Ft$ interpolates between $(-1)^{d/2} \pi a/ 2$ for the Weyl anomaly coefficients $a$ in even dimension, and $(-1)^{(d-1)/2} \log Z_{S^d}$ in odd dimensions. 
In dimensions $2,3$, and $4$ it is indeed a weak $C$-function, because so are the Weyl anomalies in $d=2,4$ and $F=-\log Z_{S^3}$ in $d=3$ \cite{Zamolodchikov:1986c,Cardy:2010fa,Komargodski:2011vj, Pufu:2016zxm,Casini:2012ei,Benedetti:2021wzt}. %
Though it appears to be a strong $C$-function with respect to perturbative RG flows \cite{Pannell:2025ixz}, there are holographic counterexamples to $\Ft$ as a strong $C$-function in $d=3$ \cite{Taylor:2016kic,Ghosh:2018qtg}.
That is not a problem, as we are only interested in $\Ft$ as a weak $C$-function. %
However, the conditions for $\Ft$ to be a weak $C$-function in continuous dimension are not clear; there are trivial counterexamples in any $d$ from the flow between generalized free field theories (precisely those of \cref{sec:GFFflow}) with scaling dimensions that violate the unitarity bound \cite{Benedetti:2021wzt}.
Nonetheless, in the following we shall show that $\Ft$ can be used in generic dimension to determine the IR limit of certain large-$N$ theories via an extremization principle. %

\subsection{Generalized free fields and \FttextOrPDF} \label{sec:GFFsAndFt}

We briefly review generalized free fields (GFFs), following \cite{Benedetti:2021wzt}. These are just free fields with arbitrary propagators.
If a theory containing GFFs is conformal, we call it a conformal generalized free field theory, or a long-range massless Gaussian theory; we discussed a simple example of the flow between two GFFs in \cref{sec:GFFflow}.
A theory of GFFs is a mean field theory (MFT), as all correlators are simply sums of products of two-point functions. 
Recall that it is a generic result that MFT is the leading contribution in large-$N$ theories, due to factorization\footnote{
In fact, when solving QFTs by expanding in small parameters like $N$ or $\epsilon$, we often find GFFs to leading order, with small corrections proportional to our expansion parameter \cite{Bissi:2022mrs}.
This makes sense, as the GFFs can be solved exactly, and are therefore a useful place to start a perturbative expansion.
}. %
Once we have a particular mean field theory CFT, it is straightforward to Weyl map it to the sphere; this enables our calculation of $\Ft$.

\subsubsection{Conventions for representations}

As a short technical note: we use the following conventions for the representations of the fields in our Euclidean QFTs. We denote a field in the $\SO(d)$ representation $\rho_\Phi$ by $\Phi_{\mu_1 \cdots \mu_s}$, where $\mu_i$ are the generalized $\SO(d)$ indices -- that is either vectorial or spinorial. This field may also transform in a representation $R_\Phi$ of some finite internal symmetry group $\Gglobal$ (i.e. $\dim \Gglobal$ does not scale with $N$), which we assume throughout to be unbroken.
Then we say that the field transforms in the representation
\begin{equation}\begin{aligned}
\rhoext_\Phi=\rho_\Phi \times R_\Phi \text{ of }\SO(d)\times \Gglobal.
\end{aligned}\end{equation} 
For example, for an $\gO(N)$ vector of complex Dirac fermions $\psi_i$ we have
\begin{equation}\begin{aligned}
\rhoext_{\psi} = (\text{Dirac fermion of }\SO(d), \text{vector of } \gO(N))
\end{aligned}\end{equation} 
so that the real dimension of $\rhoext_\psi$ is
\begin{equation}\begin{aligned}
\dim \rhoext_{\psi} = \dim_{\mathbb{R}}\rho_\psi \times\dim R_\psi=  2 \Tr[\spinid] \times N.
\end{aligned}\end{equation}
The relevant data for a melonic CFT are the Lorentz representations $\rho_\Phi$ and the dimensions of the global symmetry representations $\dim R_\Phi$ -- other details of the $\Gglobal$ representations do not matter.

\subsubsection{Generalized free fields}

The standard free bosonic field, with a local action, has scaling dimension $\Delta=\frac{d-2}{2}$. By giving up locality, we can equally consider the case of a free field with arbitrary $\Delta$. Then the action in flat space is given by
\begin{equation}\begin{aligned}
S &=\half \int \odif[d]{x} \, \phi(x) (-\partial^2)^{\frac{d}{2}-\Delta} \phi(x) = \half \int \frac{\odif[d]{p}}{(2\pi)^d} \, \tilde{\phi}(p) (p^2)^{\frac{d}{2}-\Delta} \tilde{\phi}(-p).
\end{aligned}\end{equation} 
We will prefer to use the following bilocal form, which contains the explicitly non-local implementation of the operator $(-\partial^2)^{\frac{d}{2}-\Delta}$,
\begin{equation}\begin{aligned}
S &= \lim_{r\to 0} \half \int_{\abs{x-y}>r} \odif[d]{x} \, \odif[d]{y}\, \phi(x) G_\phi^{-1}(x,y) \phi(y),
\end{aligned}\end{equation} 
though we hide the limit in all subsequent computations\footnote{We also hide various contact terms which serve only to make the correlator well-defined at coincident points \cite{Fraser-Taliente:2025udk}.}
The inverse propagator is
\begin{equation}\begin{aligned}\label{eq:inverseProp}
G_\phi^{-1}(x,y) \equiv \frac{c(d-\Delta)}{\abs{x-y}^{2(d-\Delta)}}, \quad c(\Delta) \equiv \frac{1}{2^{d-2\Delta} \pi^{d/2}} \frac{\Gamma(\Delta)}{\Gamma(\frac{d}{2}-\Delta)},
\end{aligned}\end{equation} 
which manifestly gives a conformal correlator. 
The propagator is then
\begin{equation}\begin{aligned} \label{eq:flatSpaceGFFpropagator}
G_\phi(x,y) = \frac{c(\Delta)}{\abs{x-y}^{2\Delta}},
\end{aligned}\end{equation} 
which can be shown by Fourier transforming twice.
Due to the quadratic form of the action, all higher-point correlators of a theory consisting solely of GFFs can be found by Wick contractions, in the usual manner for a free theory.

We now Weyl map \eqref{eq:flatSpaceGFFpropagator} to the sphere of radius $R$, using the coordinates of the stereographic projection.  %
Recall from \cref{sec:WeylConf} that the $d$-dimensional sphere $S^d$ with a single point removed is Weyl equivalent to Euclidean space. %
Hence, the sphere is said to be conformally flat. %
This makes the map of the flat-space CFT to the sphere CFT straightforward (up to subtleties associated with the curvature couplings, which we can neglect for GFFs \cite{Pufu:2016zxm}. Specifically, we will always assume that the correct curvature couplings are always present in the action, even in the flat space case, where they disappear, as the curvature is zero). %
In stereographic coordinates, the sphere metric is 
\begin{equation}\begin{aligned} \label{eq:sphereMetric}
g_{\mu\nu} = \Omega(x)^2 \delta_{\mu\nu}, \quad \quad \Omega(x) \equiv \frac{2 R}{(1+x^2)^\half},
\end{aligned}\end{equation} 
which is manifestly a Weyl transformation away from flat space. The chordal distance $s(x,y)$ is
\begin{equation}\begin{aligned}
s(x,y) \equiv \Omega(x)^\half\Omega(y)^\half \abs{x-y}.
\end{aligned}\end{equation} 
To Weyl map the propagator, we simply replace $\abs{x-y} \to s(x,y)$ in \eqref{eq:flatSpaceGFFpropagator}:
\begin{equation}\begin{aligned} \label{eq:sphereGFFpropagator}
G_\phi(x,y)|_{S^d} = \frac{c(\Delta)}{s(x,y)^{2\Delta}}.
\end{aligned}\end{equation} 
The same applies for fields in other Lorentz representations. However, we note that to reproduce \eqref{eq:sphereGFFpropagator}, the sphere action will generically need additional non-minimal couplings to the geometry\footnote{For example, for the conformally coupled scalar of dimension $\Delta= \frac{d-2}{2}$, the sphere action is $$S_{\phi}=\int_{S^d} \odif[d]{x} \sqrt{g} \, \half \phi \left(-\partial^2 + \frac{d-2}{4(d-1)} R\right) \phi.$$}. 
\mccorrect{We assume that the correct couplings required for an uplift to sphere of the CFT data $\{(-\Delta_i, \rho_i), C_{ijk}\}$ are always present, thus sidestepping this subtlety.}

\subsubsection{Free energy}
\begin{figure}
\centering
    \begin{subfigure}{0.5\textwidth}
\includegraphics[width=\textwidth]{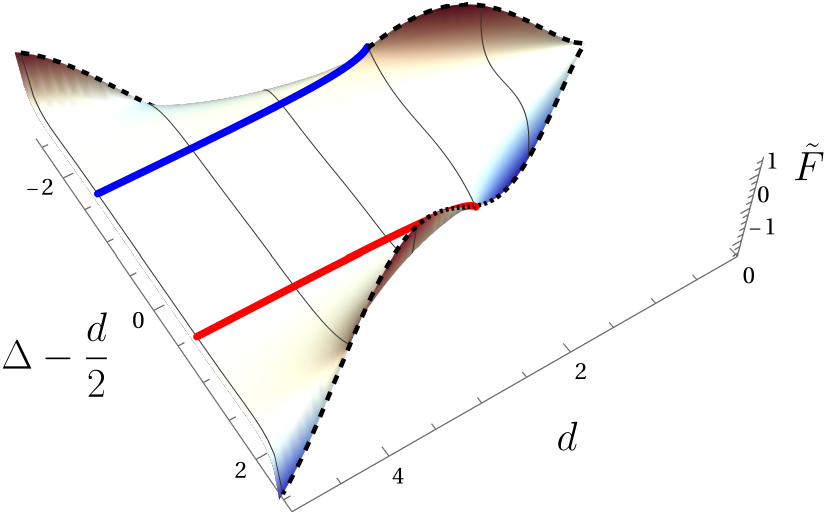}
    \caption{$\Ft$ for a free boson.}
    \end{subfigure}%
    \begin{subfigure}{0.5\textwidth} 
\includegraphics[width=\textwidth]{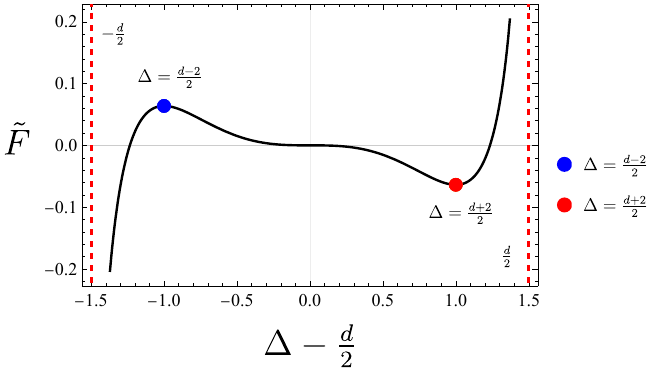}
    \caption{Cross-section for $d=3$.}
    \label{fig:Ftdeq3}
    \end{subfigure}
\caption{$\Ft_b(\Delta)$ for a generalized free scalar field, \eqref{eq:Ftboson}, shown as a surface for $0+\epsilon<\Delta<d-\epsilon$. The function is even around $\Delta=\frac{d}{2}$. $\Ft(\Delta)$ looks schematically like \cref{fig:Ftdeq3} ($d=3$), for all values of $d$: there is always a local maximum at the scaling dimension of the free field $\frac{d-2}{2}$ (in blue); a stationary point at $\frac{d}{2}$; and a local minimum for the shadow of the free field $\frac{d+2}{2}$ (in red). 
There is also always a log-divergence for $\Delta$ approaching $0,d$, which is shown with a dashed line.}
\label{fig:Ftboson3D}
\end{figure}

The free energy of a generalized free bosonic field $\phi$ on the sphere is
\begin{equation}\begin{aligned}
F_\phi &= -\log Z_{\phi,S^d}= \log \int \cD \phi\, \exp(- \int_{x,y}\half \phi(x) G_\phi^{-1}(x,y) \phi(y)) \\
&= \half\log\det G_\phi^{-1} = \half \Tr\log G_\phi^{-1},
\end{aligned}\end{equation} 
where we have defined $\Dd{\phi}$ to cancel constant factors, and $G_\phi^{-1}$ is the inverse sphere propagator. For a field $\Phi$ of arbitrary statistics, we need only modify this to $F_\Phi=(-1)^{\mathrm{F}^\Phi} \half \Tr\log G_\Phi^{-1} \equiv \half \mathrm{Str}\log G_\Phi^{-1}$ (taking $\mathrm{F}^\Phi=0$ for bosons and $1$ for fermions). Hence,
\begin{subequations} \label{eq:FtBosFerm}
\begin{equation}\begin{aligned}
\Ft_{\Phi} = -\sin(\pi d/2) \, \half \,\mathrm{Str} \log G_{\Phi}^{-1}(x,y)
\end{aligned}\end{equation} 
for any real field $\Phi$ with sphere propagator $G_{\Phi}(x,y)$. 

\mccorrect{The computation of $\Ft_\Phi$ when $G_\Phi$ is the conformal sphere propagator is a standard result from the AdS/CFT literature for any Lorentz representation of $\Phi$, and we provide a compact derivation} in \cref{app:FtComputation}.
For a free real scalar boson of dimension $\Delta$, we have
 \begin{equation}\begin{aligned} \label{eq:Ftboson}
\Ft_b(\Delta, \text{scalar}) \equiv \frac{\pi}{\Gamma(d+1)}\int_{\frac{d}{2}}^{\Delta} \odif{\Delta'} \, \frac{\Gamma (\Delta' ) \Gamma (d-\Delta' )}{\Gamma \left(\frac{d}{2}-\Delta' \right) \Gamma \left(\Delta' -\frac{d}{2}\right)},
 \end{aligned}\end{equation} 
 where we have used $\Ft(\Delta=d/2)=0$. The behaviour of this function in continuous dimension for $0<\Delta<d$ is shown in \cref{fig:Ftboson3D}. %
For a complex Dirac fermion of dimension $\Delta$, %
\begin{equation}\begin{aligned}
\Ft_f(\Delta, \text{Dirac f.}) &\equiv %
-2\Tr\spinid \frac{\pi}{\Gamma (d+1)}\int_{\frac{d}{2}}^{\Delta} \odif{\Delta'} \frac{\Gamma \left(\Delta' +\frac{1}{2}\right) \Gamma \left(d-\Delta' +\frac{1}{2}\right)}{\Gamma \left(\frac{d}{2}-\Delta' +\frac{1}{2}\right) \Gamma \left(\Delta' -\frac{d}{2}+\frac{1}{2}\right)},
\end{aligned}\end{equation} 
\end{subequations}
where we have used
\begin{equation*}
\dim\rhoext_\psi = \dim \rho_\psi \times \dim R_\psi = \dim(\text{Dirac fermion}) = 2 \Tr\spinid.
\end{equation*}
The trace structure of $\Ft_\Phi$ makes it easy to see the following two facts. First, the results \eqref{eq:FtBosFerm} should never depend on the normalization of the field $\Phi$ -- indeed they do not, because $\Tr\log(C\delta^d(x-y))= 0$, for any constant $C$. Second, it is trivial that $\Ft_\Phi$ for a GFF $\Phi$ in a representation $\rhoext=\rho \times R$ of $\SO(d) \times \Gglobal$ is %
\begin{equation}\begin{aligned}
\Ft_\Phi =\Ft(\Delta, \rhoext) = \dim R \times \Ft(\Delta, \rho).
\end{aligned}\end{equation} 
For example, a complex scalar ($\Gglobal=\gU(1)$) has $\Ft(\Delta, \text{complex scalar}) = 2 \Ft(\Delta, \text{scalar})$.
Finally, as demonstrated in \cref{fig:Ftboson3D}, for all $d$, $\Ft_b(\Delta)$ for the scalar GFF has the following properties. $\Ft$ hits a maximum for the free field, $\Delta=\frac{d-2}{2}$, with a corresponding minimum for the shadow of the free field at $\frac{d+2}{2}$ -- and a stationary point in between at $\frac{d}{2}$. The same applies for the fermion, with free field $\Delta = \frac{d-1}{2}$.
Hence, the free field values are always local maxima of $\Ft$ for GFFs, and certainly absolute maxima within the range given by the free field value and its shadow. %

\subsection{\FttextOrPDF-maximization in superconformal field theories} \label{sec:Fmaximization}%

We now review the $F$- and $a$-maximization procedures in SCFTs for the simple case of chiral superfields perturbed by a relevant operator; as mentioned in the introduction,
we unify these procedures by considering $\Ft$-maximization in general dimension \cite[\S 5]{Giombi:2014xxa} \cite{Minahan:2015any}. 
\mccorrect{The main result of this work is that the procedure for these chiral superfields is the same as that for melonic theories.}

Take a UV supersymmetric QFT in $d$-dimensions with four supercharges\footnote{This corresponds to $\cN=1$ in $d=4$, $\cN=2$ in $d=3$, and $\cN=(2,2)$ in $d=2$.}, containing free chiral superfields $\{X\}$, in some general global symmetry representations. 
Then perturb by some superpotential $W(\{X\})$ that preserves the supercharges and global symmetry.  %

Now flow the theory to the deep IR; assuming that there is an SCFT there, given certain caveats (see \cref{sec:FmaxLimitations}), the $R$-symmetry generator in the IR is a linear combination of the original $UV$ generator and any other abelian symmetry generators. 
The supersymmetry enforces the relationship 
\begin{equation}\begin{aligned} \label{eq:DeltaRrel}
\Delta_X = \frac{d-1}{2} R_X
\end{aligned}\end{equation} 
between the scaling dimensions and the $R$-charges.
We can find the IR scaling dimensions by extremizing the total $\Ft$ for a collection of generalized free chiral superfields with \textit{trial} scaling dimensions $\Delta_{X}$
\begin{equation}\begin{aligned}
\Ft = \sum_{\text{chirals } X} \tilde{\mathcal{F}}(\Delta_{X}), \quad \tilde{\mathcal{F}}(\Delta) \equiv \Ft(\Delta, \text{chiral superfield}), \label{eq:chiralFtSumRepeated}
\end{aligned}\end{equation} 
subject to a constraint. 
This constraint is implied by insisting that we have supersymmetry and an $R$-charge in the IR, which fixes the $R$-charge of the superpotential to be two. %
This constraint can be rewritten suggestively in terms of the scaling dimensions $\Delta_{X}$ using the relationship \eqref{eq:DeltaRrel}: assuming a perturbing superpotential of the form
\begin{equation}\begin{aligned}
 S_{\text{int}} \supset \int \odif[d]{x} \odif[2]{\theta} \, W, \quad W \equiv \sum_m g_m\prod_{\text{chirals }X} X^{q_{X}^m}, %
\end{aligned}\end{equation} 
then, for each monomial, we have the constraint
\begin{equation}\label{eq:chiralsConstraintsRepeated}
[\odif[d]{x} \, \odif[2]{\theta}] + \sum_{\text{chirals } X} q^m_{X} \Delta_{X}  =0, \quad [\odif[d]{x} \, \odif[2]{\theta}] = -(d-1).
\end{equation}
The relationship \eqref{eq:DeltaRrel} between $R$-charge and scaling dimension holds only for chiral primary operators, and so does not hold for $W$ (see around \cite[(2.7)]{Cachazo:2002ry}). 
Thus, this identity does not mean that $\int \odif[2]{\theta} \,W$ has scaling dimension $d$ (as at the IR fixed point there is no such marginal operator); nor does it mean that the integrated operator $\int \odif[d]{x} \odif[2]{\theta} \,W$ has scaling dimension $0$ -- but it is a helpful mnemonic.

Each chiral superfield contains a complex scalar, a complex auxiliary field, and a complex Dirac fermion; so $\tilde{\cF}(\Delta_X)$ is defined by summing up the contributions of each of these components,
\begin{equation}\begin{aligned}
\tilde{\cF}(\Delta) &\equiv \Ft_b(\Delta, \, \Cplx \text{ scalar}) + \Ft_f(\Delta+\thalf, \text{Dirac fermion}) +  \Ft_b(\Delta+1, \, \Cplx \text{ scalar}),
\end{aligned}\end{equation} 
up to a convention-dependent additive constant which drops out of all computations. To ensure supersymmetry, we take $\Tr\spinid=2$ fixed, regardless of dimension: this is the dimensional reduction scheme, a standard procedure for analytically continuing $3$d supersymmetry away from $d=3$ \cite{Giombi:2014xxa}. %

\mccorrect{Thus, as in the melonic case, we extremize $\Ft$ for a mean field theory, subject only to the constraint} \eqref{eq:chiralsConstraintsRepeated}. 
If the SCFT is also unitary, it turns out that the extremum is also a maximum; whether this holds in any sense for the melonic theories as well is not yet clear. %

\subsubsection{Simple example}

Consider the example of $N+1$ chiral superfields, and a potential $W =\frac{\lambda}{2} X \sum_{i=1}^N Z_i Z_i$. To find the exact IR scaling dimensions, we maximise
\begin{equation}\begin{aligned}
\Ft = \tilde{\cF}(\Delta_X) + N \tilde{\cF}(\Delta_Z), \text{ subject to }\Delta_X + 2\Delta_Z = d-1,
\end{aligned}\end{equation} 
with \text{no} $N$-subleading corrections!

\subsubsection{Limitations of supersymmetric \FttextOrPDF-maximization} \label{sec:FmaxLimitations}

The principal limitation of $\Ft$-maximization here is that we must assume that the flow ends at an SCFT without any accidental symmetries arising in the IR. 
If this is not the case, two things can go wrong:
\begin{enumerate}
\item Supersymmetry breaks, and we can flow to a symmetry-breaking phase (which may not be a CFT).
\item The $R$-charge can mix with any accidental symmetries that arise in the IR. 
\end{enumerate}
We can see this in a slight modification of the example above. Consider the UV perturbation
\begin{equation}\begin{aligned}\label{eq:SUSYUVperturb}
W = g_1 X \sum_{i=1}^N Z_i Z_i + g_2 X^3.
\end{aligned}\end{equation} 
This triggers an RG flow, which preserves a $\gO(N) \times \mathbb{Z}_3$ flavour symmetry and a unique $\gU(1)_R$ symmetry, under which all the fields have $R$-charge $2/3$. However, for $N>2$, at least, the coupling $g_2$ is thought to run to zero; the flavour symmetry is then enhanced to $\gO(N) \times \gU(1)$, and so the IR $R$-charges, and so scaling dimensions, are not determined by the naive constraint $3 \Delta_X = d-1$ that would come from a non-zero $g_2$ \cite{Pufu:2016zxm}. 
\mccorrect{However, this is consistent with the numerical observation that $\Ft$ from $\Ft$-maximization applied to }\eqref{eq:SUSYUVperturb} \mccorrect{in the case $g_2=0$ is strictly larger than $\Ft = (N+1) \tilde{\cF}(\tfrac{d-1}{3})$ for all $N$ in $2<d<4$.}

These two phenomena have precise parallels in the case of the melonic-type theories: essentially, the former corresponds to breaking of the large-$N$ symmetry, and the latter corresponds to the fact that when we have multiple melonic interactions $g_m$, any of them can either be tuned to zero, or could run to zero; in the latter case, the corresponding monomial is irrelevant. %

\section{Fundamental claim} \label{sec:fundamentalClaim}

\begin{mdframed}
\textbf{The complicated IR structure of melonic field theories, with arbitrary numbers of fields and interactions, can be reformulated as a constrained $\Ft$-extremization problem.} 
The defining data for these melonic theories is
\begin{enumerate}
    \item a list of $n_f$ fundamental fields and their representations (Lorentz representation and global symmetry representations -- $(\phi, \rho_{\phi}, R_{\phi})$);
    \item an $n_m \times n_f$ matrix of integers -- $q^m_{\phi}$, determined from the schematic form of the melonic-dominant potential $V = \sum_{m=1}^{n_m} g_m \prod_{\phi} \phi^{q^m_\phi}$. 
\end{enumerate}
Then, we can compute the IR scaling dimensions by extremizing the free energy $\Ft=-\sin(\pi d/2)F$ of the mean field theory consisting of a collection of generalized free fields
\begin{subequations} \label{eq:FmaxSummary}
\begin{equation}
\Ft(\{\Delta_{\phi}\}) = \sum_{\text{fields } \phi} \Ft_{\phi}(\Delta_{\phi}, \rhoext_{\phi}) = \sum_{\text{fields } \phi} \Ft_{\phi}(\Delta_{\phi}, \rho_{\phi}) \times \dim R_{\phi},
\end{equation} 
with respect to the trial scaling dimensions $\Delta_{\phi}$, subject to the melonic constraints
\begin{equation} \label{eq:melConstrsIn4}
\sum_{\text{fields } \phi} q^m_{\phi} \Delta_{\phi} -d =0
\end{equation} 
for each of the couplings $g_m$. %
This extremization will typically give a discrete infinity of solutions -- consistency with the UV description then requires that the scaling dimensions of the fields must be greater than their free (UV) scaling dimensions.
Thus,
\begin{equation} \label{eq:DelGreaterThanFree}
\Delta_{\phi} > \Delta^\mathrm{free}_{\phi},
\end{equation} 
\end{subequations}
where for dynamical scalars $\Delta_\phi^\mathrm{free} = \frac{d-2}{2}$; for fermions $\Delta_\psi^\mathrm{free} = \frac{d-1}{2}$; and for auxiliary scalars $\Delta_X^\mathrm{free} = \frac{d}{2}$.
Otherwise, the free behaviour will dominate in the IR; of course, for the interacting fields \eqref{eq:DelGreaterThanFree} also corresponds to the unitarity bound in $d \ge 2$.
Therefore, only $\{\Delta_{\phi}\}$ lying inside a polyhedron in $\mathbb{R}^{n_f}$, which we call the \textit{IR wedge}, are valid solutions\footnote{Solutions satisfying $\Delta_{\phi} < \Delta^\mathrm{free}_{\phi}$  -- lying in the \textit{UV wedge} -- are also consistent as UV CFTs; however, we will usually neglect them.}.
In the case of equality in \eqref{eq:DelGreaterThanFree}, we find the long-range models, discussed in \cref{sec:LRmodels}.

Note that it is possible for the constraint system to be over-determined (if $n_m > n_f$). In that case, to find an IR solution, some of the couplings $g_m$ must either be set to zero, or run to zero. In the latter case, IR consistency demands that $\sum_{\phi} q^m_{\phi} \Delta_{\phi} > d$. We will typically assume that we keep only the $g_m$s that are non-zero at the fixed point.
\end{mdframed}
Other than the substitution of $[\odif[d]{x} \,\odif[2]{\theta}]=-(d-1)$ with $[\odif[d]{x}]=-d$, this is precisely identical to the supersymmetric $\Ft$-maximization of \cref{sec:Fmaximization}, except that, as far as we know, the solution need not only be a maximum to exist as a formal CFT. %

Incorporating the constraints \eqref{eq:melConstrsIn4} into $\Ft$ via Lagrange multipliers $\mathfrak{g}_m'$,
\begin{equation}\begin{aligned}
\frac{1}{N}\Ft(\{\Delta_{\phi}, \mathfrak{g}_m'\}) \equiv \sum_{\text{fields } \phi} \Ft_{\phi}(\Delta_{\phi}) +\sum_{\text{melons } m} \mathfrak{g}_m' \left(\sum_{\text{fields } \phi} q^m_{\phi} \Delta_{\phi} -d\right).
\end{aligned}\end{equation} 
We have suppressed the dependence on the representations. We will show in the next section that this construction is exactly an expansion of the sphere free energy of the melonic CFT to leading order in $N$, where $\mathfrak{g}_m'$ is proportional to the running coupling constant squared (that is, including the field renormalizations).
Extremizing this quantity with respect to the $\Delta_{\phi}$s and $\mathfrak{g}_m'$s then determines the conformal scaling dimensions of the theory. The Lagrange multipliers enforcing the constraint can therefore be given a precise interpretation as being the squared running coupling constants; this is precisely the conjecture of Kutasov \cite{Kutasov:2003ux,Barnes:2004jj,Amariti:2011xp} in the case of $a/F$-maximization in SCFTs.

The final results are a function only of the discrete data of the integers $q^m_{\phi}$ and the symmetry representations of the fields. 
In generic dimension, this $\Ft$-extremization procedure must be done numerically, typically yielding multiple possible IR vacua; we now demonstrate this with an explicit example.

\subsection{Explicit example: the melonic quartic Yukawa model}\label{sec:FextrQYuk}

\begin{figure}
    \centering
    \includegraphics[width=0.7\textwidth]{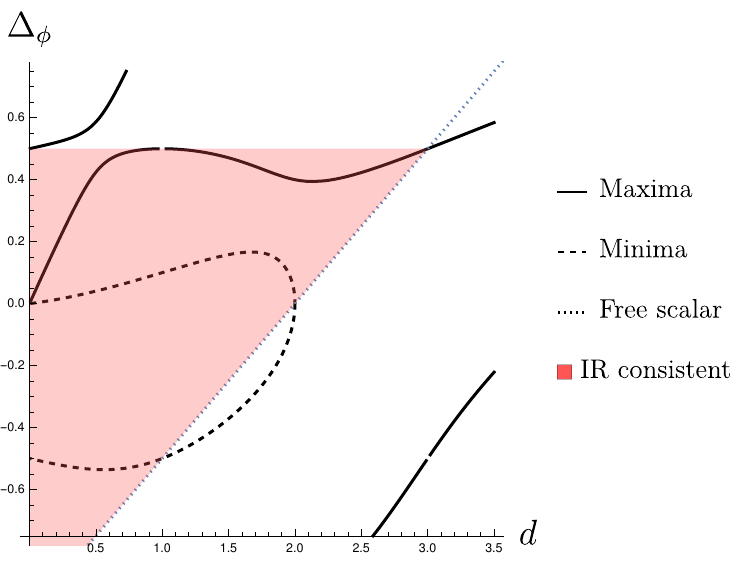}
    \caption{The scaling dimension of $\phi$ for the melonic quartic Yukawa theory \eqref{eq:lmelonicLagrangian} in continuous dimension; this follows from extremizing $\Ft(\Delta_\phi, \Delta_\psi, \mathfrak{g}')$ in \eqref{eq:lmelonicFt}. $\Delta_\psi$ can be found by the constraint $2\Delta_\phi + 2\Delta_\psi = d$. The red region indicates the IR-consistent wedge $\Delta_\phi \ge \frac{d-2}{2}$ and $\Delta_\psi \ge \frac{d-1}{2}$; this theory is free in the upper corner of this wedge, for $d=3$. Solid lines are maxima of the constrained $\Ft(\Delta_\phi) =\Ft_b(\Delta_\phi)+ \Ft_f(\tfrac{d-2\Delta_\phi}{2})$; dashed lines are minima. }
    \label{fig:phi2psibarpsiIRwedge}
\end{figure}

Consider the melonic quartic Yukawa model (studied in detail in \cref{chap:3dyuk}, based on \cite{Fraser-Taliente:2024rql}): it is the theory of $N$ Dirac fermions\footnote{As in the supersymmetric case, we take the $\gamma_\mu$s to have dimension $2$, regardless of $d$.} and $N$ bosons that is marginal in $d=3$, with schematic Lagrangian
\begin{equation}\begin{aligned} \label{eq:lmelonicLagrangian}
\cL_{QY} = \half \phi(-\partial^2)\phi +\bar\psi (-\slashed{\partial}) \psi + \half g\phi\phi\bar\psi\psi.
\end{aligned}\end{equation} 
Note that we have suppressed the particular melonic mechanism here, which could be a disorder average, as in SYK-like \cite{Prakash:2022gvb} or tensorial \cite{Fraser-Taliente:2024rql} theories.
The scalar potential has also been tuned to zero. Therefore, we extremize 
\begin{equation}\begin{aligned} \label{eq:lmelonicFt}
\Ft(\Delta_\phi, \Delta_\psi, \mathfrak{g}')/N = \Ft(\Delta_\phi, \text{scalar}) + \Ft(\Delta_\psi, \text{Dirac fermion}) + \mathfrak{g}' (2\Delta_\phi + 2\Delta_\psi -d)
\end{aligned}\end{equation} 
with respect to the $\Delta$s and the Lagrange multiplier $\mathfrak{g}'$, while also requiring $\Delta_\phi > \frac{d-2}{2}$ and $\Delta_\psi > \frac{d-1}{2}$.  Conveniently, we are taking derivatives with respect to $\Delta$, so do not need to do the integrals in \eqref{eq:FtBosFerm}.
After a numerical extremization procedure we obtain the results given in \cref{fig:phi2psibarpsiIRwedge}, where the IR wedge (evidently bounded by the scaling dimension of a free scalar) is shown in red, $\frac{d-2}{2} < \Delta_\phi < \half$. Since we are extremizing with respect to only a single variable $\Delta_\phi$, we only have a single constraint; thus all extrema are either maxima or minima. The line descending here from the free theory ($\Delta_\phi = \frac{d-2}{2}, \Delta_\psi = \frac{d-1}{2}$) that exists in $d=3$ is then indeed a maximum; however, for $d \le 2$, we also have two lines of minima. 

For certain lines, at some integer dimensions $d_0$, we note the absence of solutions -- despite the existence of perturbative solutions in $d_0-\epsilon$; this is indicated by breaks in the lines.
This typically arises due to $\odv{F_\phi}{\Delta_\phi}\to 0,\infty$ for one of the fields $\phi$ as $\Delta_\phi$ approaches some $\Delta_\text{div}$ \cite{Schaub:2024rnl}. %
A similar problem is encountered in \cite{Biggs:2023mfn}, albeit in a $d=1$ melonic model, and it indicates symmetry breaking -- but can nonetheless be solved by carefully taking the limit $\Delta_\phi \to \Delta_\text{div}$.
It would be interesting to understand the higher-dimensional analogue here.

We comment on this model further in \cref{sec:QuarticYukawaFull}.

\section{\FttextOrPDF-extremization from the 2PI effective action} \label{sec:FtMaxFrom2PI}

This mechanism is a straightforward consequence of the 2PI formalism, which we first briefly review. We aim to reach a simple expression \eqref{eq:Fin2PI} for the sphere free energy $F$ as a function of $G=\expval{\varphi(x) \varphi(y)}^\mathrm{conn}$, the full propagator. For more details on the 2PI formalism, see \cite{Benedetti:2018goh} and references therein.

\subsection{Reminder of the 2PI formalism}

The two-particle-irreducible (2PI) effective action $\Gamma[\phi,G]$ is defined as the double Legendre transform of the generating functional $\mathbf{W}[j,k]$ with respect to $j$, a one-point source, and $k$, a two-point source. Explicitly, for the theory of a scalar field with action $S[\varphi]$,
\begin{equation}\begin{aligned}
\mathbf{W}[j,k] \equiv \log \int \cD \varphi \, \exp\left(- S[\varphi] + \int_x j(x) \varphi(x) + \int_{x,y} \half \varphi(x) k(x,y) \varphi(y)\right).
\end{aligned}\end{equation} 
This is precisely (minus) the sourced free energy. Now, we define the expectation value of the field $\varphi$ in the presence of the sources
\newcommand{\bfPhi}{\Phi}
\begin{equation}\begin{aligned} \label{eq:classicalFieldDef}
\bfPhi(x) = \expval{\varphi(x)}_{j,k} = \fdv{\mathbf{W}}{j(x)};
\end{aligned}\end{equation} 
and the full connected propagator%
\begin{equation}\begin{aligned} \label{eq:GDef}
\mathbf{G}(x, y) &= \expval{\varphi(x) \varphi(y)}^{\mathrm{conn}}_{j,k} = \expval{\varphi(x)\varphi(y)}_{j,k} -\expval{\varphi(x)}_{j,k}\expval{\varphi(y)}_{j,k}\\
& = 2 \fdv{\mathbf{W}}{k(x,y)} - \fdv{\mathbf{W}}{j(x)}\fdv{\mathbf{W}}{j(y)},
\end{aligned}\end{equation} 
also in the presence of the sources. The (double) Legendre transform with respect to both $j$ and $k$ is implemented almost as usual \cite{Zia:2008LegendreTransform}, giving the 2PI effective action\footnote{We overload notation and use the same symbol to refer to a given object, regardless of whether it is the argument of a functional or a functional itself. Hence, we have either $(j,k,\bfPhi[j,k],\mathbf{G}[j,k])$ or $(j[\bfPhi, \mathbf{G}],k[\bfPhi, \mathbf{G}], \bfPhi, \mathbf{G})$; regardless of which we take, \eqref{eq:GammaAndWLT} holds.}: %
\begin{equation}\begin{aligned} \label{eq:GammaAndWLT}
\Gamma[\bfPhi, \mathbf{G}] + \mathbf{W}[j,k] = \int_x j(x) \bfPhi(x) + \half \int_{x, y} (\mathbf{G}(x, y) + \bfPhi(x) \bfPhi(y)) k(x, y).
\end{aligned}\end{equation} 
Hence, in the unsourced case $(j=0,k=0)$, $\Gamma$ coincides precisely with the free energy. The reason for the additional $\half$ and $\bfPhi(x) \bfPhi(y)$ in the second Legendre transform is to ensure that we get the connected $\mathbf{G}$ defined in \eqref{eq:GDef}, rather than the disconnected propagator.

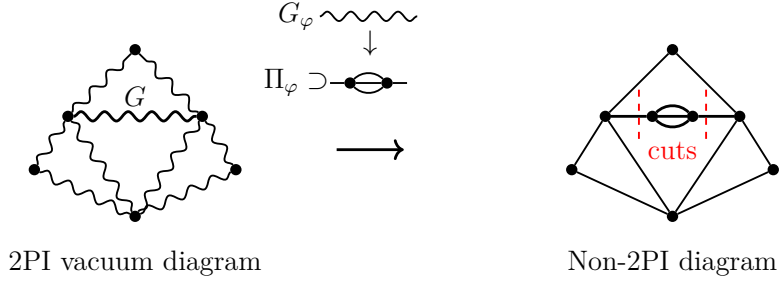
\begin{figure}
    \centering
    \begin{tikzpicture}[
    thick,
    vertex/.style={draw, circle, fill=black, inner sep=0pt, minimum size=4pt},
    Gprop/.style={decorate, decoration={snake, amplitude=2pt}, draw=black, thick},
    regular/.style={draw=black},
    label distance=0.5cm
]
\begin{scope}[xshift=0cm]
    \node[vertex] (v1) at (0,0) {};
    \node[vertex] (v2) at (2,0) {};
    \node[vertex] (v3) at (1,1) {};
    \node[vertex] (v4) at (1,-1.5) {};
    \node[vertex] (v5) at (-0.5,-0.8) {};
    \node[vertex] (v6) at (2.5,-0.8) {};
    
    \draw[Gprop,very thick] (v1) -- (v2);
    \draw[Gprop] (v1) -- (v3);
    \draw[Gprop] (v2) -- (v3);
    \draw[Gprop] (v1) -- (v4);
    \draw[Gprop] (v2) -- (v4);
    \draw[Gprop] (v5) -- (v1);
    \draw[Gprop] (v6) -- (v2);
    \draw[Gprop] (v5) -- (v4);
    \draw[Gprop] (v6) -- (v4);
    
    \node at (1,-2.2) {2PI vacuum diagram};
    \node at (1,0.3) {$G$};
\end{scope}

\draw[->, very thick] (4,-0.5) -- (5,-0.5);
\begin{scope}[yshift=1.5cm]
    \draw[Gprop] (3.75,0) -- (5.2,0); %
    \node at (3.4,0) {$G_\varphi$};
    \node at (3.4,-0.99) {$\Pi_\varphi \supset $}; %
    \node at (4.5,-0.4) {$\downarrow$};
    \node (b1) at (3.75,-1) {};
    \node (b2) at (5.2,-1) {};
    \node[vertex] (s1) at (4.2,-1) {};
    \node[vertex] (s2) at (4.75,-1) {};
    \draw[regular] (b1) -- (s1) -- (s2);
    \draw[regular] (s1) to[bend left=45] (s2);
    \draw[regular] (s1) to[bend right=45] (s2);
    \draw[regular] (s2) -- (b2);
\end{scope}

\begin{scope}[xshift=8cm]
    \node[vertex] (w1) at (0,0) {};
    \node[vertex] (w2) at (2,0) {};
    \node[vertex] (w3) at (1,1) {};
    \node[vertex] (w4) at (1,-1.5) {};
    \node[vertex] (w5) at (-0.5,-0.8) {};
    \node[vertex] (w6) at (2.5,-0.8) {};
    
    \node[vertex] (s1) at (0.7,0) {};
    \node[vertex] (s2) at (1.3,0) {};
    
    \draw[regular, very thick] (w1) -- (s1);
    \draw[regular, very thick] (s1) to[bend left=45] (s2);
    \draw[regular, very thick] (s1) -- (s2);
    \draw[regular, very thick] (s1) to[bend right=45] (s2);
    \draw[regular, very thick] (s2) -- (w2);
    
    \draw[regular] (w1) -- (w3);
    \draw[regular] (w2) -- (w3);
    \draw[regular] (w1) -- (w4);
    \draw[regular] (w2) -- (w4);
    \draw[regular] (w5) -- (w1);
    \draw[regular] (w6) -- (w2);
    \draw[regular] (w5) -- (w4);
    \draw[regular] (w6) -- (w4);
    
    \draw[red, dashed, thick] (0.5,0.4) -- (0.5,-0.4);
    \draw[red, dashed, thick] (1.5,0.4) -- (1.5,-0.4);
    
    \node at (1,-2.2) {Non-2PI diagram};
    \node[red] at (1,-0.5) {cuts};
\end{scope}
\end{tikzpicture}
    \caption{Replacing a propagator for $\varphi$ in a 2PI diagram by a contribution to the field self-energy $\Pi_\varphi$ yields a diagram which automatically disconnects on cutting two lines. Hence, all such diagrams are not 2PI.}
    \label{fig:2PIdiagrams}
\end{figure}

Solving the theory corresponds to finding $\expval{\varphi}$ and $\expval{\varphi(x) \varphi(y)}^\mathrm{conn}$ in the unsourced case: we call these the classical field $\phi$ and the full two-point function $G$ respectively,
\begin{equation}\begin{aligned}
\phi \equiv \bfPhi|_{j=0,k=0}, \quad G \equiv \mathbf{G}|_{j=0,k=0}.
\end{aligned}\end{equation} 
The Legendre transform relations are that, considered as functionals of $\bfPhi(x)$ and $\mathbf{G}(x,y)$, $j[\bfPhi, \mathbf{G}]$ and $k[\bfPhi, \mathbf{G}]$ solve
\begin{subequations}\label{eq:LegendreTransformRelations}
\begin{align}\label{eq:EoMsForVEVs}
\fdv{\Gamma[\bfPhi , \mathbf{G}]}{\bfPhi(x)} &= j(x) + \int_y k(x,y) \bfPhi(y) , \\
\fdv{\Gamma[\bfPhi, \mathbf{G}]}{\mathbf{G}(x,y)} &= \half k(x,y).
\end{align}
\end{subequations} 
Therefore, the equations of motion for $\phi$ and $G$ are
\begin{equation}\begin{aligned}
j[\phi, G]=0, \,  k[\phi, G]=0 \, \implies \, \fdv{\Gamma[\phi, G]}{\phi}  = 0, \, \fdv{\Gamma[\phi, G]}{G} = 0.
\label{eq:varGeq0}
\end{aligned}\end{equation} 
In the melonic case, $\fdv{\Gamma}{\phi}=0$ is trivially solved by assuming no symmetry breaking, $\expval{\varphi}=0$; $\fdv{\Gamma}{G}=0$ will give the Schwinger-Dyson equation for the bilocal field $G$, which is the usual route to solve a melonic field theory. It is then a standard result that the 2PI effective action $\Gamma[\phi, G]$ for a scalar field theory is given by
\begin{subequations}\label{eq:2PIeffectiveActionScalarBoth}
\begin{equation}\label{eq:2PIeffectiveActionScalar}
\Gamma[\phi, G] = S[\phi] + \frac{1}{2} \Tr \ln G^{-1} + \half \Tr C^{-1} G  + \Gamma_{2}[\phi, G].
\end{equation}
Here:
\begin{itemize}
    \item $S[\phi]$ is the classical action, evaluated for the classical field $\phi$.
    \item $C^{-1}(t,t')$ is the (matrix) inverse free propagator for $\varphi$.
    \item As usual, we treat $G(x,y)$ and $C(x,y)$ as matrices indexed by $x$ and $y$, and take the matrix logarithm.
    \item $\Gamma_{2}[\phi, G]$ is (minus) the sum of all of the two-particle-irreducible vacuum graphs. These are all graphs that do not disconnect when cutting open any two edges. 
    Crucially, the Feynman rules are slightly modified: instead of the free $\varphi$ propagator, we use $G$. In the symmetric case where $\phi=0$, the vertices used are precisely the same as those in the original action (this is not the case if $\phi\neq 0$, i.e. if the field gets a VEV).  
\end{itemize}
This means that we never need to consider self-energy insertions: they are resummed automatically by the fact that we are using $G$! To see this, take any 2PI diagram, and replace one of the $\varphi$ propagators by a diagram that would contribute to the self-energy $\Pi_\varphi$ of that field. This is demonstrated in \cref{fig:2PIdiagrams}; clearly, in that case we can cut the two edges surrounding the $\Pi_\phi$ insertion and disconnect the diagram; thus it does not contribute to $\Gamma_{2}$.

The following schematic expression \cite{Benedetti:2018goh} is a useful aide-mémoire. If the original action was $S[\varphi]$, then we can write %
\begin{equation}\begin{aligned} \label{eq:integralFormForGamma}
e^{-\Gamma[\phi,G]} = e^{-S[\phi] - \half \Tr[C^{-1} G]} \int_{\mathrm{2PI}} \cD\varphi\, e^{-\half \varphi G^{-1} \varphi - S_{\mathrm{int}}[\phi, \varphi]},
\end{aligned}\end{equation} 
\end{subequations}
where the subscript indicates that when we do the perturbative expansion of the functional integral, we keep only the 2PI graphs. Here, $S_\mathrm{int}[\phi,\varphi]$ is the interacting part of $S[\phi + \varphi]$.

This formalism trivially generalizes to the multi-field case, with $\phi_i = \expval{\varphi_i}$ and $G_i = \expval{\varphi_i \varphi_i}^{\text{conn}}$; additionally, it is purely combinatorial, and therefore must also apply to a QFT on a sphere. Thus, we have all the ingredients we need to apply it to calculating the sphere free energy of any QFT, which we note from \eqref{eq:GammaAndWLT} is just
\begin{subequations}\label{eq:FeqGextremum}
\begin{equation}\begin{aligned} \label{eq:Fin2PI}
F \equiv -\log Z_{S^d} = \Gamma[\phi_i, G_i]|_{k_i=0, j_i=0, \text{ on }S^d},
\end{aligned}\end{equation} 
where $k=0$, $j=0$ just means that we evaluate $\Gamma$ for the $\phi_i$ and $G_i$s that solve \cref{eq:EoMsForVEVs},
\begin{equation}\begin{aligned}
\forall i \quad \fdv{\Gamma}{\phi_i} = 0, \quad \fdv{\Gamma}{G_i} = 0.
\end{aligned}\end{equation} 
\end{subequations}
\textbf{That is, the (sphere) free energy of any QFT is precisely the extremum of the 2PI action $\Gamma$, with respect to the propagators $G_i$ and field VEVs $\phi_i$.}

\subsection{Application of the 2PI formalism}\label{sec:apply2PI}

In the following, we calculate the sphere free energy in the IR, which is assumed to be conformal. Thus, we evaluate \eqref{eq:Fin2PI} for large radius $R$, such that the contribution of the UV propagator is negligible. We also assume no symmetry breaking of any kind, and so drop $\phi =\expval{\varphi}=0$ and $S[\phi]=0$ in \cref{eq:2PIeffectiveActionScalarBoth,eq:FeqGextremum}. In this case, we now find that for a melonic theory of schematic form $S_\text{int} = \sum_m g_m \prod_\phi \phi^{q^m_\phi}$, \textbf{only $n_m$ diagrams appear in $\Gamma_2[G]$}: these are the complete melons
\begin{equation}\label{eq:sumMelons}
\Gamma_{2}[\{G_{\phi_i}\}] = -\half\sum_m \frac{g_m^2}{S_m} \int_{x,y}\vcenter{\hbox{\begin{tikzpicture}[anchor=base,baseline]
  \pgfmathsetmacro{\L}{4}
  \coordinate (L) at (-\L,0);
  \coordinate (R) at (\L,0);
  
  \draw[black, decorate, decoration={snake, amplitude=.4mm, segment length=2mm}]  (L) to[bend left=90,looseness=1.25] (R);
  \draw[black, decorate, decoration={snake, amplitude=.4mm, segment length=2mm}]  (L) to[bend left=75,looseness=1.20] (R);
  \draw[loosely dotted, thick] (0, {0.65*\L}) to (0, {0.55*\L});
  \node at (0.4, {(0.58*\L)}) {$q_{\mathcal{O}}$};
  \draw[black, decorate, decoration={snake, amplitude=.4mm, segment length=2mm}]  (L) to[bend left=60] (R);
  
  \draw[black, thick] (L) to[bend left=50] (R);
  \draw[black, thick] (L) to[bend left=40, looseness=1] (R);
  \draw[loosely dotted, thick] (0, {0.32*\L}) to (0, {0.22*\L});
  \node at (0.8, {(0.25*\L)}) {$q_{\Phi}$};
  \draw[black, thick] (L) to[bend left=25, looseness=0.7] (R);
  
  \node at (-\L-0.3, 0) {$x$};
  \draw[loosely dotted, thick] (0, {0.09*\L}) to (0, {-0.09*\L});
  \node at (\L+0.3, 0) {$y$};
  
  \draw[black, dashed] (L) to[bend right=40, looseness=1] (R);
  \draw[loosely dotted, thick] (0, {-0.23*\L}) to (0, {-0.33*\L});
  \node at (0.4, {-(0.29*\L)}) {$q_{\psi}$};
  \draw[black, dashed] (L) to[bend right=25, looseness=0.7] (R);
  \draw[black, dashed] (L) to[bend right=70] (R);
  \draw[black, dashed]  (L) to[bend left=90,looseness=1.25] (R);
\end{tikzpicture}}}. %
\end{equation}
Typically, we would have had to resum the melonic insertions on each leg -- but the 2PI formalism has done this automatically. Now, for convenience, we use our freedom to rescale $g_m$ to remove $S_m$: that is, in the remainder of the paper, we normalize the couplings $S_\text{int} = \sum_m g_m/s_m \prod_\phi \phi^{q^m_\phi}$, with $s_m$ such that $S_m =1$\footnote{For real fields and a conventional normalization of the interaction couplings, $S_\text{int} = \sum_m g_m \prod_\phi \phi^{q^m_\phi}/q^m_\phi!$, and so  $S_m= \prod_\phi q^m_\phi!$. In that case, setting $s_m^2 = \prod_\phi q^m_\phi!$ removes $S_m$ from \eqref{eq:sumMelons}.}. 
Additionally, we work \textit{throughout} this section with the bare fields.
Since we only want to find scaling dimensions in the limit of strong bare coupling, which when calculated non-perturbatively do not depend on field normalizations, this is not only possible but simpler.

We note that there are certain drawbacks to the 2PI reformulation of the SYK model or tensor models, explained in \cite{Benedetti:2018goh}, but these only apply at subleading orders in $N$, and so can be neglected here.

\subsubsection{2PI effective action for an SYK-like theory}

To illustrate this formalism, consider the 2PI effective action for an SYK-like theory: that is, for a melonic-type theory in $d=1$ with a single fermionic field $\psi_i$, we obtain %
\begin{equation}\label{eq:GammaSYK}
\frac{1}{N}\Gamma_{\text{SYK}}[G] = -\frac{1}{2}\Tr \ln G^{-1} - \frac{1}{2}\Tr C^{-1}G - \half J_m^2 \int_{t,t'} G(t,t')^q.
\end{equation}
Here:
\begin{itemize}
    \item $G(t, t') \delta_{ij}\equiv \expval{\psi_i(t) \psi_j(t')}$; so we have assumed no symmetry breaking.
    \item $N$ would be replaced by $N=M^q$ if we were studying an $\gO(M)^q$-tensor model, rather than a disorder-averaged model. All terms shown on the RHS are therefore order $N^0$.
    \item $C^{-1}(t,t') = \delta(t-t')\partial_t$ is the (matrix) inverse of the bare propagator.
    \item $J^2_m$ is the effective bare coupling for a complete melon\footnote{As mentioned above, we have rescaled $J_m^2$ for convenience, so it differs by: $q$ from the usual SYK normalization, $J^2_m = J_\mathrm{SYK}^2/q$; and $q!$ from the usual Feynman-diagrammatic normalization, $J^2_m = J_{F}^2/q!$.}.
    \item The minus signs in front of the trace terms compared to \eqref{eq:2PIeffectiveActionScalar} arise because of the fermionic character of $\psi$.
\end{itemize}
The result \eqref{eq:GammaSYK} is well known, and directly derivable in disorder-averaged theories, via a change of variables in the replica method \cite{Benedetti:2018goh,Maldacena:2016hyu,Gross:2016kjj,Kitaev:2017awl}; it was shown in \cite{Benedetti:2018goh} that it also applies to the tensor models, though it is more complicated to derive. Adding more fermions leads to the generalized SYK model of \cite{Gross:2016kjj}; but we can now also generalize it to arbitrary dimensions, fields, and melonic interactions.

\subsubsection{2PI effective action for an arbitrary melonic theory}

Consider a melonic theory with $n_f$ fields $\{\Phi\}$, each of which has some suppressed indexed structure $\Phi_i$ associated with a melonic mechanism. Take $n_m$ melonic couplings $g_m$, so the UV perturbation is schematically
\begin{equation}
    V=\sum_m \frac{g_m}{s_m} \cO_m, \quad \cO_m \simeq \prod_\Phi \Phi^{q^m_\Phi},
\end{equation}
for $s_m$ chosen for convenience such that we get no additional factors in the following. 
Thanks to the melonic dominance, we can immediately write down the following $d$-dimensional 2PI effective action for $G_\Phi(x,y) \delta_{ij} \equiv \expval{\Phi_i(x)\Phi_j(y)}$ to LO in $N$:
\begin{equation}\label{eq:G2PIforArbMelonic}
\frac{1}{N}\Gamma[\{G_\Phi\}] \equiv \sum_\Phi \left(\half \Str\ln G_\Phi^{-1} +  \half \Str C^{-1}_{\Phi} G_\Phi \right) -\half\sum_m g_m^2 \int_{x,y} \prod_{\Phi} G_\Phi(x,y)^{q_{\Phi}^m}.
\end{equation}
Though its form should be clear as a generalization of \eqref{eq:GammaSYK}, we comment:
\begin{itemize}
    \item We sum over the $n_f$ dynamical fields $\Phi$, not accounting for the $N$ additional copies present due to the melonic mechanism. Thus, for the SYK or quartic tensor model, the sum only has a single contribution, $\{\Phi\} = \{\psi\}$.
    \item The supertrace $\Str X_\Phi = (-1)^{\mathrm{F}^\Phi} \Tr X_\Phi$ provides a $-1$ factor for fermionic fields ($\mathrm{F}^\Phi$ is $0$ for bosons and $1$ for fermions).
    \item Since we have assumed no symmetry breaking, $R_\Phi$ then only appears in the $\Tr\id_R$ implicit in the first term -- so we see immediately that only the dimension of $R_\Phi$ matters. Hence, all fields here are assumed to be real; complex fields are implicitly accounted for by the $\dim_{\mathbb{R}} R_\Phi$ in the trace, which provides the necessary $2$ for a complex boson.
    \item The UV propagator $C$ that appears in the second term plays no role in the IR, assuming $\Delta_\Phi > \Delta_\Phi^{\text{free}}$, and we drop it from $\Gamma$ hereafter: it only serves to specify the IR wedge \eqref{eq:DelGreaterThanFree}.  
    However, when we treat the case of the long-range theories in \cref{sec:LRmodels}, we must keep it.
    \item Generically, $g_m^2$ could be a homogeneous quadratic polynomial in the bare melonic MST couplings, as in \cite{Prakash:2017hwq, Fraser-Taliente:2024rql}. 
    \item The final term in \eqref{eq:G2PIforArbMelonic} is the large-$N$ evaluation of 
    \begin{equation}\label{eq:intIs2pt}
    -\half \int_{x,y} \expval{V_x V_y}_{\mathrm{GFF}} = -\half \sum_m \frac{g_m^2}{s_m^2}\int_{x,y} \expval{\cO_m(x) \cO_m(y)}_{\mathrm{GFF}}.
    \end{equation}
    By the subscript, we mean that we evaluate this correlator in a generalized free field theory where the $\Phi$ propagators are the unknown function $G_\Phi$. We have chosen $s_m^2$ to cancel the factor of $\prod_\Phi q^m_\Phi!$ coming from the Wick contractions of the various $\Phi$s (in the real case), so we can ignore them. 
    \eqref{eq:intIs2pt} is manifestly a scalar. 
    However, for fields in non-trivial representations of $\SO(d)\times \Gglobal$, $G_\Phi$ has indices, which must therefore be contracted; but such contractions only lead to further rescalings of $g_m$ by constants, and so we can also neglect them.
\end{itemize}
The form of \eqref{eq:intIs2pt} is entirely expected from our knowledge of double-trace deformations in a large-$N$ theory. 
In a mean field theory, the correlator of an exponential can be solved exactly: hence, for $V$ a product of single-trace operators in a large-$N$ theory $T$, up to subleading terms in $N$
\begin{equation}
\expval{\exp\left(-\int_x V_x\right)}_T \simeq \exp\left(-\half \int_{x,y} \expval{V_x V_y}_T\right)
\end{equation}
a result that follows directly from the Wick contractions of GFFs \cite{Hartman:2006dy}. Of course, confirming that this is what happens appears to require the detailed melonic mechanism. %

Naturally, the stationary point conditions $\fdv{\Gamma}{G_\Phi}=0$ are just the two-point Schwinger-Dyson equations. In fact, this is the easiest way to construct them\footnote{Of course, when performing the variation, we must remember that $G_\Phi(x,y)=(-1)^{\mathrm{F}^\Phi} G_\Phi(y,x)$ are not independent \cite{Benedetti:2018goh}.}. %
We obtain precisely the leading-$N$ two-point function SDE derived diagrammatically in \cref{app:diagrammaticProof}:
\begin{equation}\begin{aligned} \label{eq:2PIEoM} 
    0 = (-1)^{\mathrm{F}^\phi}\dim(\rhoext_\phi)\, G_\phi^{-1}(x,y) + \sum_m q^m_\phi g_m^2 [G_\phi(x,y)]^{q^m_\phi -1} \prod_{\Phi \neq \phi} [G_\Phi(x,y)]^{q^m_{\Phi}}.
\end{aligned}\end{equation} 

\subsection{The fundamental claim proved I}

We now show how our fundamental claim arises.

In the language of \cite{Karateev:2018oml}, %
it is a standard result of CFT that a primary $\Phi$ has a unique two-point structure with an operator $\Phi^\dagger$, with scaling dimension $\Delta_\Phi$, transforming in the conjugate reflected representation $\rho^\dagger$; this representation is the complex conjugate of $\rho$ when working in Lorentzian signature. We suppress the global symmetry group and its indices. %
This two-point function is uniquely determined up to an arbitrary constant $\cZ_\Phi$\footnote{To avoid potential confusion: $\cZ_\Phi$ here is the full two-point function normalization for the bare fields.
If at the end of the calculation we wanted to unit-normalize our bare fields $\Phi_0$: we define renormalized fields by $\Phi_0 \equiv \sqrt{Z_\Phi} \Phi_{\mathrm{R}}$, as in \cref{sec:0dQFT}; %
then if we choose $Z_\Phi \equiv 1/\cZ_\Phi$, $\expval{\Phi_{\mathrm{R}} \Phi_{\mathrm{R}}}$ is unit-normalized.}:
\begin{equation}\begin{aligned} \label{eq:conformalG}
G_\Phi(x,y) = \expval{\Phi_{\mu_1 \cdots \mu_s}(x) (\Phi^\dagger)^{\nu_1 \cdots \nu_s}(y)} = \cZ_\Phi [\twoPt_{\Phi}(x-y)]\indices{_{\mu_1 \cdots \mu_s}^{\mu_1 \cdots \mu_s}}\,  \id_{R_\Phi}.
\end{aligned}\end{equation} 
Here, $\twoPt_\Phi$ depends only on the data of the primary, i.e. the scaling dimension $\Delta_\Phi$ and the $\SO(d)$ representation $\rho$. We suppress the $\Gglobal$ indices %
in the following.  

The quantity to be extremized, $\Ft$, is then defined to be the function given by the 2PI action \eqref{eq:G2PIforArbMelonic}, evaluated on the sphere with the conformal $G_\Phi$s: 
\begin{equation}\begin{aligned} \label{eq:FtEqGamma}
\Ft(\{\Delta_\Phi\}, \{\cZ_\Phi\}, \{g_m\}) \equiv -\sin(\pi d/2)\Gamma[\{G_\Phi\}]|_{S^d, \text{ conformal }G_\Phi}.
\end{aligned}\end{equation} 
It is regulated with a $-\sin \pi d/2$, just as $F$ was regulated in \eqref{eq:Ftdef}, making it demonstrably finite for all $d$ and generic $\Delta$. 
If the IR limit on the sphere is a CFT, then the constraints of conformal symmetry tell us the exact form of the $G_\Phi(x,y)$s, up to two numbers $\Delta_\Phi$ and $\cZ_\Phi$. The functional extremization problem on $\Gamma$ then becomes a function extremization problem on $\Ft$. 
To find the sphere propagators $\twoPt_\Phi(x,y)$s, we need only Weyl map the textbook two-point functions, as we did to find \cref{eq:sphereGFFpropagator}. 

Without loss of generality, we assume that all of the melons are IR relevant (if they are not, they simply disappear from the calculation entirely). Then, evaluating \eqref{eq:FtEqGamma} to leading order in $N$, we find
\begin{equation}\begin{aligned} \label{eq:LOfreeEnergy}
\frac{1}{N} \Ft(\{\Delta_\Phi, \mathfrak{g}_m\}) =\sum_\Phi \Ft_{\Phi} + \sum_m \frac{\mathfrak{g}_m}{(2R)^{2\mm{m}}} \Mtm(\mm{m}) %
+ O\left(N^{-1}\right).
\end{aligned}\end{equation} 
We have defined the following quantities.
\begin{itemize}
    \item As in \eqref{eq:FtBosFerm}, $\Ft_\Phi$ is $\Ft$ evaluated for a generalized free field of dimension $\Delta_\Phi$ in the representation $\rhoext_\Phi =\rho_\Phi \times R_\Phi$; it is independent of $\cZ_\Phi$, and linear in $\dim R_\Phi$.%
    \item A convenient combination of scaling dimensions associated with each melon $m$ is
    \begin{equation}\begin{aligned}
        \mm{m} & \equiv  \sum_\Phi q^m_\Phi \Delta_\Phi-d.
    \end{aligned}\end{equation} 
    The melonic constraints are $\mm{m}=0$.
    \item The renormalized squared couplings are defined to be
    \begin{equation}\begin{aligned}
        \mathfrak{g}_m  \equiv g_m^2 \prod_\Phi \cZ_\Phi^{q^m_\Phi},
    \end{aligned}\end{equation}
    and have mass dimension $-2\mm{m}$.
    Since only the fields, not the couplings, are renormalized to leading order in $N$, we find the renormalized couplings by simply stripping off the field renormalization from the bare parameter $g_m$ in the action -- and we square for convenience.
    Note that in the IR the $\cZ_\Phi$s therefore appear in $\Ft$ solely through $\mathfrak{g}_m$, and when the melonic constraints are satisfied, the $\mathfrak{g}_m$s are dimensionless.
    \item The dimensionless function $\Mtm$ is proportional to the complete melon on the sphere:
    \begin{equation}\begin{aligned}\label{eq:MtmDef}
    \Mtm(\mm{m}) &\equiv [-\sin (\pi d/2)] \, \left(-\thalf\right) (2R)^{2\mm{m}} \int \odif[d]{x} \odif[d]{y} \, \frac{\sqrt{g(x)} \sqrt{g(y)}}{s(x,y)^{2(\mm{m} +d)}}.
    \end{aligned}\end{equation}
    This complete melon integral is easily evaluated, using the homogeneity of the sphere to fix one point to zero \cite{Benedetti:2021wzt, Pufu:2016zxm, Cardy:1988cwa}: %
\begin{equation}\begin{aligned}
 \Mtm(\mm{m}) &= \frac{\pi ^d \sin \left(\frac{\pi  d}{2}\right) \Gamma \left(\frac{d}{2}\right) \Gamma \left(-\frac{d}{2}-\mm{m}\right)}{2 \Gamma (d) \Gamma (-\mm{m})}.
\end{aligned}\end{equation}  %
    This function has zeros for $\mm{m}=0, 1,2, \ldots$ in any $d$\footnote{Without the $\sin$, this integral is non-zero for even $d$ for these values of $\mm{m}$ (which is a scheme-dependent fact \cite{Gerchkovitz:2014gta}); the presence of $\sin(\tfrac{\pi  d}{2})$ allows us to ignore this issue.}.
    \item $f_\Phi^\mathrm{free}$ is the contribution from the free (UV) propagator, which is proportional to $\cZ_\Phi R^{2(\Delta_\Phi^{\text{free}}- \Delta_\Phi})$.
    Assuming $\Delta_\Phi > \Delta_\Phi^{\text{free}}$, this will be dropped in the IR. %
\end{itemize}

The factor of $-\sin(\pi d/2)$ in the definition of $\Ft$ serves to ensure that $\Mtm'(0)$ does not have poles for even integer $d$. Since $\Ft_\Phi$ for generalized free fields is also finite, the entire $\Ft$ is indeed finite in all $d$ by construction. %

\subsection{The fundamental claim proved II}

Extremizing \eqref{eq:LOfreeEnergy} with respect to the $\Delta_\Phi$s and $\cZ_\Phi$s gives
\begin{subequations}\label{eq:FtildeVariation}
\begin{align}
\forall \, \cZ_\Phi: \quad%
&\frac{1}{\cZ_\Phi}\sum_m q^m_\Phi \mathfrak{g}_m\Mtm\left(\mm{m}\right)(2R)^{-2\mm{m}} = 0,\label{eq:ZphiVariation}\\
\forall \, \Delta_\Phi: %
\quad&  \odv{\Ft_\Phi}{\Delta_\Phi} + \sum_m q^m_\Phi \mathfrak{g}_m \Mtm'\left(\mm{m}\right) (2R)^{-2\mm{m}} =0, \label{eq:deltaPhiVariation} %
\end{align}\end{subequations}
where we have immediately used \cref{eq:ZphiVariation} to cancel the term $\propto \log R$ that otherwise would appear in \cref{eq:deltaPhiVariation}, and dropped the $N$-subleading terms. %

The extrema of $\Ft$ satisfying $\mm{m}=0$ for all $m$ correspond to the extrema of the functional $\Gamma$ that are independent of $R$, and therefore can be consistently mapped to flat space. There exist other solutions to \eqref{eq:ZphiVariation}, but they lead to terms of different order in $R$ in \cref{eq:deltaPhiVariation}; the dependence of the solutions on $R$ implies that we do not have a consistent IR solution (which should be $R$-independent for large $R$). Contributions from any $g_m$s with $\mm{m} >0$ will not survive the IR limit, and therefore the associated melonic constraint will not be applied. As mentioned above, we will typically neglect this possibility.

To make contact with the usual analysis of the full Schwinger-Dyson equations, it is clear that any such solutions with $R$-dependence do not satisfy \eqref{eq:2PIEoM}; that is, they are not extrema of the full $\Gamma[G]$, but only of the conformal slice of $\Gamma$, which we have defined to be $\Ft(\Delta_\Phi, \cZ_\Phi)$.
It is only the $R$-independent extrema of $\Ft$ that give $G$s that extremize $\Gamma[G]|_{S^d}$. 
We can therefore expand \eqref{eq:LOfreeEnergy} around $\mm{m}=0$, using
\begin{equation}\begin{aligned}\label{eq:MtmExpanded}
\Mtm(\mm{m}) (2R)^{-2 \mm{m}} = \frac{\pi^{d+1}}{\Gamma(d+1)} \mm{m} + O(\mm{m}^2).
\end{aligned}\end{equation} 
Substituting \eqref{eq:MtmExpanded} into \eqref{eq:deltaPhiVariation}, we see that the functional extremization problem is, in the IR, equivalent to extremizing the function
\begin{equation}\begin{aligned}
\frac{1}{N}\Ft(\{\Delta_\Phi, \mathfrak{g}_m'\})  = \sum_\Phi \Ft_\Phi + \sum_m \mathfrak{g}'_m \left(\sum_\Phi q^m_\Phi \Delta_\Phi-d\right)
\end{aligned}\end{equation}  
with respect to the Lagrange multipliers
\begin{equation}\begin{aligned}
\mathfrak{g}_m' = \frac{\pi^{d+1}}{\Gamma(d+1)} g_m^2 \prod_\Phi \cZ_\Phi^{q^m_\Phi}.
\end{aligned}\end{equation} 
Hence, the extremum of $\Ft(\{\Delta\})$ corresponds precisely to the actual value of $\Ft_\mathrm{CFT}$ for this CFT, and is just
\begin{equation}\begin{aligned}
\Ft_\mathrm{CFT} = \sum_\Phi \Ft_\Phi|_{\Delta_\Phi}.
\end{aligned}\end{equation} 
The complete melon has been regulated to zero; this can be understood as the fact that the trace of the identity is zero in continuous dimension, $\Tr \id =0$ \cite{Benedetti:2021wzt} -- and the melon diagram at criticality is proportional to $\Tr \id$.
Naturally, all of this agrees with the equations obtained by a diagrammatic analysis in the appendix, \eqref{eq:melonicsummary}. 

This is the fundamental claim \eqref{eq:FmaxSummary}: the melonic interaction precisely implements the (linear) melonic constraint in the $\Ft$-extremization procedure. 
The large-$N$ limit protects the form of the potential, so no additional terms are generated, and this is consistent. %

Thus, to conclude in words: the conformal IR solution of a melonic QFT can be found by extremizing the sum of the free energies of a collection of generalized free fields, subject to a marginality constraint $\mm{m}=0$ (or $\mathfrak{g}_m=0$) for each melon, and lying within the IR wedge \eqref{eq:DelGreaterThanFree}. 
The constraint ensures that said melon survives in the IR. 

\begin{mccorrection}

Any non-Gaussian behaviour in the correlators of the fundamental fields is $N$-subleading, which is clear from taking derivatives with respect to $G_{ij}$ of the 2PI action.
Hence, the correlators of the fundamental fields at separated points can be described by a mean field theory with constrained extremal $\Ft$ to leading order in $N$.
\end{mccorrection}

\section{Long-range models} \label{sec:LRmodels}

In \cref{sec:apply2PI} we specified that the solutions $\{\Delta_\Phi\}$ must lie within the IR wedge in order for the contribution of the UV propagator $\Str C^{-1}_\Phi G_\Phi$ to drop out of the 2PI effective action.
However, if instead of using standard free scalar fields and adding an interaction, we start with particular GFFs from the beginning, we can straightforwardly access \textit{lines} of (nonlocal) melonic fixed points: these are called the long-range models, as opposed to the models with standard kinetic terms, which are called short-range models.
It is easy to extend the treatment above to these models, and so we do so in this section. %

The general idea is that we tune the free kinetic term such that the free propagator $C_\Phi$ matches the scaling of the conformal propagator $G_\Phi$ that is picked out by a given short-range model:
\begin{equation}
\Delta_\Phi^\mathrm{free} = \Delta_\Phi \rvert_\mathrm{SR}.
\end{equation}
This makes the theory conformal at all scales, not just in the IR like in short-range models. 
This was first done with cSYK \cite{Gross:2017vhb} for general $q$, then for bosonic tensor models for $q=4,6$ \cite{Benedetti:2019eyl,Benedetti:2019ikb}, and more recently the Amit-Roginsky model, which implements $q=3$ \cite{Benedetti:2020iku}.
In \cite{Benedetti:2020yvb}, a perturbative proof was given of the conformal symmetry of these models at the IR fixed point; this was done by embedding the fields in a $d' = d+p$, $p=2-2\zeta$ dimensional space, but localising the interaction to the $d$-dimensional space.

Long-range models in the multi-field case are conceptually slightly more complicated. We need to identify a short-range fixed point, and then we choose to promote that short-range fixed point to a long-range model; this is done by modifying the free two-point function to make it match the scaling dimensions of the short-range model; the free propagator is then marginal in the IR. 
From then on, all of the scaling dimensions are fixed, as they are part of the long-range theory's definition (rather than being a result found in the IR).

\subsection{Solving the long-range two-point function}

Once we have fixed some UV kinetic terms, we are done, as we have already assembled everything required to solve the theory: we need only keep the UV propagators in \eqref{eq:G2PIforArbMelonic} this time. Thus, there is an additional contribution to $\Ft$ in \eqref{eq:FtEqGamma} for each field, namely
\begin{equation}
-\sin(\pi d/2)\half\Str C_\Phi^{-1} G_\Phi = - \frac{\cZ_\Phi}{\cZ_\Phi^{\text{free}}} \frac{(-1)^{\mathrm{F}^\Phi} \dim \rhoext_\Phi}{\cN_{\Phi,\Delta_\Phi^{\text{free}}}} \Mtm(\Delta_\Phi - \Delta_\Phi^\text{free}) (2R)^{-2(\Delta_\Phi - \Delta_\Phi^\text{free})}
\end{equation}
where we have used the identity \eqref{eq:inverseIsShadow}. 
But we recognize from \eqref{eq:plancherelDef} precisely the derivative of a GFF $\Ft$, $\Ft_\Phi'(\Delta_\Phi^\text{free})$:
\begin{equation}
- \frac{\cZ_\Phi}{\cZ_\Phi^{\text{free}}} \Ft_\Phi'(\Delta_\Phi^\text{free}) \frac{\Gamma(d+1)}{\pi^{d+1}} \Mtm(\Delta_\Phi - \Delta_\Phi^\text{free}) (2R)^{-2(\Delta_\Phi - \Delta_\Phi^\text{free})}
\end{equation}
Consider some candidate IR solution $\{\Delta_\Phi\}$ chosen from the known solutions to the short-range model. If $\Delta_\Phi < \Delta_\Phi^\text{free}$, this kinetic term would dominate in the IR, and so the interaction term is irrelevant -- so this would not work. If $\Delta_\Phi > \Delta_\Phi^\text{free}$, it is irrelevant, so we return to the case of the usual short-range model. However, if we tune $\Delta_\Phi^\text{free}$ to equal the desired IR $\Delta_\Phi$, then this defines the long-range model. 

Hence, in the effective action, we can expand $\Mtm$ around zero, and keep only the leading part. Conveniently, $\frac{\Gamma(d+1)}{\pi^{d+1}} \Mtm(x) = x + O(x^2)$, and so
\begin{equation}
  -\sin(\pi d/2)\half\Str C_\Phi^{-1} G_\Phi =- \frac{\cZ_\Phi}{\cZ_\Phi^{\text{free}}} \Ft_\Phi'(\Delta_\Phi^\text{free}) (\Delta_\Phi - \Delta_\Phi^\text{free})
\end{equation}
up to higher-order terms. Extremizing $\Ft$ with the addition of these terms, we find that, unsurprisingly, we are forced to set $\Delta_\Phi$ to the values that we tuned, and we remain only with a set of equations for the $\cZ_\Phi$s:
\begin{equation}
\text{For each field $\Phi$: } \left(1- \frac{\cZ_\Phi}{\cZ_\Phi^{\text{free}}}\right)\odv{\Ft_\Phi(\Delta_\Phi)}{\Delta_\Phi} =- \frac{\pi^{d+1}}{\Gamma(d+1)} \sum_{m} q^m_\Phi  \mathfrak{g}_m
\end{equation}
Manifestly, the short-range solution for $\mathfrak{g}_m$ is recovered in the limit where all $\cZ_\Phi \to 0$. This is a strong coupling limit, with $\lambda_b \to \infty$, $\cZ_\Phi \to 0$ such that $\mathfrak{g}_m \to$ a constant, the short-range value. 
We forget about the UV propagator, and so necessarily arrive at the short-range case.

This procedure might appear similar to simply changing the value of the $\dim \rhoext_\Phi$. Not so: if we vary $\dim \rhoext_\Phi$ in the short-range models, we modify the scaling dimensions. That is not possible here, as we have fixed those $\Delta_\Phi$s from the start; this is made obvious by seeing that varying $\rhoext$ keeps the theory local, whereas varying the bare scaling dimension from its canonical value makes the theory non-local (in particular, there is no local stress-energy tensor). The differences between the CFTs for the various values of the bare couplings can be seen in the dimensions of the operators appearing in the OPE of the fundamental fields \cite{Benedetti:2020rrq,Benedetti:2024mqx}.

\subsection{The long-range melonic \texorpdfstring{$\phi^q$}{phi\^q} model}

Let us illustrate the short-range to long-range transition for a single-field $q$-tensor, with melonic interaction $\sim \lambda_b \phi^q/s_m$. We therefore choose the free kinetic term such that $\expval{\phi_x \phi_0}_{\lambda=0} = \cZ^\text{free}/\abs{x}^{2d/q}$. As usual for a melonic theory, the only parameter actually appearing here is $\lambda_b^2$; hence, we are motivated to define a pair of squared couplings that account for the field normalization:
\begin{equation}
\mathfrak{g}=\lambda_b^2 \cZ^q, \quad \mathfrak{l}_b = \lambda_b^2 (\cZ^\text{free})^q.
\end{equation}
As usual, since to leading order in $N$ the coupling is not renormalized, it makes sense to call these the renormalized and bare parameters, respectively.

This theory is conformal for all values of $\lambda_b$. Since $\cZ^\text{free}$ is just a constant, we can equally parametrize these CFTs by our new bare coupling $\mathfrak{l}_b$, and we will justify this particular choice later, with equation \eqref{eq:LRparamrelation}. Defining a critical coupling value 
\begin{equation}
\mathfrak{g}_c = -\frac{1}{q}\frac{\Gamma(d+1)}{\pi^{d+1}} \odv{\Ft_\phi}{\Delta}\big|_{\Delta=d/q}, \text{ we find } \, \frac{\cZ}{\cZ^\mathrm{free}} = \left(1 - \frac{\mathfrak{g}}{\mathfrak{g}_c}\right).
\end{equation}
We can immediately interpret $\mathfrak{g}_c$ as the value of $\mathfrak{g}$ that gives the short-range model. The minus sign makes sense in the bosonic case: for the $\phi^q$ model, $\mathfrak{g}_c >0$ precisely in the renormalizable range $0<d<\frac{2q}{q-2}$\footnote{This is the usual $d< 6$ for $q=3$, $d<4$ for $q=4$, and $d<3$ for $q=6$. For \cite{Benedetti:2019eyl,Benedetti:2020iku} ($q=3$, $q=4$)%
, their $g_c$ is related to ours by $\mathfrak{g}_c= \frac{g_c^2}{q \cF_{2\Delta,0}^q}$, for $\cF_{\lambda, s} \equiv i^{-s}2^{d-\lambda}\pi^{d/2} \frac{\Gamma(\frac{d+s-\lambda}{2})}{\Gamma(\frac{\lambda+s}{2})}$ \cite{Karateev:2018oml}. %
}. Then we have 
\begin{equation}\begin{aligned}\label{eq:LRparamrelation}
\mathfrak{l}_b = \frac{\mathfrak{g}}{\left(1-\frac{\mathfrak{g}}{\mathfrak{g}_c}\right)^q},
\end{aligned}\end{equation} 
which indeed justifies the choice of parameter $\mathfrak{l}_b$. We note the similarity of this reparametrization to that of the toy model of GFF flow in \eqref{eq:GFFcouplings}.
\begin{figure}
    \begin{subfigure}{0.5\textwidth}
  \centering
\begin{tikzpicture}[scale=2]
\begin{axis}[
    xmin=-0.4, xmax=0.9,
    y=3cm/5,
    x=3cm,
    clip=false, %
    axis lines=middle,
    xmajorticks=false,
    ymajorticks=false,
    xlabel=$\mathfrak{l}_b$,
    ylabel=$\mathfrak{g}$,
    every axis x label/.style={at={(1,0.32)}, anchor=west}, every axis y label/.style={at={(0.3,1)}, anchor=south}
]
\addplot [->,thick, domain=4:2.2066, samples=50, blue, dashed] ({x/(1-x)^4}, {x}) node[right] {$\scriptstyle\infty$};
\addplot [thick, domain=-0.3333:0.28, samples=50, blue,->] ({x/(1-x)^4}, {x}) node[right] {$\scriptstyle\infty$};
\addplot [thick, domain=-2:-0.3333, samples=50, dashed] ({x/(1-x)^4}, {x});
\node[gray] at (-0.1,1) {$\mathfrak{g}_c$};
\draw[thick, densely dotted, gray] (0,1) -- (1.2,1);
\node at (1.4, 1) {SR};
\draw[gray] (1.4,1) ellipse ({0.5/5} and 0.5);
\draw[thick, densely dotted, gray] (-0.105,-1/3) -- (0, -1/3);
\node[gray] at (0.1,-1/3-0.1) {$\mathfrak{g}_0$}; %
\end{axis}
\end{tikzpicture}
  \caption{Even $q$ ($q=4$)}
  \label{fig:LRccsQEven}
\end{subfigure}%
\begin{subfigure}{0.5\textwidth}
  \centering
\begin{tikzpicture}[scale=2]
\begin{axis}[
    xmin=-0.4, xmax=0.7,
    y=3cm/5,
    x=3cm,
    clip=false, %
    axis lines=middle,
    xmajorticks=false,
    ymajorticks=false,
    xlabel=$\mathfrak{l}_b$,
    ylabel=$\mathfrak{g}$,
    every axis x label/.style={at={(1,0.32)}, anchor=west}, every axis y label/.style={at={(0.3,1)}, anchor=south}
]
\addplot [->,thick, domain=4:2.4066, samples=50, blue, dashed] ({x/(1-x)^3}, {x}) node[left] {$-\scriptstyle\infty$};
\addplot [thick, domain=-0.5:0.28, samples=50, blue,->] ({x/(1-x)^3}, {x}) node[right] {$\scriptstyle\infty$};
\addplot [thick, domain=-2:-0.5, samples=50, dashed] ({x/(1-x)^3}, {x});
\node[gray] at (-0.1,1) {$\mathfrak{g}_c$};
\draw[thick, densely dotted, gray] (0,1) -- (0.8,1);
\node at (1, 1) {SR};
\draw[gray] (1,1) ellipse ({0.5/5} and 0.5);
\draw[thick, densely dotted, gray] (-0.105,-0.5) -- (0, -0.5);
\node[gray] at (0.1,-0.5-0.1) {$\mathfrak{g}_0$}; %
\end{axis}
\end{tikzpicture}
  \caption{Odd $q$ ($q=3$)}
  \label{fig:LRccsQOdd}
\end{subfigure}
    \caption{Schematic plot of the dependence of the renormalized (squared) coupling $\mathfrak{g}$ on the bare coupling $\mathfrak{l}_b$, showing the three different branches. The branch assumed to be physical, since it includes the free theory, is indicated with solid blue; the other branches are dashed. The two graphs are schematically correct for any odd or even values of $q$ for $d$ with $\mathfrak{g}_c>0$, but the plots are for $q=4,3$. In the limit of strong (positive) bare coupling, we have $\mathfrak{g} \to \mathfrak{g}_c$, which gives SR, the usual short-range bosonic $\phi^q$ model.}\label{fig:LRcouplings}
\end{figure}
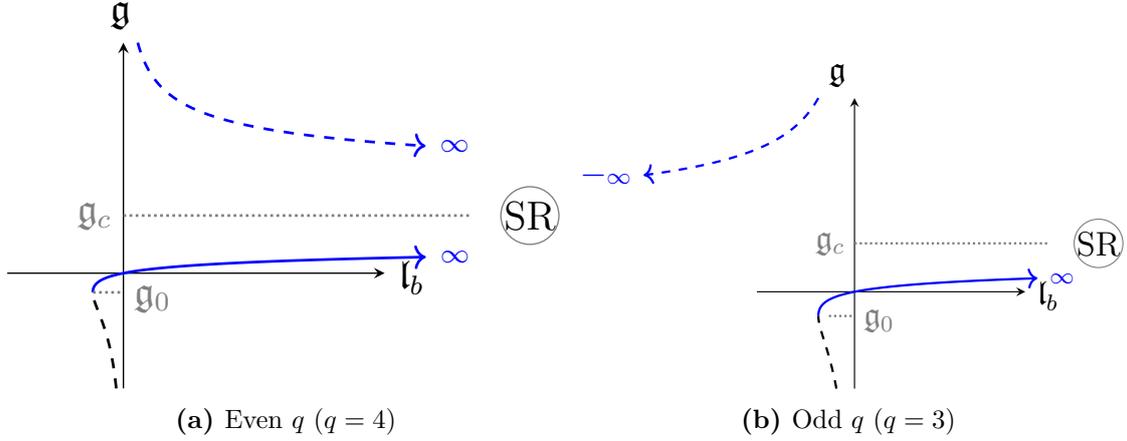

Let us now assume that we are in the renormalizable dimensional range, so $\mathfrak{g}_c >0$. The relationship \eqref{eq:LRparamrelation} is then shown for even and odd $q$ in \cref{fig:LRcouplings}. The figure shows that it is only invertible for $\mathfrak{g}$ between $\mathfrak{g}_0 \equiv -\mathfrak{g}_c (q-1)^{-1}$ and $\mathfrak{g}_c$. Now, $\lambda_b$ must have some actual physical value. 
If we require that value to be real (so $\lambda_b$ could be pure imaginary), then from a renormalization group point of view, the range of renormalized couplings $\mathfrak{g}$ that are real and consistent is between $\mathfrak{g}_0$ and $\mathfrak{g}_c$, inclusive. Hence, $\cZ/\cZ^\text{free}$ lies between $0^+$ and $\frac{q}{q-1}$.  Naturally, this includes the free theory of GFFs at $\lambda_b = \mathfrak{g} =0$, where $\cZ = \cZ^\text{free}$. What about $\mathfrak{l}_b < \mathfrak{l}_0 \equiv \mathfrak{l}_b|_{\mathfrak{g}=\mathfrak{g}_0} = - \frac{g_c}{q} \left(\frac{q-1}{q}\right)^{q-1}$ (i.e. $\cZ = \frac{q}{q-1}$)? For odd $q$, this is just on the other branch; however, for even $q$, this corresponds to a complex $\mathfrak{g}$. 

These manipulations make it completely clear that as we take the strong coupling limit of the short-range model, we forget about the UV propagator and return to the usual short-range model: that is, $\mathfrak{g}$ asymptotes to $\mathfrak{g}_c$ from below as $\mathfrak{l}_b \to \infty$ (and $\cZ \to 0^+$).
Accordingly, the bilinear spectrum of the long-range CFT in the strong-coupling limit is guaranteed to match that of the short-range CFT.
Away from this strong-bare-coupling limit, we will have a non-local theory that lacks a local stress tensor in its spectrum (see \cite[\S 2]{Fraser-Taliente:2026gdh} for a definition of non-local CFT in this context).

\section{The pattern: critical large-\texorpdfstring{$n$}{n} vector models} \label{sec:vectorModels}

\begin{mccorrection}

The general melonic theories will have some possibly unfamiliar features.
In this section, we will consider the critical $\gO(n)$ vector model, which we can use to demonstrate that said features are already present in this more familiar context. 
This is possible, as at the conformal fixed point of this model we find that the first corrections to the scaling dimensions of the fundamental fields (given in \cite{Vasiliev:1981dg, Goykhman:2019kcj,Fraser-Taliente:2025udk}) are exactly encoded by $\Ft$-extremization.

The features that consideration of the critical vector model will help explain are: the multiple vacuum solutions; the collision and hence disappearance (complexification) of these vacua as $d$ is varied; and the missing solutions at certain integer values of $d$. 
To fully justify these steps, we will promote the vector model to an SYK-like large-$N$ model.
Taking $N\to\infty$, the theory is melonic for all values of $n$; thus by varying $n$, we interpolate from a standard large-$n$ vector theory to a more typical melonic theory.%
\end{mccorrection}

\subsection{The vector model and the disordered vector model}

After introducing an auxiliary field $\sigma$, the Lagrangian of the $\tfrac{\lambda_b}{8n}(\phi_I \phi_I)^2$ model is
\begin{equation}\begin{aligned} \label{eq:vectorOn}
  \cL_{\gO(n)} = \sum_{I=1}^n \half \phi\indices{_I} C^{-1} \phi_{I}  - \frac{1}{2 \lambda_b} (\sigma \sigma) - \sum_{I=1}^n \frac{1}{2\sqrt{n}} \sigma\phi\indices{_I} \phi_{I},
  \end{aligned}\end{equation} 
where $C^{-1}$ is the inverse propagator of the conformally coupled free scalar\footnote{The conformally coupled free scalar has $C^{-1} = -\partial^2 + \frac{d-2}{8(d-1)} \mathcal{R}$, for $\mathcal{R}$ the Ricci scalar. On the sphere, $\mathcal{R}|_{S^d}= d(d-1)/R^2$ \cite{Tarnopolsky:2016vvd}.}.
The leading corrections to the two-point functions at criticality (i.e. tuning the mass to vanish) are
\begin{equation}\begin{aligned}
\Pi_\phi = \frac{1}{n} \verticalcenter{\begin{tikzpicture}\begin{feynman}
  \vertex (a)  at (0,0);
  \vertex (b) at (1,0);
  \vertex (c) at (2,0);
  \vertex (d) at (3,0);
  \diagram* {
    (a) -- [scalar] (b),
    (b) -- [boson, half left, looseness=1.5] (c),
    (c) -- [scalar, half left, looseness=1.5] (b),
    (c) -- [scalar] (d)
  };\end{feynman}\end{tikzpicture}} + O(n^{-2}), \quad 
\Pi_\sigma= \frac{1}{2} \, \verticalcenter{\begin{tikzpicture}\begin{feynman}
  \vertex (a) at (0,0);
  \vertex (b) at (1,0);
  \vertex (c) at (2,0);
  \vertex (d) at (3,0);
  \diagram* {
    (a) -- [boson] (b),
    (b) -- [scalar, half left, looseness=1.5] (c),
    (c) -- [scalar, half left, looseness=1.5] (b),
    (c) -- [boson] (d)
  };\end{feynman}\end{tikzpicture}} + O(n^{-1}) \label{eq:vectorModelLO}
\end{aligned}\end{equation} 
\mccorrect{which are indeed the simplest kind of melon. 
Hence, we say that if a $\sigma$ field has been introduced, then the first corrections to the critical theory are melonic, with melonic data}
\begin{equation}\begin{aligned} \label{eq:OnMelonicData}
q_\phi =2, \quad q_\sigma = 1, \quad \frac{\dim \rhoext_\phi}{\dim \rhoext_\sigma} = n.
\end{aligned}\end{equation}
Indeed, $\Ft$-extremization does provide the correct leading values of the anomalous dimensions.
To aid our later comparison with the melonic models, we want to make this theory exactly melonic for all values of $n$. We can achieve this by adding a melonic mechanism: a disordered coupling constant\footnote{As usual, a suitable generalization of the Amit-Roginsky model \cite{Benedetti:2020iku}, or a tensor model, would also give the same results without the disorder average.} $g_{aij}$, following \cite{Chang:2021wbx,Shen:2023srk}. Promoting both $\phi_I$ and $\sigma$ to $N$-component fields, \eqref{eq:vectorOn} becomes the \textit{disordered vector model}
\begin{equation}\begin{aligned}
\cL_{\text{disordered }\gO(n)} = \sum_{i=1}^N \sum_{I=1}^n \half \phi\indices{^i_I} C^{-1} \phi\indices{^i_I}  - \sum_{a=1}^N\frac{1}{2\lambda_b} (\sigma^a \sigma^a) - \sum_{a,i,j=1}^N  \sum_{I=1}^n \frac{1}{2\sqrt{n}} g_{aij} \sigma^a \phi\indices{^{i}_I} \phi\indices{^j_I},
\end{aligned} \end{equation}
where we disorder average the coupling $g_{aij}$. We then take $N\to \infty$, while keeping $n$ finite. The melonic data \eqref{eq:OnMelonicData} is identical, and so $2\Delta_\phi + \Delta_\sigma = d$ holds as an exact statement for finite $n$ as well (up to $1/N$ corrections). 
\mccorrect{In this sense, the conformal IR of this disordered vector theory is a consistent extension for finite $n$ of the large-$n$ physics of the critical vector model}. 

Assuming (as usual) unbroken symmetry, and using \eqref{eq:FmaxSummary}, we find that in the IR of the disordered vector model, to leading order in the melonic mechanism's $N$, but exactly in $n$, we have $2\Delta_\phi + \Delta_\sigma = d$ and
\begin{equation}\begin{aligned} \label{eq:polesForVector}
\frac{\Ft_b'(\Delta_\sigma)}{\Ft_b'(\Delta_\phi)} = \frac{n}{2} = \frac{\Gamma (2 \Delta_\phi ) \Gamma (d-2 \Delta_\phi ) \Gamma \left(\frac{d}{2}-\Delta_\phi \right) \Gamma \left(\Delta_\phi -\frac{d}{2}\right)}{\Gamma (\Delta_\phi ) \Gamma \left(\frac{d}{2}-2 \Delta_\phi \right) \Gamma (d-\Delta_\phi ) \Gamma \left(2 \Delta_\phi -\frac{d}{2}\right)},
\end{aligned}\end{equation} 
with the IR wedge defined by $\Delta_\phi > \frac{d-2}{2}, \, \Delta_\sigma > \dotwo$.

We now make the connection to the standard large-$n$ vector model by \textit{also} taking the large-$n$ limit. 
Since $\Ft_b'(\tfrac{d-2}{2})=0$, we have a solution for $\Delta_\phi = \frac{d-2}{2} + \frac{\gamma_{\phi,1}}{n}+O(1/n^2)$, $\Delta_\sigma =2 + O(1/n)$, with
\begin{equation}\begin{aligned}\label{eq:gammaPhiFromFtextr}
\gamma_{\phi,1} =\frac{2\Ft_b'(2)}{\Ft_b''(\tfrac{d-2}{2})} = \frac{-2 \Gamma (d-2)}{\Gamma \left(2-\dotwo\right) \Gamma \left(\dotwo-2\right) \Gamma \left(\dotwo-1\right) \Gamma \left(\frac{d+2}{2}\right)}, %
\end{aligned}\end{equation}

which is exactly the standard result for the anomalous dimension of the field $\phi_I$ \cite{zinn-justin_quantum_2002}. For $2<d<4$ it is positive, and so lies in the IR wedge.
\mccorrect{We note that at large $n$, only the $\sigma$ field has a large anomalous dimension; however, for both fields the first contributions to the anomalous dimensions come from the melon diagram (despite being order $n^{-1}$ and $n^0$ for $\phi_I$ and $\sigma$ respectively).}

There is one slight complication: at this stage, the correction to $\Delta_\sigma$ in the large-$n$ limit for the original vector model is undetermined, as there $\Delta_\sigma = d-2\Delta_\phi$ only up to $O(1/n)$ corrections. However, the actual value of $\Delta_\sigma$ would be calculable by studying the spectrum of bilinears appearing in the OPE $\phi \times \phi$. Nonetheless, in the case of the disordered vector model we find the result described in \cite{Chang:2021wbx}, of a one-parameter family that interpolates from the critical $\gO(n)$ model at $n\to\infty$ (dual to higher spin AdS gravity) to a theory with a classical string dual. %

\subsection{Further solutions to \FttextOrPDF-extremization} \label{sec:furtherSolutions}

\begin{figure}[ht]
\centering
    \includegraphics[width=0.8\textwidth]{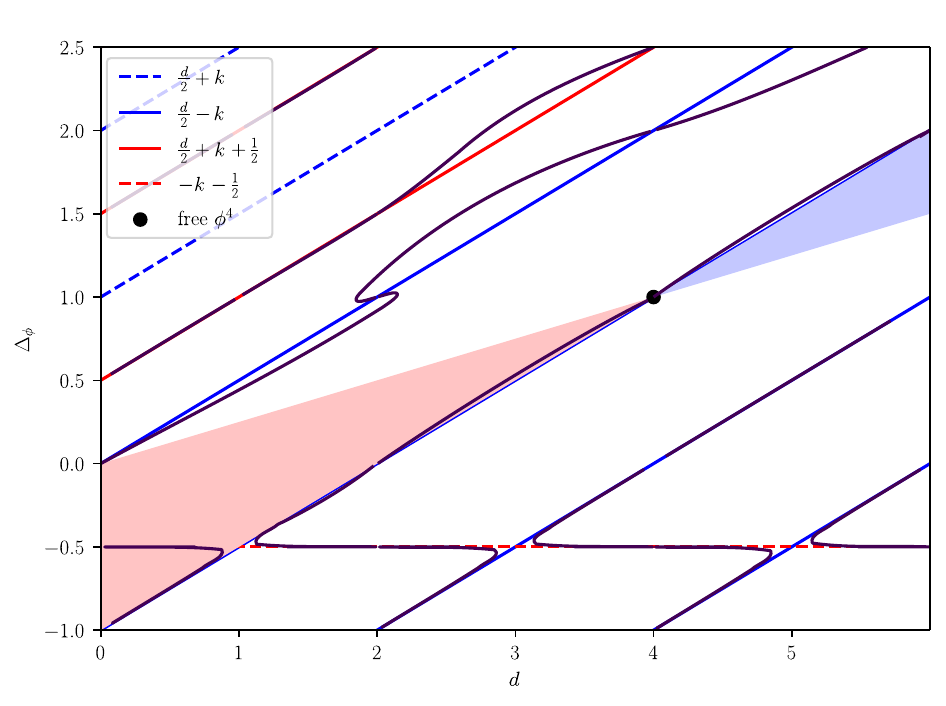}
    \caption{Large (but finite) $n$ contours superimposed in black onto the poles of \eqref{eq:polesForVector}. Note the absence of the $\Delta_\phi \sim \dotwo + k, k>0$ contours; this is due to the complexity of the anomalous dimensions around those poles. The curious twisted behaviour of the $\Delta_\phi \sim \dotwo$ line is due to the triple pole. The line descending from the free theory should be in the IR consistent (red) region only for $2<d<4$. We have taken $n=200$, but note that the size of the perturbation around $\Delta_\phi = \frac{d-2}{2}$ has been manually enhanced for clarity, to indicate in which dimensional range the scaling dimensions lie in the regions of IR and UV validity.}
    \label{fig:otherBranchesVectorModelFinite}
\end{figure}

The story above gives the usual solution of the $\gO(n)$ vector model, but evidently \cref{eq:polesForVector} has many other solutions for $\Delta_\phi$ in generic $d$: these are shown in \cref{fig:otherBranchesVectorModelFinite} by the black contours. 
Note that for $d>1$, only one of the contours lies within the IR wedge \eqref{eq:DelGreaterThanFree}, shown in red (for completeness we also show the UV wedge in blue). This is the standard solution \eqref{eq:gammaPhiFromFtextr}.
However, we are always allowed to modify the free propagators of $\phi$ and $\sigma$ to those of some other local GFF -- that is, some expression like $\phi (-\partial^2)^k \phi$ for integer $k$ -- possibly at the cost of unitarity (and we note that the implementation of this operator is more complicated on the sphere) \cite{Benedetti:2021wzt}. 
This modifies the UV scaling dimension $\Delta^{\mathrm{free}}_{\phi,\sigma}$, which changes the IR wedge. Then, the modified Lagrangian can access these extra solutions in its conformal limit.

Let us understand where these additional solutions come from in the large-$n$ limit; these lessons will transfer to the generic melonic case. The gamma function has no zeros. Therefore, in the large-$n$ limit, all solutions to \eqref{eq:polesForVector} must come from near the poles $\Delta^0_\phi$ of the gamma functions in the numerator. Of course, these poles must not coincide with poles in the denominator; expanding about them with $\Delta_\phi \equiv \Delta_\phi^0 + \gamma_\phi$, we can determine the leading-order anomalous dimensions for the fixed points. The order in $n$ of the correction then depends on the multiplicity of the pole. A pole of multiplicity $m$ at some $\Delta_\phi^0(d)$ leads to a leading-order equation
\begin{equation}\begin{aligned}
\frac{f(d)}{\gamma_\phi^m} +O(n^0) = \frac{n}{2} \quad \implies \quad \gamma_\phi \propto \frac{1}{n^{\frac{1}{m}}};
\end{aligned}\end{equation} 
this may be complex, depending on $m>1$ and the sign of $f(d)$ (which is always real, as $\Gamma(x)$ is real for real $x$). Thus, in practice, some perturbative poles will not give physical IR CFTs, even in integer $d$. 
The locations and multiplicities of such poles for integer $k$ are
\begin{equation*}\begin{aligned}
\begin{matrix}
     & \textbf{multiplicity} & \Delta_\phi^0 & \textbf{ values } & \gamma_\phi \textbf{s}\\ \hline
    \text{(i)} & \text{triple pole} &\frac{d}{2}& \frac{d}{2} & \text{ one $\R$, two $\Ibb$ } \propto \frac{i}{n^{1/3}} \\
     \text{(ii)} & \text{double pole }& \frac{d}{2} + k, k>0 &\quad \frac{d}{2} + 1,\frac{d}{2} + 2,\ldots & \text{ for d<4, two $\Ibb$ } \propto \frac{i}{n^{1/2}} \\
     \text{(iii)} & \text{single pole }&\frac{d}{2} - k, k>0&\quad \frac{d}{2}, \frac{d}{2}- 1,\ldots & \text{ real } \propto \frac{1}{n}\\
     \text{(iv)} & \text{single pole }&\frac{d}{2} + k + \half, k \ge 0&\quad \frac{d}{2} + \half, \frac{d}{2} + \frac{3}{2},\ldots & \text{ real } \propto \frac{1}{n}\\
     \text{(v)} & \text{single pole }&-k -\half, k \ge 0 &\quad -\half, \, -\frac{3}{2}, \ldots & \text{ real } \propto \frac{1}{n}
\end{matrix}
\end{aligned}\end{equation*}
These five cases can be written compactly as 
\begin{equation}
    \text{(i-iii) } \Delta^0_\phi = \frac{d}{2} \pm k \quad \text{ and }\quad \text{(iv-v) } \Delta^0_\sigma = \frac{d}{2} \pm \left(\frac{d}{2} +2 k +1 \right), \text{ for } k \ge 0. %
\end{equation}%
To make it clear that the solutions are perturbations around the pole lines, they are shown in colour in \cref{fig:otherBranchesVectorModelFinite}. %
We make the following comments:
\begin{itemize}
    \item To ensure similarity to the actual large-$n$ vector model, we have chosen $n=200$ for the ratio of degrees of freedom of $\phi$ and $\sigma$. This ratio is tunable, and, as shown in the tables, all anomalous dimensions are order $1/n^k$. For small $n$, the solutions depend strongly on $n$.
    \item For example, we have one real and two complex solutions around $\Delta_\phi =\frac{d}{2}$:
\begin{equation}\begin{aligned}
\Delta_\phi = \frac{d}{2} + \frac{\alpha}{n^{1/3}} + O(1/n^{2/3}), \quad \alpha^3 = \frac{\Gamma (d)}{\Gamma \left(-\frac{d}{2}\right) \Gamma \left(\frac{d}{2}\right)^3},
\end{aligned}\end{equation} 
which gives the distinct twisted structure.
\item There are no visible solutions for $\Delta^0_\phi = \frac{d}{2} + k$, $k=1,2$, since they are complex.
\item We note that for $d=1$ exactly, for $\Delta_\phi>0$, there is only one real solution, with $\Delta_\phi = \frac{d}{2} + O(1/n^{1/3})|_{d=1}$. %
However, there are an infinite number of real solutions perturbatively in $d=1+\epsilon$, which is why the contours appear unbroken. %
\item The fact that $\gamma_\phi$ is positive for $2<d<4$ means that the $\Delta_\phi = \frac{d-2}{2} + \gamma_\phi$ solution is consistent in this region. We observe the presence of a second consistent IR solution (that is, one lying within the IR wedge) for $\Delta_\phi = -\half + O(1/n)$ for $d<1$, and hence $\Delta_\sigma = d+1 + O(1/n)$, that seemingly descends from the free $d=1$ boson. %
\item We can immediately identify the solutions around the poles at $\Delta^0_\phi = \frac{d}{2} - k$ in the large-$n$ limit as being the Wilson-Fisher CFT for the real $\Box^{k}$ CFT (with kinetic term $\phi(-\partial^2)^k \phi$), which is dual to the minimal type-A$_k$ higher spin gravity \cite{Sun:2020ame,Bekaert:2013zya,Brust:2016zns}. 
The other solutions are less clear, but could be reached by modifying the UV kinetic terms of $\sigma$ and/or $\phi$ to $\Box^{\pm k}$, making them free higher-derivative scalars \cite{Brust:2016gjy}; this was done, for example, in \cite{Fei:2014yja,Giombi:2019upv} to access the critical $\gO(N)$ CFT in the IR of a QFT for $4<d<6$. 
These solutions appear to be associated with the cases where the UV kinetic term of one of the fields is in an exceptional representation of the global conformal group (specifically, where $F_{\phi/\sigma}'(\Delta)= 0,\infty$) \cite{Dobrev:1977qv,Schaub:2024rnl}.
This also means that there is no momentum space representation of the two-point functions in flat space, as the Fourier transform either diverges or is zero \cite{Bzowski:2015pba}. %
\end{itemize}

This analysis extends immediately to other large-$n$ vector models, such as all of the vectorial models found in \cite{Jepsen:2023pzm}, or the Gross-Neveu model \cite{Zinn-Justin:1991ksq}, where we would take
\begin{equation}\begin{aligned}
\frac{\dim \rhoext_\psi}{\dim \rhoext_\sigma} =2 \Tr[\spinid]\times n,
\end{aligned}\end{equation} 
and again solve to leading order in $n$. The result matches the computations of \cite{Zinn-Justin:1991ksq,Goykhman:2020tsk}, and %
changing the $\dim \rhoext$s, also solves the chiral and non-abelian \cite{Gracey:2018qba,Gracey:2021yyl} extensions. 
Hence, we have a simple example of how, as promised, only the dimension of complicated finite symmetry representations matters. 
The analysis performed here is identical to the \enquote{simple loop} approximation of \cite{Vasiliev:1981sf}. 
In that paper, QCD was studied without taking a large-$N$ limit, but using the uncontrolled approximation of only taking the melonic diagrams (the \enquote{simple loops} are just the simplest kind of melon, as above). 
The authors then studied the other lines of vacua that we have noted here.%

We have shown that the IR of the disorder-averaged vector model indeed gives a consistent extension of the leading-$n$ physics of the critical vector model ($n\to \infty$) to finite values of $n$. 
Of course, the melonic structure of the standard large-$n$ vector model does not persist to the next order in $n$: if we attempt to compute the subleading terms in $\Ft$, along the lines of \cite{Tarnopolsky:2016vvd,Fraser-Taliente:2025udk}, we find non-melonic diagrams. 
This means that when solving for the IR, we no longer have the neat interpretation of the constrained extremization of $\Ft(\Delta_\phi,\Delta_\sigma)$ for two GFFs, $=N\Ft_b(\Delta_\phi) + \Ft_b(\Delta_\sigma)$. 
However, recently a related phenomenon was observed: if we analytically continue the vector model in $\Delta_\phi$, we find the so-called long-range vector models, which have a free energy $\Ft^\mathrm{LR}_{\gO(N)}(\Delta_\phi)$.
The standard (short-range) CFT is then the CFT which maximizes $\Ft(\Delta_\phi)$ \cite{Fraser-Taliente:2025udk}.

The large-$n$ vector models provide simple examples in which we can observe these characteristic features. We turn next to the melonic models, in which case the equivalent of the parameter $n$ is decreased to order one, and so the anomalous dimensions are also order one. This makes it harder to see directly the reason for, for example, the disappearing contour lines (the generation of complex anomalous dimensions), but they occur for reasons identical to those shown here.
Likewise, the additional lines of solutions in the other melonic models can be thought of as coming from (large) perturbations around theories with modified kinetic terms. %

\section{Melonic models: some examples} \label{sec:melonicModels}

For a generic melonic model, the anomalous dimensions are order one, and so it can be harder to interpret what happens as we change $d$ -- this is why we began with the large-$n$ critical vector models in \cref{sec:vectorModels}. In this section, we demonstrate how the characteristic features identified in the critical vector models also hold here:
\begin{enumerate}
    \item There are multiple conformal vacua, only some of which lie inside the IR wedge, for theories with $n_f>n_m$. As in the vector case, each line arises from a possibly large perturbation around the pole of a gamma function, which reflects a scaling coming from a UV kinetic term.
    \item At certain values of $d$, pairs of real solutions collide and become a complex conjugate pair: thus we have disappearing lines of solutions.
    \item There are gaps in the solution contours at certain integer values of $d$. 
    \item The existence and location of real solutions depend strongly on the ratios of $\dim\rhoext_\phi$s.
\end{enumerate}
We use three multi-field models as our example: first, two single-interaction models, and then a multi-interaction model, which also possesses supersymmetric vacua. We do not discuss the single-field melonic models (or more generally, the theories with $n_f=n_m$), as they have only a single solution, $\Delta = d/q$, directly from the constraint(s).

\subsection{Two fields, one interaction: the quartic Yukawa model} \label{sec:QuarticYukawaFull}
\begin{figure}
    \centering
    \includegraphics[width=0.6\linewidth]{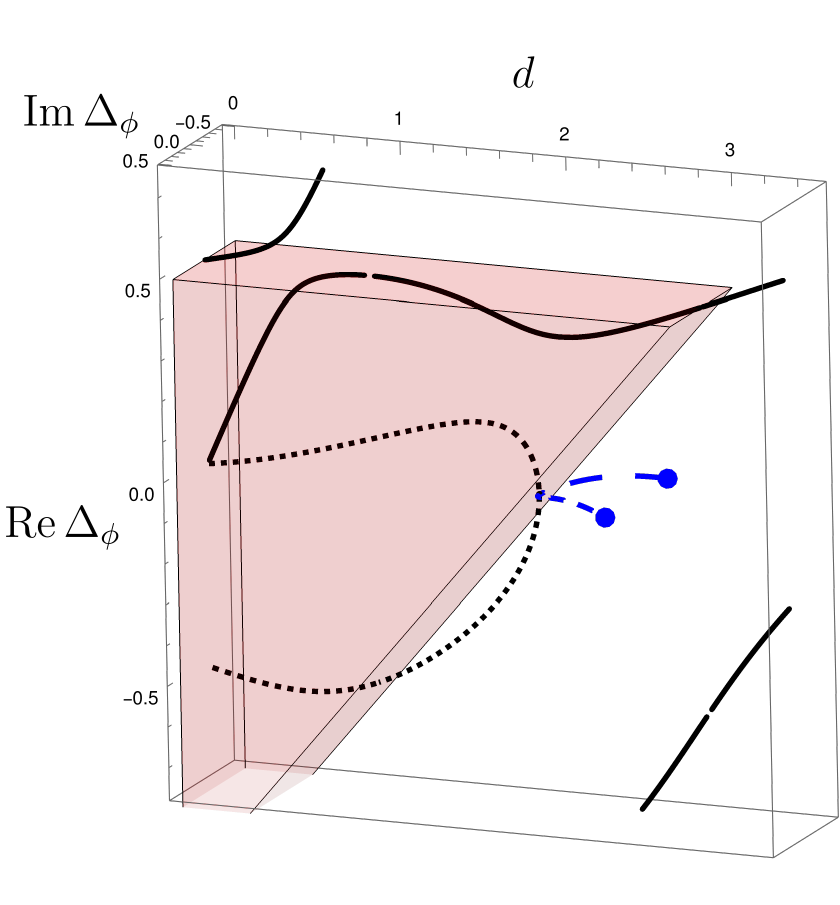}
    \caption{Plot of solutions for the scaling dimension $\Delta_\phi$ in the $\phi^2 \bar\psi \psi$ melonic model. $\Delta_\psi$ is obtained from the melonic constraint $2\Delta_\phi + 2\Delta_\psi=d$. Complex solutions are blue and dashed, and leave the volume of the plot at the blue dots. The real slice of this figure corresponds to \cref{fig:phi2psibarpsiIRwedge}. On the real axis, solid lines represent maxima of $\Ft$; dashed lines represent minima. The gaps in the contours are real, and indicate missing solutions in $d=1$ and $d=3$ here.} %
    \label{fig:QuarticYukawaSolutionPlot3D}
\end{figure}

In \cref{fig:QuarticYukawaSolutionPlot3D}, we give the contour plot of the solution space for the $\phi^2 \bar\psi \psi$ melonic model, tuned so that
\begin{equation}\begin{aligned} \label{eq:QYdimrat}
\frac{\dim \rhoext_\psi}{\dim \rhoext_\phi} = 2 \Tr \spinid = 4.
\end{aligned}\end{equation} 
Unlike \cref{fig:phi2psibarpsiIRwedge}, we also include here the (blue, dashed) complex solutions; these collide at $d=2$ on the free scalar line, giving the two real solutions that exist for $d \le 2$. 
The details of the contours and the occurrence of complexification are highly dependent on the ratio \eqref{eq:QYdimrat}: this will be demonstrated in \cref{sec:SDEanalysis}.

\subsection{Three fields and one interaction: the Popović model}

\begin{figure}
    \centering
    \begin{subfigure}{0.65\textwidth}
    \includegraphics[width=\textwidth]{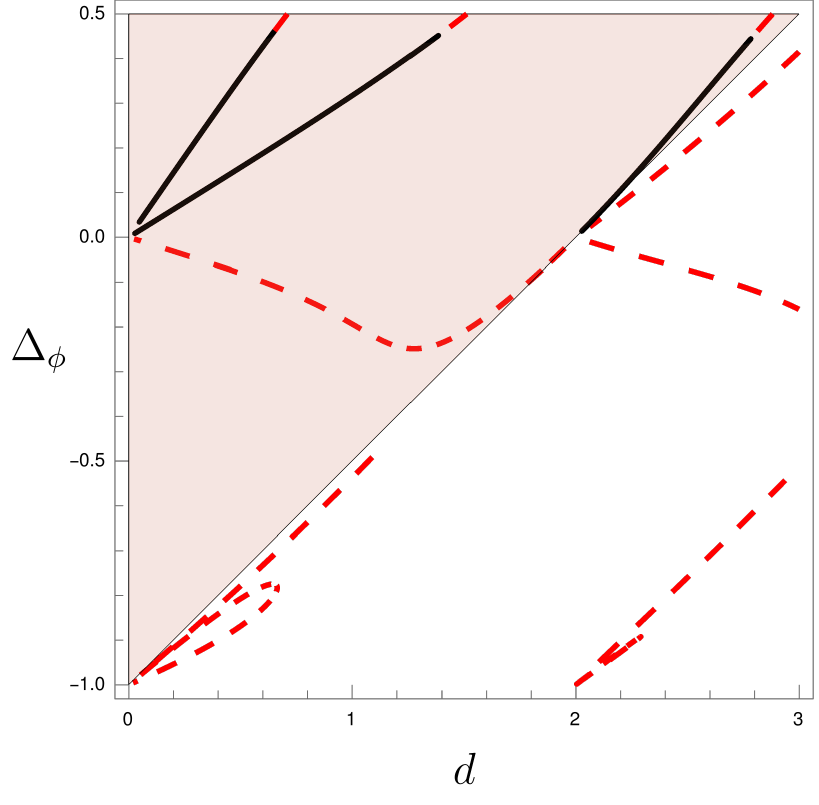}
    \caption{Contour plot of $\Delta_\phi$ against $d$.} %
    \label{fig:PopN5r2}
    \end{subfigure}%
    \vspace*{0.3cm}
    \begin{subfigure}{0.65\textwidth}
    \vspace*{\fill}
    \includegraphics[width=\textwidth]{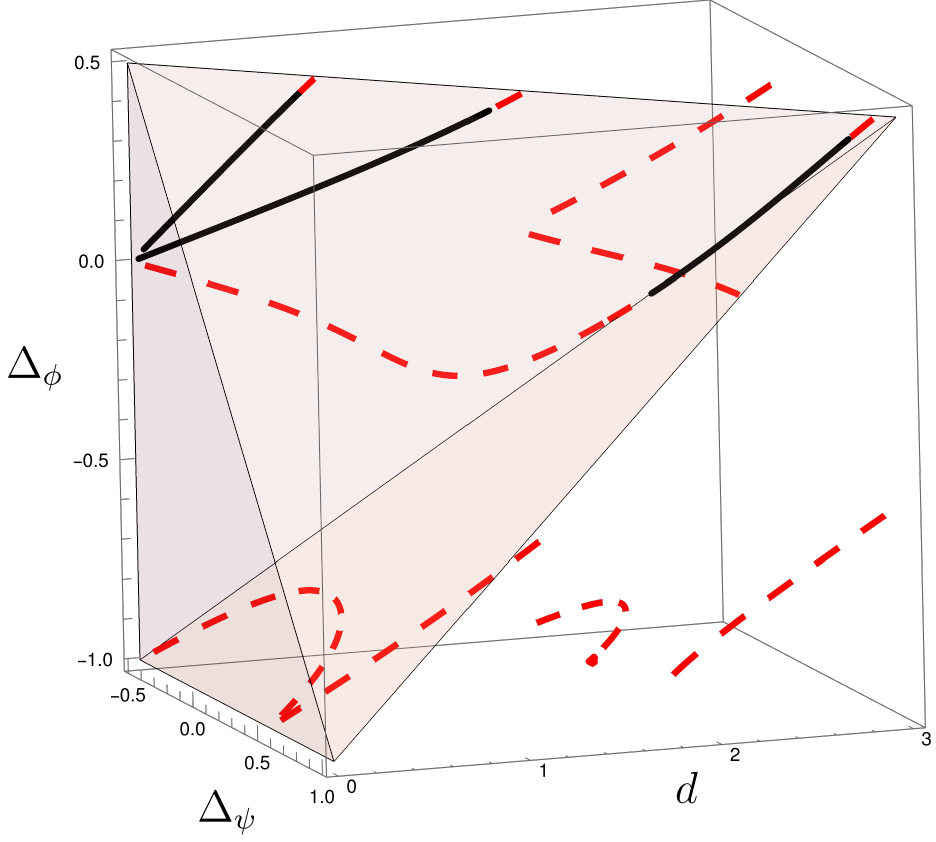}
    \vspace*{\fill}
    \caption{Angle view, showing 3d structure.}
    \end{subfigure}
    \caption{Scaling dimension solutions of the melonic Popović model \eqref{eq:melPopLag} for various $d$, illustrated for $n=5$, $\Tr[\spinid]=2$; we see a complex network of conformal vacua. The IR wedge is now an IR tetrahedron, which is shaded: black lines lie inside it; the dashed red lines lie outside. Unlike before, we do not indicate the maxima and minima. The rightmost black line in \cref{fig:PopN5r2} is the solution $\Delta_\phi = \frac{d-2}{2} + O(1/n), \Delta_\psi = \frac{d-1}{2} + O(1/n)$. The two left-hand black lines are $\Delta_\phi = \frac{d}{2} \pm O(1/n^\half), \Delta_\psi = \frac{d-1}{2} + O(1/n)$.} 
    \label{fig:Popoviccontours}
\end{figure}

In the case of more fields, we find a higher-dimensional generalization of the above behaviour. 
\Cref{fig:Popoviccontours} shows the IR solutions for a melonic version of the Popović model \cite{Popovic:1977cq}. 
With melonic mechanism suppressed, the Lagrangian for this is
\begin{subequations}\label{eq:melPopLag}
\begin{equation}\begin{aligned}
\cL_{\text{Popović}} \sim \bar\phi_I (-\partial^2) \phi_I  + \bar{\psi}_I(-\slashed{\partial}) \psi_I - \bar\chi \chi + g_0(\bar\chi \phi_I \psi_I  + \bar\psi_I \phi^\dagger_I \chi),
\end{aligned}\end{equation} 
where $I$ runs from $1$ to $n$, and $\chi$ is a fermionic auxiliary field\footnote{Note that the Popović model is an $\gO(n)$ vector model where the singlet field is $\chi \propto \phi_I \psi_I$ instead of the usual $\phi_I \phi_I$.} which becomes dynamical. 
We choose arbitrarily
\begin{equation}\begin{aligned}
\frac{\dim \rhoext_\psi}{\dim \rhoext_\phi} = \Tr[\spinid]= 2,\quad \frac{\dim \rhoext_\phi}{\dim \rhoext_\chi} = n = 5.
\end{aligned}\end{equation} 
\end{subequations}
The melonic mechanism constrains $\Delta_\phi + \Delta_\psi + \Delta_\chi =d$, and so eliminating $\Delta_\chi$, the extrema of $\Ft$ lie in the three-dimensional space $(d,\Delta_\phi,\Delta_\psi)$ shown in the figure.
The IR wedge becomes an IR tetrahedron
\begin{equation}\begin{aligned}
\Delta_\phi > \frac{d-2}{2},\quad \Delta_\psi > \frac{d-1}{2}, \quad \Delta_\chi > \frac{d}{2}.
\end{aligned}\end{equation} 
We still find a discrete set of vacua in each $d$, only some of which, drawn in black, lie within the IR tetrahedron.

\subsection{Multi-interaction melonics; a supersymmetric model} \label{sec:SUSYmelons}

As described above, we can consider multiple melonic interactions, i.e. $n_m>1$.
One natural case where these arise is in the supersymmetric melonic theory of a single chiral field and its supersymmetry-breaking deformations, which we now turn to.
Recall that the standard supersymmetric $\Ft$-extremization trivially collides with the melonic $\Ft$-extremization.
The advantage of the melonic theory is that we are also permitted to consider the vacua with broken SUSY.
We now foreshadow the discussion of \cref{sec:hlamprismaticSDEfound}, where this model will be discussed further.
We take the melonic-type theory of a single complex scalar superfield with four supercharges, as given in \cite{Lettera:2020uay}.
As usual, we suppress the melonic mechanisms, which can be found in \cite{Murugan:2017eto,Chang:2018sve,Popov:2019nja,Lettera:2020uay}.
The form of the superpotential makes the $\Ft$-extremization trivial,
\begin{equation}\begin{aligned} \label{eq:QYauxSUSYSol}
W[\Phi] \sim g \Phi^4, \quad  \Delta_\Phi=\frac{d-1}{4}.
\end{aligned}\end{equation} 
However, we can also consider breaking up the superfield into its components in the usual way, 
\begin{equation}\begin{aligned}
\Phi= \phi + \theta \psi - \theta^2 X. %
\end{aligned}\end{equation} 
In this case, %
we obtain a potential of schematic form
\begin{equation}\label{eq:susyBrokenV}
V(\phi, \psi, X) = \rho (X \phi^3 + X^\dagger (\phi^\dagger)^3) + \lambda \phi^\dagger \phi \bar\psi \psi,
\end{equation}
where supersymmetry gives the relation $\rho \sim \lambda$. 
This is a complex version of the full quartic Yukawa model, with auxiliary field, of \cref{sec:SDEanalysis} (\cite{Fraser-Taliente:2024rql}), which was defined as $h\lambda_\text{prismatic}$ there. 

\begin{figure}
    \centering
    \includegraphics[width=0.8\linewidth]{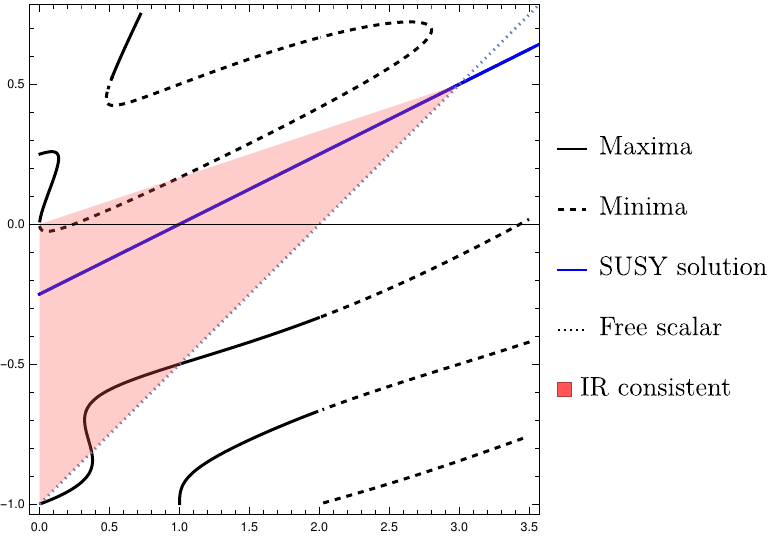}
    \caption{The scaling dimension $\Delta$ of $\Phi$ for the supersymmetry-breaking vacua of the melonic theory of a single complex scalar superfield with four supercharges in $0<d<3.5$; i.e. with potential \eqref{eq:susyBrokenV}. The SUSY-preserving solution \eqref{eq:QYauxSUSYSol} is the line $\Delta = \frac{d-1}{4}$. 
    We indicate whether the given lines are maxima or minima of the function $\Ft(\Delta_\phi)=\sum_i \Ft_i$ (though not necessarily of the full free energy).} %
    \label{fig:hlamprismatic-r2}
\end{figure}

Let us now consider perturbing by the SUSY-breaking operator \eqref{eq:susyBrokenV}; that is, we allow $\rho$ and $\lambda$ to vary independently. 
We take $\dim \rhoext_X= \dim \rhoext_\phi$ and $\dim \rhoext_\psi = 2 \dim \rhoext_\phi$. 
Since we have two interactions, there are now two melonic constraints,
\begin{equation}\begin{aligned}
\Delta_X + 3 \Delta_\phi = d \, \text{ and } \, 2\Delta_\phi + 2\Delta_\psi = d,
\end{aligned}\end{equation} 
and so the extrema are only in the space $(d,\Delta_\phi)$, which we plot in \cref{fig:hlamprismatic-r2}. The supersymmetric solution \eqref{eq:QYauxSUSYSol}, where $\Delta_\phi = \Delta_\psi - \thalf= \Delta_X-1$, is the straight line in the figure. In addition, the component melonic approach has given non-perturbative access to all the SUSY-breaking vacua; however, in this case the additional vacua are only IR consistent for $d<1$.

\begin{mccorrection}
    \subsection{Limitations of melonic \FttextOrPDF-extremization: do we actually reach the melonic CFT?} \label{sec:MelonicFmaxLimitations}
    
    We end this section by summarising the limitations of melonic $\Ft$-extremization.
    They are similar to those discussed in \cref{sec:FmaxLimitations} for supersymmetric $\Ft$-maximization, with the additional limitation that we do not yet have a principle for determining the nature of the extremum of $\Ft$: whether it is maximized or minimized.
    In the case of multiple vacua in the IR, this would allow us to determine which one is realised in the IR.

    The main limitation is that we have assumed that the Lagrangian QFT that we write down actually flows to the conformal IR CFT found by $\Ft$-extremization.
    Of course, if this does not occur, it is still possible to consider that CFT as a (possibly non-unitary and complex) CFT defined by its CFT data $\{(-\Delta_i,\rho_i),C_{ijk}\}$.
    The IR wedge is the most basic general constraint on whether the flow occurs; however, there are other possible obstructions.
    As discussed in \cite{Murugan:2017eto}, %
    the self-energy is in fact UV divergent, which we have ignored by using analytic continuation in $\Delta$ and $d$.
    In a physical system, we would need to tune the more relevant interactions/counterterms in order to hit such a critical point, which is not guaranteed to be possible. 

    The indications of instability that might generally arise are then the following:
    \begin{itemize}
        \item Complex scaling dimensions can indicate an instability of the theory, as the true IR is not conformal; we comment on how this can be diagnosed by calculation of the spectrum of operators in the OPE of the fundamental fields in \cref{sec:windowsOfStability}. 
        \item We have assumed that there exists a stable vacuum; this is certainly not clear for the tensor models, where the interaction terms have no positivity properties. 
        Additionally, when considered on the sphere, the quadratic curvature couplings $\sim \mathcal{R}\phi_i \phi_i$ are positive, and therefore serve to make the perturbative vacuum metastable \cite{Giombi:2019upv}. 
        Hence, perturbatively we may indeed land on the CFT with the data described by the large-$N$ limit.
        Non-perturbative instabilities remain, due to the negative directions in the interacting potential, but detecting them will require a different approach to that of the strict large-$N$ limit \cite{Murugan:2017eto,Herderschee:2024zlc}.
    \end{itemize}
    Nonetheless, in these cases we can still consider these non-unitary CFTs as a formal solution to the SDEs \cite{Giombi:2017dtl}, as mentioned above.
    \begin{itemize}
        \item Finally: the breaks in the lines of conformal solutions in various integer dimensions, visible in \cref{fig:otherBranchesVectorModelFinite,fig:QuarticYukawaSolutionPlot3D} also seem to indicate that the true IR in these dimensions lies in a symmetry breaking phase \cite{Biggs:2023mfn}; in this case not even a formal CFT exists.
    \end{itemize}
\end{mccorrection}

\begin{subappendices}
\section{Two-point functions and computing \FttextOrPDF}

\subsection{Conformal two-point functions and conventions}

Given a conformal field $\phi$ with conformal data $(-\Delta_\phi, \rhoext_\phi = \rho_\phi \times R_\phi)$, the inverse of its two-point function $\twoPt_\phi$, if it exists, is defined by
\begin{equation}\begin{aligned}
\int \odif[d]{x} \, \twoPt_\phi (w-x)\indices{_{\mu_1 \cdots \mu_s}^{\nu_1 \cdots \nu_s}} \, [\twoPt_\phi^{-1}(x-y)]\indices{_{\nu_1 \cdots \nu_s}^{\sigma_1 \cdots \sigma_s}} &\equiv \delta^d(w-y) \, \hat{\delta}\indices{_{\mu_1\cdots \mu_s}^{\sigma_1 \cdots \sigma_s}}.
\end{aligned}\end{equation} 
We can write this in the shorthand form
\begin{equation}\begin{aligned}
\int \odif[d]{x} \, \twoPt_\phi (w-x) \cdot \twoPt_\phi^{-1}(x-y) &\equiv \delta^d(w-y)\, \id_{\rho_{\phi}},
\end{aligned}\end{equation} 
where $\id_{\rho_{\phi}} \equiv\hat{\delta}\indices{_{\mu_1\cdots \mu_s}^{\sigma_1 \cdots \sigma_s}}$ is the appropriate identity for the Lorentz representation $\rho_{\phi}$. 
Henceforth, as in the main text, we shall suppress all indices, including the $\Gglobal$ indices of the representation $R_\phi$.

Now define the unique shadow field $\tilde\phi$ to be the field with scaling dimension $\tilde{\Delta}_\phi \equiv d-\Delta_\phi$, transforming in the \textit{reflected representation}\footnote{The reflected representation of $\rho$ is defined by $\rho^R(g) \equiv \rho(RgR^{-1})$, where $R \in \gO(d)$ is a reflection in any direction. 
In odd dimensions, $\rho^R$ and $\rho$ are equivalent, because $\gO(d)$ factorizes into $\SO(d) \times \mathbb{Z}_2$ as $-1$ is a reflection matrix.  
In even dimensions, to obtain $\rho^R$, we swap the weights associated with the two spinor representations.} $\rho_{\phi}^{\mathrm{ref}}$. 
The operator $\tilde{\phi}^\dagger$ then has scaling dimension $d-\Delta_\phi$ and $\SO(d)$ representation $\rho^\star$ (the dual of $\rho$).
The inverse propagator and the shadow field propagator transform identically under conformal symmetry. Since both are unique, we must have%
\begin{equation}\begin{aligned}
\twoPt_\phi^{-1}(x-y) &=\frac{1}{\cN_{\phi}} \twoPt_{\tilde\phi}(x-y), %
\label{eq:inverseIsShadow}
\end{aligned}\end{equation} 
for some $\cN_\phi$; this must be a purely representation-theoretical quantity (since $D_\phi$s are unit-normalized) that can be calculated explicitly by taking the inverse in momentum space. 
This result is a generic identity for CFTs, provided that $\cN_{\phi} \neq \infty,0$, which occurs for operators transforming in the exceptional series of conformal group representations. 
$\cN_\phi$ can also be found using \cite{Karateev:2018oml} %
\begin{equation}\begin{aligned}\label{eq:plancherelDef}
\frac{\dim_{\mathbb{R}}(\rho_\phi)}{\cN_\phi} = (-1)^{\mathrm{F}^\phi} \Omega_d \, \mu(\phi),
\end{aligned}\end{equation} 
where $\Omega_d=2^d \vol\SO(d)$ is a constant that drops out of all computations, $(-1)^{\mathrm{F}^\phi}$ just gives a minus sign for fermionic reps,
and $\mu(\phi)$ is the Plancherel measure of $\SO(d+1,1)$ for the conformal representation of $\phi$.
For comparison with the main text, we note that $\Ft$ for a generalized free field is defined by
\begin{equation}\begin{aligned} \label{eq:FtEqIntMu}
\Ft(\Delta_\phi, \rho_\phi) &= \frac{\pi^{d+1}}{\Gamma(d+1)}\int_{\frac{d}{2}}^{\Delta_\phi} \odif{\Delta'} \,(-1)^{\mathrm{F}^\phi} \Omega_d \, \mu(\phi|_{\Delta'}),
\end{aligned}\end{equation} 
where for a scalar field, $\cN_\phi^{-1} = c(\Delta)c(d-\Delta)$, with $c(\Delta)$ from \eqref{eq:inverseProp}.
If $\phi$ transforms in a nontrivial global symmetry representation $R_\phi$, we also multiply this by a factor of $\dim R_\phi$, so
\begin{equation}
\Ft(\Delta_\phi, \rhoext_\phi) = \Ft(\Delta_\phi, \rho_\phi) \times \dim R_\phi.
\end{equation}

\begin{mccorrection}
  \subsection{Computation of \FttextOrPDF for GFFs} \label{app:FtComputation}

We can compute the sphere free energy for a real conformal GFF $\Phi$ of arbitrary dimension $\Delta$ and representation $\rho$, which has a propagator $G_\Phi =\cZ_\Phi D_\Phi$.
The normalization $\cZ_\Phi$ does not contribute, since $\Tr \delta^{(d)}(x-y) =0$ in DREG, so assuming that $\Tr \log G_\Phi^{-1} = - \Tr \log G_\Phi$, we find
\begin{equation}
F_\Phi \equiv (-1)^{\mathrm{F}^\Phi}\half \Tr \log G_\Phi^{-1} = -(-1)^{\mathrm{F}^\Phi}\half \Tr \log D_\Phi.
\end{equation}
This calculation is a standard result from the AdS/CFT literature for any Lorentz representation of $\Phi$ \cite{Sun:2020ame,Benedetti:2021wzt,Harribey:2022esw}. %
We provide here an arbitrary $d$ calculation which demonstrates the relation to $\cN_\Phi(\Delta,\rho)$ defined in \eqref{eq:inverseIsShadow}.
We begin by differentiating: %
\begin{align}
	F_\Phi' = \odv{F_\Phi}{\Delta} &= -\half (-1)^{\mathrm{F}^\Phi} \Tr G_\Phi^{-1} \odv{G_\Phi}{\Delta} = -\half(-1)^{\mathrm{F}^\Phi}\odv{}{\alpha}\left(\Tr G_\Phi^{-1} G_{\Phi,\Delta_\Phi=\alpha}\right) \big\rvert_{\alpha=\Delta},
\end{align}
where we repeatedly swap the order of derivatives and integrals, assuming everything to converge: we are therefore implicitly working in the range of $d$ where this is convergent as the UV cutoff is taken to infinity and then analytically continuing back up.
Explicitly writing out the position space trace,
\begin{align}
	F'_\Phi &= -\half \frac{1}{\cN_\Phi} \odv{}{\alpha} \int_{x,y} \tr_\rho \Big[D_\Phi^{-1}(x,y) \underbrace{(-1)^{\mathrm{F}^\Phi}D_{\Phi,\alpha}(y,x)}_{D_{\Phi,\alpha}(x,y)}\Big]  \Big\rvert_{\alpha=\Delta}\\
  &= -\half \frac{\tr_\rho(\id_\rho)}{\cN_\Phi} \odv{}{\alpha} \int_{x,y} \frac{1}{s(x,y)^{2(d-\Delta+\alpha)}} \Big\rvert_{\alpha=\Delta},
\end{align}
where we have used the known behaviour of conformal $D_\Phi(x,y)$s under $x \leftrightarrow y$.
The fact that we do not need to consider the suppressed indices, being guaranteed to end up with only an $\id_\rho$, is by definition of the inverse.
Using our convenient function $\Mtm$ from \eqref{eq:MtmDef},
\begin{equation}\begin{aligned}
\odv{\Ft_\Phi}{\Delta} &= \frac{\dim_\mathbb{R}(\rho)}{\cN_\Phi} \odv{}{\alpha} (2R)^{-2(\alpha-\Delta)} \Mtm(\alpha-\Delta) \big\rvert_{\alpha=\Delta}\\
&= \frac{\dim_\mathbb{R}(\rho)}{\cN_\Phi} \Mtm'(0).
\end{aligned}\end{equation}
 Applying the expansion \eqref{eq:MtmExpanded}, we find exactly \eqref{eq:FtEqIntMu} for any conformal GFF:
 \begin{equation}
 \Ft_\Phi(\Delta,\rho) = \frac{\pi^{d+1}}{\Gamma(d+1)} \int_{\frac{d}{2}}^{\Delta} \odif{\Delta'} \frac{\dim_\mathbb{R}(\rho)}{\cN_{\Phi,\Delta'}}.
 \end{equation}
We have assumed here that in DREG $\Ft_\Phi(\Delta=\tfrac{d}{2})=0$, which follows immediately from the identity $\Tr \log G_\Phi^{-1} = - \Tr \log G_\Phi$.

\section{The regularized volume of hyperbolic space} \label{sec:volHd}

The regularized volume of Euclidean hyperbolic space is straightforwardly obtained, following section 2.1 of \cite{Diaz:2007an}. %
Taking the metric on $d+1$-dimensional hyperbolic space to be $\odif{s^2} = 4R^2 (1-x^2)^{-2} \odif{x^2}$, where $r^2 \equiv \abs{x}^2 \le 1$, we have 
\begin{align}
\vol \mathbb{H}^{d+1} = \int\odif[d]{x} \, \sqrt{g} = \int_{\abs{x}\le 1} \frac{ \odif[d]{x} (2R)^{d+1}}{(1-x^2)^{d+1}} = (2R)^{d+1} \vol S^d \,\int_0^1 \frac{\odif{r}\, r^d}{(1-r^2)^{d+1}}.
\end{align}
This converges for $-1<\Re(d) <0$. 
Evaluating it there, and analytically continuing back up to the $d$ of interest (which is precisely dimensional regularization), we find that the dimensionally regularized volume of hyperbolic space is
\begin{equation}
\vol \mathbb{H}^{d+1} = \pi^{d/2} \Gamma(-d/2) R^{d+1}.
\end{equation}
The manifestation of the Weyl anomaly is that this quantity diverges in any even $d$.

We could instead choose the metric to be $\odif{s^2} = 4 (1-x^2/R^2)^{-2} \odif{x^2}$. 
Then,
\begin{equation}
\vol \mathbb{H}^{d+1} = \int\odif[d]{x} \, \sqrt{g} = 2^{d+1} \vol S^d \, \int_0^R \frac{\odif{r}\, r^d}{(1-r^2/R^2)^{d+1}}.
\end{equation}
For comparison, consider the $(d+1)$-sphere volume, with metric $\odif{s^2} = 4 (1+x^2/R^2)^{-2} \odif{x^2}$.  %
This differs by a sign and the limit of the integral, but is absolutely convergent
\begin{equation}
\vol S^{d+1} = 2^{d+1} \vol S^d \, \int_0^\infty \frac{\odif{r}\, r^d}{(1+r^2/R^2)^{d+1}} = \frac{2 \pi^{\dotwo+1}}{\Gamma(\dotwo +1)} R^{d+1}.
\end{equation}
Combining these two, we indeed see that when evaluated with the same $R$,
\begin{equation}
\half \frac{\vol S^{d+1}}{\vol \mathbb{H}^{d+1}} = - \sin(\pi d/2).
\end{equation} %
\end{mccorrection}
\section[Diagrammatic proof of constrained \FttextOrPDF-extremization]
{\texorpdfstring{Diagrammatic proof of constrained\\ $\Ft$-extremization}{Diagrammatic proof of constrained F~-extremization}} %
\label{app:diagrammaticProof}

In the following, we demonstrate how $\Ft$-extremization can be recovered by a more standard Feynman-diagrammatic calculation of the two-point Schwinger-Dyson equations in flat space. 
These manipulations, although less enlightening, demonstrate that there is no sleight of hand occurring in \cref{sec:FtMaxFrom2PI}.
\subsection{Two-point Schwinger-Dyson equations}

As is usual for conformal field theories, we will work in position space, where the two-point functions of an arbitrary field $G_\phi=\cZ_\phi D_\phi$ (defined in \eqref{eq:conformalG}) satisfy the usual Schwinger-Dyson equation
\begin{equation}\begin{aligned}%
\vcenter{\hbox{\begin{tikzpicture}
  \begin{feynman}[every blob={/tikz/fill=gray!30}]
    \vertex[small,rectangle, draw=black,fill=gray!30]  (m) at (0,0) {$\mathcal{Z}_\phi D_\phi$};
    \vertex (a) at (-1,0) ;
    \vertex (b) at ( 1,0);
    \diagram* {
      (a) --[scalar] (m) --[scalar] (b),
      };
  \end{feynman}
\end{tikzpicture}}}
\quad &= \quad \vcenter{\hbox{\begin{tikzpicture}
  \begin{feynman}
  \vertex[small, rectangle, draw=black] (m) at (0,0) {$C^{\text{free}}_\phi$};
    \vertex (a) at (-1,0) ;
    \vertex (b) at ( 1,0);
    \diagram* {
      (a) --[scalar] (m) --[scalar] (b),
      };
  \end{feynman}
\end{tikzpicture}}} \quad+ \quad
\vcenter{\hbox{\begin{tikzpicture}
    \begin{feynman}
    \vertex (a) at (-1,0) ;
     \vertex[small, rectangle,draw=black] (m) at (0,0) {$C^{\text{free}}_\phi$};
   \vertex[large,blob,fill=gray!30] (centreblob2) at (1.5,0) {$\Pi_\phi$};
   \vertex[small,rectangle, draw=black,fill=gray!30] (end) at (3,0) {$\mathcal{Z}_\phi D_\phi$};
    \vertex (e) at (4,0);
    \diagram* {
      {[edges={scalar}] (a) -- (m)  -- (centreblob2) -- (end) -- (e)}
      };
    \end{feynman}
  \end{tikzpicture}}}\\
\cZ_\phi \twoPt_\phi &= C^{\text{free}}_\phi + C^{\text{free}}_\phi \star \Pi_\phi \star \cZ_\phi \twoPt_\phi.
\end{aligned}\end{equation}
Here, $C^{\text{free}}_\phi(x,y)$ is the bare propagator of the field, assumed also to be conformal with scaling dimension $\Delta_\phi^\mathrm{free}$; $\Pi_\phi(x,y)$ is the one-particle-irreducible (1PI) self-energy for the field $\phi$; and $\star$ indicates convolution of these two-index objects.
Convolving with the inverses, we find 
\begin{equation}\begin{aligned} \label{eq:inversesRelationship}
[C_\phi^{\text{free}}]^{-1} = \frac{1}{\cZ_\phi} [\twoPt_{\phi}]^{-1} + \Pi_\phi.
\end{aligned}\end{equation} 
We assume that the free propagator drops out. For an IR or UV CFT, this means that each field must satisfy
\begin{equation}\begin{aligned} \label{eq:UVIRscaling}
\text{IR: }\Delta_\phi > \Delta_\phi^{\text{free}}; \qquad \text{UV: }\Delta_\phi < \Delta_\phi^{\text{free}}.
\end{aligned}\end{equation} 
The case of equality, $\Delta_\phi = \Delta_\phi^{\text{free}}$, leads to the long-range melonic models \cite{Gross:2017vhb,Benedetti:2019rja,Benedetti:2020rrq,Benedetti:2024mqx,Benedetti:2021wzt,Shen:2023srk}, which are briefly discussed in \cref{sec:LRmodels}.
Note that for canonical free field kinetic terms, any UV CFT must violate the unitarity bounds.

\subsection{Melonic theories}

We will specify a melonic-type theory with schematic interaction Lagrangian\footnote{Note that $\tilde{g}_m$ is the coupling constant with the conventional Feynman-diagrammatic normalizing factor associated with a melon. 
This differs from the coupling normalization commonly used for SYK.}
\begin{equation}\begin{aligned}
\sum_{m}^{n_m} \tilde{g}_m \prod_\Phi \frac{\Phi^{q^m_\Phi}}{q^m_\Phi!}.
\end{aligned}\end{equation}  
As discussed in the main text, we assume that there is an unspecified underlying mechanism (typically a disorder average or tensorial structure) that enforces the large-$N$ melonic dominance.
It is standard that the self-energy $\Pi_\phi$ of each field can then be resummed to
\begin{equation}\begin{aligned}
\Pi_\phi(x,y) =\sum_m \vcenter{\hbox{
\begin{tikzpicture}[anchor=base,baseline]
  \pgfmathsetmacro{\L}{4}
  \coordinate (L) at (-\L,0);
  \coordinate (R) at (\L,0);
  
  \draw[black, decorate, decoration={snake, amplitude=.4mm, segment length=2mm}]  (L) to[bend left=90,looseness=1.25] (R);
  \draw[black, decorate, decoration={snake, amplitude=.4mm, segment length=2mm}]  (L) to[bend left=75,looseness=1.20] (R);
  \draw[loosely dotted, thick] (0, {0.65*\L}) to (0, {0.55*\L});
  \node at (0.4, {(0.58*\L)}) {$q^m_{\Phi_i}$};
  \draw[black, decorate, decoration={snake, amplitude=.4mm, segment length=2mm}]  (L) to[bend left=60] (R);
  
  \draw[black, thick] (L) to[bend left=50] (R);
  \draw[black, thick] (L) to[bend left=40, looseness=1] (R);
  \draw[loosely dotted, thick] (0, {0.32*\L}) to (0, {0.22*\L});
  \node at (0.8, {(0.25*\L)}) {$(q^m_{\phi} -1)$};
  \draw[black, thick] (L) to[bend left=25, looseness=0.7] (R);
  
  \node at (-\L-0.3, 0) {$x$};
  \draw[loosely dotted, thick] (0, {0.09*\L}) to (0, {-0.09*\L});
  \node at (\L+0.3, 0) {$y$};
  
  \draw[black, dashed] (L) to[bend right=50] (R);
  \draw[black, dashed] (L) to[bend right=40, looseness=1] (R);
  \draw[loosely dotted, thick] (0, {-0.23*\L}) to (0, {-0.33*\L});
  \node at (0.4, {-(0.29*\L)}) {$q^m_{\Phi_k}$};
  \draw[black, dashed] (L) to[bend right=25, looseness=0.7] (R);
\end{tikzpicture}
}},
\end{aligned}\end{equation} 
where the propagator on each leg is the full resummed propagator $\cZ_\Phi D_\Phi$. Each diagram has symmetry factor $\prod_\Phi q^m_\Phi!/q^m_\phi$, so in the scaling limit \eqref{eq:inversesRelationship} becomes
\begin{subequations}
\begin{equation}\begin{aligned}
\frac{-1}{\cZ_\phi} [\twoPt_\phi(x-y)]^{-1} =\frac{1}{\dim(\rhoext_\phi)} \sum_{m} q^m_\phi \tilde{\mathfrak{g}}_m [\cZ_\phi \twoPt_\phi (x-y)]^{q^m_\phi -1} \, \prod_{\Phi \neq \phi} [\cZ_\Phi \twoPt_{\Phi}(x-y)]^{q^m_{\Phi}},
\end{aligned}\end{equation} 
where
\begin{equation}\begin{aligned}
\tilde{\mathfrak{g}}_m = \tilde{g}_m^2 \prod_\Phi \frac{\dim(\rhoext_\Phi)}{q^m_\Phi!}
\end{aligned}\end{equation} \label{eq:2ptSDEfull}
\end{subequations}
There are no signs from the fermions, as every $(-1)$ from a fermion loop cancels with a $(-1)$ from flipping the propagator $D_\Phi(x-y)= (-1)^{\mathrm{F}^\Phi} D_\Phi(y-x)$.
Dimensional analysis of \eqref{eq:2ptSDEfull} tells us immediately that the continuous data of this melonic CFT, being the scaling dimensions, are forced to obey the following equality for each melon $m$:
\begin{equation}\begin{aligned}
\sum_{\Phi} q^m_{\Phi} \Delta_\Phi = d. \label{eq:dimSumRuleGeneral}
\end{aligned}\end{equation} 
By \eqref{eq:inverseIsShadow}, the right-hand side must transform in the shadow representation of the field $\phi$, that is $\rhoext_{\tilde{\phi}}$. For this reason, we do not need to keep track of the Lorentz indices: the various contractions must end up giving an identity $\id_{\rho_\phi}$ on the right-hand side; likewise, since the symmetry group is assumed unbroken, the $\Gglobal$ indices marshal themselves into a $\id_{R_\phi}$.

As before, the $D_\phi$s are unit-normalized, and therefore we must have
\begin{equation}\begin{aligned}
[\twoPt_\phi (x-y)]^{q^m_\phi -1} \, \prod_{\Phi \neq \phi} [\twoPt_{\Phi}(x-y)]^{q^m_{\Phi}} & = \twoPt_{\tilde\phi}(x-y).
\end{aligned}\end{equation} 
Plugging that in to \eqref{eq:2ptSDEfull}, we find
\begin{equation}\begin{aligned}
[\cZ_\phi \twoPt_\phi(x-y)]^{-1} &= -\frac{1}{\dim \rhoext_\phi} \sum_{m} q^m_\phi \mathfrak{g}_m \twoPt_{\tilde\phi}(x-y),
\end{aligned}\end{equation} 
where we have defined a renormalized coupling constant for each melon
\begin{equation}\begin{aligned}
\mathfrak{g}_m \equiv \tilde{\mathfrak{g}}_m \left(\prod_{\Phi}^{\text{melon}} \cZ_\Phi^{q_\Phi}\right) = g_m^2 \prod_\Phi \frac{\cZ_\Phi^{q_\Phi^m}}{q^m_\Phi!}.
\end{aligned}\end{equation} 
Then using the identity \eqref{eq:inverseIsShadow}, for each field $\phi$ we obtain
\begin{equation}\begin{aligned}
\frac{\dim\rhoext_\phi}{\cN_\phi} =- \sum_{m} q^m_\phi  \mathfrak{g}_m. \label{eq:otherGeneralRule}
\end{aligned}\end{equation} 
Thus, if we have $n_m$ melons and $n_f$ fields $\{\phi\}$ in $\SO(d)\times \Gglobal$ representations $\rhoext_{\phi}$, we have $n_m$ equations from \cref{eq:dimSumRuleGeneral} and $n_f$ equations from \cref{eq:otherGeneralRule}. Therefore, generically we can find a solution for the unknown $\mathfrak{g}_m$s and $\Delta_{\phi}$s. 

Recall that we required $\Delta_{\phi} > \Delta^{\text{free}}_{\phi}$ in order to obtain consistent IR scaling. This picks out a polyhedron of allowed scaling dimensions in $\mathbb{R}^{n_f}$.
However, we can always tune the free scaling dimension of the fields by modifying the kinetic terms; and so we ignore this condition. %
Using \eqref{eq:FtEqIntMu}, we can eliminate $\cN_\phi$ in favour of $\odv{\Ft_\phi}{\Delta_\phi}$, and so derive:
\begin{subequations}
\begin{empheq}[box=\widefbox]{align}
\text{Given a set of melonic data: }& \{m: \,\, \prod_\Phi \Phi^{q^m_\Phi} \}\\
\text{For each melon: }& \sum_{\Phi}^{\text{melon}} q^m_{\Phi} \Delta_\Phi = d\\
\text{For each field $\phi$: }&  \odv{\Ft_\phi(\Delta_\phi)}{\Delta_\phi} =- \frac{\pi^{d+1}}{\Gamma(d+1)} \sum_{m} q^m_\phi  \mathfrak{g}_m \label{eq:melonicsummaryDiagrammatic}
\end{empheq}\label{eq:melonicsummary}
\end{subequations}
These equations give a complete solution for the conformal melonic limit, assuming appropriate IR (UV) scaling \eqref{eq:UVIRscaling}. 
We have assumed here that all melons are IR-relevant; if any melon $m$ is not, $\sum_{\Phi} q^m_\Phi \Delta_\Phi > d$, so the associated monomial operator is irrelevant, and the melon will drop out in the IR. 
These equations \eqref{eq:melonicsummary} are manifestly recoverable from the $\Ft$-extremization of \eqref{eq:FmaxSummary}, and therefore provide a conventional diagrammatic proof of $\Ft$-extremization. 
It is notable that we did not need to put in the factor of $-\sin(\pi d/2)$ by hand: we only needed to multiply \eqref{eq:otherGeneralRule} by the monotonic factor $\pi^{d+1}/d!$ to obtain \eqref{eq:melonicsummaryDiagrammatic}, and hence it was $\Ft$, rather than $F$, that naturally jumped out at us.

\end{subappendices}

\chapter{The quartic Yukawa model and general features of melonic CFTs}\label{chap:3dyuk}

\section{Introduction}

In this chapter, based on \cite{Fraser-Taliente:2024rql}, we wish to study the additional structure that these multi-field melonic theories have. Accordingly, we choose to study non-perturbatively the complete set of (symmetry-unbroken) large-$N$ melonic fixed points of the quartic Yukawa model; namely, the tensorial generalization of a Dirac fermion and real scalar with interaction $\phi^2 \bar{\psi}\psi+\phi^6$.
With two distinct types of field and two distinct interactions, this theory is especially rich -- hence our interest in it.

This theory has distinct types of fixed point, with and without fermions.
We will show that these fixed points match to those of Wilson-Fisher type found by a perturbative analysis around the upper critical dimension, $d=3-\epsilon$. This gives access to the flows between fixed points, which cannot be seen non-perturbatively via the methods of \cref{chap:fextr}.
Models of this kind have much richer structure than the melonic theories containing a single (vector/tensor) field. 
We will observe a complex network of fixed points as the dimension is varied, and study their stability and unitarity, both perturbatively and non-perturbatively.
Further investigating the spectrum of bilinears of this CFT, we will draw these observations together to suggest general features of the spectra, stability, and unitarity of melonic CFTs with the variation of $d$ and number of degrees of freedom.

\subsection{Chapter summary}

In \cref{sec:modelanalysis} we comment on the general features of the non-tensor $\phi^2 \bar\psi \psi + \phi^6$ model. 
We present the results of a perturbative analysis in $d=3-\epsilon$ of this theory to third order in the coupling constants, in \cref{sec:betas}, both in full generality and specialising to the large-$N$ melonic limit; here we will identify the apparent line of fixed points in both the bosonic and fermionic sectors.
In \cref{sec:SDEanalysis} we then use the large-$N$ melonic limit to analyse non-perturbatively the conformal field theories arising at the fixed points of the model, including a matching to the supersymmetric theory; we highlight a number of features which arise in the simpler setting of \hprismatic, before proceeding to the fermionic theories. 
In \cref{sec:bilinears}, this is then further developed in the specific case of the \lammelonic fixed point, where we use diagonalisation of the four-point kernel to obtain the exact spectrum $\{\Delta\}$ of bilinear operators in the OPE of the fundamental fields via a Bethe-Salpeter-like equation; this diagonalisation is exact in the melonic limit.
First we explain the computation, and then in \cref{sec:bilinearsCalculationResults} we study the properties of the spectrum so obtained.
A number of useful and technical results are provided in the appendices, particularly some concerning the $\phi^2 \bar{\psi}\psi +\phi^6$ model.
We postpone the conclusion to \cref{sec:3dyukOutlook}.

\subsection{Fermionic tensor models and the quartic Yukawa model}

Previously the tensorial Gross-Neveu model was studied in $d=3$ \cite{Delporte:2020rce}.
This model is non-renormalizable at finite $N$, but becomes renormalizable in the large-$N$ limit.
The theories at finite and large $N$ are fundamentally different, which makes it less likely that interesting structure persists to finite $N$.
This motivates looking for a theory of fermions in the large-$N$ limit where the precursor theory is renormalizable at finite $N$.
However, the obvious step of studying a Yukawa-like interaction $\lambda \phi \bar{\psi}\psi+ g \phi^4$ is not possible for a rank-3 tensor model\footnote{This is because there are no invariants in $\phi_{abc} \bar\psi_{def} \psi_{ghi}$ that could go in an action. 
We note that there is a disorder-averaged approach to this melonic $d\le 4$ theory \cite{Prakash:2022gvb}.
However, this model loses the RG flow information and the connection to the $\phi^6$ and prismatic bosonic tensor models.
Analogous models have also been studied in a supersymmetric quantum mechanical context \cite{Marcus:2018tsr}.}, and so we move on to what we call the \textit{quartic Yukawa model}. %

\subsubsection{The quartic Yukawa model}

We consider a theory with two tensor fields -- one real scalar $\phi_{abc}$ and one Dirac fermion $\psi_{abc}$ -- with an interaction term given by the tensorial generalization of
\begin{equation}\begin{aligned}
V(\phi, \psi) = \frac{\lambda}{2} \phi^2 \bar\psi \psi + \frac{h}{6!} \phi^6.
\end{aligned}\end{equation}
This is the unique extension of the tensorial sextic bosonic scalar field theory \cite{Giombi:2018qgp,Benedetti:2019rja,Harribey:2021xgh} to a theory containing fermions, which is still renormalizable in $d\le 3$. This Lagrangian without the $\phi^6$ potential was first studied -- in the vector large-$N$ limit -- by Popović in 1977 \cite{Popovic:1977cq} (commented on in \cref{sec:Popovic}); with the additional potential, it was studied in exactly three dimensions in \cite{Dilkes:1997vc,McKeon:1999vx}, and then with the addition of a Chern-Simons coupling in \cite{Jack:2016utw}; a non-tensor version with the scalar potential was briefly considered in \cite{Herzog:2022jlx}; with Majorana spinors instead, it gives the supersymmetric Wess-Zumino model $(\Phi^4)$ in 2+1 dimensions \cite{DeWolfe:2019etx}. We will analyse the melonic fixed points both perturbatively in $d=3-\epsilon$ and via the non-perturbative Schwinger-Dyson equations. Additionally, as we will see, identifying non-perturbatively all perturbative symmetry-unbroken fixed points requires the introduction of a non-dynamical auxiliary scalar field $X_{abc}$. %

\subsubsection{Results of the analysis of the quartic Yukawa model}

In this chapter, our analysis will find the known fixed points of the bosonic sector (the melonic \cite{Benedetti:2019rja,Harribey:2021xgh} and prismatic \cite{Giombi:2018qgp}), but will also uncover three new fermionic generalizations of these melonic fixed points. We will see in perturbation theory an apparent collision of these fixed points for a particular dimension of the gamma matrices, but it will be resolved by a non-perturbative analysis. An apparent line of fixed points in both the fermionic and bosonic sectors (in the same direction for both), will also be found, but we will be unable to establish if it is just an artifact of the order of the perturbative calculation.
Focusing on the simplest of these new CFTs, which we refer to by \lammelonic{}, we will investigate its spectrum as a function of $d$. We will frequently draw comparison to the simpler sextic prismatic fixed point (which we here call \hprismatic), and so various results will also be presented for it.

Specifically, we will consider the reality of the scaling dimensions of the spectrum, as a probe of stability and, separately, unitarity. 
As we mentioned in \cref{sec:continuousD}, for non-integer-$d$, at high scaling dimension, we expect the so-called evanescent operators/negative-norm states to appear, making the CFT non-unitary \cite{Hogervorst:2015akt,DiPietro:2017vsp,Ji:2018yaf}; these operators disappear in integer $d$. 
Nonetheless, it is interesting to study the extent to which we can consider these theories to be unitary (in the sense of having properties usually associated with unitarity), just as in the case of the bosonic melonic field theories \cite{Giombi:2017dtl,Benedetti:2019ikb} -- precisely because they are the dimensional continuation of theories that become unitary in integer dimension \cite{Fei:2015oha}. %

\section{General comments on the quartic Yukawa model} \label{sec:modelanalysis}

\subsection{Our model, \texorpdfstring{$\phi^2 \bar{\psi}\psi+\phi^6$}{phi\^2 psibar psi plus phi\^6}}

The most general renormalizable theory of a Dirac fermion and a real scalar field in Euclidean $d\le 3$ is described by the following Lagrangian
\begin{equation}\begin{aligned} \label{eq:Neq1Lagrangian}
\cL&=-\bar{\psi}\left(\slashed{\partial}+M\right) \psi + \frac{1}{2}\phi (-\partial^2 + m^2) \phi + V_{\mathrm{int}}(\phi, \psi),\\
V_{\mathrm{int}}(\phi,\psi) &\equiv \frac{\lambda}{2} \phi^{2} \overline{\psi} \psi + \frac{g}{4 !} \phi^{4} + \frac{h}{6!}\phi^6.
\end{aligned}
\end{equation}
We have: suppressed counterterms $Z_i$ and the RG scale $\mu$ for convenience; assumed a $\mathbb{Z}_2$ symmetry to remove $\phi$-odd terms; used the usual positive $(+++\cdots)$ signature in the conventions of \cite{zinn-justin_quantum_2002} (see \cref{app:calcNotes} for more details).
The (complex) Dirac fermions are used because they exist in any dimension, unlike Majorana fermions.
The real scalar fields are used partly for simplicity, and partly to ensure that we can straightforwardly access the prismatic fixed points. 
The fixed points for a complex scalar field will be straightforward to obtain from the general results below; we need only add an index of $\SO(2)\cong \gU(1)$.

It will prove convenient, for the purposes of accessing the full range of IR CFTs (specifically, those of prismatic type), to also consider the addition of a non-dynamical real auxiliary field $X$; this is just as in the standard $\phi^4$ vector model \cite{Moshe:2003xn}.
\begin{equation}
\cL_{\mathrm{aux}} = \half X^2 + V_{\mathrm{int,aux}}(X,\phi),  \quad V_{\mathrm{int,aux}}(X,\phi) = \frac{\rho}{3!} X \phi^3
\end{equation}
Since $X$ enters only quadratically, it can of course be integrated out exactly, leading only to a shift in the value of $h$.

\subsection{Comments on the model}
We make the following observations about this model as a non-tensorial quantum field theory.
\begin{itemize}
    \item As in the case of the Wilson-Fisher fixed point, which exists in $d=4-\epsilon$ for $\phi^4$, we expect to be able to find a fixed point of the renormalization group in $d=3-\epsilon$.
    The renormalized coupling constants should have a perturbative expansion around zero for $\epsilon \ll 1$; this is done in \cref{sec:vectorModelAnalysis}. 
    In standard perturbative QFT, this analysis is trustworthy for small $\epsilon$, with the fixed point colliding with the trivial fixed point in $d=3$ exactly. However, in the case of the melonic theories, it is possible to exactly solve for the scaling dimensions of these theories for all values of $d$, while also matching on to the perturbative $\epsilon \ll 1$ expansion.
    \item It is possible to consistently set $\lambda=0,h\neq 0$ or $\lambda \neq 0, h=0$.
    That $\lambda=0$ is allowed is obvious, since then the fermions are non-interacting; that $h=0$ does not flow if $\lambda\neq 0$ is more surprising, and occurs because the one-loop $\lambda^3$ contribution to $\expval{\phi^6}$ is not divergent (even though if working in exactly $3$d $\Tr(\gamma_\mu \gamma_\nu \gamma_\rho) \propto \epsilon_{\mu\nu\rho} \neq 0$). 
    \item Any value of $M\neq 0$ explicitly breaks parity symmetry in $3$d. %
    \item To sidestep the confusion of fermions in non-integer dimensions (reviewed in \cite{Pannell:2023tzc,Jack:2023zjt}), we will only deal with Dirac fermions, which are well-defined in any integer dimension. We will follow the standard approach of leaving the dimension of the gamma matrices as a free parameter, $T \equiv \Tr[\spinid]$. 
    The ratio of the number of fermionic degrees of freedom to the number of bosonic degrees of freedom will also prove a useful parametrization, $r\equiv 2T$.
    We discuss this point further in \cref{sec:rparameter}.
\end{itemize}

\subsection{Indexology}

Now, let us consider the $\cal{N}$-vector version of the model. Sprinkling $O(\cN)$ indices:
\begin{align}\label{eq:NvectorLagrangian}
\begin{split}
\cL&=-\bar{\psi}_I\left(\slashed{\partial} \delta_{IJ}+M_{IJ}\right) \psi_J 
+ \frac{1}{2} \phi_I (-\partial^2 \delta_{IJ} + m^{2}_{IJ}) \phi_J + V_{\mathrm{int}}(\phi,\psi)\\
 V_{\mathrm{int}}(\phi,\psi) &= \frac{\lambda_{(IJ)KL}}{2} \phi_I \phi_J \overline{\psi}_K \psi_L+\frac{g_{(IJKL)}}{4 !} \phi_I\phi_J\phi_K\phi_L + \frac{h_{(IJKLMN)}}{6!}\phi_I\phi_J\phi_K\phi_L\phi_M\phi_N
\end{split}\\
 \cL_{\mathrm{aux}} &=\frac{1}{2} X_I X_I + \frac{\rho_{I(JKL)}}{3!} X_I \phi_J \phi_K \phi_{L}\label{eq:auxLagrangian}
\end{align}
Once again, note that integrating out the non-dynamical $X$ simply leads to a redefinition of $h_{(IJKLMN)}$. 
Thus, for the perturbative analysis, we do not need to deal with the additional complication of a separate coupling $\rho$.

Depending on the precise values chosen for the coupling constants, the symmetry group in these cases may be $\gU(N)^3$ instead: however, we need only ensure that we pick the faithfully acting subgroup of a product of three general linear groups -- if we account for the symmetry factors, the melonic limit is unchanged.

We will be using the term \textit{superindex} to refer to a grouped set of three indices of $\gO(N)^3$. That is, $\phi_I = \phi_{i_r i_b i_g}$, with $i_r,i_g,i_b=1,\ldots,N$.

\section{\texorpdfstring{$3-\epsilon$}{3-epsilon} beta functions at large \NPDF in the melonic limit} \label{sec:betas}

We first perform a standard perturbative analysis of the multi-field theory, for finite $N$, and completely arbitrary couplings. For a similar analysis of the marginal theory of scalars and fermions in $4$d ($\phi \bar\psi \psi + \phi^4$), see \cite{Osborn:2020cnf,Pannell:2023tzc}. The Feynman loop integrals are standard (see \cref{app:loopIntegrals}), except for being in $d=3$; note, of course, that diagrams with different tensor structures may have identical momentum structure. %

\subsection{Vector beta functions in \texorpdfstring{$d=3-\epsilon$}{d=3-epsilon}} \label{sec:vectorBetaFunctions}

We calculate the beta functions and field anomalous dimensions for the Lagrangian \eqref{eq:NvectorLagrangian} in $\overline{\mathrm{MS}}$ scheme.
In the following: Greek indices are dummy indices which are summed over; $F$ and $G$ are the indices of an anti-fermion and fermion respectively; the Latin $B_i$ indices indicate the index of a boson, which must be symmetrized over with weight one; we set $s=1/(8\pi)$. Then, to the indicated order in the marginal coupling constants\footnote{Non-marginal coupling constants have been set to zero, as they could be obtained from these calculations via the \textit{dummy field method} of \cite{Martin:1993zk} (pedagogically reviewed in \cite{Schienbein:2018fsw}). We do not, for example, use $g_{IJKL}$ here, as to find $\epsilon$-perturbative fixed points, it, like the field masses, must be tuned to zero.} $\lambda_{(IJ)KL}$ and $h_{(IJKLMN)}$, we find

\begin{subequations}
{\allowdisplaybreaks
\begin{align}
    \begin{split}
    \label{eq:vectorBetas3mepsLam}
    &\beta[\lambda]_{B_1 B_2 FG} =-\epsilon  \lambda_{B_1 B_2 FG}+ \frac{1}{3} s^4 h_{B_1 \beta \gamma \delta \varepsilon \zeta } h_{B_2 \beta \gamma \delta \varepsilon \mu } \lambda_{\zeta \mu FG}\\
    &\quad +2\lambda_{B_1 \beta FG}\left[\frac{s^2}{3} T\lambda_{B_2 \zeta \eta \theta } \lambda_{\beta \zeta \theta \eta } + \frac{s^4}{90}  h_{B_2 \kappa \gamma \delta \varepsilon \zeta } h_{\beta \kappa \gamma \delta \varepsilon \zeta}\right]\\ 
    &\quad+ \frac{s^2}{3}  \left[\lambda_{B_1 B_2 F\delta } \lambda_{\varepsilon \zeta \delta \theta } \lambda_{\varepsilon \zeta \theta G}+\lambda_{B_1 B_2 \gamma G} \lambda_{\varepsilon \zeta \gamma \theta } \lambda_{\varepsilon \zeta \theta F}\right]\\
    &\quad+\frac{s^2}{3} \left(6 T \lambda_{B_1 \beta \gamma \delta } \lambda_{B_2 \zeta \delta \gamma } \lambda_{\beta \zeta FG}+6 \lambda_{B_1 B_2 \gamma \delta } \lambda_{\varepsilon \zeta \delta G} \lambda_{\varepsilon \zeta F\gamma }\right.\\
    &\quad \quad\left.+12 \lambda_{B_1 \beta \gamma G} \lambda_{\beta \zeta F\eta } \lambda_{B_2 \zeta \eta \gamma }+12 \lambda_{B_1 \beta F\delta } \lambda_{B_2 \zeta \delta \theta } \lambda_{\beta \zeta \theta G}\right) +O(\lambda^4,\ldots)|_{\text{sym } B_i},
    \end{split}\\
    \begin{split}
    \label{eq:vectorBetas3mepsh}
    &\beta[h]_{B_1 B_2B_3B_4B_5B_6} = -2 \epsilon  h_{B_1 B_2B_3B_4B_5B_6}+\frac{20}{3} s^2 h_{B_1 B_2 B_3\eta \nu \xi }h_{B_4 B_5 B_6\eta \nu \xi } \\
    &\quad +6 h_{B_1 B_2B_3B_4B_5\nu } \left[\frac{s^2}{3} T \lambda_{B_6\rho \tau \sigma } \lambda_{\nu \rho \sigma \tau } +\frac{s^4}{90}  h_{B_6\rho \sigma \tau \upsilon \varphi } h_{\nu \rho \sigma \tau \upsilon \varphi }\right]\\
    &\quad +5 h_{B_1 B_2B_3B_4\nu \xi } (6 s^2 T \lambda_{B_5\nu \sigma \tau } \lambda_{B_6\xi \tau \sigma }+s^4 h_{B_5\nu \sigma \tau \upsilon \varphi } h_{B_6\xi \sigma \tau \upsilon \varphi })\\
    &\quad -\tfrac{15}{2} \pi ^2 s^4 h_{B_1B_2\beta \eta \nu \xi } h_{B_3B_4\beta \eta \upsilon \varphi } h_{B_5B_6\nu \xi \upsilon \varphi }-80 s^4 h_{B_1B_2B_3\eta \nu \xi } h_{B_4B_5\eta o\chi \psi } h_{B_6\nu \xi o\chi \psi }\\
    & \quad - 360 s^2 T \left(\lambda_{B_1B_2\gamma \delta } \lambda_{B_3\zeta \delta \theta }\lambda_{B_4B_5\theta \kappa } \lambda_{B_6\zeta \kappa \gamma } +\lambda_{B_1B_2\gamma \delta } \lambda_{B_3B_4\delta \kappa }\lambda_{B_5\zeta \kappa \eta } \lambda_{B_6\zeta \eta \gamma }  \right)\\
    &\quad +O(h^4,\lambda^5,\ldots) |_{\text{sym } B_i},
    \end{split}\\
    &\gamma^\phi_{B_1B_2} = \frac{s^2}{3} T \lambda_{B_1\beta \gamma \delta }\lambda_{B_2\beta \delta \gamma } + \frac{s^4}{90} h_{B_1\beta \gamma \delta \varepsilon \zeta } h_{B_2\beta \gamma \delta \varepsilon \zeta } +O(\lambda^3,\ldots) |_{\text{sym } B_i}, \label{eq:generalVecGammaPhi}\\
    &\gamma^\psi_{FG} = \frac{s^2}{3} \lambda_{\alpha \beta F\delta } \lambda_{\alpha \beta \delta G}+O(\lambda^3,\ldots) \label{eq:generalVecGammaPsi},
\end{align}
}\label{eq:generalVectorResults}

\end{subequations}
where we can isolate the contributions of the anomalous dimensions to the beta functions in square brackets. Note that in this formulation, to obtain the $N=1$ beta functions, we need only drop all the index structure -- for example, $\beta_\lambda = -\epsilon \lambda + \frac{1}{45} s^2 (s^2 h ( 15 h \lambda + h \lambda) + \ldots) + \ldots$, etc. This is not a complete four-loop calculation, as we have not calculated the $O(\lambda^4)$ contributions to the anomalous dimension, or the $O(\lambda^5)$ contributions to $h$ which appear at four loops.

We have three checks on our results here.
\begin{enumerate}
    \item The anomalous dimensions precisely match the leading order conformal calculation of \eqref{eq:conformalAnalysisDims}.
    \item We were able to reproduce some finite-$N$ two-loop beta functions calculations: those of the bosonic sector matched appendix A of \cite{Giombi:2018qgp}; and, other than one discrepancy (see \cref{app:JackPooleDiscrepancy}), the full theory matched the $d=3$ results of \cite{Jack:2016utw} to the order calculated there. %
    Likewise, the bosonic sector matches the vectorial calculation of \cite{Pisarski:1982vz}, up to their definition of $\gamma_\varphi \equiv 2 \gamma_\phi$; it also matches \cite{Jepsen:2020czw} and \cite{Hager:2002uq}, although we note that the latter contains errors at six-loop order \cite{TrenoginEtAl}.
    \item The anomalous dimensions at the fixed points agree (to four loops) with the non-perturbative melonic analysis of the Schwinger-Dyson equations of \cref{sec:SDEanalysis}, which is a completely orthogonal calculation. This should happen, as the scaling/anomalous dimensions of physical operators $\gamma_{\cO}$ evaluated at fixed points are physical and so scheme-independent: $\Delta_{\cO} = \Delta_{\cO,\mathrm{free}} + \gamma_{\cO}$. 
    Of course, away from fixed points, this is no longer true.
\end{enumerate}

\subsection{A simple example of tensor contractions}

As described above, we can calculate these quantities for the tensor model by breaking each superindex $I=1,\ldots, \cN=N^3$ into three separate indices $(ijk)$, each transforming under a separate $\gO(N)^3$. Taking the melonic limit then requires substituting suitable combinations of delta functions for each of the coupling constants. 
It is then an exercise in (automated) index contractions to evaluate each of the tensor beta functions above, and to decompose them to find the beta functions of each $\gO(N)^3$-invariant coupling constant. Taking the large-$N$ limit, assuming no symmetry breaking of $\gO(N)^3$, we obtain the required results for the large-$N$ melonic theory.

We will now illustrate this, taking a rank two (matrix) model of scalars for simplicity, because the procedure is identical but more amenable to compact presentation. Consider a matrix model with fields $\phi_{ab},\psi_{ab}$, each transforming in the $\Box\times \Box$ of $\gO(N)\times \gO(N)$. This has $\kappa_{IJKL}=\kappa_{(i_r i_b)(j_r j_b)(k_r k_b) (l_r l_b)}$\footnote{These brackets do not mean symmetrization.}. Note that capital letters indicate a superindex of $\gO(N)^2$, $I=(i_r i_b)$, and so both the red and blue indices. We also impose the symmetry under \textit{colour-averaging}, which is switching $r \leftrightarrow b$. Then, to evaluate an example index contraction $\beta[\kappa]_{B_1 B_2 FG} \supset \alpha \kappa_{B_1 \alpha F\delta}\kappa_{\alpha B_2 \delta G}$, we first must find the symmetric tensorial form of $\kappa_{IJKL}$:
\begin{equation}\begin{aligned}
&V(\phi,\psi)=\frac{\kappa_{IJKL}}{2} \phi_I \phi_J \textcolor{red}{\bar{\psi}_K} \textcolor{gray}{\psi_L} \equiv \frac{\kappa_{\mathrm{dt}}}{2} \times \frac{1}{2} \left(\phi_{ab} \phi_{cb} \bar\psi_{ad}\psi_{cd} +\phi_{ab} \phi_{ac} \bar\psi_{db}\psi_{dc}\right)\\ %
&\implies \kappa_{IJKL}=\kappa_{(i_r i_b)(j_r j_b)(k_r k_b) (l_r l_b)} = \frac{\kappa_{\mathrm{dt}}}{2}(\delta_{i_r k_r} \delta_{i_b j_b}\delta_{j_r l_r} \delta_{k_b l_b}+ \delta_{i_r j_r} \delta_{i_b k_b} \delta_{j_b l_b} \delta_{k_r l_r})|_{\text{sym }I\leftrightarrow J}\\
& \equiv \frac{\kappa_{\mathrm{dt}}}{2} O^{dt}_{(i_r i_b)(j_r j_b)(k_r k_b) (l_r l_b)}  \label{eq:exampleKappa}
\end{aligned}\end{equation}
Given an example expression for the beta function,
\begin{equation}\begin{aligned} \label{eq:exampleBeta}
&\beta[\kappa]_{B_1 B_2 F G} =\alpha \kappa_{B_1 \alpha F\delta}\kappa_{\alpha B_2 \delta G} |_{\text{sym } B_1 \leftrightarrow B_2},
\end{aligned}\end{equation}
we can then substitute in \eqref{eq:exampleKappa} expression for $\kappa$, and perform the index contractions:
\begin{equation}\begin{aligned}
&\beta[\kappa]_{(i_r i_b)(j_r j_b)(k_r k_b) (l_r l_b)} = \kappa_{(i_r i_b)(j_r j_b)(\gamma_r \gamma_b) (\delta_r \delta_b)}  \kappa_{(k_r k_b)(l_r l_b)(\delta_r \delta_b) (\gamma_r \gamma_b)}|_{\text{sym}}\\
&=\alpha \frac{N \kappa_{\mathrm{dt}}^2}{8} O^{dt}+ \text{other structures}\\
&=\beta[\kappa_{\mathrm{dt}}] O^{dt} + \cdots,
\end{aligned}\end{equation}
where $|_{\text{sym}}$ now indicates that we perform $\frac{1}{4!} \sum_{\text{perms}}$ over the $2!$ permutations of the 2 multi-indices $I=(i_r,i_b),J=(j_r,j_b)$. %

Thus, if \eqref{eq:exampleBeta} were the general beta function, the coupling constant defined by $\kappa_{\mathrm{dt}}$ would flow according to the beta function
\begin{equation}\begin{aligned}
\beta[\kappa_{\mathrm{dt}}]= \alpha \frac{N \kappa_{\mathrm{dt}}^2}{8}
\end{aligned}\end{equation}
These delta function combinations $O^i$ are much more conveniently represented visually:
\begin{equation}\begin{aligned}
V(\phi,\psi)=\frac{\kappa_{IJKL}}{2} \phi_I \phi_J \textcolor{red}{\bar{\psi}_K} \textcolor{gray}{\psi_L}= \frac{\kappa_{\mathrm{dt}}}{2} \times \frac{1}{2} \left(\begin{tikzpicture}[baseline={([yshift=-.5ex]current bounding box.center)}]
\draw[color=red] (1,0) -- (1,1) (0,1) -- (0,0);
\draw[color=blue] (1,1) -- (0,1) (0,0) -- (1,0);
\fill[color=black] (0,0) circle (0.05) (1,0) circle (0.05); 
\fill[color=gray] (1,1) circle (0.05);
\fill[color=red] (0,1) circle (0.05);
\end{tikzpicture}+\begin{tikzpicture}[baseline={([yshift=-.5ex]current bounding box.center)}]
\draw[color=blue] (1,0) -- (1,1) (0,1) -- (0,0);
\draw[color=red] (1,1) -- (0,1) (0,0) -- (1,0);
\fill[color=black] (0,0) circle (0.05) (1,0) circle (0.05); 
\fill[color=gray] (1,1) circle (0.05);
\fill[color=red] (0,1) circle (0.05);
\end{tikzpicture}\right).\\
\end{aligned}\end{equation}
Here, red lines indicate contraction of the first index, and blue lines indicate contraction of the second index between two fields. $\phi_I$ is indicated by a black dot; $\bar{\psi}_K$ a red dot; $\psi_J$ a grey dot. A red line between $\phi_I$ and $\phi_J$ corresponds to a delta function $\delta_{i_r j_r}$, etc. Thus, these graphical depictions map to an analytical expression of tensors. When we introduce the third index on these fields below, we shall use the natural generalization of this notation, with additional green lines.

\newpage
\subsection{The \texorpdfstring{$\gO(N)^3$}{O(N)\^3} model and its beta functions} \label{sec:findingPerturbativeBetas}

\begin{figure}[h]
\includegraphics[width=\textwidth]{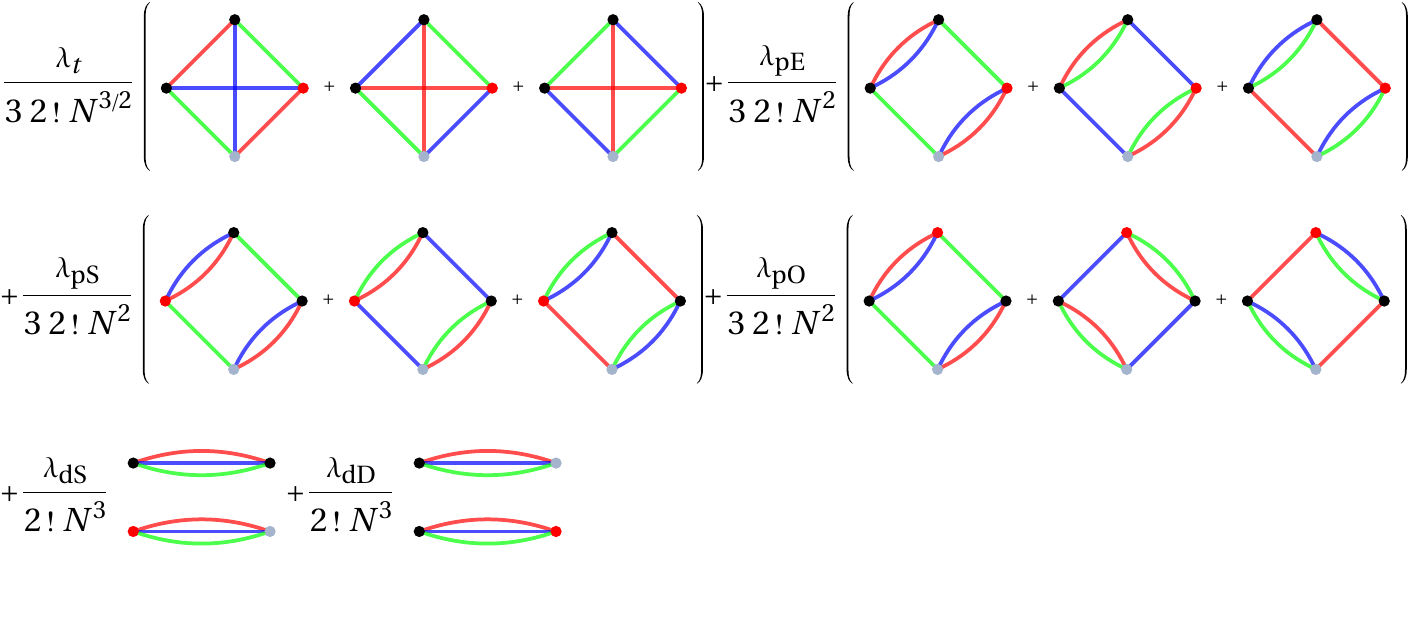}
\includegraphics[width=\textwidth]{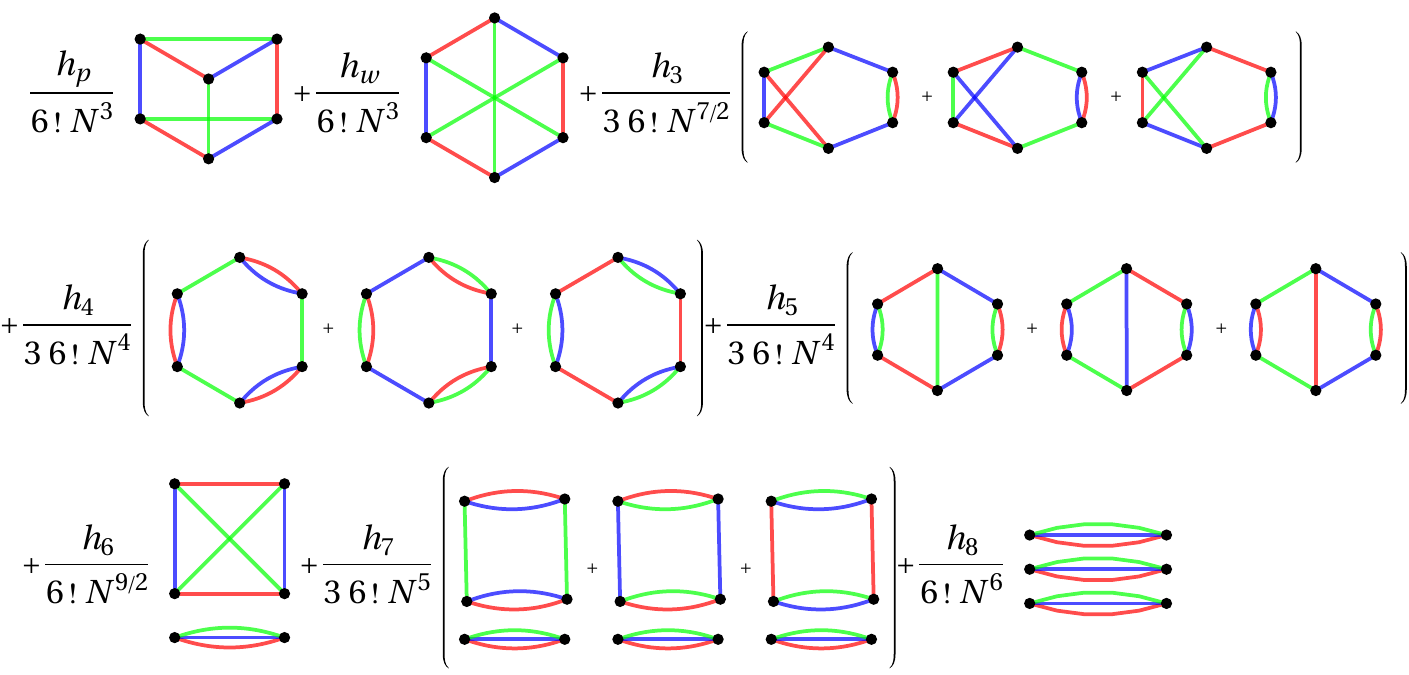}
\caption{The interaction terms $V_{\mathrm{int}}$ for the $\gO(N)^3$ $\phi^2 \bar{\psi}\psi + \phi^6$ model. Black dots indicate $\phi_{abc}$s; red dots $\bar{\psi}_{abc}$s; grey dots $\psi_{abc}$s. Red, green, and blue lines indicate contractions between the first, second, and third index. Note the symmetry of this potential under the interchange of the three colours.} \label{fig:potentialWithH}
\end{figure}

The marginal sector of the potential for the tensorial version of the $\phi^2 \bar{\psi}\psi$ model that we will use is depicted graphically in \cref{fig:potentialWithH}. 
By marginal, we mean that the masses and $g_{IJKL}$ have been pre-emptively set to zero: we keep only the operators that have dimension $\propto \epsilon$ in the $d=3-\epsilon$ free theory.
This is the most general marginal potential in $d=3$ that satisfies an $\gO(N)^3 \times S_3$ symmetry, where the $S_3$ corresponds to the permutation symmetry of each of the $\gO(N)$ groups (i.e. the three colours). Without the $S_3$, for example, we would need a different coupling constant for each of the three 3-colourings of the $\lambda_t$ invariant \begin{tikzpicture}[baseline={([yshift=-.5ex]current bounding box.center)}]
\draw[color=black] (0.4,0) -- (0.4,0.4) (0,0.4) -- (0,0) (0.4,0.4) -- (0,0.4) (0,0) -- (0.4,0) (0,0) -- (0.4,0.4) (0,0.4) -- (0.4,0);
\fill[color=black] (0,0) circle (0.05) (0.4,0) circle (0.05); 
\fill[color=gray] (0.4,0.4) circle (0.05);
\fill[color=red] (0,0.4) circle (0.05);
\end{tikzpicture} visible in the figure.

\noindent A full analytic expression is given in \cref{app:potential}.
In addition, these results can be used to reproduce the results of the complex sextic \cite{Benedetti:2019rja} and real prismatic \cite{Giombi:2018qgp} tensor models; see \cref{app:potentialcomparison} %
for details.

Applying the general formulae of \eqref{eq:generalVectorResults}, the field anomalous dimensions are 
\begin{subequations}
\begin{align}
    \gamma_\phi &=\frac{\lambda_t^2 s^2}{18}  T + \frac{s^4}{97200} \left(3 \hp^2+9 \hw^2-50 \lambda_t^4 T (22 T+20)\right) + O(\lambda_t^5,\ldots), \label{eq:anomDimPhiGeneralPert}\\
    \gamma_\psi &= \frac{\lambda_t^2 s^2}{18} - \frac{\lambda_t^4 s^4}{486} (1+8 T)+O(\lambda^5,\ldots), \label{eq:anomDimPsiGeneralPert}
\end{align}
\end{subequations}
with the field scaling dimensions at any fixed points being $\Delta_\phi= \frac{d-2}{2} + \gamma_\phi$, $\Delta_\psi= \frac{d-1}{2} + \gamma_\psi$. The only couplings that contribute to the two-point function of the fields, and therefore this anomalous dimension, in the large-$N$ limit, are $\{\lambda_t, \hp, \hw\}$; these are the \textit{maximally single trace (MST) couplings}.
These are defined by being the only invariants in \cref{fig:potentialWithH} that are \textit{maximally single trace}, in that each of the 2-colored subgraphs obtained by forgetting the third colour are connected. 
In this way, they are similar to a single-trace
invariant in a matrix model \cite{Benedetti:2020seh}.
\begin{itemize}
    \item $\lambda_t$ is the tetrahedral interaction, MST as in the standard bosonic $\phi^4$ \cite{Giombi:2017dtl}.
    \item $\hp$ is the prismatic interaction, MST as in the prismatic model \cite{Giombi:2018qgp}.
    \item $\hw$ is the wheel graph (which is also the complete bipartite graph $K_{3,3}$, MST just as in the bosonic $\phi^6$ model \cite{Benedetti:2019eyl}).
\end{itemize}

These three have beta functions that depend only on each other; if these three beta functions are zero, then the fixed point values of the remaining coupling constants are determined. Up to $O(\lambda_t^5,\dots)$, we have the MST coupling beta functions
\begin{equation}\begin{aligned}
\beta[\lambda_t]&=-\epsilon  \lambda_t+\frac{1}{9}  (T+1) \lambda_t^3 s^2+ \frac{\lambda_t}{3}\left(\frac{\hp^2+3\hw^2}{5400} -\frac{\lambda_t^4}{162} (T (11 T+10) + 16 T +2) \right)s^4,\\
\beta[\hp]&=-2 \hp \epsilon + \hp  \left(\frac{1}{3} T \lambda_t^2 +\frac{\hp}{90} \right)s^2+ \hp\left(\frac{\hp^2+3 \hw^2}{5400}-\frac{\lambda_t^4}{162} T (11 T+10)\right)s^4,\\
\beta[\hw]&=-2 
\hw \epsilon +\hw\left(\frac{1}{3} T \lambda_t^2\right) s^2+ \hw\left(\frac{\hp^2+3 \hw^2}{5400}-\frac{ \lambda_t^4}{162} T (11 T+10)\right)s^4.%
\end{aligned}\end{equation}
The full set of 14 beta functions for the marginal coupling constants is presented in \cref{app:betas}%
\footnote{For the large-$N$ melonic theory, the momentum structure of the surviving diagrams means that it is simple enough to calculate all leading-$N$ corrections to four loops. We therefore can calculate to order $\sim h^3, \lambda^5$, rather than the $\sim h^3,\lambda^3$ general results given in \cref{sec:vectorBetaFunctions}. Thus, we indicate with $O(s^5)$ the order of 5-loop contributions, since when working in $d=3-\epsilon$, each loop order brings a factor of $s=1/(8\pi)$.}.

We solve for the Wilson-Fisher-like fixed points of the flow ($\beta_i =0$) perturbatively in $\epsilon$ for $d =3-\epsilon$; we can trust the results for $\epsilon \ll 1$. %

The trivial and pure bosonic fixed points, where the fermions decouple (i.e. $\lambda_i=0$ exactly), are well known and correspond to the free theory, the sextic bosonic \cite{Benedetti:2019rja}, and the prismatic models \cite{Giombi:2018qgp}:
\begin{equation}\begin{aligned}\label{eq:bosonicFixedPoints}
\mathrm{trivial}:& \quad \text{ all zero}\\
\hmelonic:&\quad  s^2 \hw = \pm 60s^2   \sqrt{\epsilon} \implies \gamma_\phi = \epsilon/3\\
\hprismatic:&\quad s^2  \hp = (180 \epsilon - 540 \epsilon^2) \implies \gamma_\phi = \epsilon^2 %
\end{aligned}\end{equation}
These formulae for the fixed points are perturbative solutions in $\epsilon$, truncated to the order shown. The fixed points with interacting fermions ($\lambda_i \neq 0$) are the following:
\begin{equation}\begin{aligned} \label{eq:fermionicFixedPoints}
\hlammelonic:&\, s \lambda_t  = \pm_1 \sqrt{3}\sqrt{\epsilon}, \,  s^2  \hw = \pm_2 30 i  \sqrt{2T-4} \sqrt{\epsilon}, \text{ (independent signs)}\\
& \implies \gamma_\phi = \epsilon/3, \, \gamma_\psi = \epsilon/6\\
\lammelonic:& \, s \lambda_t = \pm \left(\tfrac{3 \sqrt{\epsilon}}{\sqrt{T+1}} + \tfrac{(11T^2 +26T +2) \epsilon^{3/2}}{4(T+1)^{5/2}}\right) \\
& \implies \gamma_\phi = \tfrac{\epsilon}{2} \tfrac{T}{T+1} + \tfrac{T(5T-8)}{12(T+1)^3} \epsilon^2, \quad \gamma_\psi = \tfrac{\epsilon}{2} \tfrac{1}{T+1}- \tfrac{T(5T-8)}{12(T+1)^3} \epsilon^2 \\
\hlamprismatic:& \, s \lambda_t = \pm \left(\tfrac{3 \sqrt{\epsilon}}{\sqrt{T+1}} + \tfrac{(4T^2 +19T -5) \epsilon^{3/2}}{2(T+1)^{5/2}}\right),\, s^2 \hp = -\tfrac{90(T-2)s^2}{T+1}\epsilon \\
& \implies \gamma_\phi = \tfrac{\epsilon}{2} \tfrac{T}{T+1} + \tfrac{(T-1)(2T-3)}{3(T+1)^3} \epsilon^2, \quad \gamma_\psi = \tfrac{\epsilon}{2} \tfrac{1}{T+1} - \tfrac{(T-1)(2T-3)}{12(T+1)^3} \epsilon^2 
\end{aligned}\end{equation}
We note that:
\begin{enumerate}
\item The particular names that we have given these fixed points will be justified during the SDE analysis in \cref{sec:SDEanalysis}, where we shall also re-derive these results to all orders in $\epsilon$. This will confirm that the anomalous dimensions in \hmelonic and \hlammelonic are exactly linear in $\epsilon$, and so have no higher order corrections. We re-iterate that the anomalous dimensions at fixed points are physical and thus must be scheme-independent. The values of the coupling constants, however, are not.
\item The precise order in $\epsilon$ that we can calculate to depends on the particular fixed point; for example, in $h\lambda_{\mathrm{melonic}}$, we are only able to calculate to $\sqrt{\epsilon}$ order, because the four-loop result for the $h$s only reaches order $h^3$. 
\item At the fermionic fixed points, $\gamma_\phi + \gamma_\psi=\epsilon/2$ to the order known. 
From \cref{sec:SDEanalysis}, we expect this to be true to all orders in $\epsilon$. 
This is a non-trivial check, as the implied cancellation of terms between \eqref{eq:anomDimPsiGeneralPert} and \eqref{eq:anomDimPhiGeneralPert} occurs only at the fixed point.

\end{enumerate}

\subsection{Values of the other coupling constants} \label{sec:othercouplingconstants}

Solving the remaining 11 beta functions requires non-zero values for some coupling constants. We summarise in the following table, where we give each coupling constant value at the fixed point to all known orders in $\epsilon=3-d$. All coupling constants not shown or blank are zero to the order calculated. First, the bosonic fixed points:
\begin{center}
$\begin{array}{l|l|l}
 \mathrm{} & \hmelonic & \hprismatic \\
 \hline
 \hp & \mathrm{} & 180 \epsilon -540 \epsilon ^2 \\
 \hw & \pm 60 \sqrt{\epsilon } & \mathrm{} \\
 h_4 & \mathrm{} & -90 \pi ^2 \epsilon ^2 \\
 h_5 & -270\pm (-540) \sqrt{\epsilon } & 2160 \epsilon ^2+1080 \epsilon  \\
 h_7 & \frac{2700}{7}  \pm 540 \sqrt{\epsilon }& 540 \epsilon -540 \left(\pi ^2-83\right) \epsilon ^2 \\
 h_8 &-\frac{1090}{7}  \pm (-180) \sqrt{\epsilon } & -180 \left(\pi ^2-126\right) \epsilon ^2 \\
\end{array}$
\end{center}
The fermionic fixed points: %
\begin{center}
$\begin{array}{l|l|l|l}
 \mathrm{} & \lammelonic & \hlammelonic & \hlamprismatic\\
 \hline
 \lambda_t & \pm 3 \sqrt{\frac{\epsilon }{1+T}} & \pm_1 \sqrt{3} \sqrt{\epsilon } & \pm 3 \sqrt{\frac{\epsilon }{1+T}} \\
 \hp & \mathrm{} & \mathrm{} & -\frac{90 (T-2) \epsilon }{T+1} \\
 \hw & \mathrm{} & \pm_2 30 i \sqrt{2} \sqrt{(T-2) \epsilon } & \mathrm{} \\
 h_5 & \mathrm{} & -\frac{405 (T-2)}{T-3} - 9 h_w & \frac{1080 (T-2) \epsilon }{(T+1) (3 T-2)} \\
 h_7 & \mathrm{} & \frac{2430 (T-2) (2 T-5)}{(T-3) (8 T-21)} + 9 h_w & -\frac{270 (T-2) (9 T+2) \epsilon }{(T+1) (3 T-2) (3 T-1)} \\
 h_8 & \mathrm{} & -\frac{5 (T-2) \left(808 T^2-3978 T+4905\right)}{(T-3) (2 T-5) (8 T-21)} - 3h_w & \frac{1620 (T-2) T (2 T+1) \epsilon }{(T+1) (3 T-2) (3 T-1) (5 T-1)} \\
\end{array}$
\end{center}
We now can make the following observations, recalling first that $T=2$ is a distinguished value, being the minimal dimension of the gamma matrices in $d=3$:
\begin{enumerate}
\item In \hlammelonic, $\pm_1$ can be chosen independently of $\pm_2$.
\item In \hmelonic, $h_{5,7,8}$ are not small, just as was found in \cite{Benedetti:2019rja}. The same is true for \hlammelonic, $h_{5,7,8}$, unless $T$ is taken near $2$. This might make us doubt the validity of the perturbative approach. However, there is still a possibility of the perturbative series being trustworthy if these large coupling constants always appear together with the MST $\lambda_t, h_{1,2}$. In \cref{sec:SDEanalysis} we will identify these same fixed-points, but in a way that ignores the non-MST coupling constants.
\item All of the fermionic fixed points seem (to this order) to reduce to the same fixed point for $T=2$. However, our non-perturbative analysis in \cref{sec:SDEanalysis} will show these fixed points to be distinct even at $T=2$. Therefore, this collision is an artefact of our inability to calculate to higher loop order. In particular, it is unrelated to supersymmetry.
\item If we were interested in calculating the $O(1/N)$ corrections here, we would have to take care with the way we take the large-$N$ expansion. 
This is because beyond optimal scaling the large-$N$ and small $\epsilon$ expansions do not commute \cite{Harribey:2021xgh, Jepsen:2023pzm}: we would need to first take $\epsilon N\to\infty$, and then $\epsilon\to 0$.
\end{enumerate}

\subsection{Scaling dimensions of interaction terms} \label{sec:stabmats}

We can calculate the conformal dimensions of the interaction terms by calculating the eigenvalues of the stability matrix \eqref{eq:stabMat}. The stability matrices themselves are too large to reproduce, so we give only the eigenvalues here. Each of these eigenvalues equals $\Delta_\cO-d$, where $\Delta_\cO$ is the scaling dimension of one of the marginal $\gO(N)^3$ singlet operators in the theory; these are linear combinations of the 14 operators appearing in the potential of \cref{fig:potentialWithH}. Where it is easy to do so, we also indicate the direction in the space of coupling constants that a given eigenvalue corresponds to (to leading order). 

First, for the bosonic theories, where the $\lambda$s do not mix with the $h$s:
\begin{subequations}
{\scriptsize
\begin{align} %
\hmelonic &: h\mathrm{s}: \left\{h_8: 30 \epsilon ,14 \epsilon ,10 \epsilon , 6 \epsilon ,2 \epsilon ,0,\left(
\begin{array}{cc}
 4 \epsilon  & 1 \\
 0 & 4 \epsilon  \\
\end{array}
\right)\right\} + \lambda \mathrm{s}\mathrm{: } \left\{\frac{29 \epsilon }{3},\frac{5 \epsilon }{3},-\frac{\epsilon }{3},-\frac{\epsilon }{3},-\frac{\epsilon }{3},-\frac{\epsilon }{3}\right\} \label{eq:hmelonicEigs}\\
\hprismatic &: h\mathrm{s}: \{6 \epsilon ,2 \epsilon ,2 \epsilon,-2 \epsilon ,-2 \epsilon ,-2 \epsilon ,-2 \epsilon ,-2 \epsilon \} + \lambda \mathrm{s}\mathrm{: } \left\{-\epsilon ,-\epsilon ,-\epsilon , -\epsilon , -\epsilon ,-\epsilon\right\}.
\end{align}
}

Next, for the theories with coupled fermions, where the two types do mix:
{\scriptsize
\begin{align}
\lammelonic &: \Bigg\{h_8: \frac{2 (5 T-1) \epsilon }{T+1},\frac{2 (3 T-1) \epsilon }{T+1}, \lambda_{d_S}:  3 \epsilon ,\frac{2 (2 T-1) \epsilon }{T+1},\frac{2 (2 T-1) \epsilon }{T+1},2 \epsilon ,\frac{(3 T-2) \epsilon }{T+1},\frac{6 \epsilon }{T+1},\epsilon ,\\
& \quad \quad \quad \frac{2 (T-1) \epsilon }{T+1},\frac{2 \epsilon }{T+1},\frac{(T-2) \epsilon }{T+1},\frac{(T-2) \epsilon }{T+1},-\frac{2 \epsilon }{T+1}\Bigg\} \notag\\
\hlammelonic &: \left\{2 \epsilon ,\frac{7-2 T+\sqrt{4 (T-2)^2 + 9}}{3} \epsilon ,\frac{2 \epsilon }{3},\frac{1}{3} (7-2 T) \epsilon ,0,-2 (T-3) \epsilon ,-\frac{4}{3} (T-3) \epsilon ,-\frac{2}{3} (T-3) \epsilon ,\right. \\
& \quad \quad \quad  \left.-\frac{2 \epsilon }{3}, \frac{7-2 T-\sqrt{4 (T-2)^2 + 9}}{3} \epsilon ,(11-4 T) \epsilon ,\frac{2}{3} (21-8 T) \epsilon ,2 (5-2 T) \epsilon ,6 (5-2 T) \epsilon \right\}\notag\\
\hlamprismatic &: \bigg\{\frac{2 (5 T-1) \epsilon }{T+1},\frac{2 (3 T-1) \epsilon }{T+1},3 \epsilon ,\frac{2 (2 T-1) \epsilon }{T+1},2 \epsilon ,\frac{(3 T-2) \epsilon }{T+1},\frac{6 \epsilon }{T+1},\frac{6 \epsilon }{T+1},\epsilon ,\frac{2 \epsilon }{T+1},\\
& \quad \quad \quad \frac{2 \epsilon }{T+1},\frac{(T-2) \epsilon }{T+1},-\frac{(T-2) \epsilon }{T+1},-\frac{2 \epsilon }{T+1}\bigg\}\notag.
\end{align}
}
\end{subequations}
\noindent All the fixed points that differ only by $\pm_i$ have the same eigenvalues: this suggests that the seemingly distinct fixed points related by switching those signs describe the same CFT. We might have expected this from the diagrammatic expansion, where all diagrams are constructed out of melons, and hence the MST couplings always appear squared.

Some of these operators are marginally irrelevant ($\Delta_{\cO}>d$), and some are marginally relevant ($\Delta_{\cO} < d$) -- in the case of the fermionic fixed points, which are which depends on $T$: therefore these CFTs are saddle points of RG flow.

\begin{enumerate}
\item Despite \hmelonic being the theory of a real scalar field, in the strict large-$N$ limit it is identical to the complex sextic bosonic model of \cite{Benedetti:2019rja}, as mentioned above. 
Indeed, the $\gO(N)$ invariants that do not descend from $\gU(N)$ invariants (see \cref{app:potentialcomparison}) are zero in this theory. 
The stability matrix is not diagonalisable in this case, but has a Jordan normal form \eqref{eq:hmelonicEigs} that shows it to be a logarithmic CFT\footnote{See, for example \cite{Cardy:2013rqg,Hogervorst:2016itc}. 
The same phenomenon also occurs in the strict large-$N$ limit of the Fishnet Conformal Field Theories (FCFTs) of \cite{Kazakov:2022dbd}.} at the fixed point, and so automatically non-unitary, to leading order in $N$. 
Working to higher orders in the coupling constants will not modify this. However, at subleading orders in $N$, this is lifted, due to new $h^2,h^3$ terms in the beta functions \cite{Harribey:2021xgh}. 
\item \hlammelonic has a diagonalisable stability matrix, unlike its bosonic cousin \hmelonic.
\item The zero eigenvalues in the case of both \hmelonic and \hlammelonic are in the same direction $-\hp + 3 h_5 - 3 h_7 + h_8$, indicating the presence of a similar marginal operator in both; hence these CFTs appear to sit on a line of fixed points. 
However, this free parameter would appear at an order higher than we were able to calculate, so this degeneracy could be lifted at the next order. 
This deformation would have broken the $\gU(N)$ symmetry of the complex sextic model (see \cref{app:potentialcomparison} for details), and therefore was invisible to \cite{Benedetti:2019rja}. 
We discuss this line further in \cref{sec:3dyukOutlook}.
\item None of the other theories are obviously non-unitary; indeed, for all values of $T$, even $T<0$, all of these anomalous dimensions for all fermionic fixed points are real; likewise for $\epsilon<0$. This is particularly notable in the case of \hlammelonic, which has complex coupling constants. This observation agrees with what is seen in the bilinear analysis of \lammelonic in \cref{sec:bilinears}, where at least perturbatively close to $d=3-\epsilon$, the bilinear spectrum is fully real.
\end{enumerate}

\subsection{The flow system for the three MST couplings}

\begin{figure}
    \centering
    \includegraphics[width=0.8\textwidth]{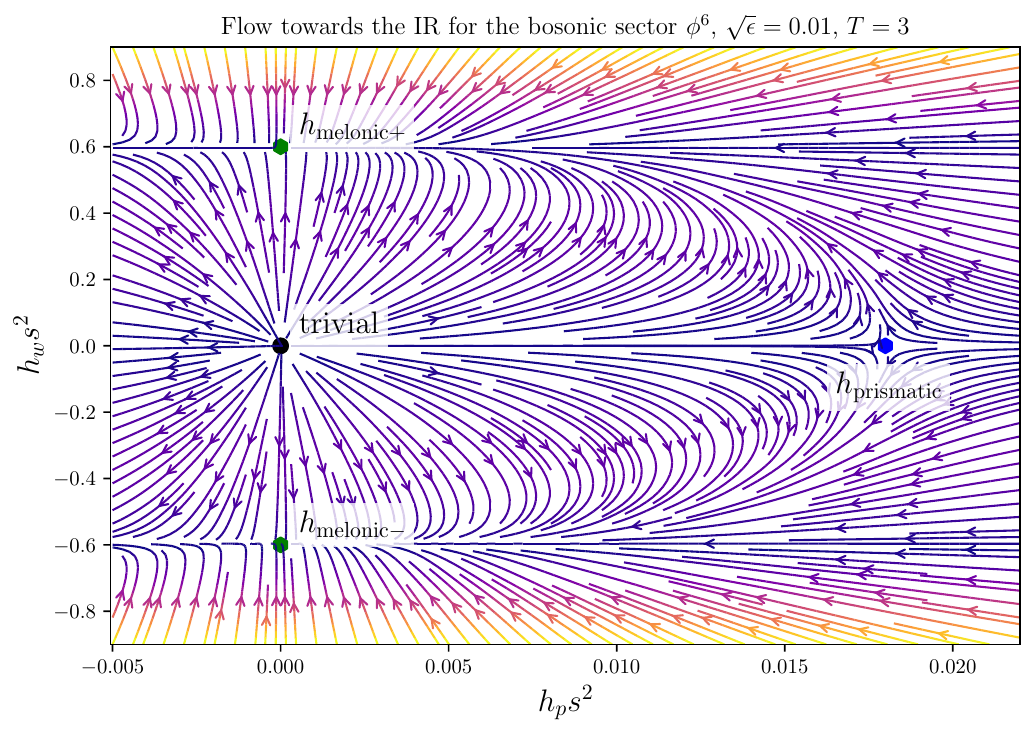}
    \caption{Flow towards the IR for the bosonic sector of the theory -- when the fermions are decoupled. We expect the two \hmelonic fixed points to represent the same CFT, as they only differ by the sign of the coupling constant. All non-trivial fixed points shown here are saddle points.}
    \label{fig:flowSystemForBosonicSector}
\end{figure}

\begin{figure}
    \centering
    \includegraphics[width=0.8\textwidth]{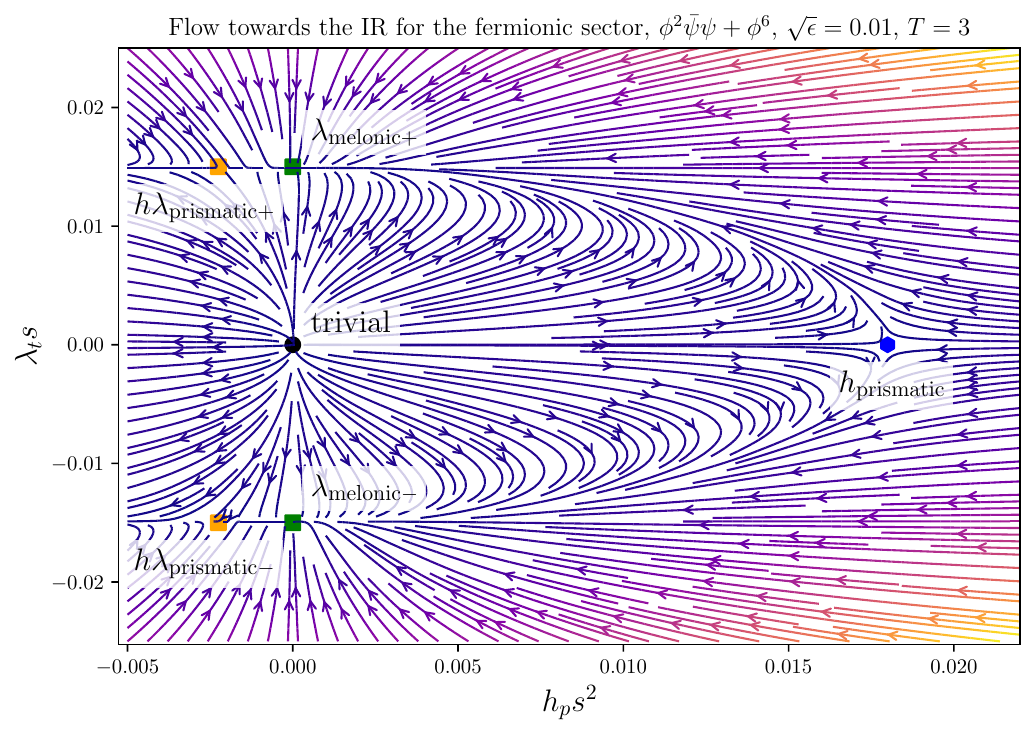}
    \caption{Flow towards the IR for the fermionic sector of the theory; this plot shares its $x$-axis with \cref{fig:flowSystemForBosonicSector}, and also displays both the trivial fixed point and \lammelonic. We once again expect the two \lammelonic fixed points to refer to the same CFT; likewise for \hlamprismatic. All non-trivial fixed points shown here are saddle points.}
    \label{fig:flowSystemForFermionicSector}
\end{figure}

To understand the relationship between these CFTs in the space of coupling constants, it is convenient to reduce to the MST coupling constants $(\lambda_t, \hp, \hw)$, and consider their flow under the renormalization group towards the IR. 
For concreteness, in the following plots of the flow we take $\sqrt{\epsilon} =1/100$. In \cref{fig:flowSystemForBosonicSector} we show the bosonic sector: the flow between the existing sextic bosonic models in the literature, \hmelonic \cite{Benedetti:2019rja,Harribey:2021xgh} and \hprismatic \cite{Giombi:2018qgp}. 
In \cref{fig:flowSystemForFermionicSector}, we show the sector with coupled fermions with $T=2$, which avoids the collision of the perturbative coupling constants visible in \cref{eq:fermionicFixedPoints} at $T=2$. 
As shown by the signs of the eigenvalues, all the non-trivial fixed points here are saddle points of RG flow; IR-stable to some deformations, but IR-unstable to others. The fate of the fixed points perturbed in the unstable directions is unknown.

We are unable to illustrate the location of \hlammelonic on this plot, due to the imaginary nature of $\hw$ at that fixed point. However, if we restrict to the subspace $\hp=0$, we can write down the beta functions for the squared coupling constants $H_w=\hw^2$ and $\Lambda_t=\lambda_t^2$, up to $O\left(\Lambda_t^4,\ldots\right)$ terms
\begin{equation}\begin{aligned}
\beta[\Lambda_t]|_{\hp=0}&= -2\epsilon  \Lambda_t +\frac{2}{9}  (T+1) \Lambda_t^2 s^2 + \frac{\Lambda_t}{3} \left(\frac{H_w}{900}-\frac{\lambda_t^2}{81} (T (11 T+10) + 16 T+2) \right)s^4\\
\beta[H_w]|_{\hp=0}&=-4 \epsilon H_w + \frac{2}{3} T H_w   \Lambda_t s^2+ \left(\frac{H_w^2}{900}-\frac{\lambda_t^2}{81} H_w T (11 T+10) \right)s^4.
\end{aligned}\end{equation}
The fixed points here occur for real values of these squared coupling constants -- albeit negative in the case of $\hlammelonic$:
\begin{equation}\begin{aligned}
\hmelonic &: \, s^4 H_w= 3600\epsilon + O(\epsilon^2),\\
\lammelonic &: \, s^2 \Lambda_t =  \frac{9 \epsilon }{T+1}+\frac{3 (T (11 T+26)+2) \epsilon ^2}{2 (T+1)^3}+O\left(\epsilon ^3\right)\\
\hlammelonic &: \,  s^4 H_w = -1800(T-2)\epsilon + O(\epsilon^2), s^2 \Lambda_t = 3\epsilon + O(\epsilon^2)
\end{aligned}\end{equation}
The reason that we could not do this for $\hp$ is due to the quadratic term appearing in the beta function; this is because $\hp$ is only non-zero at the fixed points of prismatic-type theories, where it is not true that the coupling constants only appear squared. In such theories, the actual melonic coupling is $\rho X \phi^3$, so only $\rho^2$ appears. However, since $\rho^2 \sim \hp$, an $\hp^2$ term appears in the beta function.

\begin{figure}[H]
    \centering
    \includegraphics[width=0.8\textwidth]{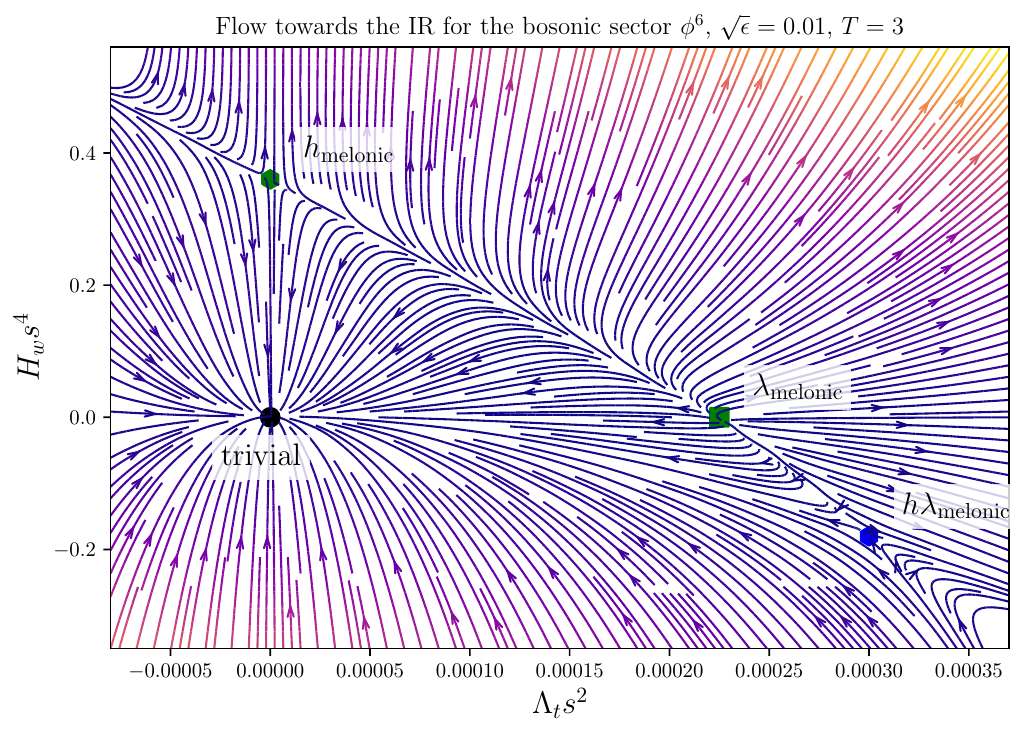}
    \caption{Flow towards the IR for the melonic-type theories, where we visualise the flow of the value of the coupling constants squared, to avoid the imaginary nature of $h_w$ for the \hlammelonic theory.}
    \label{fig:flowSystemForSquaredCCs}
\end{figure}

Tuning $T$ moves both $\lammelonic$ and $\hlammelonic$, with an apparent collision occurring for $T \to 2$; a collision that we will discover to be an artifact of perturbation theory in \cref{sec:hlamprismaticSDEfound}. Increasing $T$ seems to move \lammelonic{} along the line separating \hmelonic and \hlammelonic, towards the latter.

In the next section we look at these same fixed points from a complementary point of view: that of the Schwinger-Dyson equations at large-$N$, which can calculate the scaling dimensions of the fundamental fields and singlet bilinears to all orders in $\epsilon$. 
However, the price we pay is having to sit precisely at the fixed points, therefore losing any information about the RG flow, or the values of the non-MST coupling constants.

\section{Schwinger-Dyson equation analysis} \label{sec:SDEanalysis}

In this section, we use the Schwinger-Dyson equations (SDEs) to investigate the fixed points non-perturbatively, assuming that the scaling symmetry at the fixed points is promoted to full conformal symmetry. 
This enhancement of symmetry is common, but not guaranteed. Various assumptions are required for a proof of this, including unitarity and locality \cite{Nakayama:2013is}, and that the conformal limit is taken before the large-$N$ limit \cite{Benedetti:2020yvb}. 
We shall use standard momentum-space Schwinger-Dyson equations, though we note that we would obtain identical results via $\Ft$-extremization. 
The equivalence of the two procedures was proven in \cref{app:diagrammaticProof}, and indeed, in this section we are just performing a momentum-space version of that general analysis.

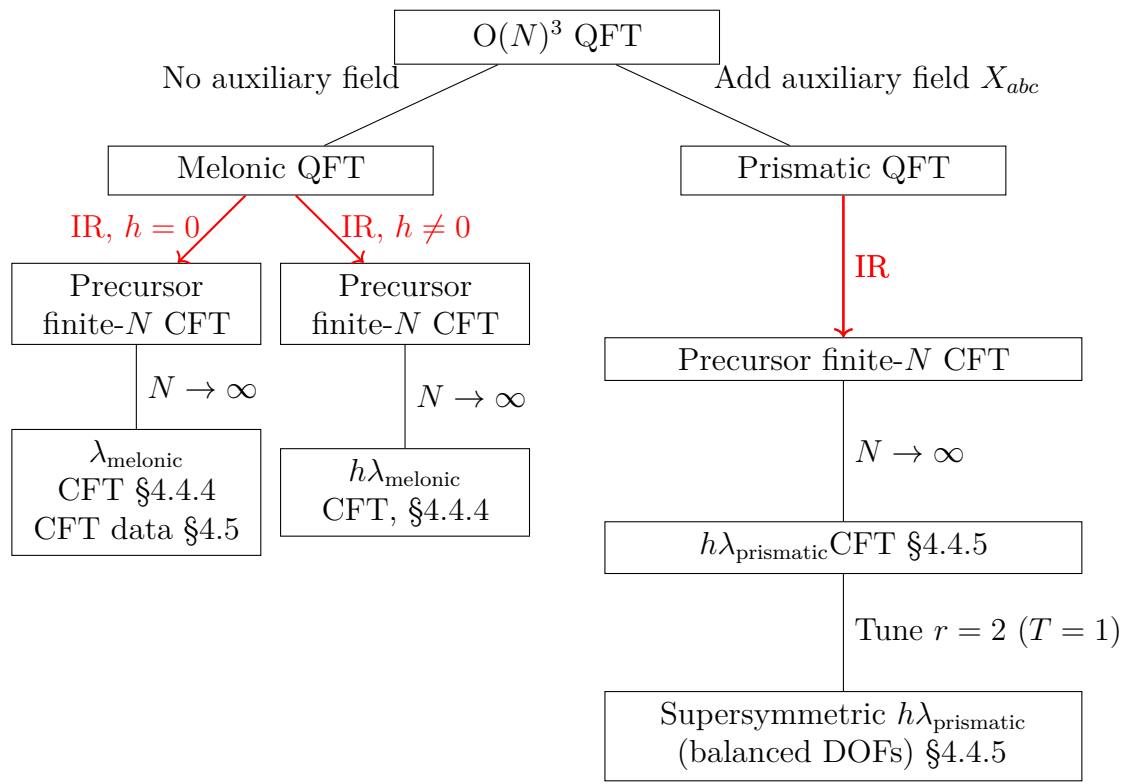
\begin{figure}[h]
\centering
\begin{tikzpicture}[node distance=2.5cm, auto]
    \node (theory) [rectangle, draw, text width=4cm, text centered] {$\gO(N)^3$ QFT};
    \node (prismatic) [rectangle, draw, text width=4cm, text centered, below right of=theory, xshift=2cm, yshift=0cm] {Prismatic QFT};
    \node (melonic) [rectangle, draw, text width=4cm, text centered, below left of=theory, xshift=-2cm, yshift=0cm] {Melonic QFT};
    \node (lamMelonic1) [rectangle, draw, text width=3cm, text centered, below left of=melonic, yshift=0cm] {Precursor finite-$N$ CFT};
    \node (hlamMelonic1) [rectangle, draw, text width=3cm, text centered, below right of=melonic, yshift=0cm] {Precursor finite-$N$ CFT};
    \node (hlprismatic1) [rectangle, draw, text width=6cm, text centered, below of=prismatic, yshift=0cm] {Precursor finite-$N$ CFT};
    \node (lamMelonic2) [rectangle, draw, text width=3cm, text centered, below of=lamMelonic1, yshift=0cm] {\lammelonic CFT \S\ref{sec:lammelonicSDEfound}\\CFT data \S\ref{sec:bilinears}};
    \node (hlamMelonic2) [rectangle, draw, text width=3cm, text centered, below of= hlamMelonic1, yshift=0cm] {\hlammelonic CFT, \S\ref{sec:lammelonicSDEfound}};
    \node (hlprismatic2) [rectangle, draw, text width=6cm, text centered, below of=hlprismatic1, yshift=0cm] {\hlamprismatic CFT \S\ref{sec:hlamprismaticSDEfound}};
    \node (hlSUSYic) [rectangle, draw, text width=6cm, text centered, below of=hlprismatic2, yshift=0cm] {Supersymmetric \hlamprismatic\\(balanced DOFs) \S\ref{sec:qGenAndSUSYprismatic}};

    \path (theory) edge node {Add auxiliary field $X_{abc}$} (prismatic);
    \path (theory) edge node[above left] {No auxiliary field} (melonic);
    \path (melonic) edge[red, ->, thick] node[left] {IR, $h=0$} (lamMelonic1);
    \path (melonic) edge[red,->, thick] node[right] {IR, $h\neq 0$} (hlamMelonic1);
    \path (prismatic) edge[red, ->,thick] node {IR} (hlprismatic1);
    \path (lamMelonic1) edge node {$N\to\infty$} (lamMelonic2);
    \path (hlamMelonic1) edge node {$N\to\infty$} (hlamMelonic2);
    \path (hlprismatic1) edge node {$N\to\infty$} (hlprismatic2);
    \path (prismatic) edge[red, ->,thick] node {IR} (hlprismatic1);
    \path (hlprismatic2) edge node {Tune $r=2$ ($T=1$)} (hlSUSYic);
\end{tikzpicture} %
\caption{Map of interacting fixed points in the theory in generic dimension (that is, ignoring the theories with free fermions, \hmelonic{} and \hprismatic{}). The CFT we obtain after the IR limit is determined by the initial coupling constants. We also ignore all symmetry-breaking possibilities. Adding the auxiliary field does not change the theory, but does allow us to access new fixed points non-perturbatively.} \label{fig:FPstructure} %
\end{figure}

A similar analysis to what will follow is provided for the short-range and long-range $\gO(N)^3$ bosonic tensor models in the series of papers \cite{Benedetti:2019eyl, Benedetti:2019ikb, Benedetti:2020yvb}. 
However, in the present case we will observe a more complex structure of fixed points, of which we give a cartoon in \cref{fig:FPstructure}. We stress in the cartoon that we are always taking the IR limit before taking the $N\to\infty$ limit, to avoid any contributions from unwanted diagrams (discussed in section 6 of \cite{Benedetti:2020yvb}). These exact solutions will match precisely the Wilson-Fisher-like fixed points found perturbatively in $d=3-\epsilon$ in \cref{sec:betas}.

By ignoring the auxiliary field in the SDEs, we are assuming that of the MST couplings, $h_w \neq 0$ and $h_p =0$. On the other hand, with the auxiliary field, we allow for $h_p\neq 0$: this gives us access to the prismatic-type fixed points via the SDEs.

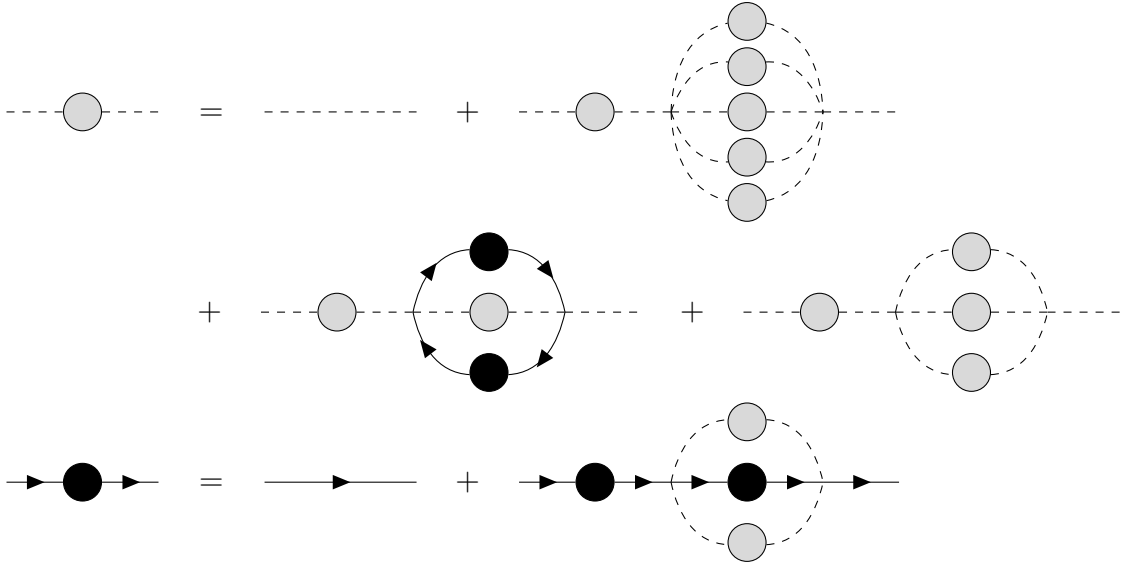
\begin{figure}[H]
  \begin{align*}%
\vcenter{\hbox{\begin{tikzpicture}
  \begin{feynman}[every blob={/tikz/fill=gray!30,/tikz/inner sep=2pt}]
    \vertex[small, blob] (m) at (0,0) {};
    \vertex (a) at (-1,0) ;
    \vertex (b) at ( 1,0);
    \diagram* {
      (a) --[scalar] (m) --[scalar] (b),
      };
  \end{feynman}
\end{tikzpicture}}}
\quad &= \quad \vcenter{\hbox{\begin{tikzpicture}
  \begin{feynman}
    \vertex (a) at (-1,0) ;
    \vertex (b) at ( 1,0);
    \diagram* {
      (a) --[scalar] (b),
      };
  \end{feynman}
\end{tikzpicture}}} \quad+ \quad\vcenter{\hbox{\begin{tikzpicture}
    \begin{feynman}[every blob={/tikz/fill=gray!30,/tikz/inner sep=2pt,/tikzfeynman/small}, every edge=scalar]
   \vertex[small,blob] (m) at (0,0) {};
    \vertex (a) at (-1,0) ;
    \vertex (start) at ( 1,0);
    \vertex[small, blob] (centreblob0) at ( 2,1.2) {};
    \vertex[small, blob] (centreblob1) at ( 2,0.6) {};
    \vertex[small, blob] (centreblob2) at ( 2,0) {};
    \vertex[small, blob] (centreblob3) at ( 2,-0.6) {};
    \vertex[small, blob] (centreblob4) at ( 2,-1.2) {};
    \vertex (end) at ( 3,0);
    \vertex (e) at ( 4,0);
    \diagram* {
      (a) -- (m) -- (start) -- (centreblob2) -- (end) -- (e),
      (start) --[quarter left, out=30] (centreblob0) --[quarter left,in=150] (end),
      (start) --[quarter left, out=30] (centreblob1) --[quarter left,in=150] (end),
      (start) --[quarter right, out=-30] (centreblob3) --[quarter right,in=210] (end),
      (start) --[quarter right, out=-30] (centreblob4) --[quarter right,in=210] (end)
      };
    \end{feynman}
  \end{tikzpicture}}}\\ %
  &+\quad \vcenter{\hbox{\begin{tikzpicture}
    \begin{feynman}[every blob={/tikz/fill=gray!30,/tikz/inner sep=2pt,/tikzfeynman/small}]
   \vertex[small,blob] (m) at (0,0) {};
    \vertex (a) at (-1,0) ;
    \vertex (start) at ( 1,0);
    \vertex[small, blob,black] (centreblob1) at ( 2,0.8) {};
    \vertex[small, blob] (centreblob2) at ( 2,0) {};
    \vertex[small, blob,black] (centreblob3) at ( 2,-0.8) {};
    \vertex (end) at ( 3,0);
    \vertex (e) at ( 4,0);
    \diagram* {
      {[edges={scalar}] (a) -- (m) -- (start) -- (centreblob2) -- (end) -- (e)},
      (start) --[fermion, quarter left, out=30] (centreblob1) --[fermion, quarter left,in=150] (end),
      (start) --[anti fermion, quarter right, out=-30] (centreblob3) --[anti fermion,quarter right,in=210] (end)
      };
    \end{feynman}
  \end{tikzpicture}}} \quad +\quad \vcenter{\hbox{\begin{tikzpicture}
    \begin{feynman}[every blob={/tikz/fill=gray!30,/tikz/inner sep=2pt,/tikzfeynman/small}, every edge=scalar]
   \vertex[small,blob] (m) at (0,0) {};
    \vertex (a) at (-1,0) ;
    \vertex (start) at ( 1,0);
    \vertex[small, blob] (centreblob1) at ( 2,0.8) {};
    \vertex[small, blob] (centreblob2) at ( 2,0) {};
    \vertex[small, blob] (centreblob3) at ( 2,-0.8) {};
    \vertex (end) at ( 3,0);
    \vertex (e) at ( 4,0);
    \diagram* {
      (a) -- (m) -- (start) -- (centreblob2) -- (end) -- (e),
      (start) --[quarter left, out=30] (centreblob1) --[quarter left,in=150] (end),
      (start) --[quarter right, out=-30] (centreblob3) --[quarter right,in=210] (end),
      };
    \end{feynman}
  \end{tikzpicture}}}\\% Fermion SDE:
\vcenter{\hbox{\begin{tikzpicture}
  \begin{feynman}[every blob={/tikz/fill=gray!30,/tikz/inner sep=2pt}]
    \vertex[small, blob,black] (m) at (0,0) {};
    \vertex (a) at (-1,0) ;
    \vertex (b) at ( 1,0);
    \diagram* {
      (a) --[fermion] (m) --[fermion] (b),
      };
  \end{feynman}
\end{tikzpicture}}}
\quad &= \quad \vcenter{\hbox{\begin{tikzpicture}
  \begin{feynman}
    \vertex (a) at (-1,0) ;
    \vertex (b) at ( 1,0);
    \diagram* {
      (a) --[fermion] (b),
      };
  \end{feynman}
\end{tikzpicture}}} \quad+ \quad
\vcenter{\hbox{\begin{tikzpicture}
    \begin{feynman}[every blob={/tikz/fill=gray!30,/tikz/inner sep=2pt,/tikzfeynman/small}]
   \vertex[small,blob,black] (m) at (0,0) {};
    \vertex (a) at (-1,0) ;
    \vertex (start) at ( 1,0);
    \vertex[small, blob] (centreblob1) at ( 2,0.8) {};
    \vertex[small, blob,black] (centreblob2) at ( 2,0) {};
    \vertex[small, blob] (centreblob3) at ( 2,-0.8) {};
    \vertex (end) at ( 3,0);
    \vertex (e) at ( 4,0);
    \diagram* {
      {[edges={fermion}] (a) -- (m) -- (start) -- (centreblob2) -- (end) -- (e)},
      (start) --[scalar, quarter left, out=30] (centreblob1) --[scalar, quarter left,in=150] (end),
      (start) --[scalar, quarter right, out=-30] (centreblob3) --[scalar,quarter right,in=210] (end)
      };
    \end{feynman}
  \end{tikzpicture}}}
  \end{align*}
\caption{Graphical SDE for the two-point functions in $\phi ^2 \bar{\psi}\psi$ theory, the black blobs denote full fermion propagators $F(p)$, and the grey blobs denote full boson propagators $B(p)$. We omit diagrams that vanish in dimensional regularization, as the fields are massless.}
\label{on3-2pt-sde-3dyuk}
\end{figure}

In the strict large-$N$ limit, melonic dominance enables the complete resummation of the SDEs to all orders in the coupling constants: indeed, for the purposes of this section, we could now forget about the tensorial origin of this theory, and imagine it as coming from an SYK-like theory with disorder, since we do not consider subleading corrections.
Then we obtain the SDEs shown graphically in \cref{on3-2pt-sde-3dyuk}. %
\noindent To each vertex we assign a generic coupling to represent some combination of bare $\gO(N)^3$-invariant coupling constants: $\lambda$ to the boson-fermion vertex, $h$ to the boson 6-point vertex, and $g$ to the boson 4-point vertex. For example, $\lambda^2 =\frac{\lambda_{t,\text{bare}}^2}{6}$, though their precise forms will not matter.
Denoting by $F(p)$ and $F_0(p)$ the full and free fermion propagators, and by $B(p)$ and $B_0(p)$ the full and free boson propagators, we can divide through by $B_0(p)B(p)$ or $F_0(p)F(p)$ to obtain the Euclidean space SDEs

\begin{subequations}\label{eq:SDEsWithoutX}
\begin{align}
B_0(p)^{-1} & = B(p)^{-1} + \Sigma_B(p), \\
\Sigma_{B}(p) &= \frac{h^2}{5!} \int_{k,l,m,n} B(k+l+m+n+p)B(k)B(l)B(m)B(n) \notag\\
&+ \frac{\lambda^2}{1} \int_{k,l} (-1)\Tr\left[B(k+l+p)F(l)F(k)\right]\\
&+ \frac{g^2}{3!} \int_{k,l} B(k+l+p)B(l)B(k),\\ %
F_0(p)^{-1} & = F(p)^{-1} + \Sigma_F(p)  = F(p)^{-1} + \frac{\lambda^2}{1}  \int_{k,l} F(k+l+p)B(l)B(k). \label{eq:fermionSDEWithoutX}
\end{align}
\end{subequations}
We neglect snail-like diagrams that vanish in DREG, as we have tuned the renormalized masses and $\phi^4$ couplings to zero by assumption.
By assuming the conformal form $\sim 1/\abs{x}^{2\Delta}$ for the full two-point functions, and taking an IR limit, where the free propagators are negligible, we exactly determine the scaling dimensions of the fundamental fields $\phi$ and $\psi$. 
We will find that all of our results depend only on the ratio of bosonic to fermionic degrees of freedom $r$. 
We note parenthetically that in the low energy/strong coupling limit (where we ignore the free propagator), the truncated SDEs have a large set of local symmetries, as described in \cite{Choudhury:2017tax,Benedetti:2018goh,Chang:2018sve}.

\subsection{Auxiliary field theory}

The motivation to introduce the auxiliary scalar $X_{abc}$ comes from the existence of perturbative solutions with $h\sim \epsilon$: \hprismatic and \hlamprismatic. These do not appear as solutions of \eqref{eq:SDEsWithoutX}; so, by comparison with the componentwise Schwinger-Dyson equations of the supersymmetric tensor model \cite{Popov:2019nja}, we conclude that the auxiliary field is necessary to obtain the SDEs for the prismatic-type QFTs.

This is simply another quadratic tensor interaction, and therefore does not modify the combinatorics of the large-$N$ limit: it simply adds a new SDE for $X_{abc}$ and an extra term for the $\phi$ SDE:
\begin{figure}[H]
  \begin{align*}%
\vcenter{\hbox{\begin{tikzpicture}
  \begin{feynman}[every blob={/tikz/pattern color=black,/tikz/inner sep=2pt}]
    \vertex[small, blob,pattern=north east lines,pattern color=black] (m) at (0,0) {};
    \vertex (a) at (-1,0) ;
    \vertex (b) at ( 1,0);
    \diagram* {
      (a) --[ghost] (m) --[ghost] (b),
      };
  \end{feynman}
\end{tikzpicture}}}
\quad &= \quad \vcenter{\hbox{\begin{tikzpicture}
  \begin{feynman}
    \vertex (a) at (-1,0) ;
    \vertex (b) at ( 1,0);
    \diagram* {
      (a) --[ghost] (b),
      };
  \end{feynman}
\end{tikzpicture}}} \quad+ \vcenter{\hbox{\begin{tikzpicture}
    \begin{feynman}[every blob={/tikz/fill=gray!30,/tikz/inner sep=2pt,/tikzfeynman/small}]
   \vertex[small,blob,pattern=north east lines] (m) at (0,0) {};
    \vertex (a) at (-1,0) ;
    \vertex (start) at ( 1,0);
    \vertex[small, blob] (centreblob1) at ( 2,0.8) {};
    \vertex[small, blob] (centreblob2) at ( 2,0) {};
    \vertex[small, blob] (centreblob3) at ( 2,-0.8) {};
    \vertex (end) at ( 3,0);
    \vertex (e) at ( 4,0);
    \diagram* {
      (a) --[ghost] (m) --[ghost] (start) --[scalar] (centreblob2) --[scalar] (end) --[ghost] (e),
      {[edges={scalar}]
      (start) --[quarter left, out=30] (centreblob1) --[quarter left,in=150] (end)},
      {[edges={scalar}]
      (start) --[quarter right, out=-30] (centreblob3) --[quarter right,in=210] (end)},
      };
    \end{feynman}
  \end{tikzpicture}}}\\
\Sigma_{B,\mathrm{aux}}(p) &= \Sigma_B(p) + \quad \vcenter{\hbox{\begin{tikzpicture}
    \begin{feynman}[every blob={/tikz/fill=gray!30,/tikz/inner sep=2pt,/tikzfeynman/small}]
   \vertex[small,blob] (m) at (0,0) {};
    \vertex (a) at (-1,0) ;
    \vertex (start) at ( 1,0);
    \vertex[small, blob] (centreblob1) at ( 2,0.8) {};
    \vertex[small, blob,pattern=north east lines, pattern color=black] (centreblob2) at ( 2,0) {};
    \vertex[small, blob] (centreblob3) at ( 2,-0.8) {};
    \vertex (end) at ( 3,0);
    \vertex (e) at ( 4,0);
    \diagram* {
      (a) --[scalar] (m) --[scalar] (start) --[ghost] (centreblob2) --[ghost] (end) --[scalar] (e),
      {[edges={scalar}]
      (start) --[quarter left, out=30] (centreblob1) --[quarter left,in=150] (end)},
      {[edges={scalar}]
      (start) --[quarter right, out=-30] (centreblob3) --[quarter right,in=210] (end)},
      };
    \end{feynman}
  \end{tikzpicture}}}
  \end{align*}
\label{auxSDEsAdded}
\end{figure}
\begin{equation}\begin{aligned}\label{eq:auxSDEsAdded}
A_0(p)^{-1} &= A(p)^{-1} + \Sigma_A(p)=A(p)^{-1} + \frac{\rho^2}{3!} \int_{k,l} B(k+l+p) B(l)B(k)\\
\Sigma_{B,\mathrm{aux}}(p) &= \Sigma_B(p) + \frac{\rho^2}{2!} \int_{k,l} A(k+l+p) B(l)B(k)
\end{aligned}\end{equation}
\noindent $A(p),A_0(p)$ denote the full and bare auxiliary field propagators, $\rho^2$ here stands for some quadratic combination of the $\gO(N)^3$ invariant coupling constants that make up the 12-index coupling constant $\rho_{I(JKL)}$.

\subsection{Momentum scaling analysis of SDEs without auxiliary field}

Let us begin by studying the case without the auxiliary scalar.
Assuming, as discussed above, that the fixed point possesses full conformal symmetry, and so is a CFT, we make the ansatz  
\begin{align}
B(p) &= \frac{B}{(p^2)^b}, \qquad  F(p) = \frac{F \slashed{p}}{(p^2)^{f + \half}} %
\end{align}
for the momentum space two-point functions. The bare two-point functions are
\begin{align}
 B_0(p) = \frac{B_0}{p^2 + m_0^2}\, ,\qquad
F_0(p) = \frac{F_0}{\slashed{p} +M_0}\,.
\end{align}
$M_0$ plays no role in the subsequent analysis (due to parity symmetry) and can consistently be set to zero. The SDEs then take the form
\begin{align}
    (p^2+m_0^2) BB_0^{-1} &= p^{2b}+\lambda^2F^{-2}c_1\,p^{2(d-b-2f)}+h^2B^{-4}c_2\, p^{2(2d-5b)}+g^2B^{-2}c_3\,p^{2(d-3b)}\, ,\label{eqn:SDEsubsB}\\
    (p^2)^\half FF_0^{-1}&= p^{2f}+\lambda^2B^{-2}c_4\,p^{2(d-2b-f)}\, ,\label{eqn:SDEsubsF}
\end{align}
where $c_i$, $i=1\ldots 4$ are computable coefficients (see \cref{app:loopIntegrals}, or in full generality the procedure explained in \cref{chap:fextr}). From demanding non-trivial IR scaling ($f<\frac{1}{2}$ and/or $b<1$), where we drop the free propagators, we conclude that:
\begin{enumerate}
    \item \label{SDEhmelonic}$\lambda =0$ gives the real sextic bosonic tensor model plus a free tensor fermion; we identify this theory as \hmelonic \cite{Benedetti:2019rja}:
\begin{equation}\label{eq:stdBosonicPhi6Sol}
    \gamma_\phi = \frac{\epsilon}{3}, \quad \gamma_\psi = 0, \quad \Leftrightarrow \quad \Delta_\phi = \frac{d}{6}, \quad \Delta_\psi = \frac{d-1}{2}
\end{equation}
This requires $d<3$ for validity of the IR solution. If $d>3$, we have a UV solution that breaks the unitarity bound for a scalar.
\item $\lambda\ne 0$ implies, from \eqref{eqn:SDEsubsF}, that  \begin{align} 2b+2f = d\,. \label{eqn:fbconstraint}
\end{align}
Consequently $g$ can be ignored and set to zero. To see this, note that if $g$ is relevant then, from \eqref{eqn:SDEsubsB},  $b=d/4$ and hence $f=d/4$ -- which is not a consistent IR solution if $d>2$. There are then two possibilities: \begin{enumerate}
    \item \label{SDElmelonic}$\lambda\ne 0$, $h=0$, gives solutions satisfying \eqref{eqn:fbconstraint} if and only if $d<3$; solving for the coefficients will then pick out a particular solution.
    \item \label{SDEhlmelonic}$\lambda\ne 0$, $h\ne 0$ implies from \eqref{eqn:SDEsubsB} that  $b \le d/3$ and therefore from  \eqref{eqn:fbconstraint} that $f \ge d/6$, so these solutions exist only if $d\le 3$. Note that the  $b=d/3$, $f=d/6$ solution has qualitatively different behaviour from the others as there are three terms of the same order on the r.h.s. of \eqref{eqn:SDEsubsB}. %
\end{enumerate}
\end{enumerate}
Unitarity has not entered into these considerations; we have only demanded consistent IR scaling of the equations.

As a warm-up, we now locate the bosonic fixed points that we found perturbatively earlier (in \eqref{eq:bosonicFixedPoints}): both of these are known, and correspond to the sextic bosonic tensor model and the prismatic tensor model. 
For convenience, we now switch to scaling dimensions instead of the momentum space powers $b,f$: $b=1-\gamma_\phi$ and $f=1/2 - \gamma_\psi$; $\Delta_\phi = \frac{d-2}{2} + \gamma_\phi$ and $\Delta_\psi = \frac{d-1}{2} + \gamma_\psi$. 

\subsection{Bosonic fixed points; or, \texorpdfstring{\hprismatic}{h-prismatic} and an introduction to the general characteristics of a multi-field melonic CFT} \label{sec:generalCharacteristicsPrismatic}

The first bosonic fixed point that we found perturbatively, $h_{\text{melonic}}$, with $\gamma_\phi= \epsilon/3$, is well understood as the \textit{sextic melonic CFT}\footnote{See \cite{Benedetti:2019rja} for a complex version, which is identical up to a modification to the spectrum of bilinears: in the real case, the odd-spin bilinears do not exist.}, and is trivially identified in the SDEs analysis without auxiliary field above as case \ref{SDEhmelonic}, \eqref{eq:stdBosonicPhi6Sol}. Thus, the non-perturbative value of the scaling dimension is linear in $d$: $\Delta_\phi = d/6$; so this theory with a single field has only a single solution, which is straightforward.

To identify the second bosonic fixed point, \hprismatic, we need the SDEs with auxiliary field \eqref{eq:auxSDEsAdded}\footnote{We note again that the theory with an auxiliary field is identical at the quantum level to the theory without auxiliary field, except for different values of the coupling constants $\lambda_i, h_i$. Due to these, different terms dominate in the SDEs.}. %
Indeed, if the above analysis is performed without fermions, then by solving the SDEs about $d=3-\epsilon$, we obtain:
\begin{align}
  &\frac{\Gamma (\Delta_\phi ) \Gamma (d-\Delta_\phi )}{\Gamma \left(\frac{d}{2}-\Delta_\phi \right) \Gamma \left(\Delta_\phi -\frac{d}{2}\right)}-\frac{3 \Gamma (3 \Delta_\phi ) \Gamma (d-3 \Delta_\phi )}{\Gamma \left(\frac{d}{2}-3 \Delta_\phi \right) \Gamma \left(3 \Delta_\phi -\frac{d}{2}\right)} = 0 \label{eq:prismatic2Pt}\\
  \implies &\gamma _{\phi } = \epsilon ^2-\frac{20 \epsilon ^3}{3}+\frac{1}{9} \left(472+3 \pi ^2\right) \epsilon ^4+O\left(\epsilon ^5\right). \label{eq:hprismaticPerturbative3meps}
\end{align}
\begin{figure}[ht]
\centering
\includegraphics[width=0.9\textwidth]{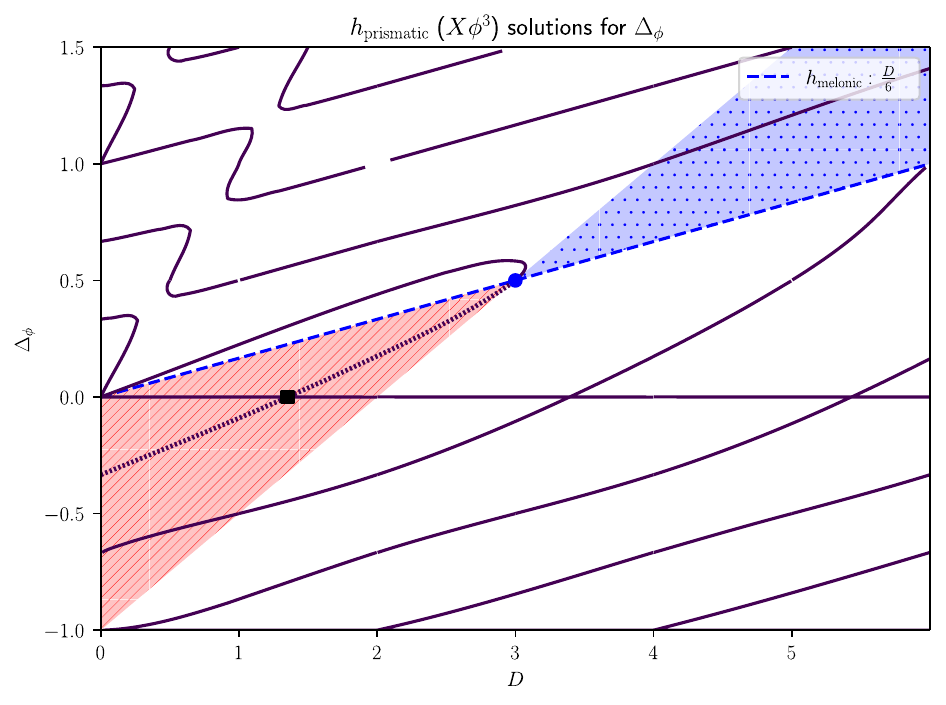}
\caption{Adding an auxiliary field $X$ to the sextic bosonic theory, we solve for \hprismatic{}, the sextic prismatic theory. The contours indicate the $\Delta_\phi$s which solve the SDEs, and $\Delta_X = d-3\Delta_\phi$. The region of IR validity \eqref{eq:IRscalingPrismatic} is shaded in ruled red; this also corresponds to the region where the anomalous dimensions of both $\phi$ and $X$ are positive (satisfying unitarity bounds). For $d>3$, we have the unitarity-bound-violating UV fixed point region in dotted blue. The standard \hmelonic{} solution is shown for comparison as a dashed blue line, $\Delta_\phi = d/6$. The free theory in $d=3$ is marked with a blue dot; the line of solutions \eqref{eq:hprismaticPerturbative3meps} descending from it in $d=3-\epsilon$ is indicated with a finely dashed line. For $d \simeq 1.353$, we have $\Delta_\phi=0$, marked with a black square; this will be discussed in \cref{sec:zeroScaling}.}
\label{fig:Bosonic6SDEdimPhi}
\end{figure}
This is precisely identical to the results of the \textit{prismatic bosonic tensor model} of \cite{Giombi:2018qgp}, which was also studied in \cite[\S 3.2]{Jepsen:2023pzm} (and the combinatorics are identical to the $d=0$ theory studied in \cite{Krajewski:2023tbv}). 
The full solutions of the allowed $\Delta_\phi$ for every $d$ are plotted below in figure \ref{fig:Bosonic6SDEdimPhi}. For comparison, we indicate with the dashed blue line the \hmelonic theory.

Now, the solutions here demonstrate a number of features which we expect to be generic for multi-field melonic CFTs, which reflect the results of \cref{chap:fextr}.
Therefore, for emphasis, we provide these as a list:
\begin{itemize}
\item \textbf{The IR/UV wedges} The range of validity for an IR solution to \eqref{eq:prismatic2Pt} is 
\begin{equation}\label{eq:IRscalingPrismatic}
\gamma_\phi >0\text{ and }\gamma_X = \epsilon - 3\gamma_\phi  >0,
\end{equation}
indicated with the left red striped region, which we term the \textit{IR wedge}. 
Alternatively, swapping the signs of both of these conditions gives the non-unitary UV fixed point indicated with the second blue dotted region, the UV wedge.
Note that positivity of all anomalous dimensions (for canonical kinetic terms) is also a necessary condition for unitarity of the theory.
Therefore, any solution outside the IR wedge must describe a non-unitary CFT (regardless of any concerns about evanescent operators).
\item \textbf{Infinite branches of solutions} We observe multiple different branches for almost every value of $d$ -- in fact, an infinite number. Those within the two wedges signal distinct vacua of the theory \cite{Chang:2021wbx}; those outside the wedges might seem to have no interpretation at all. However, this is not correct. 
The seemingly inaccessible lines of non-unitary fixed points can in fact become accessible, if we modify the kinetic terms for the scalar fields to be $\sim \int \phi (-\partial^{2})^{\pm k} \phi$  (with suitable tuning of the more relevant kinetic terms) for arbitrary $k$ \cite{Benedetti:2019eyl}, as discussed in \cref{sec:furtherSolutions}. 
This allows us to modify the range of IR/UV validity, and we obtain typically manifestly non-unitary CFTs.
\item \textbf{Collision of two lines of fixed points} Occasionally, we appear to have fewer solutions than we might expect. For example, increasing $d$ from around $d=3$, two solutions collide and disappear at around $d \simeq 3.074$. $\Gamma(z)^\star = \Gamma(z^\star)$ then means that we have a pair of conjugate $\Delta_\phi$s; these will be illustrated graphically for \lammelonic in \cref{fig:Yuk46SDEdimPhiR4-3D}. This represents an instability of the theory \cite{Benedetti:2021qyk}, and we will comment on it further in \cref{sec:windowsOfStability}.
\item \textbf{Disappearing solutions at exceptional values of the dimension $d$ and scaling dimension $\Delta_\phi$} We see breaks in the scaling dimension at certain \textit{exceptional values of the dimension}. For example, in even dimensions we find only a finite number of solutions to \eqref{eq:prismatic2Pt}, because the ratios of gamma functions become a rational function of $\Delta_\phi$. For example, in $d=2$ the infinite number of SDE solutions truncates to only two,
 \begin{equation}
 \Delta_\phi = \frac{1}{13}(4 \pm \sqrt{3}).
 \end{equation}
Likewise, in even dimensions, we find $\lfloor \frac{2d +3}{3} \rfloor$ solutions to the scaling dimension equation, some of which may be complex.
However, for all of these $d$s there are still perturbative fixed points in $d=2n\pm \epsilon$, which is why the contours only appear to have a small break in them. 
Another exceptional value is $\Delta_\phi=\half$ for $d=1$, where similarly we observe a perturbative solution $\Delta_\phi = \half + \frac{\epsilon}{6} + O(\epsilon^2)$, but no solution for exactly $\epsilon =0$. The nature of these exceptional fixed points is not understood.
\item \textbf{Zero scaling dimensions} 
The analysis that led to \eqref{eq:hprismaticPerturbative3meps} must fail for certain values of $d$. 
For example, $\Delta_\phi=0$ implies a logarithmic two-point function; however, blindly evaluating \eqref{eq:hprismaticPerturbative3meps} with this value, we find that this is a solution for all values of $d$. 
In the IR region, this occurs for the branch descending from $d=3$ at $d_c \simeq 1.35$, as mentioned in \cite{Giombi:2018qgp}, as well as for the branch that always has $\Delta_\phi=0$.
This produces singularities in the dimensions of scalar bilinears, as we will see in \cref{sec:zeroScaling}, indicating that our approach to this theory breaks down. 
\end{itemize}
We have now established in a simple multi-field model the various elements we will require: the IR wedge, the infinite number of solutions in general dimension, the breakdown of the conformal analysis at exceptional values of $d$ and $\Delta_\phi$, and the associated breaks in the contours; we are now ready to introduce fermions.

\subsection{Fermionic fixed points: SDEs without auxiliary field}

We now turn to the case where the fermions are coupled, where $\lambda \neq 0$, but for the moment we neglect the auxiliary field.

\subsubsection{\texorpdfstring{$\lambda_{\text{melonic}}$}{lambda-melonic} and \texorpdfstring{$h\lambda_{\text{melonic}}$}{hlambda-melonic} fixed points} \label{sec:lammelonicSDEfound}

\begin{figure}
\centering
\begin{subfigure}{\textwidth}
  \centering
\includegraphics[width=0.8\textwidth]{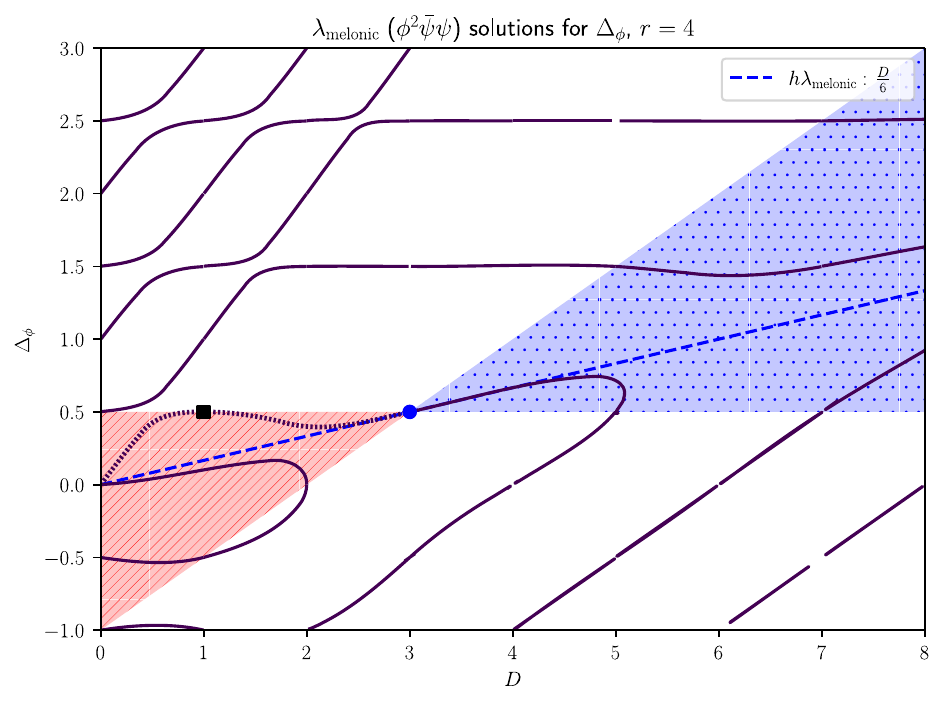}
  \caption{$r=4$, the branch used in the scaling dimensions analysis below.}
  \label{fig:Yuk46SDEdimPhiR4}
\end{subfigure}
\begin{subfigure}{\textwidth}
  \centering
\includegraphics[width=0.8\textwidth]{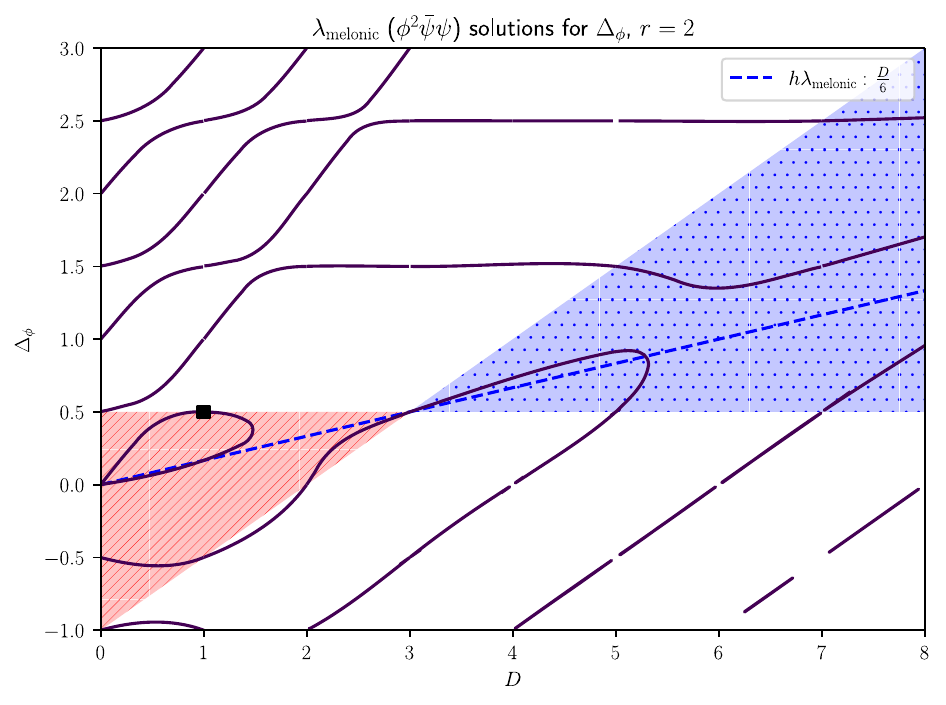}
  \caption{$r=2$ presented for comparison.}
  \label{fig:Yuk46SDEdimPhiR2}
\end{subfigure}
  \caption{$\phi$ scaling dimension for \lammelonic for $r=4,2$. 
  The $r=4$ branch that we will use in the \lammelonic{} bilinear analysis of \cref{sec:bilinearsCalculationResults} is finely dotted; it descends from the $d=3$ free theory, which is marked with a blue dot. 
  The wedge of validity of the IR scaling solution to the SDEs is shaded in ruled red, and also corresponds to the regions where the scaling dimensions of $\phi,\psi$ satisfy the unitarity bounds. We give for comparison the \hlammelonic scaling dimension, $\Delta_\phi = \frac{d}{6}$ as a blue dashed line.  Also, note that in $d=1+ \epsilon$ for all values of $r$ we find $\Delta_\phi = \half- O(\epsilon^2)$, $\Delta_\psi =0 + O(\epsilon)$, marked with a black square. This solution does not exist in $d=1$ exactly, however. The IR wedge of \cref{fig:Yuk46SDEdimPhiR4} is shown in more detail and including the solutions for complex $\Delta_\phi$ in \cref{fig:Yuk46SDEdimPhiR4-3D}.}
  \label{fig:Yuk46SDEdimPhiBothRs}
\end{figure}

Setting $2b+2f=d$, the fermion SDE becomes
\begin{equation}\begin{aligned}
    &1+\frac{\lambda^2 B^2 F^2}{2  (4\pi)^d}\frac{\Gamma \left(\frac{d}{2}-b\right)^2}{\Gamma (b)^2} \frac{  \Gamma \left(b-\frac{d}{2}+\frac{1}{2}\right) \Gamma \left(b+\frac{1}{2}\right) }{\Gamma \left(\frac{d}{2} - b + \half\right) \Gamma \left(d-b+\frac{1}{2}\right)} = 0
\end{aligned}\end{equation}
This can be substituted into the boson SDE, which yields:
\begin{equation}\begin{aligned}
&\frac{B^6 h^2\Gamma (5 b-2 d)\Gamma \left(\frac{d}{2}-b\right)^5 }{ 5! (4\pi)^{2 d} \Gamma (b)^5 \Gamma \left(\frac{5 d}{2}-5 b\right)}(p^2)^{2 d-6 b}+\frac{2^{d-4 b} r \Gamma (-b) \Gamma (2 b) \Gamma \left(d-b+\frac{1}{2}\right)}{\Gamma \left(b-\frac{d}{2}+\frac{1}{2}\right) \Gamma \left(b+\frac{d}{2}\right) \Gamma (d-2 b)}=-1 \label{eq:SDEsForFermionic}%
\end{aligned}\end{equation}
A consistent solution requires either $h=0$ or  $2d-6b=0$.  We will discover that the former gives precisely the \lammelonic{} fixed point identified above. We plot the solutions and regions of validity for $h=0$ in \cref{fig:Yuk46SDEdimPhiR4,fig:Yuk46SDEdimPhiR2} for two different values of $r$. In \cref{sec:bilinears}, we will focus on the $r=4$ line of $\Delta_\phi$ solutions descending from the free theory in $d=3$, indicated in \cref{fig:Yuk46SDEdimPhiR4}.

\begin{figure}
  \centering
\includegraphics[width=0.8\textwidth]{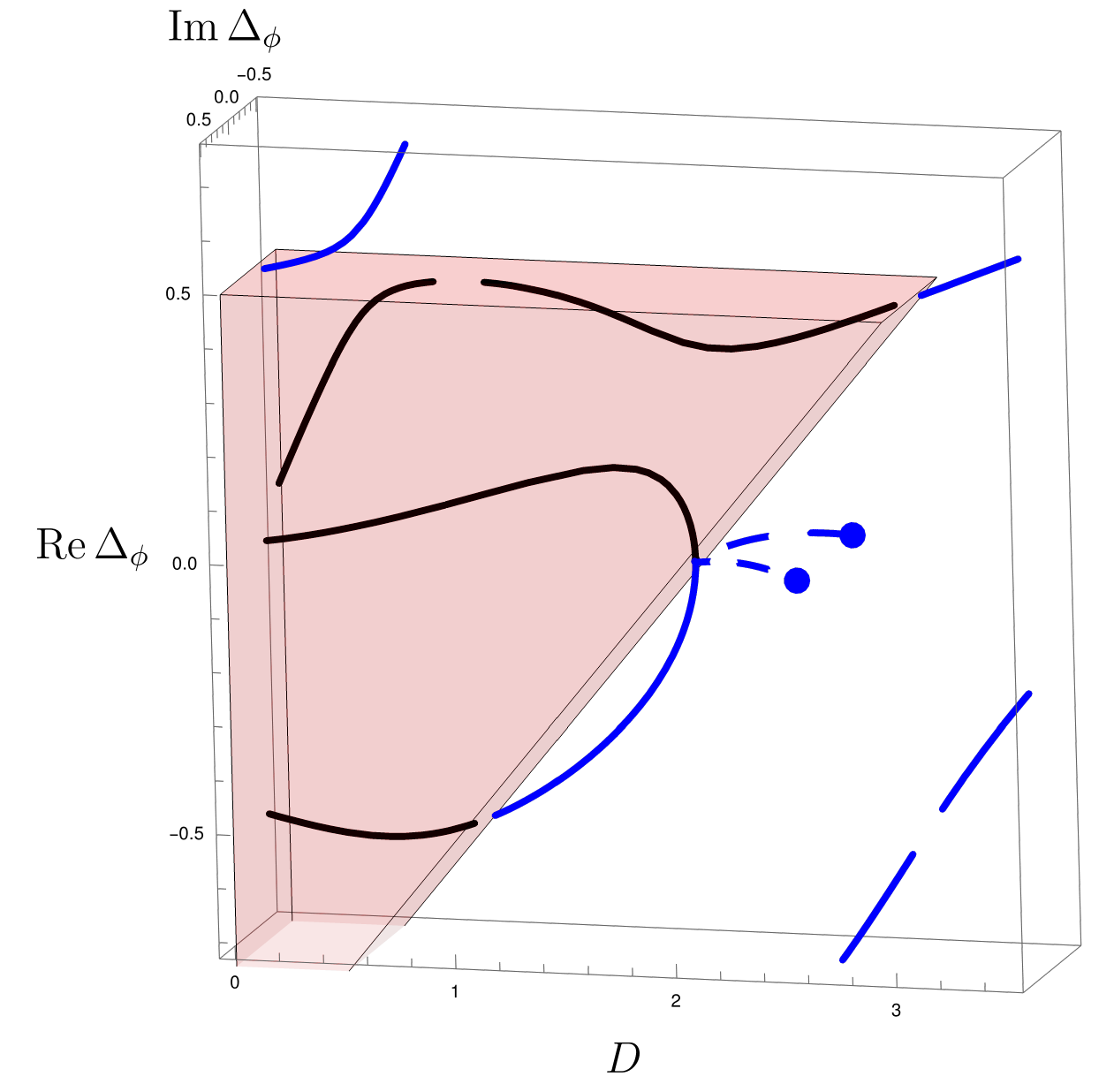}
  \caption{$\phi$ scaling dimension for \lammelonic for $r=4$; this is a reproduction of the IR wedge of \cref{fig:Yuk46SDEdimPhiR4}, but including complex $\Delta_\phi$s. Black lines lie inside the IR wedge; blue lines lie outside it. The two dashed lines in the centre are complex, and leave the plotting region at points marked with blue dots.}
  \label{fig:Yuk46SDEdimPhiR4-3D}
\end{figure}

Expanding $\gamma_\phi$ for $d=3-\epsilon, \Delta_\phi = \frac{d-2}{2}+\gamma_\phi$ in $\epsilon$ gives
\begin{align}\label{eq:gammaphiSDEsol}
\begin{split}
\gamma _{\phi } &= {\scriptstyle \frac{r \epsilon }{2(r+2)}+\tfrac{r (5 r-16) \epsilon ^2}{6 (r+2)^3}+\frac{r \left(2 r (r (17 r+164)-472)-224-3 \pi ^2 (r-2) (r+2)^2\right) \epsilon ^3}{36 (r+2)^5}}\\
&{\scriptstyle +\frac{r \left[\splitfrac{\scriptstyle 3 \pi ^2 (r (r (5 r-82)+200)-64) (r+2)^2-378 (r-2) \zeta(3) (r+2)^4}{\scriptstyle +4 r (r (r (r (62 r+1069)+5839)-17384)-7136)-4096}\right] \epsilon ^4}{216 (r+2)^7}+O(\epsilon ^5)},
\end{split}
\end{align}
which exactly matches the $O(\epsilon^2)$ $\phi_{\mathbb{R}}\psi_{\mathbb{C}}$ perturbative result in \eqref{eq:fermionicFixedPoints} for $r=2T$. For later reference, for $r=4$ at fourth order 
\begin{equation}\label{eq:lamOnlyFPGammaPhiEp4}
    \gamma_\phi = \frac{\epsilon }{3}+\frac{\epsilon ^2}{81}+\frac{\left(428-27 \pi ^2\right) \epsilon ^3}{8748}+\left(-\frac{7 \zeta (3)}{108}-\frac{4 \pi ^2}{2187}+\frac{6299}{59049}\right) \epsilon ^4+O\left(\epsilon ^5\right).
\end{equation}
Note that this $d=3-\epsilon$ fixed point path, shown with a finely dotted line in \cref{fig:Yuk46SDEdimPhiR4}, leaves the IR wedge for $d>3$. However, as mentioned above, we can modify this by changing the UV kinetic term to some non-local $\int \phi (-\partial^{2})^\zeta \phi$, as in \cite{Benedetti:2019eyl,Fraser-Taliente:2025udk}, at the cost of locality.
We note the presence of various features that first appeared in \hprismatic in \cref{sec:generalCharacteristicsPrismatic}. We have already mentioned the IR/UV wedges, and the infinite number of solutions in generic dimension are manifest in \cref{fig:Yuk46SDEdimPhiR4,fig:Yuk46SDEdimPhiR2}.
\begin{itemize}
\item  \textbf{Collision of two lines of fixed points}: once again, we have a collision of two lines of fixed points, that occurs, for example, around $d\sim 5$ for $r=2,4$, and at $d\simeq 1.48$ for $r=2$. At these points the actual solutions complexify, as is demonstrated for the latter in \cref{fig:Yuk46SDEdimPhiR4-3D}.
\item \textbf{Disappearing solutions at exceptional values of the dimension} In integer dimensions, the ratio of gamma functions in \eqref{eq:SDEsForFermionic} becomes a rational function of $\Delta$, and so we obtain $d$ solutions for even $d$, and $(d+1)/2$ solutions in odd $d$.
\item We have in $d=1$ a combination of both the \textbf{disappearing solutions at exceptional values of $\Delta_\phi$} and the \textbf{zero scaling dimensions}. That is, we have perturbative solutions $\Delta_\phi = 1/2 - O(\epsilon^2)$ and so $\Delta_\psi = 0 + O(\epsilon)$ in $d=1+\epsilon$, but no such solution for $\epsilon =0$. This will lead to the singularities in the dimensions of the scalar bilinears, as we will see in \cref{sec:zeroScaling}; however, as we know from \hprismatic, these two phenomena should not be confused, just because they occur together here. In fact, for example, in the $\phi^p \bar\psi \psi$ theory, we always have a break in the contour for $\Delta_\phi = 1/2 - O(\epsilon^2)$, even though $\Delta_\psi = \half(d-p\Delta_\phi) \neq 0$ \cite{Fraser-Taliente:2024hzv}.
\end{itemize}

However, we now also have a second melonic coupling: as discussed above, $b=d/3$ also gives a solution for $h\neq 0$. If the fermions are not free, i.e. $\lambda \neq 0$, then $2f+2b=d$ enforces $f=d/6$, which is indeed precisely the \hlammelonic fixed point that we found perturbatively in \cref{sec:betas}. At this fixed point, the $b=d/3,\Delta_\phi = d/6$ scaling dimension matches up exactly with the scaling dimension found in the \textit{complex} sextic bosonic tensor model \cite{Benedetti:2019rja}, despite the presence of the interacting fermions. Of course, the scaling behaviours of the respective SDEs are identical regardless of the fixed point, so the complex nature of that field, and the presence of fermions, are irrelevant. 

We also note here that \lammelonic and \hlammelonic have manifestly different scaling dimensions beyond leading order in $\epsilon$. 
Though a perturbative analysis is not feasible here, it is likely simply that the value of the $h$s at the fixed point $h_i^* \propto \half (r-4)( \epsilon + g(r) \epsilon^2) + f(r) \epsilon^3$, where $f(4)\neq 0$. Thus, the $h^2$ term in $\gamma_\phi$ contributes only at $\epsilon^4$ order when $r=4$, so these fixed points which appeared to be identical at leading order in $\epsilon$ are actually different. This resolves the apparent collision of \lammelonic and \hlammelonic{} noticed perturbatively in \cref{sec:othercouplingconstants}. It still remains to deal with \hlamprismatic, but we will do so in \cref{sec:hlamprismaticSDEfound}.

\subsubsection{Majorana fermions, or other variations} \label{sec:rparameter}

We demonstrated in \cref{chap:fextr} that the seemingly arbitrary choice of a real scalar and complex fermion is not relevant, as any other choice can simply be subsumed (in the strict large-$N$ limit) by appropriate variations of $r$. 
For example, we can repeat the SDE analysis above with Majorana fermions, or other alternative fermions. 
The Feynman rules \cite{Denner:1992me} are straightforward in the context of our limited set of diagrams; we simply modify the symmetry factors. 
If we label the real or complex scalar field by $\phi_{\mathbb{R}/\mathbb{C}}$, and the Majorana or Dirac fermion by $\psi_{\mathbb{R}/\mathbb{C}}$, then the symmetry factors associated with the diagrams of \cref{on3-2pt-sde-3dyuk} (excluding the $\phi^4$ melon) are
\begin{center}
\begin{tabular}{L|LLL|L}
  & S_{FBB} & S_{BFF} & S_{B^5}  & \text{equivalent to }$r=$\\ \hline
 \phi_{\mathbb{R}} \psi_{\mathbb{{R}}} & 2 & 2 & 5! & T \\
 \phi_{\mathbb{R}} \psi_{\mathbb{{C}}} & 2 & 1 & 5! & 2T\\
 \phi_{\mathbb{C}} \psi_{\mathbb{{R}}} & 1 & 2 & 2! \cdot 3! & T/2\\
 \phi_{\mathbb{C}} \psi_{\mathbb{{C}}} & 1 & 1 & 2! \cdot 3! & T
\end{tabular}.
\end{center}
If the $h$ couplings are irrelevant, i.e. we are at the \lammelonic{} fixed point, then the solution depends only on the ratio $r=T=S_{FBB}/S_{BFF}$, and the result is the same as above in \eqref{eq:gammaphiSDEsol}. This case shows no particularly interesting features, unlike the situation when the auxiliary field is included, analysed below. If $h$ is relevant, then we know that we have the completely constrained $b = d/3$ fixed point. Changing the value of $r$ does not modify this at all, and the solution always exists.
If $\lambda$ is irrelevant, the fermions are free, and the $r$ parameter has no meaning.

In fact, we will see in the $\lambda_{\text{melonic}}$ model that we can interpret $r=2T$ as the ratio of the number of real fermionic degrees of freedom in the field $\psi$ to the number of real bosonic degrees of freedom in $\phi$.
This can be made obvious by the fact that if we add a $\gU(N_f)$ fermion symmetry (i.e. we have $N_f$ tensor fermion fields $\psi_{abc,K}, K=1,...N_f$) and likewise for $N_b$ tensor boson fields, the effect in the equations of the model is exactly the same\footnote{The Feynman rules say each fermion loop gives a factor of $(-1)\times T$; if we have a vector of $N_f$ fermions, which only interact via $\gU(N_f)$-symmetric terms like $\sum_K \lambda_t \delta^{t,p,dt}_{abc,def,hij,klm} \phi_{abc} \phi_{def} \bar\psi_{hij,K} \psi_{klm, K}$, then each loop also contributes a factor of $N_f$. } as modifying $r=\frac{2T N_f}{N_b}$; for this reason we can at least formally modify $r$ to whatever value we like. 

This also includes both negative and non-integer values of $r$.
Combined with the fact that $T$ always comes with a factor of $(-1)$ due to fermion loops, we can interpret Dirac fermions of negative $T$ as $\Sp(\abs{T}/T_d)$-symmetric commuting fermions, via the well-known formal relation $\SO(-N) \simeq \Sp(N)$\footnote{At least up to ordering of operators in scattering, which is irrelevant here; see \cite{Gurau:2022dbx,Keppler:2023lkb,Keppler:2023qol}.}, where $\abs{T}$ is the free parameter that we vary, and $T_d$ is the dimension of the minimal complex Dirac spinor in $d$ dimensions. 
As was discussed in \cref{sec:solvingCFTs}, a mathematical framework justifying non-integer values of $N_{f,b}$ and hence $r$ is presented in \cite{Binder:2019zqc}; see also \cite{Jepsen:2020czw}.
Note that as $r$ is smoothly varied from $r=4$ to $r=2$, at $r\simeq 3.07$, the line of theories that we will use splits: this means that there is no valid IR solution for a dimensional window around $d \simeq 1.8$, as is shown in \cref{fig:Yuk46SDEdimPhiBothRs}. 
In the language of \cite{Kaplan:2009kr}, for $r=2$ the two IR fixed points annihilate with each other around $d=1.48$ as $d$ increases from $1$.

\subsection{The SDEs with auxiliary field} \label{sec:hlamprismaticSDEfound}

We can consider the set of SDEs with auxiliary field, \eqref{eq:auxSDEsAdded}; ansatzing the IR propagator to be 
\begin{equation}
    \expval{X_I(p)X_J(-p)}=A(p)\delta_{IJ} \equiv \frac{A}{p^{2a}} \delta_{IJ},
\end{equation}
we require only $a>0$ for IR consistency.
This modifies only the bosonic sector of the SDEs, and in the deep IR we have:%
\begin{subequations}
\begin{align}
0 & = F(p)^{-1} +  \frac{\lambda^2}{2}  \int_{k,l} F(k+l+p)B(l)B(k),\\
0 & = B(p)^{-1} + \frac{h^2}{5!} \int_{k,l,m,n} B(k+l+m+n+p)B(k)B(l)B(m)B(n) \notag\\
&+\frac{\rho^2}{2} \int_{k,l} A(k+l+p)B(l)B(k) + \frac{\lambda^2}{1} \int_{k,l} (-1)\Tr\left[B(k+l+p)F(l)F(k)\right], \\
0 &= A(p)^{-1} + \frac{\rho}{3!} \int_{k,l} B(k+l+p)B(k)B(l).
\end{align}
\end{subequations}
Therefore, we obtain once again the usual $2f+2b=d$, along with $a+3b=d$. We again can either take $h\neq 0$, or $h=0$. 
In the former case, we force $\Delta_\phi =d/6$, which forces $\Delta_X=d/2$, which is the free field scaling dimension; this gives $\rho^2 A B^3 =0$, and so we find \hlammelonic again. 
In the latter case we can solve for $B^2 F^2$ and $AB^3$, which yields the result
\begin{equation}
\frac{\csc (\pi  b) \Biggl[\splitfrac{3 (d-3 b) \sin (3 \pi  b-\pi  d) \Gamma \left(3 b-\frac{d}{2}\right) \Gamma \left(\frac{3 d}{2}-3 b\right)}{-r \Gamma \left(b+\frac{1}{2}\right) \cos \left(\pi  b-\frac{\pi  d}{2}\right) \Gamma \left(d-b+\frac{1}{2}\right)}\Biggr]}{b \Gamma \left(\frac{d}{2}-b\right) \Gamma \left(b+\frac{d}{2}\right)}=-1.
\end{equation}
This can be solved perturbatively to any order desired. Taking $b=1-\gamma_\phi$ and $d=3-\epsilon$, we obtain a match to the $\hlamprismatic$ perturbative analysis:
\begin{equation}\label{eq:auxFieldSRFC}
\gamma_\phi = \tfrac{r \epsilon }{2(r+2)}+ (r-2)\left(\tfrac{4 (r-3) \epsilon ^2}{3 (r+2)^3}-\tfrac{\left(r \left(3 \pi ^2 (r+2)^2-8 r (5 r+99)+3056\right)-3840\right) \epsilon ^3}{36 (r+2)^5}+O(\epsilon^4)\right).
\end{equation}
Note that for $r=4$, this equals the \lammelonic{} scaling dimension calculated in \eqref{eq:gammaphiSDEsol} up to order $\epsilon^3$. They then end up differing at order $\epsilon^4$: note the 6326 here, compared to 6299 in equation \eqref{eq:lamOnlyFPGammaPhiEp4}:
\begin{equation} \gamma_\phi = \frac{\epsilon }{3}+\frac{\epsilon ^2}{81}+\frac{\left(428-27 \pi ^2\right) \epsilon ^3}{8748}+\left(-\frac{7 \zeta (3)}{108}-\frac{4 \pi ^2}{2187}+\frac{\mathbf{6326}}{59049}\right) \epsilon ^4+O\left(\epsilon ^5\right).
\end{equation}
So, the auxiliary field solution is again different at order $\epsilon^4$, for the same reason as before. So all three of the fixed points which apparently collide at $r=2T=4$ in the perturbative analysis of \eqref{eq:fermionicFixedPoints} are in fact distinct. 

\begin{figure}
\begin{subfigure}{\textwidth}
  \centering
\includegraphics[width=0.8\textwidth]{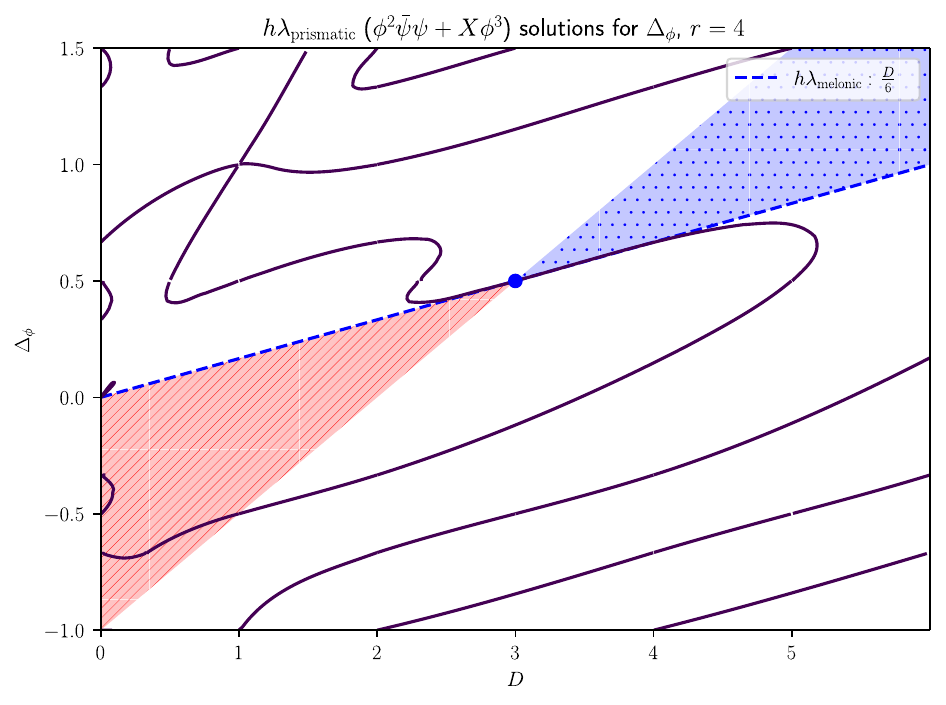}
  \caption{$r=4$}
  \label{fig:hlamprismatic2ptSDEs-r4}
\end{subfigure}
\begin{subfigure}{\textwidth}
  \centering
\includegraphics[width=0.8\textwidth]{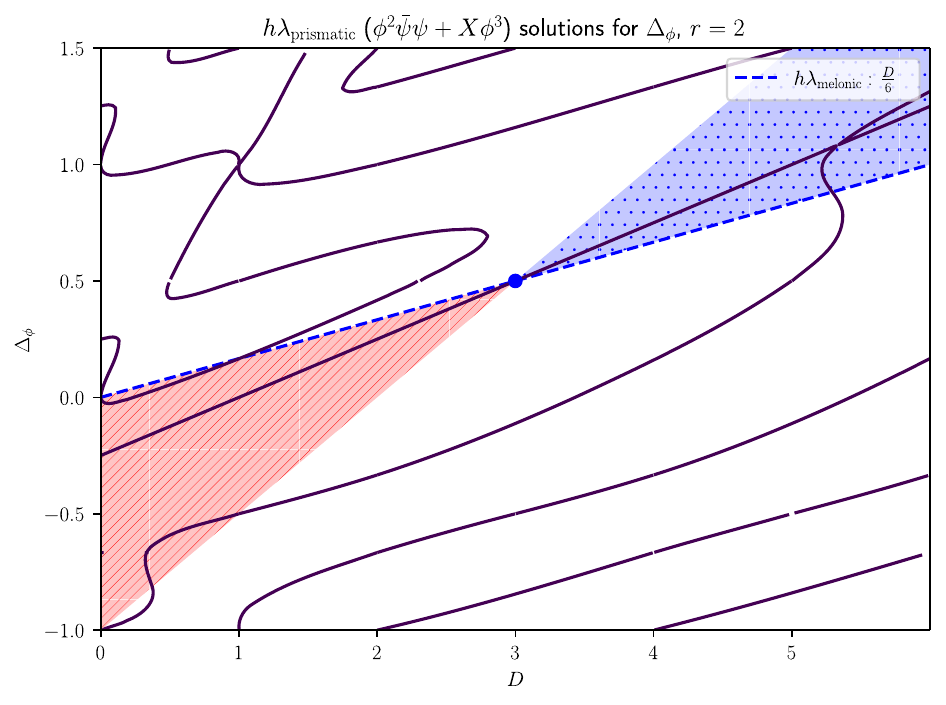}
  \caption{$r=2$, which contains the putative supersymmetric theory with $\Delta_\phi = \frac{d-1}{4}$ referred to in \cref{sec:qGenAndSUSYprismatic}, as well as other non-supersymmetric IR solutions.}
  \label{fig:hlamprismatic2ptSDEs-r2}
\end{subfigure}
\caption{$\phi$ scaling dimension for the prismatic theory, \hlamprismatic{}, with $r=4,2$ again; that for \hlammelonic{} is also illustrated for comparison. The regime of validity of the IR scaling solution to the SDEs is shaded in dashed red. 
The dotted blue region captures the manifestly non-unitary UV solutions to the SDEs.} \label{fig:prismaticScalingDimensions}
\end{figure}

In \cref{fig:prismaticScalingDimensions} we demonstrate the space of fixed points. Once again, red indicates the region of validity, and \hlammelonic is again indicated with the blue dashed line. For $r=0$, clearly \hlamprismatic{} reduces to \hprismatic{} (plus a free fermion CFT).

\subsubsection{Variation of \texorpdfstring{$r$}{r}, \texorpdfstring{$q$}{q}-generalizations and the supersymmetric solution} \label{sec:qGenAndSUSYprismatic}

As before, it is trivial to solve the above for general symmetry factors, which correspond to the following theories. Note that $\psi_{\mathbb{R}}$ corresponds to Majorana and $\psi_{\mathbb{C}}$ corresponds to Dirac fermions, although we leave $T$ general.

Note that if $r=2$, we obtain
\begin{equation}
\gamma_\phi = \gamma_\psi = \gamma_F = \epsilon/4
\end{equation}
exactly, to all orders.  
This corresponds to the scaling dimension of the supersymmetric solution in the $\phi_{\mathbb{R}} \psi_{\mathbb{R}}$  case ($\cN=1$ scalar superfield) and $\phi_{\mathbb{C}} \psi_{\mathbb{C}}$  cases ($\cN=2$ scalar superfield); both of these have $\Tr[\spinid]=2$, the expected dimension of the spinor space in $3$d. For this value of $r$, if we include the auxiliary field, the number of off-shell fermionic degrees of freedom then matches the number of off-shell bosonic degrees of freedom $N^3+N^3$, as it should for a supersymmetric theory. Of course, for the theory $\phi_{\mathbb{R}} \psi_{\mathbb{C}}$, this would require $\Tr[\spinid]=1$, which is not a physically permitted dimension of the spinor space; nonetheless, we still appear to have supersymmetry.

Continuous dimension is, in general, incompatible with supersymmetry; however, we have performed a variant of the dimensional reduction scheme used in loop calculations of supersymmetric theories: we have kept $T$ fixed and continued in $d$, in order to consider a supersymmetric theory in general dimension \cite{Giombi:2014xxa}.

Assuming either real scalar fields and Majorana fermions, or complex scalar fields and Dirac fermions, adding in an auxiliary field, we can use the superspace formalism: that is, we combine the three fields into a (real or complex) superfield $\Phi$. The superfield formalism enforces that the dimensions of each of the three components of the superfield ($\phi$, $\psi$, $X$) differ by $1/2$, and hence the anomalous dimensions of each must be equal \cite{Popov:2019nja,Lettera:2020uay}. Thus, in a manner very similar to the standard $4$d quartic bosonic model, we define the superfield propagator to be $P(p)$, and find the IR SDEs to be $P(p)^{-1} \propto \int_{k,l} \Tr(P(k+l+p) P(k) P(l))$. Therefore, simple dimensional analysis enforces $\gamma_\Phi = \epsilon/4 = \gamma_\phi = \gamma_\psi = \gamma_X$ -- precisely the result of \eqref{eq:auxFieldSRFC}. 
In fact, this can be done with general coupling $\Phi^q$ to find that the superfield gets scaling dimension $\Delta_\Phi = \frac{d-2}{2} + \gamma_\Phi = \frac{d-1}{q}$. Motivated by this, these SDEs with auxiliary field easily generalize, with or without fermions. We simply set the auxiliary field to be $X=\phi^{q-1}$, and perform the same set of integrals. If we include fermions this is a generalization of the aforementioned $q$-generalized supersymmetric tensor models \cite{Popov:2019nja,Lettera:2020uay} to non-matching bosonic and fermionic degrees of freedom. Without the auxiliary field, and without fermions, this is the $q$-generalization of the original tensor model studied in \cite{Giombi:2017dtl}.

\section{Spectrum of bilinears for the \texorpdfstring{\lammelonic}{lambda-melonic} fixed point} \label{sec:bilinears}

The simplest of these new fermionic fixed points is \lammelonic{}, which we now investigate in more detail by computing the spectrum of the set of $\gO(N)^3$ singlet bosonic operators that appear in the conformal OPEs $\phi_{abc} \times \phi_{abc}$ and $\bar\psi_{abc} \times \psi_{abc}$. 
For these melonic theories, this is a standard computation using the Schwinger-Dyson equation, which at large $N$ we can write down exactly \cite{Polchinski:2016xgd,Benedetti:2023mli}. 
In the conformal limit this Schwinger-Dyson equation can be integrated analytically, and the computation of the spectrum reduces to finding the scaling dimensions $\Delta$ such that a particular kernel matrix $k(\Delta)^a_{b}$ has a unit eigenvalue.

We shall find results that are analytic in spin; however, it is enlightening to consider explicitly the form of the spin-zero operators appearing in this OPE. They should be 
$\phi_{abc} (\partial^2)^n \phi_{abc}$, $\bar\psi_{abc} (\partial^2)^n \psi_{abc}$, and $\bar\psi_{abc} (\partial^2)^n \slashed{\partial} \psi_{abc}$; of course, operators of definite scaling dimensions will be a linear combination of these.

There are four main observations to be made about this part of the bilinear spectrum for the \lammelonic fixed point:
\begin{enumerate}
\item Near the free theory and near $d=0$ the operators in the spectrum exactly match the known scaling dimensions. Additionally, we have cross-checks via the perturbative computations of the mass operators.
\item There are windows of stability of the parameters $r$ and $d$, outside which operators have a complex scaling dimension; these appear to be characteristic of tensor models.
\item The spectrum contains conserved operators at spins $s=0,1,2$. These are, respectively, a redundant operator, the $\gU(1)$ current, and the stress tensor.
\item Divergences appear in the spectrum when the scaling dimension of one of the fundamental fields goes to zero; in this model, this occurs at $d=1$, just as occurs for \hprismatic in $d\simeq 1.35$.
\end{enumerate}

We will restrict ourselves to the subset of the operator spectrum consisting of bosonic operators $\cO_{(\mu_1 \cdots \mu_s)}$ that are traceless symmetric tensors in $\SO(d)$. These transform in the traceless symmetric spin-$s$ representations $\rho$, and so are labelled by Dynkin label $(s,0,0,\ldots)$, and exist in any $d$.  They take the schematic form\footnote{We will not need the exact forms, which can be found in e.g. \cite{Hikida:2016cla}.}
\begin{equation}\begin{aligned}
\cO_{s} \sim [(\partial^2)^x \partial_{(\mu} \cdots \partial_{\mu_i} \phi][\partial_{\mu_{i+1}} \cdots \partial_{\mu_s)} (\partial^2)^y \phi] -\text{traces}
\end{aligned}\end{equation}
We will not consider any other representations (thus ignoring any fermionic operators), as in that case analytic continuation in the dimension and Dynkin label entries is less clear.

We will begin with an overview of the pure bosonic calculation in \cref{sec:bilinearCalculationOverview}. %
Moving to the \lammelonic{} theory, we demonstrate the elaborations necessary for a fermionic theory in \cref{sec:bilinearsCalculationDetails}, and end with observations and comparison to \hprismatic in \cref{sec:bilinearsCalculationResults}.

\subsection{Spectrum for the bosonic tensor model with \texorpdfstring{$V(\phi)=\phi^q$}{V(phi)=phi\string^ q}} \label{sec:bilinearCalculationOverview}

First, we recall the determination of the spectrum for a purely bosonic CFT, recapping the salient points of \cite{Giombi:2017dtl}. It is informative to see the dependence of the melonic spectrum on the degree of the potential, $q$, which we keep general. 
To identify the operators $\cO_{\Delta,\rho}$, of scaling dimension $\Delta$ and $\SO(d)$ representation $\rho$, appearing in the OPE $\phi\times\phi$, we consider the Schwinger-Dyson equation for the three-point function $v(\cO)=\expval{\phi(x)\phi(y)\cO_{\Delta,\rho}}$. %
The spectrum consists of those operators $\cO_{\Delta,\rho}$ for which this Schwinger-Dyson equation is self-consistent.

\subsubsection{The Schwinger-Dyson equation for the three-point function}

The self-consistent Schwinger-Dyson equation for the three-point function $v(\cO)$ takes the form \cite{Gross:2016kjj,Liu:2018jhs}
\begin{figure}[H]
  \begin{align*}%
  \vcenter{\hbox{
\begin{tikzpicture}
    \begin{feynman}
        \vertex [blob] (a) at (0,0) {$v$};
        \vertex (b) at (-1,0.75) {$x$};
        \vertex (c) at (-1,-0.75) {$y$};
        \vertex (d) at (1.5,0) {$z$};
        \diagram* {
            (a) -- (b),
            (a) -- (c),
            (a) -- (d),
        };
    \end{feynman}
\end{tikzpicture}
}}
\, &= \,
\vcenter{\hbox{
\begin{tikzpicture}
    \begin{feynman}
        \vertex [blob] (a) at (0,0) {$v$};
        \vertex (y1) at (-1,0.75);
        \vertex (y2) at (-1,-0.75);
        \vertex (d) at (1.5,0) {$z$};
        \vertex (x) at (-3.0, 0.75) {$x$};
        \vertex (y) at (-3.0,-0.75) {$y$};
        \diagram* {
            (a) -- (y1) -- (x),
            (a) -- (y2)--(y),
            (a) -- (d),
        };
    \end{feynman}
\end{tikzpicture}
}}
\, + \, \vcenter{\hbox{
\begin{tikzpicture}
    \begin{feynman}
        \vertex [blob] (a) at (0,0) {$v$};
        \vertex (y1) at (-1,0.75);
        \vertex (y2) at (-1,-0.75);
        \vertex (d) at (1.5,0) {$z$};
        \vertex (x) at (-3.0, 0.75) {$x$};
        \vertex (y) at (-3.0,-0.75) {$y$};
        \diagram* {
            (a) -- (y1) -- (x),
            (a) -- (y2)--(y),
            (a) -- (d),
        };
        \draw[draw=gray,fill=gray] (-1,0.75) rectangle ++(-1.0,-1.5);
    \end{feynman}
\end{tikzpicture}
}}
  \end{align*}
\label{fig:3ptFuncVisualSDE}
\end{figure}
\noindent where the grey block represents the 2PI connected four-point Bethe-Salpeter kernel. %
The first term on the r.h.s., which is the free-field contribution to the kernel, %
drops out in the conformal IR. The tensor structure at the \lammelonic fixed point is identical to that of the quartic model, so the argument of \cite{Giombi:2017dtl} transfers exactly. %
The complete non-perturbative connected Bethe-Salpeter kernel is (illustrated for $q=4$):
\begin{figure}[H]
  \begin{align*}%
\vcenter{\hbox{\begin{tikzpicture}
    \begin{feynman}[every blob={/tikz/pattern color=black,/tikz/inner sep=2pt}]
    \vertex (a) at (-1,0) ;
    \vertex (b) at ( 1,0);
    \vertex (c) at (-1,1);
    \vertex (d) at ( 1,1);
    \draw[draw=gray,fill=gray] (-0.5,0) rectangle ++(1.0,1.0);
    \diagram* {
      (a) -- (b),
      (c) -- (d)
      };
  \end{feynman}
\end{tikzpicture}}}
\quad &= 
\quad \half \quad \vcenter{\hbox{\begin{tikzpicture}
    \begin{feynman}[every blob={/tikz/pattern color=black,/tikz/inner sep=2pt}]
    \vertex (a) at (-1,0) ;
    \vertex (b) at (1.5,0);
    \vertex (m1) at (0,0);
    \vertex (m2) at (0,1);
    \vertex (c) at (-1,1);
    \vertex (d) at (1.5,1);
    \foreach \point in {(-0.7,0), (-0.7,1), (0.22,0.5), (-0.22, 0.5)} {
        \node [circle, fill, inner sep=3pt] at \point {};
    }
    \diagram* {
      (a) -- (m1)-- (b),
      (m1)-- [quarter left](m2),
      (m1)-- [quarter right](m2),
      (c) -- (m2)-- (d),
      };
  \end{feynman}
\end{tikzpicture}}} + \half \quad \vcenter{\hbox{\begin{tikzpicture}
    \begin{feynman}[every blob={/tikz/pattern color=black,/tikz/inner sep=2pt}]
    \vertex (a) at (-1,0) ;
    \vertex (b) at (1.5,0);
    \vertex (b1) at (2.5,1);
    \vertex (m1) at (0,0);
    \vertex (m2) at (0,1);
    \vertex (c) at (-1,1);
    \vertex (d) at (1.5,1);
    \vertex (d1) at (2.5,0);
    \foreach \point in {(-0.7,0), (-0.7,1), (0.22,0.5), (-0.22, 0.5)} {
        \node [circle, fill, inner sep=3pt] at \point {};
    }
    \diagram* {
      (a) -- (m1)-- (b)--(b1),
      (m1)-- [quarter left](m2),
      (m1)-- [quarter right](m2),
      (c) -- (m2)-- (d)--(d1),
      };
  \end{feynman}
\end{tikzpicture}}}
  \end{align*}
\label{fig:3ptFuncVisualSDEMelonic}
\end{figure} %
\noindent where, as before, the blobs on lines indicate full resummed propagators. %
In the IR limit, and in position space, we get
\begin{equation}\begin{aligned}\label{eq:3ptSDEnobasis}
v(\cO)(x,y,z) = %
\int_{w_1,w_2} K((x,y),(w_1,w_2))\,  v(\cO)(w_1,w_2, z)\, ,
\end{aligned}\end{equation}
where, for general $q$,
\begin{align}
    K((x,y),(x_a,x_b)) = (q-1)\lambda^2 \times \half & \left[G(x,x_a)  G(y,x_b) G(x_a,x_b)^{q-2} \right. \nonumber \\ &\qquad\left. + G(x,x_b) G(y,x_a) G(x_b,x_a)^{q-2}\right]\, . \label{eqn:4ptKernel}
\end{align}
$K$
is, %
in general, a complicated function but in the conformal limit, we know exactly the form of the two-point function $G$. The resulting integral in \eqref{eqn:4ptKernel} is tractable, allowing us to determine precisely the three-point functions $v$, up to overall factors; this gives us the scaling dimensions.

\subsubsection{Conformal structures}

In a conformal theory, $v$ is simply determined by the three-point coefficients $c_{\phi\phi\cO}^a$:
\begin{equation}\begin{aligned}
\expval{\phi(x)\phi(y) \cO(z)} = \sum_a c_{\phi\phi\cO}^a \expval{\phi(x)\phi(y) \cO(z)}^a \equiv  \sum_a c_{\phi\phi\cO}^a \, v^a(\Delta,\rho)(x,y,z)\, . \label{eqn:three-pt}
\end{aligned}\end{equation}
Here, the index $a$ runs over the possible conformal structures that can appear in the OPE of two scalars and an operator $\cO$ in $SO(d)$ representation $\rho$; it is a group-theoretical exercise to enumerate these \cite{Kravchuk:2016qvl}. %
The action of the kernel on each of these structures is restricted %
by conformal invariance to take the form
\begin{equation}\begin{aligned}\label{eq:3ptSDE}
\int_{w_1,w_2} K((x,y),(w_1,w_2))\,  v^a(\cO)(w_1,w_2,z) \equiv k^a{}_b\,  v^b(\cO)(x,y,z),
\end{aligned}\end{equation}
defining the matrix $k^a{}_b$ %
which is a function only of the conformal representations of $\phi$ ($\Delta_\phi$, scalar) and $\cO$ ($\Delta$, $\rho_s$).
Finding the scaling dimensions in the IR limit for the melonic theory from \eqref{eq:3ptSDEnobasis} is then identical to solving the matrix eigenvalue problem
\begin{equation}\begin{aligned}
\det(k^a{}_b - \delta^a{}_b) = 0\,.
\end{aligned}\end{equation}

For traceless bosonic spin-$s$ operators $[\cO_{\Delta,\rho_s}]_{\mu_1 \ldots \mu_s}$ %
only one conformal structure is allowed in \eqref{eq:3ptSDE}; 
therefore $k^a{}_b=k\,\delta^a{}_b$.
Introducing a generic null vector $\xi^\mu$,
we can remove all the Lorentz indices by considering the operator $\cO_{\Delta,\rho_s} (x;\xi)= [\cO_{\Delta,\rho_s}(x)]_{\mu_1 \ldots \mu_s} \xi^{\mu_1} \ldots \xi^{\mu_s}$, and the corresponding correlator
\begin{equation}\begin{aligned}
& v^1_{\phi_1\phi_2\cO_{s}}(x_1,x_2,x_3;\xi) =  \expval{\phi_1(x_1) \phi_2(x_2) \cO_{\Delta,\rho_s}(x_3; \xi)}  \\ &=\frac{(\mathcal{X}_3 \cdot \xi)^{s}}{(x_{12}{}^2)^{\frac{1}{2}(\Delta_1+\Delta_2-\Delta+s)}(x_{23}{ }^2)^{\frac{1}{2}(\Delta+\Delta_2-\Delta_1-s)}(x_{31}{ }^2)^{\frac{1}{2}(\Delta+\Delta_1-\Delta_2-s)}},
\end{aligned}\end{equation}
where
\begin{equation}\begin{aligned}
x_{ij} = x_i - x_j, \quad \mathcal{X}_{3 \mu}=\frac{(x_{31})_\mu}{x_{31}{ }^2} -\frac{(x_{32})_\mu}{x_{32}{ }^2}  .
\end{aligned}\end{equation}
Then we exploit the fact that \eqref{eq:3ptSDE} applies for any $z$ to take $z\to\infty$ and work with the index-free eigenvector %
\begin{equation}\begin{aligned}
v^1_{\phi\phi\cO_{s}}(x_{12};\xi)=\lim_{x_3 \to \infty} (x_3{}^2)^{\Delta} v^1_{\phi\phi\cO_{s}}(x_1,x_2,x_3;\xi) = \frac{(\xi \cdot x_{12})^s}{\abs{x_{12}}^{2 \Delta_\phi - \Delta + s}}\, ,
\end{aligned}\end{equation}
where $\Delta_1=\Delta_2=\Delta_\phi$.

The conformal scalar IR propagator is known from the two-point SDE \cite{Giombi:2017dtl}
\begin{equation}\begin{aligned}
 G(x)= \frac{B}{\abs{x}^{2\Delta_\phi}},\quad \Delta_\phi = \frac{d}{q}, \quad \lambda^2 B^q \equiv \frac{-1}{X_\phi X_{\tilde{\phi}}} = \frac{-1}{\cN_\phi},
\end{aligned}\end{equation}
 where, for scalar operators such as $\phi$ and its shadow $\tilde{\phi}$, we define for convenience
 \begin{equation}\begin{aligned}
X_\phi &= \pi^{d/2} \frac{\Gamma\left(\frac{d}{2}-\Delta_\phi\right)}{\Gamma(\Delta_\phi)} \quad \implies \quad X_{\tilde\phi} = \pi^{d/2} \frac{\Gamma\left(\frac{d}{2}-\tilde\Delta_\phi\right)}{\Gamma(\tilde\Delta_\phi)}.
\end{aligned}\end{equation}
These $X_i$s arise from the Fourier transform of a conformal propagator, albeit with some factors of $2$ and $\pi$ (that almost always cancel) removed.
We have defined the shadow dimension, the dimension of the shadowed operator, by $\Delta_{\tilde{\cO}} \equiv \tilde{\Delta}_\cO \equiv d-\Delta_{\cO}$; see \cite{Akyuz:2026oec,Fraser-Taliente2026spectrum} for more on shadow operators and these computations.
Using the kernel \eqref{eqn:4ptKernel} and $v^1_{\phi\phi\cO_{s}}(x_{12};\xi)$, we can then compute \eqref{eq:3ptSDE} to find that the eigenvalue condition, $\det(k^1{}_1-1) =0$,  is 

\begin{equation}\begin{aligned}\label{eqn:implicitdelta}
 1+(q-1) P_s \frac{Y_\phi(s,\Delta)}{Y_{\tilde\phi}(s,\Delta)} =0\, .%
\end{aligned}\end{equation}
Here we have defined the useful function $Y_\cO(s,\Delta)$ and the projector on to even spin:
\begin{equation}\begin{aligned}\label{eq:Yfuncdef}
Y_\cO (s,\Delta) &= X_{\cO} \, \Gamma\left(\Delta_\cO +\frac{s}{2} - \frac{\Delta}{2}\right) \Gamma\left(\Delta_\cO + \frac{s}{2} - \frac{\tilde\Delta}{2}\right),\\
P_s &\equiv \frac{1+(-1)^s}{2}\,.
\end{aligned}\end{equation}
We will often suppress the $\Delta$ dependence, writing $Y_\cO(s)$. The implicit equation \eqref{eqn:implicitdelta} has an infinite number of $\Delta$  solutions for fixed $d,s$; these can be found either perturbatively around the free theory (to any order desired) or exactly numerically. 
Note that there are no solutions for odd $s$; this is because $\phi$ is a real field. This reflects the standard fact \cite{Simmons-Duffin:2023cftNullPlane} that if we wish to analytically continue in $s$ we must do so for the cases of odd and even $s$ separately.

\subsection{Extension to the \texorpdfstring{\lammelonic}{lambda-melonic} fixed point} \label{sec:bilinearsCalculationDetails}

The essence of the analysis is unchanged, as the tensor structure at the \lammelonic fixed point is identical to that of the quartic model. 
However, the computations are now complicated by (1) multiple fields, including complex fermions, (2) multiple conformal structures appearing in a given conformal correlator, and (3) mixing between conformal correlators of different types. Note also that the results will in general now depend on the ratio of fermionic/bosonic degrees of freedom, $r=2T$.

The structures compatible with conformal invariance are now
\begin{equation}\begin{aligned}
\{v^a(\cO)\} \, \equiv \, \{ \langle \phi_{abc}(x_1) \phi_{abc}(x_2) \cO(x_3) \rangle^a\} \cup \{\langle \psi_{abc}(x_1) \bar\psi_{abc}(x_2) \cO(x_3)\rangle^a\},
\end{aligned}\end{equation}
 where the operators $\cO$ with non-zero $c^a_{\phi\phi\cO}$ and $c^a_{\psi\bar{\psi}\cO}$ are precisely those operators which could appear in the OPE of $\phi \times \phi$, $\psi\times\bar{\psi}$ respectively \eqref{eqn:three-pt}.
We do not here need to consider $\langle \psi \phi \cO\rangle$, which is non-zero only for fermionic $\cO$.

\subsubsection{Eigenvector candidates}

It is convenient to work in a $v^a$ basis labelled by the free field operators in $d=3$, shown in Table \ref{table:3Dfree} grouped by parity. Recall that we are considering operators in the traceless symmetric $s$-tensor representation in the index-free representation, and sending $x_3\to\infty$ in all correlators. 

The purely bosonic operators take the same form as before. 
The spin-0 bilinear operators made of fermions take the schematic form $\bar{\psi} (\slashed{\partial})^n \psi$. 
In $d=3$, for odd $n=2m+1$, this is the parity-even scalar $\bar\psi(\partial^2)^m \slashed{\partial}\psi$; for even $n=2m$ it is the parity-odd pseudoscalar $\bar{\psi} (\partial^2)^m \psi$ \cite{Giombi:2017rhm}.

The most general non-chiral (i.e. without $\gamma_5$) three-point function of two fermions and an $s=0$ scalar operator must then take the form %
\begin{equation}\begin{aligned}\label{eq:threeptpsipsibarO}
\left\langle\psi(x_1)\bar{\psi}(x_2)\mathcal{O}_\Delta(x_3)\right\rangle=\frac{c_{\phi\bar{\phi}\cO}^{P+} \left(\frac{\slashed{x}_{12}}{\abs{x_{12}}^2}\right)+c_{\phi\bar{\phi}\cO}^{P-} \left(\frac{\slashed{x}_{13}\slashed{x}_{32}}{\abs{x_{12}}\abs{x_{31}}\abs{x_{23}}}\right)}{\left|x_{31}\right|^\Delta\left|x_{12}\right|^{2 \Delta_\psi-1-\Delta}\left|x_{23}\right|^\Delta}.
\end{aligned}\end{equation}
Taking $x_3\to\infty$ gives the two eigenvectors $ v_{s=0}^{F\bar{F}_{\slashed{x}}}$ and $v_{s=0}^{F\bar{F}_{1}}(x_1, x_2)$, shown in Table \ref{table:3Dfree}. 
Generalizing to the case of $s>0$, we find one $\expval{\phi\phi\cO}$ and four $\expval{\psi\psi\cO}$ structures.%

\begin{table}[h!]
\begin{tabular}{L|L|L|L|L}
  d=3 \text{ symmetry} &\text{name} & \text{schematic form} & \text{eigenvector candidate}\\\hline
  &BB & \phi (\xi \cdot \partial)^s  (\partial)^{2n} \phi & \frac{(\xi \cdot x_{12})^s}{\abs{x_{12}}^{2\Delta_\phi - \tau}} & \\
  P\text{-even} &F\bar{F}_{\slashed{\xi}} & \bar{\psi} (\xi \cdot \partial)^{s-1} (\partial)^{2n} \slashed{\xi} \psi & \frac{(\xi \cdot x_{12})^{s-1} \slashed{\xi}}{\abs{x_{12}}^{2\Delta_\psi -1 - \tau}}& \\
  & F\bar{F}_{\slashed{x}} & \bar{\psi}(\xi \cdot \partial)^s  (\partial)^{2n} \slashed{\partial}\psi & \frac{(\xi \cdot x_{12})^s \slashed{x}_{12}}{\abs{x_{12}}^{2\Delta_\psi + 1 - \tau}}& \\\hline
  P\text{-odd} &  F\bar{F}_{1} & \bar{\psi}(\xi \cdot \partial)^s (\partial)^{2n} \psi & \frac{(\xi \cdot x_{12})^{s}}{\abs{x_{12}}^{2\Delta_\psi - \tau}}& \\
   &  F\bar{F}_{\slashed{\xi}\slashed{x}} & \bar{\psi}(\xi \cdot \partial)^{s-1} (\partial)^{2n} ( \slashed{\xi}\slashed{\partial}- \slashed{\partial}\slashed{\xi}) \psi & \frac{ (\xi \cdot x_{12})^{s-1} \, (\slashed{x}_{12} \slashed{\xi}-\slashed{\xi}\slashed{x}_{12} ) }{\abs{x_{12}}^{2\Delta_\psi  - \tau}} & \\
\end{tabular}
\caption{The basis of conformal structures for bilinears.}
\label{table:3Dfree}
\end{table}
Note that the $P$ symmetry constrains the theory to have no mixing between odd and even operators. 
Additionally, the $P$-odd kernel turns out to be diagonal in the basis of \cref{table:3Dfree}, and so the $5\times 5$ kernel $k^a{}_b$ is in fact block diagonal.  %

\subsubsection{Kernel}

Writing down the melonic kernel corresponds to drawing the forward two-particle scattering diagrams allowed in the melonic limit, and we find
\newcommand\kernelDiagram[6]{\vcenter{\hbox{\scalebox{0.5}{\feynmandiagram[small, layered layout, horizontal=a to b] {%
i1 -- [#1] a -- [#2] b,
i2 -- [#3] c -- [#4] d,
{ [same layer] a -- [#5, quarter right] c},
{ [same layer] a -- [#6,quarter left] c},
};}}}}
\begin{align}\label{eq:kernelBBFFbPCons}
    K((x_1, x_2), (x_a, x_b)) = \begin{pmatrix}
    \kernelDiagram{scalar}{scalar}{scalar}{scalar}{anti fermion}{fermion} & \kernelDiagram{scalar}{anti fermion}{scalar}{fermion}{scalar}{fermion}\\
    \kernelDiagram{anti fermion}{scalar}{fermion}{scalar}{scalar}{anti fermion} & \kernelDiagram{anti fermion}{anti fermion}{fermion}{fermion}{scalar}{scalar}
    \end{pmatrix} \equiv \frac{\lambda_t^2}{6}\begin{pmatrix}
        (-T)K_{BB\gets BB} & \mathbf{2}(-T)K_{BB\gets F\bar{F}}\\
        K_{F\bar{F}\gets BB} & \frac{1}{2} K_{F\bar{F}\gets F\bar{F}}
    \end{pmatrix}.
\end{align}
 Clearly, the off-diagonal components mean that a $BB$-type eigenvector will generically mix with an $F\bar{F}$-type, and vice versa. Note in addition that
 \begin{enumerate}
    \item The prefactor $\lambda_t^2/6$  comes from $\lim_{N\to\infty}$ of each of the tensor contractions of two $\lambda_{IJKL}$s. 
    \item It is necessary to keep track of closed fermion loops; each generates a factor of $-T$ in the evaluation of \eqref{eqn:three-pt}. These occur when the top row of $K$ is contracted with an eigenvector, either as an explicit fermion loop in the diagram for $K_{BB\gets F\bar{F}}$ or when the external fermion lines of $K$  cap off the fermionic component in an eigenvector. 
    \item The bold factor of $2$ in front of $K_{BB-F\bar{F}}$ could go on either of the off-diagonal entries; it comes from the two ways of placing an element of the kernel to complete a fermion loop in a ladder. The loop can either match or reverse the orientation of the previous fermion loop.
    \item Though there is only one boson structure, $K_{i\gets FF}$ generically will mix between the different possible fermion structures. Indeed, for example, $k^{BB}{}_{F\bar{F}_{\slashed{xi}}}\neq k^{BB}{}_{F\bar{F}_{\slashed{x}}}$. %
\end{enumerate}
We illustrate the calculation of these $k$s via \eqref{eq:3ptSDE} by calculating $k^{BB}_{F\bar{F}_i}$, which is only non-zero for $i=F\bar{F}_{\slashed{\xi}},F\bar{F}_{\slashed{x}}$:
\begin{equation}\begin{aligned}
&\hspace*{-20pt}\kernelDiagram{anti fermion}{scalar}{fermion}{scalar}{scalar}{anti fermion}\cdot v^{BB}(x_{ab}) = \frac{\lambda_t^2}{6}\int K_{F\bar{F}\gets BB}\cdot v^{BB}(x_{ab}) \\
&=\frac{\lambda_t^2}{6}\int_{x_a, x_b} F(x_{1a}) B(x_{ab}) F(x_{ab}) F(x_{b2})\frac{v^{BB}(x_{ab}) + v^{BB}(x_{ba})}{2}\\
&=\frac{\lambda_t^2}{6}\int_{x_a, x_b} F(x_{1a}) B(x_{ab}) F(x_{ab}) F(x_{b2}) P_s v^{BB}(x_{ab})\\
& \equiv \sum_i k^{BB}{}_{F\bar{F}_i}\, v^{F\bar{F}_i}(x_{12}),
\end{aligned}\end{equation}
where we have used the fact that $v^{BB}(x_{ba})=(-1)^s v^{BB}(x_{ab})$, and in the final step decomposed the result of the integral in the basis of Table \ref{table:3Dfree}. The other calculations proceed likewise; plugging these $k$s into $\det(k-1)=0$, we can eliminate $\lambda^2 B^2 F^2$ with the two-point function solution, and $T$ via  \eqref{eq:SDEsForFermionic}; this gives an implicit equation for the scaling dimension $\Delta$ of the bilinear.

\subsection{Bilinear spectrum results for \texorpdfstring{\lammelonic}{lambda-melonic} fixed point} \label{sec:bilinearsCalculationResults}

In the following, we solve numerically and plot as a function of $d$ the scaling dimensions of these $\gO(N)^3$-singlet, traceless spin-$s$, bosonic bilinears -- and then discuss the results. We select the branch of the \lammelonic{} theory that descends from the free theory in $d=3$ dimensions.

\subsubsection{Eigenvalue conditions}

\noindent \textbf{Parity-even sector}. Using again the projector onto even spins $P_s$, we have:
\begin{equation}\begin{aligned}\label{eq:PevenGeneralsSDcondition}
0=&\left(1-\frac{Y_{\psi }(s-1)}{Y_{\tilde{\psi }}(s-1)}\right) \left(1+P_s \frac{Y_{\phi }(s)}{Y_{\tilde{\phi }}(s)}+\frac{Y_{\psi }(s+1)}{Y_{\tilde{\psi }}(s+1)} -3P_s\frac{Y_{\phi }(s)}{Y_{\tilde{\phi }}(s)}\frac{Y_{\psi }(s+1)}{Y_{\tilde{\psi }}(s+1)} \right)\\
&+2 s \frac{Y_{\psi }(s-1)}{Y_{\tilde{\psi }}(s+1)} \left(\Delta _{\phi }+\left(d+s-2-\Delta _{\phi }\right) P_s \frac{Y_{\phi }(s)}{Y_{\tilde{\phi }}(s)}\right),\\
&\text{where } X_\psi \equiv -i \pi^{d/2} \frac{\Gamma\left(\frac{d}{2}-\Delta_\psi + \half \right)}{\Gamma(\Delta_\psi + \half)}.\\
\end{aligned}\end{equation}
We stress the different definition of $X_\psi$ for a fermionic field.

\noindent \textbf{Parity-odd sector}. This is in fact diagonal in the basis of Table \ref{table:3Dfree}. 
Recalling the definition \eqref{eq:Yfuncdef} of $Y_\cO(s)$, we find:
\begin{equation}\begin{aligned} \label{eq:PoddGeneralsSDcondition}
F\bar{F}_1 : \quad 0&=1+\frac{Y_\psi(s)}{Y_{\tilde\psi}(s)}, \quad \quad F\bar{F}_{\slashed{\xi}\slashed{x}} : \quad 0= 1-\frac{Y_\psi(s)}{Y_{\tilde\psi}(s)}.
\end{aligned}\end{equation}
It is clear that when $s=0$, the sector corresponding to $F\bar{F}_{\slashed{\xi}}$ decouples as expected, since $F\bar{F}_{\slashed{\xi}}$-type operators are non-local for $s=0$.

In these results, we expect a \textit{shadow symmetry} of the spectrum. 
That is, every physical operator of dimension $\Delta$, has a corresponding non-physical, non-local operator of dimension $\tilde{\Delta}\equiv d- \Delta$, called the shadow operator \cite{Giombi:2018qgp}. 
In a given CFT, we can identify which of this pair of operators is physical by analytically continuing from the free theory. 
Mathematically, these arise due to the Schwinger-Dyson equations having a symmetry under $\Delta \leftrightarrow \tilde\Delta$. 
Since $Y_{i}(s,\Delta)=Y_{i}(s,\tilde\Delta)$ is manifestly shadow symmetric, these eigenvalue conditions are also symmetric about the line $\Delta= \frac{d}{2}$, shown in the plots below in widely dashed blue. 
This is similar to the shadow symmetry observed in Chew-Frautschi plots of Regge trajectories \cite{Caron-Huot:2022eqs,Simmons-Duffin:2023cftNullPlane}.

To solve \eqref{eq:PevenGeneralsSDcondition} and \eqref{eq:PoddGeneralsSDcondition} numerically, we must input the scaling dimensions of the fundamental fields. We can eliminate $\Delta_\psi$ via $2\Delta_\phi + 2\Delta_\psi =d$; however, if we want to find the spectrum for a given $d$, we must first find $\Delta_\phi$ via the two-point SDEs, \eqref{eq:SDEsForFermionic}. Of course, there are multiple branches of $\Delta_\phi$ available at each $d$; here, we use the solution branch for $\Delta_\phi$ descending from the free theory in $d=3$, marked in \cref{fig:Yuk46SDEdimPhiR4}. 
As noted above, although this theory exits the regime of validity at $d < 1.46$, it is nonetheless interesting to look at the continuation of this path of theories down to $d=0$; although non-unitary, they could be accessed with a modified free scaling dimension.

Once again, due to the presence of $P_s$, even and odd spins are part of different Regge trajectories, and so must be analytically continued separately \cite{Simmons-Duffin:2023cftNullPlane}. 

\subsubsection{Scaling dimensions} \label{sec:bilinearsObservations}

\begin{figure}[ht]
\centering
\includegraphics[width=0.7\textwidth]{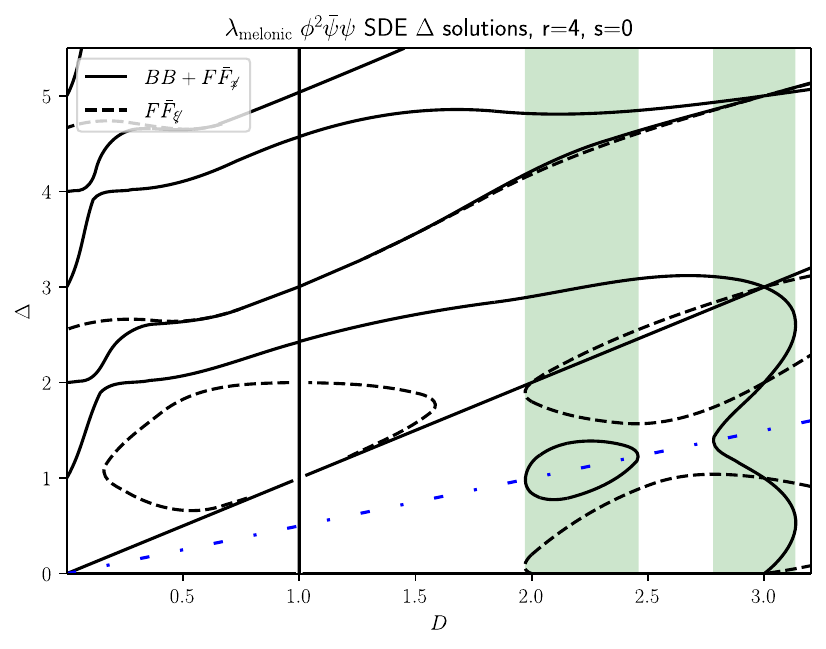}
\caption{P-even $s=0$ spectrum for $r=4$ \lammelonic.  The dimensional windows of stability, where all known local operators are real, are shown here in green. In the left-hand green window, the $F\bar{F}_{\slashed{\xi}\slashed{x}}$ (shown in \cref{fig:lmelonicSpecPoddr4S0}) has a complex scaling dimension; however, it is non-local, so  the implications for the fate of the theory are unclear. Note that the theory for $d>3$ automatically violates the unitarity bounds, since $\Delta_\phi < \frac{d-2}{2}$ (see \cref{fig:Yuk46SDEdimPhiR4}). The widely dashed blue line indicates the line of shadow symmetry. To see what happens to the disappearing solution, see the complex version of this plot in \cref{fig:lmelonicSpecPevenr4S0-3Dplot}.} %
\label{fig:lmelonicSpecPevenr4S0}
\end{figure}

\begin{figure}[ht]
\centering
\includegraphics[width=0.7\textwidth]{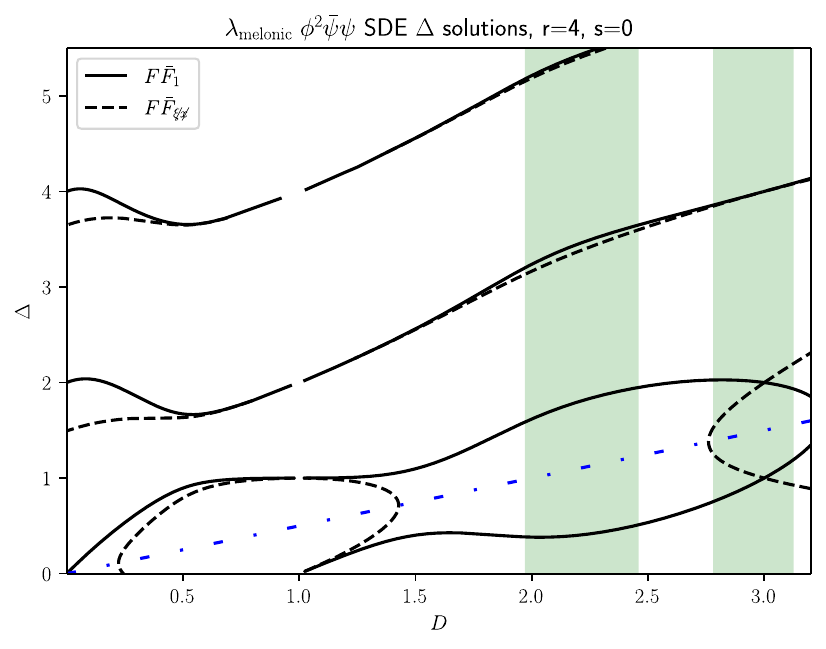}
\caption{P-odd $s=0$ spectrum for $r=4$ \lammelonic. Note that the dashed $F\bar{F}_{\slashed{\xi}}$ is again non-local for this value of $s$. There exist no solutions for $d=1$, hence the break in the line. We shade the $d$-range of stability given by \cref{fig:lmelonicSpecPevenr4S0} again in green.} %
\label{fig:lmelonicSpecPoddr4S0}
\end{figure}

\begin{figure}[ht]
\centering
\begin{subfigure}{0.5\textwidth}
  \centering
\includegraphics[width=\textwidth]{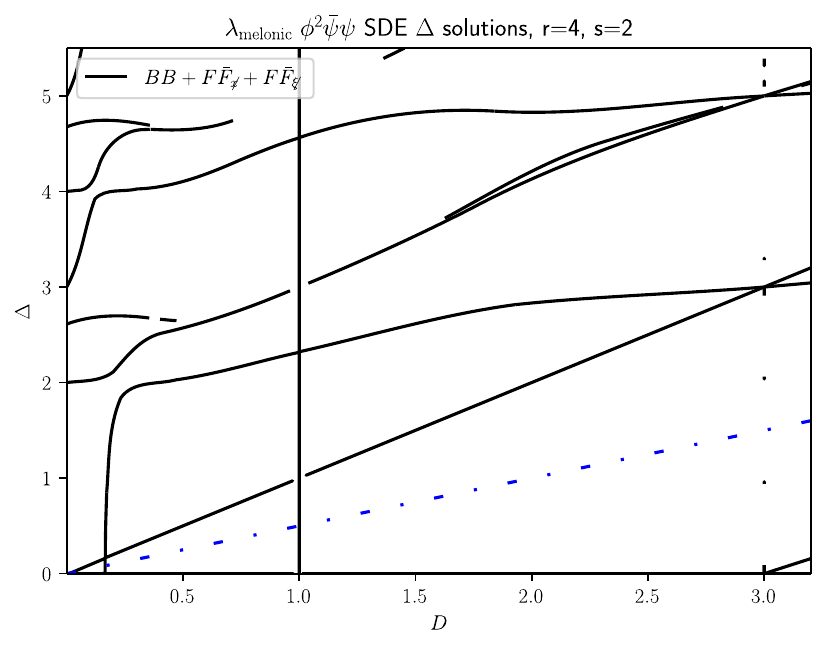}
  \caption{P-even $s=2$}
  \label{fig:lmelonicSRFCs4T2specPEven}
\end{subfigure}%
\begin{subfigure}{0.5\textwidth}
  \centering
\includegraphics[width=\textwidth]{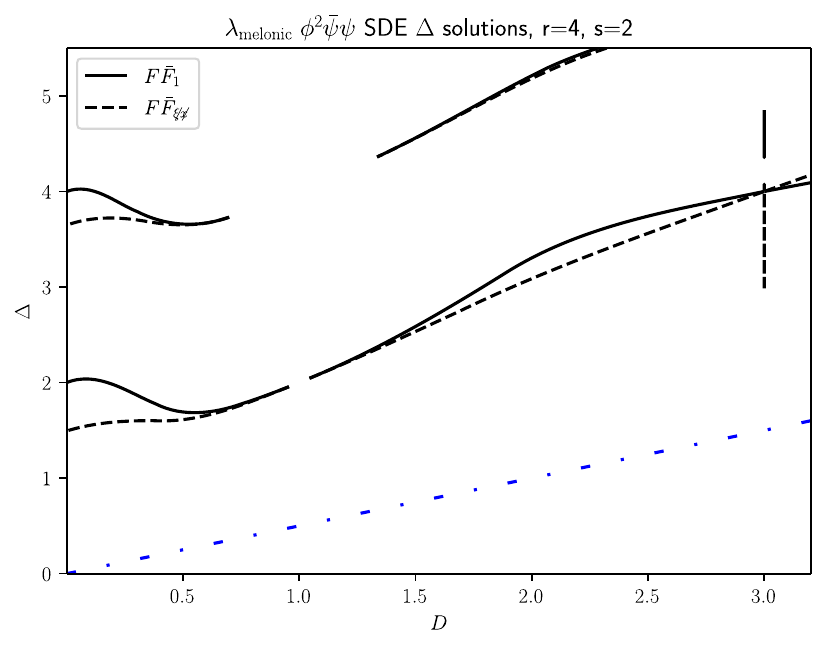}
  \caption{P-odd $s=2$}
  \label{fig:lmelonicSRFCs4T2specPOdd}
\end{subfigure}
  \caption{$s=2$ spin spectrum for $r=4$ \lammelonic{}. All operators shown here are local, including $F\bar{F}_{\xi}$, which is no longer decoupled. As usual, we show the line of shadow reflection. Note the presence of the $\Delta=d$ stress tensor in the P-even spectrum.} \label{fig:lmelonicSRFCs4T2spec}
\end{figure}

In \cref{fig:lmelonicSpecPevenr4S0,fig:lmelonicSpecPoddr4S0,fig:lmelonicSRFCs4T2spec} we display the spectrum of this theory as a function of continuous dimension $d<3$; the spectrum at a particular value of $d$ can be found by slicing the contour plot at that $d$. We restrict to the low-$\Delta$ region (around $\Delta <5$), as the high-$\Delta$ spectrum rapidly asymptotes to a known trivial form,
\begin{align}
&\text{for $BB$: }\quad \Delta=2\Delta_\phi + 2n + s+ O(1/n);\\
&\text{for $F\bar{F}_i$: }\quad \Delta = 2\Delta_\psi + 2n + s+ O(1/n).
\end{align}
i.e. those of the operators appearing in the OPE of a generalized free field theory with scaling dimensions $\Delta_\phi$ and $\Delta_\psi$, up to $O(1/n)$ corrections.
We postpone discussion of the $d=1$ region, where all $\Delta$s solve \eqref{eq:PevenGeneralsSDcondition} and no $\Delta$s solve \eqref{eq:PoddGeneralsSDcondition}, to \cref{sec:zeroScaling}.

\subsubsection{Discussion of the spectrum}

We start by identifying the elements of the physical spectrum. This is not necessarily trivial, as discussed in appendix A of \cite{Benedetti:2020iku}; however, we can exploit the fact that near the free theory in $d=3$, the operators in the spectrum must exactly match the known scaling dimensions of the physical operators. In the P-even spectrum for $s=0$, we ignore the $F\bar{F}_{\slashed{\xi}}$ operator, as it is both non-local and decoupled. We can then identify in \cref{fig:lmelonicSpecPevenr4S0} that there are five physical operators. In $d=3$, we can identify them exactly: they have dimensions $\Delta=1,3,5$ and $\Delta=3,5$, and correspond to $\bar\phi (\partial^2)^{n=0,1,2} \phi$ and $\bar\psi (\partial^2)^{n=0,1} \slashed{\partial} \psi$. Hence, the solid contour leaving $\Delta=1$ is the physical scalar mass operator $\phi^2$, and the solid contour leaving $\Delta=2$ is its shadow. For $s=2$, \cref{fig:lmelonicSRFCs4T2specPEven}, all operators shown are physical.

In the P-odd spectrum, for $s=0$, we again ignore the non-local $F\bar{F}_{\slashed{\xi}\slashed{x}}$. 
We then have three physical local operators in \cref{fig:lmelonicSpecPoddr4S0}, which we can identify as $\bar\psi (\partial^2)^{n=0,1,2} \psi$. 
This makes it clear that, in this case, the second from bottom solid contour is the fermion mass operator $\bar\psi \psi$, with the bottom one being its non-physical shadow. For $s=2$, \cref{fig:lmelonicSRFCs4T2specPOdd}, all operators are again physical.

Note that the singlet mass operators of the two fundamental fields, $\phi^2$ and $\bar\psi\psi$, are present, so we can use them to cross-check our interpretation.
Setting $d=3-\epsilon$ and expanding the eigenvalue condition for $s=0$, we find that for $\Delta=1+O(\epsilon)$ in the parity-even sector and $\Delta=2+O(\epsilon)$ for $F\bar{F}_1$ in the parity-odd sector, we exactly match the $O(\epsilon^2)$ perturbative computations of $\Delta_{\bar\phi \phi}^{\text{pert}} \equiv d+\odv{\beta_{m^2}}{m^2}|_{\lambda_*}$ and $\Delta_{\bar\psi \psi}^{\text{pert}} \equiv d+\odv{\beta_{M}}{M}|_{\lambda_*}$ at the \lammelonic fixed point (given in \eqref{eq:quadraticSDs}). 
Additionally, since this is a large-$N$ CFT, we expect the scaling dimensions of multi-trace operators to factorize.
We calculated $\Delta_{h_8}^{\text{pert}}  = [(\phi_{abc}\phi_{abc})^3]$ and $\Delta_{\lambda_{d_S}}^{\text{pert}}  = [(\phi_{abc}\phi_{abc})(\bar\psi_{abc}\psi_{abc})]$ perturbatively in \cref{sec:stabmats}, and indeed we find them to be equal to $3\Delta_{\phi^2}$ and $\Delta_{\phi^2} +\Delta_{\bar\psi \psi}$ respectively.

\subsubsection{Windows of stability} \label{sec:windowsOfStability}

It is worth first distinguishing the following three types of CFT:
\begin{enumerate}
    \item A unitary CFT with all local scaling dimensions real.
    \item A non-unitary CFT with all low-lying local scaling dimensions real.
    \item A CFT with some complex low-lying local scaling dimensions, which makes it automatically non-unitary, and potentially unstable.
\end{enumerate}
\noindent In non-integer dimension, from studies of the Wilson-Fisher fixed point, we expect all CFTs to be non-unitary due to the presence of evanescent operators (with complex scaling dimension), leading to states of negative norm \cite{Hogervorst:2015akt,DiPietro:2017vsp,Ji:2018yaf}. 
However, such operators usually have a high scaling dimension, and are thought not to immediately appear in the OPE of the fundamental field:
we see in \cref{fig:lmelonicSpecPevenr4S0} that for some ranges of dimension $d$, all the low-lying local scaling dimensions are real: namely, $d\in (1.97,2.46)$ and $(2.78, 3.14)$. 

\begin{figure}
\centering
\includegraphics[width=0.7\textwidth]{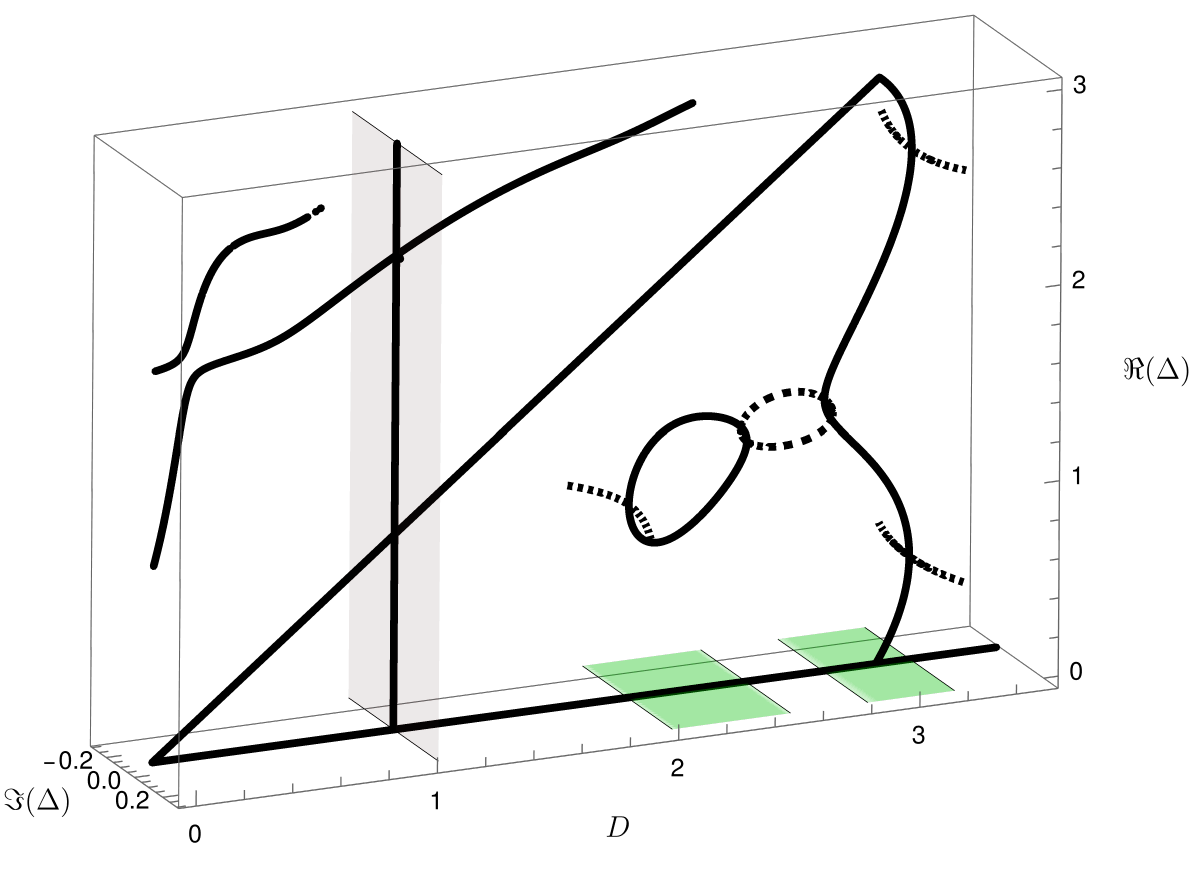}
\caption{P-even $s=0$ local spectrum for $r=4$ \lammelonic ($BB+F\bar{F}_{\slashed{x}}$), now including the complex solutions for $\Delta$. The real slice here precisely matches the lower part of \cref{fig:lmelonicSpecPevenr4S0}. Purely real solutions are solid, and complex solutions are dashed; all lines that seemingly end actually just continue on into the complex $\Delta$ plane. Once again, the windows of stability (with no imaginary solutions) are shown in green, and we shade $d=1$ to indicate that all complex values of $\Delta$ solve the equations. Note the collision of the $\phi^2$ operator with its shadow, giving a pair of operators with dimension $\Delta = \frac{d}{2} \pm i \alpha$ for $d$ outside the green region.} 

\label{fig:lmelonicSpecPevenr4S0-3Dplot}
\end{figure}

On the boundaries of these ranges, the scaling dimensions of two operators collide, and we then obtain a pair of operators with complex conjugate scaling dimensions. This is manifest for the $d\in (1.97,2.46)$ window; in the complex plot of \cref{fig:lmelonicSpecPevenr4S0-3Dplot}, we see that the $\phi^2$ operator complexifies in that $d$ range. 
We noticed in \cref{sec:generalCharacteristicsPrismatic} that the same behaviour occurs for the fundamental field in certain dimensional ranges as well.

Hence, inside the windows of stability: in integer dimension, the \lammelonic{} theory is type 1; in non-integer dimension, it is type 2. Outside the windows, it is type 3.

This phenomenon of windows of stability for the parameter values, only apparent non-perturbatively, appears to be characteristic of melonic models \cite{Giombi:2017dtl,Prakash:2017hwq,Murugan:2017eto,Giombi:2018qgp,Benedetti:2019eyl,Benedetti:2019rja,Kim:2019upg,Benedetti:2020iku,Chang:2021wbx,Prakash:2022gvb} (the windows in $d$ are demonstrated for \hprismatic{} in \cref{fig:Bosonic6SDEdimPhi}). 
However, this behaviour depends strongly on the details of the theory -- in this case, $r$ and $d$: for example, for $r=8$, the $d\in (1.97,2.46)$ window visible in \cref{fig:lmelonicSpecPevenr4S0} disappears entirely. 
Additionally, we see that $F\bar{F}_{\slashed{\xi}\slashed{x}}$ remains complexified throughout the $(1.97,2.46)$ window; however, it is a non-local operator and its interpretation is unclear. 

The complex scaling dimension of this form means that the critical points along this line are unstable. 
That is, the free energy is not minimised (not even locally) by the conformal solution, as there exist fluctuations that lower the free energy; hence the conformal vacuum is not the true vacuum of the precursor melonic QFT. For this reason, we term these as windows of stability.
\cite{Benedetti:2021qyk} proved that if there is an operator $\cO$ of the form $\Delta=d/2 + i \alpha$ in the OPE of two fundamental scalar fields, the conformal solution is unstable. 
Beyond melonic theories, such operators are often observed in large-$N$ (gauge/melonic/fishnet) CFTs more generally \cite{Kim:2019upg,Dymarsky:2005uh,Pomoni:2009joh, Grabner:2017pgm, Gorbenko:2018ncu,Benini:2019dfy}.
By the AdS/CFT duality, this is the CFT counterpart of the well-known Breitenlohner-Freedman bound in AdS: the dual field of the operator has a mass below the BF bound. 
It is conjectured that the fate of these theories after this instability is that in the true vacuum, this operator acquires a VEV, $\expval{\cO}\neq 0$ \cite{Kim:2019upg,Benedetti:2021qyk}.

It is interesting to observe that in the bosonic $\gO(N)^3$ quartic and sextic models (and their generalizations and extensions) it is also the operator that roughly corresponds to $\phi^2$ that complexifies along with its shadow \cite{Giombi:2017dtl,Benedetti:2019rja} when descending in $d$ from the marginal theory.
In all of these models, the operator that complexifies is the physical bilinear operator of lowest scaling dimension, with $\Delta < d/2$. %

Despite the non-unitarity of the CFTs, it is easy to check that for integer spins $s$ in $d<3$, the physical operators identified here all satisfy the standard $d\ge 2$ CFT unitarity bounds \cite{Benedetti:2023mli}:
\begin{equation}
    \Delta_{s=0} \ge \frac{d}{2}-1,\quad  \Delta_{s \neq 0}\ge d+s-2
\end{equation}
There is one exception: for $d=1$ \cref{fig:lmelonicSpecPevenr4S0} makes it clear that every scaling dimension is a solution to \eqref{eq:PevenGeneralsSDcondition}. 
However, this dimension is pathological in any case, as we will discuss in \cref{sec:zeroScaling}. For $d>3$, of course, $\phi$ itself manifestly violates the unitarity bound $\Delta_\phi \ge \tfrac{d-2}{2}$.

\subsubsection{Operators of protected dimension} \label{sec:protectedOperators}

We frequently observe operators of protected dimension; that is, operators which have scaling dimension fixed at $d$ or $d-1$. These are usually associated with some symmetry.

\begin{mccorrection}
In the P-even $s=0,2$ sectors, there are always operators with dimension $\Delta=d$. 
The $s=2$ operator corresponds to the conserved stress tensor. 
The $s=0$ operator is fictitious: it appears to be a vanishing mixture of $\phi \, \partial^2\,  \phi$ and $\bar\psi \slashed{\partial}\psi$ -- this occurs because, by the equations of motion, both are proportional to $\phi^2 \bar\psi \psi$; hence, a linear combination of them vanishes as an operator \cite{Bulycheva:2017ilt}.
Note that this operator is unrelated to the trace of the stress tensor (which is certainly zero as an operator), as no analogue exists in the single-field melonic theories. 
In the case of the generalized SYK model, Gross and Rosenhaus interpreted this as coming from the IR rescaling symmetry \cite{Gross:2016kjj}; in our case, this is the IR local rescaling symmetry $\psi \to A \psi, \phi \to A^{-1} \phi$, for a local function $A(x)\neq 0$. 
Indeed, in the generalized SYK model, its OPE coefficient with the fundamental fields was found to vanish, further confirming it to be a zero operator; we suspect that the same would apply here, though further calculations are required to confirm this \cite{Fraser-Taliente2026spectrum}.
\end{mccorrection}

In the P-even spin-$1$ sector, we find a bilinear that involves only the fermions (because $P_{s=1}=0$), with $\Delta = d-1$; this corresponds to the conserved $\gU(1)$ current that rotates the complex fermions, $\bar\psi \gamma^\mu \psi$. Like the rescaling symmetry, this generically conserved current is, in the deep IR, upgraded to a local $\gU(1)$ symmetry, just as in the SYK model and melonic supertensor models \cite{Davison:2016ngz,Chang:2018sve}: $\psi(x) \to \psi(x) e^{i \alpha(x)}$. In the P-odd spin-$1$ sector, we also find an $F\bar{F}_{\slashed{\xi}\slashed{x}}$-type operator with $\Delta = d$; the associated symmetry is not clear. The $s\neq 0, 2$ sectors are otherwise uninteresting, and so not plotted.

\subsubsection{Spectrum divergences when fundamental fields' dimensions hit zero} \label{sec:zeroScaling}

In exactly $d=1$ dimensions, there is no solution giving $\Delta_\phi=1/2$. However, perturbatively, in $d=1+\epsilon$, we find a solution $\Delta_\phi = \half- O(\epsilon^2)$, $\Delta_\psi =0 + O(\epsilon^2)$, for arbitrary $r$. These are marked with black squares in \cref{fig:Yuk46SDEdimPhiBothRs}. If we ignore this, and proceed anyway, taking $\epsilon \to 0$, we find that every value of $\Delta$ solves the P-even determinant condition, for any value of $s$. For the P-odd family, the determinant condition evaluates to $-1=0$ for $d=1$, and so there is a break visible in every contour along $d=1$. This missing tower of operators is just as in the $d=2$ case of the standard bosonic tensor model \cite{Giombi:2017dtl}: here we have perturbative solutions for $d=1+\epsilon$, $\Delta = 2n+ O(\epsilon)$, which do not exist at $d=1$.
 
\begin{figure}
\centering
\includegraphics[width=0.7\textwidth]{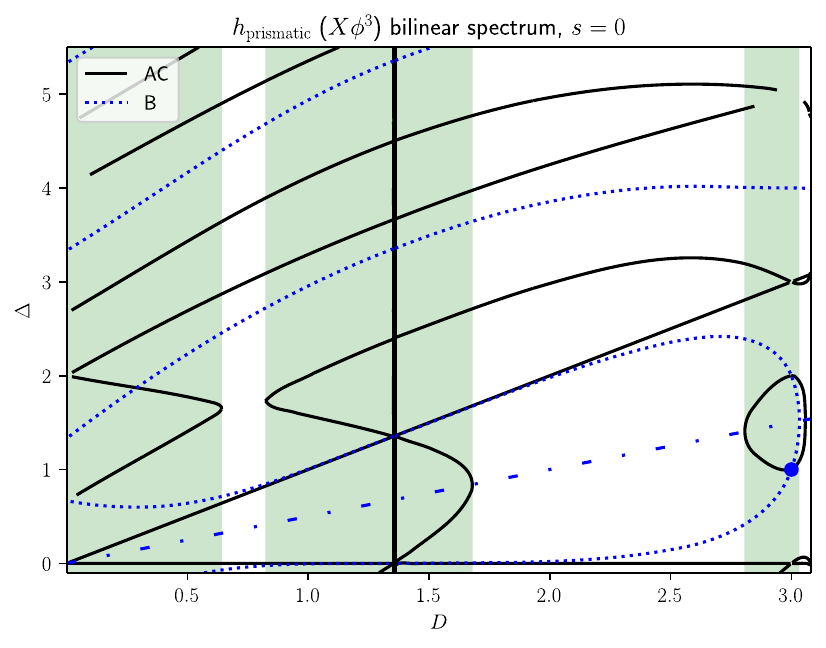}
\caption{Sextic prismatic model, \hprismatic, $s=0$ scaling dimensions \cite{Giombi:2018qgp} for the branch indicated in \cref{fig:Bosonic6SDEdimPhi}. Green shading again indicates the regions of stability; the line of shadow reflection is blue and dashed. For $d\simeq 1.353$, every $\Delta$ solves the eigenvalue equation for AC-type bilinears, and no $\Delta$s solve the B-type. The mass operator $\phi^2$ in the $d=3$ free theory is marked with a blue dot; just as with \lammelonic, it is this operator that obtains a complex scaling dimension at $d\simeq 2.81$. The bilinear $X\phi$, by contrast, is the first to obtain a complex scaling dimension if we increase $d$ to $d\simeq 3.03$.} 
\label{fig:hprismaticSpectrum}
\end{figure}

The same phenomenon also occurs for other theories when the scaling dimension of one of the fields $\phi$ hits certain integer values. Returning to the prismatic $\phi^6$ model \cite{Giombi:2018qgp}: following the family of CFTs descending from $d=3-\epsilon$ indicated in \cref{fig:Bosonic6SDEdimPhi}, while still within the IR wedge we reach $\Delta_\phi =0$ at $d_C \simeq 1.353$\footnote{$d=d_C$ solves $\psi ^{(0)}(d)+\gamma_E=\psi ^{(0)}(\frac{d}{2})-\psi ^{(0)}(-\frac{d}{2})$, with $\psi^{(0)}(x)=\mathrm{Polygamma}[0,x]$ and $\gamma_E$ Euler's constant.}. As shown in \cref{fig:hprismaticSpectrum}, for any spin, every value of $\Delta$ solves the A/C-type bilinear equations there, and no values of $\Delta$ solve the B-type bilinear equations.

Likewise, consider the other branches of \lammelonic in \cref{fig:Yuk46SDEdimPhiBothRs} -- in particular, those that descend from the free theory in $d=2n+1-\epsilon$, when they hit $d=2n-1$, for $n \in \mathbb{N}$. At those points\footnote{It is worth noting, however, that the $d=1$ evaluation of the $\Delta_\phi=+1/2$ solution branch was doubly incorrect, as at precisely this point $\Delta_\psi =0$. This is the same as the free scaling dimension $\frac{d-1}{2}$ in the two-point SDE \eqref{eq:fermionSDEWithoutX}, so in exactly $d=1$ we were not permitted to drop it in the IR.}, we have $\Delta_\phi = (2n-1)/2 + O(\epsilon^2)$ exactly, and hence $\Delta_\psi = 0+ O(\epsilon^2)$. Likewise, along the line $\Delta_\phi =0$, we also see breaks in the contours. 
Indeed, we expect this behaviour to be generic for any melonic-type CFT with multiple fundamental fields -- of course, if there is only one field, $\Delta_\phi =0$ only in $d=0$. 
One exception to this, in the case of only one field, is in the supersymmetric $\cN=1$ quartic $\gO(N)^q$ model, which, as described in \cref{sec:qGenAndSUSYprismatic}, has $\Delta_\Phi = (d-1)/q$; thus, taking the $\cN=1$ theory \cite{Popov:2019nja} we do indeed observe in $d=1$: a missing tower of operators for the BB and FF-type operators; that every value of $\Delta$ satisfies the eigenvalue condition for BF-type operators. %

Despite the presence of these spectrum divergences, there is evidence to suggest that the theories considered in the divergent dimensions may still have the missing states in the spectrum. 
\cite{Benedetti:2019ikb} used a long-range tensorial $g \phi^4$ model, which exhibits a break in the scaling dimension solutions as $d\to 2$ for arbitrary values of the marginal coupling: a character decomposition derivation of the spectrum for the free theory hinted that those states which seem to be absent are in fact still present. 
By continuity of the spectrum in $g$, this would also hold for the short-range models. 
If true, this would also apply to the other melonic $\phi^q$ models, each of which has no solutions to the kernel for $d= \mathbb{N}\frac{2q}{q-2}$.

\begin{subappendices}

\section{Potential for the tensor model} \label{app:potential}

The full analytic expression for the potential at optimal $N$ scaling \cite{Benedetti:2023mli} is as follows.
We have normalized each tensor term such that it has overall weight one: i.e. if $N=1$, then $V=\frac{\sum_i\lambda_i}{2} \phi^2 \bar{\psi}\psi +  \frac{\sum_i h_i}{6!}\phi^6$.
For compactness, we present the coupling constants that have not yet been symmetrized (with weight one) under permutation of the bosonic indices; to use these formulae in a full calculation the symmetrized versions must be used\footnote{This symmetrization is straightforward in each case: on one example $\lambda$ term, we do $r_k^j r_l^i b_k^j b_l^i g_k^j g_l^i \mapsto \frac{1}{2}(r_k^j r_l^i b_k^j b_l^i g_k^j g_l^i + (i \leftrightarrow j))$; on an example $h$ term, $r_j^i r_m^l r_n^k b_l^k b_m^i b_n^j g_l^k g_m^j g_n^i \mapsto \frac{1}{6!}(\text{same} + (6!-1) \text{ perms})$. }. We also use as shorthand $r^i_j = \delta_{i_r j_r}$ (likewise for $g$ and $b$).
\begin{subequations}\label{eq:explicitPotential}
\begin{equation}\begin{aligned}
&\lambda_{(i_r i_g i_b)(j_r j_g j_b)(k_r k_g k_b)(l_r l_g l_b)} =\Big[\\
& +\frac{\lambda_t}{3 N^{3/2}} \left(r_j^i r_l^k b_k^i b_l^j g_k^j g_l^i+r_k^i r_l^j b_j^i b_l^k g_k^j g_l^i+r_k^i r_l^j b_k^j b_l^i g_j^i g_l^k\right)\\
 & +\frac{\lambda_{\text{dD}}}{N^3} \left(r_k^j r_l^i b_k^j b_l^i g_k^j g_l^i\right)\\
 & +\frac{\lambda_{\text{dS}}}{N^3} \left(r_j^i r_l^k b_j^i b_l^k g_j^i g_l^k\right)\\
 & +\frac{\lambda_{\text{pE}}}{3 N^2} \left(r_j^i r_l^k b_j^i b_l^k g_k^j g_l^i+r_j^i r_l^k b_k^j b_l^i g_j^i g_l^k+r_k^j r_l^i b_j^i b_l^k g_j^i g_l^k\right)\\
 & +\frac{\lambda_{\text{pO}}}{3 N^2} \left(r_k^i r_l^j b_k^i b_l^j g_k^j g_l^i+r_k^i r_l^j b_k^j b_l^i g_k^j g_l^i+r_k^j r_l^i b_k^i b_l^j g_k^j g_l^i\right)\\
 & + \frac{\lambda_{\text{pS}}}{3 N^2} \left(r_j^i r_l^k b_k^j b_l^i g_k^j g_l^i+r_k^j r_l^i b_j^i b_l^k g_k^j g_l^i+r_k^j r_l^i b_k^j b_l^i g_j^i g_l^k\right)\Big]_{\text{symmetrized on $i\leftrightarrow j$}}
\end{aligned}
\end{equation}
\begin{equation}\begin{aligned}
&h_{(j_r j_g j_b)(k_r k_g k_b)(l_r l_g l_b)(m_r m_g m_b)(n_r n_g n_b)} = \Big[\\
 &+\frac{h_p}{N^3} \left(r_j^i r_m^l r_n^k b_l^i b_m^k b_n^j g_l^k g_m^j g_n^i\right)\\
 &+\frac{h_w}{N^3} \left(r_l^j r_m^i r_n^k b_l^i b_m^k b_n^j g_l^k g_m^j g_n^i\right)\\
 &+\frac{h_3}{3 N^{7/2}} \left(r_j^i r_m^l r_n^k b_l^k b_m^i b_n^j g_l^k g_m^j g_n^i+r_k^i r_m^l r_n^j b_l^i b_m^k b_n^j g_l^k g_m^j g_n^i+r_k^j r_m^l r_n^i b_l^i b_m^k b_n^j g_l^k g_m^j g_n^i\right)\\
 &+\frac{h_4}{3 N^4} \left(r_l^i r_m^k r_n^j b_l^i b_m^k b_n^j g_l^k g_m^j g_n^i+r_l^i r_m^k r_n^j b_l^k b_m^j b_n^i g_l^k g_m^j g_n^i+r_l^k r_m^j r_n^i b_l^i b_m^k b_n^j g_l^k g_m^j g_n^i\right)\\
 &+\frac{h_5}{3 N^4} \left(r_l^i r_m^k r_n^j b_l^k b_m^i b_n^j g_l^k g_m^j g_n^i+r_l^k r_m^i r_n^j b_l^i b_m^k b_n^j g_l^k g_m^j g_n^i+r_l^j r_m^k r_n^i b_l^k b_m^i b_n^j g_l^k g_m^j g_n^i\right)\\
 &+\frac{h_6}{N^{9/2}} \left(r_j^i r_l^k r_n^m b_l^k b_m^i b_n^j g_l^k g_m^j g_n^i\right)\\
 &+\frac{h_7}{3 N^5} \left(r_l^k r_m^i r_n^j b_l^k b_m^i b_n^j g_l^k g_m^j g_n^i+r_l^k r_m^i r_n^j b_l^k b_m^j b_n^i g_l^k g_m^j g_n^i+r_l^k r_m^j r_n^i b_l^k b_m^i b_n^j g_l^k g_m^j g_n^i\right)\\
 &+\frac{h_8}{N^6} \left(r_l^k r_m^j r_n^i b_l^k b_m^j b_n^i g_l^k g_m^j g_n^i\right)\Big]_{\text{symmetrized on $S_6$}}
\end{aligned}\end{equation}
\end{subequations}
The unpleasantness of the notation should make manifest the reason for the visual shorthand. %

\subsection{Potential comparison} \label{app:potentialcomparison}
To ease comparison with the other sextic models of the literature, we directly compare our real bosonic potential sector, $\{h_i\}$, to the \textit{complex} bosonic sextic potential in \cite{Benedetti:2019rja} ($\{\boldsymbol{\lambda}_i (\bar{\phi}\phi)^3\}$), and the real bosonic sextic potential in \cite{Giombi:2018qgp} ($\{\boldsymbol{g}_i \phi^6\}$). 

If we make the fermion sector free ($\lambda_i =0$), and restrict to the couplings that descend from $\gU(N)$-invariant couplings (see the table below), we obtain from our beta functions exactly the large-$N$ results of the complex sextic melonic model, up to a rescaling $h= 15 \boldsymbol{\lambda}/s^3$. 
This is despite their model being complex, and so in theory having double the number of degrees of freedom; this is likely a simple consequence of the large-$N$ limit. %

Since all $O \in \gO(N)$ also satisfy $O \in \gU(N)$, as $O^\dagger = O^T =O^{-1}$, we note that all possible $\gU(N)^3$ interaction terms are manifestly also invariant under $\gO(N)^3$. Thus, the $\gU(N)^3$ interaction terms are a subset of the $\gO(N)^3$ interaction terms, so we can identify:
\begin{center}
\begin{tabular}{|c|c|c|c|}
  \hline
  \textbf{Real sextic} & \textbf{Complex sextic $ (\bar{\phi}\phi)^3$} & \textbf{Real prismatic $\phi^6$} \\
  \hline
  $h_p/N^3$ & Forbidden by $\gU(N)$ & $\boldsymbol{g}_1/N^3$ \\
  \hline
  $h_w/N^3$ & $\boldsymbol{\lambda}_1/N^3$ & $\boldsymbol{g}_2/N^5$ \\
  \hline
  $h_3/N^{7/2}$ & Forbidden by $\gU(N)$ & $\boldsymbol{g}_5/N^4$ \\
  \hline
  $h_4/N^4$ & $\boldsymbol{\lambda}_3/N^4$ & $\boldsymbol{g}_4/N^5$ \\
  \hline
  $h_5/N^4$ & $\boldsymbol{\lambda}_2/N^4$ & $\boldsymbol{g}_3/N^4$ \\
  \hline
  $h_6/N^{9/2}$ & Forbidden by $\gU(N)$ & $\boldsymbol{g}_6/N^5$ \\
  \hline
  $h_7/N^5$ & $\boldsymbol{\lambda}_4/N^5$ & $\boldsymbol{g}_7/N^5$ \\
  \hline
  $h_8/N^6$ & $\boldsymbol{\lambda}_5/N^6$ & $\boldsymbol{h}_8/N^7$ \\
  \hline
\end{tabular}
\end{center}
We can also recover the beta functions of the prismatic models, though in \cite{Giombi:2018qgp} non-optimal scalings were chosen for the couplings, and so some possible fixed points were missed.

\section{Perturbative beta functions}
\subsection{Notes on their calculation} \label{app:calcNotes}
\subsubsection{Gamma matrices in \texorpdfstring{$(0+3)$d}{(0+3)d}} \label{app:gammaMatrices}

Our fields here are Euclidean Dirac spinors, which in $d=3$ have $2$ complex, and so $4$ real, components.
Following the conventions of (12.3) of \cite{zinn-justin_quantum_2002}, we define $\bar{\psi} = -\psi^\dagger \gamma_0$.
The gamma matrices in $(0+3)d$, with signature $(+++)$ are the Pauli matrices, $\gamma^\mu = \gamma_\mu =\sigma_\mu$ for $\mu=1,2,3$:

In precisely three dimensions, the trace rules are the following:
\begin{equation}\begin{aligned}
\Tr[\gamma^\mu] &=0, \qquad \Tr[\gamma^\mu \gamma^\nu] = \delta^{\mu\nu} \Tr \spinid = 2\delta^{\mu\nu},\\
\Tr[\gamma^\mu \gamma^\nu \gamma^\rho] &= 2i\epsilon^{\mu\nu\rho}, \qquad \epsilon^{\mu\nu\rho} \text{ s.t. } \epsilon^{123} = +1,
\end{aligned}\end{equation}
with $\epsilon^{\mu\nu\rho}$ the alternating tensor with the normalization indicated. However, it is well known that DReD is algebraically inconsistent; in particular, because different contractions of three or more $\epsilon^{\mu\nu\rho}$ factors yield different results in $d<3$ \cite{Siegel:1980qs}. Our computation of the beta functions in $d=3-\epsilon$ did not suffer from this as we have no more than one momentum at any point in our computation (as we renormalized the couplings at zero momentum), and thus no epsilon factors can ever occur, because we can always reduce any trace using the identity $\slashed{p}\slashed{p}=p^2$. Thus, as long as we take traces only at the end, no problems can arise. 
In any case, we keep $\Tr \spinid = T$ general in our computations.

\subsubsection{Lorentzian massive melonic integrals in the epsilon expansion} \label{app:massiveMelonicIntegrals}
We give here the fermionic sunrise integral, in full generality of masses and momenta, along with conventions. 
The standard Lorentzian sunrise integral, with metric signature $(+--)$ in $d=3-\epsilon$ can be found in the literature \cite{Rajantie:1996np}:
\begin{equation}
\begin{split}
&\mu^{2\epsilon}\int\frac{\mathrm{d}^d k\, \mathrm{d}^d l}{(2\pi)^{2d}} \frac{1}{(k^2-m_1^2+i \eta ) (l^2-m_2^2+i \eta ) ((k+l-p)^2-m_3^2+i \eta )}\\
&=\frac{1}{32 \pi ^2}\left[\frac{1}{\epsilon }+\log \left(\tfrac{\bar{\mu }^2}{M^2-p^2}\right)-\frac{2M}{p} \tanh ^{-1}\left(\tfrac{p+0i}{M}\right)+3+O(\epsilon ^1)\right],
\end{split}
\end{equation}
where we have defined the usual $\overline{\mathrm{MS}}$ parameter, $\bar{\mu}^2 = e^{-\gamma_E} 4 \pi \mu^2$, and the mass sum $M\equiv m_1+m_2+m_3$. Next, including fermions:
\begin{equation}
\begin{split}
&\mu^{2\epsilon}\int\frac{\mathrm{d}^{d} k\, \mathrm{d}^{d} l}{(2\pi)^{2d}} \frac{k\cdot l}{(k^2-m_1^2+i \eta ) (l^2-m_2^2+i \eta ) ((k+l-p)^2-m_3^2+i \eta )}\\
&=\tfrac{1}{192\pi^2} \Biggl[\left(p^2 -3 (m_1^2+ m_2^2-m_3^2)\right) \left(\frac{1}{\epsilon} + \log \left(\tfrac{\bar{\mu }^2}{M^2-p^2}\right)\right) \\
&\quad +\tfrac{2 \left(3 m_3 \left(m_1^2+m_2^2-p^2\right)-m_3^3+2 \left(m_1^3+m_2^3\right)\right)}{p} \tanh ^{-1}\left(\tfrac{p+0i}{M}\right)\\
&\quad -7 m_1^2-7 m_2^2+5 m_3^2-2 m_1 m_2+4 \left(m_1+m_2\right) m_3+3 p^2 +O(\epsilon^1)\Biggr]
\end{split}
\end{equation}
and 
\begin{equation}
\begin{split}
&\mu^{2\epsilon}\int\frac{\mathrm{d}^{d} k\, \mathrm{d}^{d} l}{(2\pi)^{2d}} \frac{k\cdot p}{(k^2-m_1^2+i \eta ) (l^2-m_2^2+i \eta )((k+l-p)^2-m_3^2+i \eta )}\\
&=\tfrac{1}{96\pi^2} \Biggl[p^2\left(\tfrac{1}{\epsilon }+ \log \left(\tfrac{\bar{\mu }^2}{M^2-p^2}\right)\right) +\tfrac{\left(\left(m_2+m_3\right)^3-3 \left(m_2+m_3\right) p^2-2 m_1^3-3 \left(m_2+m_3\right) m_1^2\right) }{p}\tanh ^{-1}\left(\tfrac{p+0i}{M}\right)\\
&\quad +\left(2 m_1-m_2-m_3\right)M+3 p^2+O(\epsilon^1)\Biggr].
\end{split}
\end{equation}
We have specified the branch of $\mathrm{arctanh}$ that must be taken for timelike $p = \sqrt{p^2} >M=m_1+m_2+m_3$. All divergent $2$-loop integrals in $3-\epsilon$ dimensions can be obtained from variations on this theme, although integration by parts identities may be required for the more non-trivial ones (say, with a numerator of $(k\cdot p) (l\cdot p)$).

\subsection{Full beta functions} \label{app:betas}

We present the full beta functions for the tensorial $\phi^2 \bar{\psi} \psi$ model at leading order in $N$, with all non-marginal couplings set to zero. We use $s=1/(8\pi)$ to indicate the order calculated to, since each $\lambda$ comes with an $s$ and each $h$ comes with an $s^2$. Plugging \eqref{eq:explicitPotential} into \eqref{eq:generalVectorResults}, we obtain:
{\scriptsize
\begin{align*}
\beta[\lambda_t]&=-\epsilon  \lambda_t+\frac{1}{9} s^2 (T+1) \lambda_t^3+s^4 \left(\left(\frac{h_p^2 + 3 h_w^2}{16200}\right) \lambda_t+\frac{1}{486} (-T (11 T+26)-2) \lambda_t^5\right)+O(s^5)\\
\beta[\lambda_{d_S}]&=-\epsilon  \lambda_{d_S}+\frac{2}{9} s^2 (T+1) \lambda_t^2 \left(\lambda_{p_E}+2 \lambda_{d_S}\right)\\
    & + s^4 \left(\left(h_p^2+3 h_w^2\right)\frac{ 3\lambda_{p_E}+4 \lambda_{d_S}}{4050}-\frac{2}{243} \lambda_t^4 \left(18 \lambda_{p_E}+23 \lambda_{d_S}+5 T^2 \lambda_{d_S}+56 T \lambda_{d_S}+39 \lambda_{p_E} T\right)\right)+O(s^5)\\
\beta[\lambda_{d_D}]&=-\epsilon  \lambda_{d_D}+\frac{1}{9} s^2 \lambda_t^2 \left((T+7) \lambda_{d_D}+4 \left(\lambda_{p_O}+\lambda_{p_S}\right)\right)\\
    &+ \frac{s^4}{48600} \left(\splitfrac{
(3h_p^2 + 9h_w^2 - 540h_p\lambda_t^2 - 11000\lambda_t^4 - 1100T^2\lambda_t^4 - 11600T\lambda_t^4)\lambda_{d_D}}{-(480h_p\lambda_t^2 + 8400\lambda_t^4 + 6000T\lambda_t^4)(\lambda_{p_O}+\lambda_{p_S})
}\right)+O(s^5)\\
\beta[\lambda_{p_E}]&=-\lambda_{p_E} \epsilon +\frac{2}{9} \lambda_{p_E} s^2 (T+1) \lambda_t^2+s^4 \left(\frac{h_p^2 \lambda_{p_E}}{4050}+\frac{h_w^2 \lambda_{p_E}}{1350}-\frac{2}{243} \lambda_{p_E} (T (5 T+17)+5) \lambda_t^4\right)+O(s^5)\\
\beta[\lambda_{p_O}]&=-\epsilon  \lambda_{p_O}+\frac{1}{9} s^2 \lambda_t^2 \left(\lambda_{p_O}(1+T)+2 \lambda_{p_S}\right)\\
    &+s^4 \left(\frac{\left(h_p^2+3 h_w^2\right) \lambda_{p_O}}{16200}-\frac{1}{810} h_p \lambda_{p_O} \lambda_t^2 - \frac{1}{486} \lambda_t^4 \left((T+2) (11 T+4)\lambda_{p_O} + 6 (5 T+3) \lambda_{p_S}\right)\right)+O(s^5)\\
\beta[\lambda_{p_S}]&=-\epsilon  \lambda_{p_S}+\frac{1}{9} s^2 \lambda_t^2 \left(\lambda_{p_S}(1+T)+ 2 \lambda_{p_O}\right)\\
    &+s^4 \left(\frac{\left(h_p^2+3 h_w^2\right) \lambda_{p_S}}{16200}-\frac{1}{810} h_p \lambda_{p_S} \lambda_t^2 - \frac{1}{486} \lambda_t^4 \left((T+2) (11 T+4) \lambda_{p_S}+ 6 (5 T+3) \lambda_{p_O}\right)\right)+O(s^5)\\%%
\beta[h_p]&=-2 \epsilon h_p+s^2 \left(\frac{1}{3} h_p T \lambda_t^2+\frac{h_p^2}{90}\right)+s^4 h_p \left(\frac{ h_p^2 + 3h_w^2}{5400}-\frac{1}{162} T (11 T+10) \lambda_t^4\right)+O(s^5)\\
\beta[h_w]&=-2\epsilon h_w+\frac{1}{3} h_w s^2 T \lambda_t^2+s^4 h_w \left(\frac{h_p^2 + 3 h_w^2}{5400}-\frac{1}{162} T (11 T+10) \lambda_t^4\right)+O(s^5)\\
\beta[h_3]&=-2\epsilon h_3 +s^2 \left(\frac{4}{9} h_3 T \lambda_t^2+\frac{h_p h_3}{45}\right)+s^4 h_3 \left(\frac{h_p^2 + 3 h_w^2}{5400}-\frac{1}{162} T (11 T+10) \lambda_t^4\right)+O(s^5)\\
\beta[h_4]&=-2\epsilon h_4 +s^2 \left(\frac{2}{3} T \left(h_4-10 \lambda_{p_E}^2\right) \lambda_t^2+\frac{h_3^2}{270}\right)+s^4 h_4 \left(\frac{h_p^2 + 3 h_w^2}{5400}-\frac{1}{162} T (11 T+10) \lambda_t^4\right)+O(s^5)\\
\beta[h_5]&=-2\epsilon h_5 +\frac{1}{270} s^2 \left(30 T \lambda_t^2 (6 h_p + 18 h_w + 5 h_5) +18 h_p^2+81 h_w^2+2 h_3^2-1800 T\lambda_t^2(\lambda_{p_E}^2 + 2 \lambda_t^2)\right)\\
    & +s^4 h_5 \left(\frac{h_p^2 + 3 h_w^2}{5400}-\frac{1}{162} T (11 T+10) \lambda_t^4\right)+O(s^5)\\
\beta[h_6]&=-2 \epsilon h_6+s^2 \left(\frac{2}{9} \left(h_3+3 h_6\right) T \lambda_t^2+\frac{h_p}{45}  \left(h_3+2 h_6\right)\right)+s^4 h_6 \left(\frac{h_p^2 + 3 h_w^2}{5400}-\frac{1}{162} T (11 T+10) \lambda_t^4\right)+O(s^5)\\
\beta[h_7]&=-2 \epsilon h_7+\frac{s^2}{270}  \left(\splitfrac{30T \lambda_t^2 (3 h_p + 9 h_w + 6 h_4 + 6h_5 + 8 h_7)}{+9 h_p^2+7 h_3^2+12 h_3 h_6-1800 T \lambda_t^2 (\lambda_t^2+12 \lambda_{p_E} \lambda_{d_S} +11 \lambda_{p_E}^2)}\right)\\
    &+s^4 h_7 \left(\frac{h_p^2 + 3 h_w^2}{5400}-\frac{1}{162} T (11 T+10) \lambda_t^4\right)+O(s^5)\\
\beta[h_8]&=-2\epsilon h_8 + \frac{s^2}{270}  \left(\splitfrac{30T \lambda_t^2 (h_5 + 4 h_7 + 12 h_8) +9 h_w^2+2 h_3^2+12 h_6^2+12 h_3 h_6}{-1800 T \lambda_t^2 (24 \lambda_{p_E} \lambda_{d_S} + 18 \lambda_{d_S}^2 + 5 \lambda_{p_E}^2)}\right)\\
& +s^4 h_8 \left(\frac{h_p^2 + 3 h_w^2}{5400}-\frac{1}{162} T (11 T+10) \lambda_t^4\right)+O(s^5).
\end{align*}
}
The general large-$N$ anomalous dimensions are as given in \cref{eq:anomDimPhiGeneralPert,eq:anomDimPsiGeneralPert}. For completeness, we also give the beta functions for the masses (where necessarily they must now take non-zero values):
{\scriptsize
\begin{subequations}
\begin{align}
    \beta [m^2]&=-2 m^2+\frac{4}{9} s^2 T \left(m^2-3 M^2\right) \lambda_t^2+8 M \sqrt{m^2} s^3 T \lambda_t^2 \left(\lambda_{p_E}+\lambda_{d_S}\right)\\
    & \quad +s^4 \left(\frac{2 (h_p^2 +3 h_w^2) m^2}{2025}+\frac{2}{243} T \lambda_t^4 \left(63 M^2 (T+3)-m^2 (5 T+43)\right)\right)+O(s^5)\notag \\
\beta [M]&=-M+\frac{4}{9} M s^2 \lambda_t^2+\frac{2 M}{243} \lambda_t^2 s^4 \left(648T (\lambda_{d_S}+\lambda_{p_E})^2 -13 T \lambda_t^2-23 \lambda_t^2\right)+O(s^5). \label{eq:perturbativeMbetaFunc}
\end{align}
\end{subequations}
}
Note that if we consider $\lambda \sim h \sim \epsilon$ of the same magnitude, then these beta functions are only correct to cubic order in the coupling constants. To get the full quintic corrections it would be necessary to renormalize $h$ using the $h^5$ diagrams.

At the \lammelonic{} fixed point, we can use \eqref{eq:perturbativeMbetaFunc} to compute the scaling dimensions of the mass operators; these match the non-perturbative calculations using \cref{eq:PevenGeneralsSDcondition,eq:PoddGeneralsSDcondition}:
\begin{subequations}\label{eq:quadraticSDs}
\begin{align}
\Delta_{\phi^2}^{\text{pert}} &=1+\left(3-\frac{4}{T+1}\right) \epsilon + \frac{2 T \left(6 T^2-22 T-41\right)}{3 (T+1)^3} \epsilon^2 +O\left(\epsilon ^3\right) = d+\odv{\beta[m^2]}{m^2}|_{fp}\\
\Delta_{\bar{\psi}\psi}^{\text{pert}}  &= 2+\left(\frac{4}{T+1}-1\right) \epsilon -\frac{(4 T (T+5)+42) \epsilon ^2}{3 (T+1)^3}+O\left(\epsilon ^3\right) = d+\odv{\beta[M]}{M}|_{\lambda_*}.
\end{align}
\end{subequations}

\subsection{Discrepancy with Jack and Poole} \label{app:JackPooleDiscrepancy}

In $d=3$, we can transform each Dirac fermion into two Majorana fermions via $\psi_a = \frac{1}{\sqrt{2}}(\xi_{a,1} + i \xi_{a,2})$. This keeps the kinetic terms canonical, and modifies the interaction term from $\half \lambda_{abcd} \phi_a \phi_b \bar\psi_c \psi_d = \frac{1}{4} Y_{abCD} \phi_a \phi_b \bar\xi_C \xi_D $, with $Y_{abCD} = \lambda_{abcd} \delta_{x_c x_d}$, where $C,D$ are superindices combining $c,d$ and $x_{c,d}\in \{1,2\}$. Plugging this into the results of \cite{Jack:2016utw}, ignoring the gauge couplings, we find beta functions for $Y$ and $h$. Though we have an alternate sign for $\lambda$, this does not change the beta functions. These should match ours when we substitute $T=\half T'$ in the equations of \eqref{eq:generalVectorResults}, since in $3$d a single Majorana fermion is half of a Dirac fermion:
\begin{equation}\begin{aligned}\label{eq:PooleMatching}
[\beta_{Y}]_{ab(c,x_c)(d,x_d)} = [\beta_{\lambda}]_{abcd}\, \delta_{x_c, x_d}|_{T \to \half T'}, \quad [\beta_h]_{abcdef} = [\beta_h]_{abcdef}|_{T \to \half T'}.
\end{aligned}\end{equation}
This is almost the case, except for the coefficient of $h_{B_1 B_2B_3B_4B_5\nu } \lambda_{B_6\rho \tau \sigma } \lambda_{\nu \rho \sigma \tau }$. This matches up with the coefficient $d_{3}^{(2)}$ of the tensor structure $V_3^{(2)}$. To match with our results as per \eqref{eq:PooleMatching}, we need $d_{3}^{(2)}=T' s^2$; the paper \cite{Jack:2016utw} has $d_{3}^{(2)}=2T' s^2$ (taking $T'=2$). %
Since this term contributes to all of the $\beta[h_i]$s in the melonic limit, the fact that the independent non-perturbative computations of this paper match our perturbative computations suggests that $d_{3}^{(2)}=T' s^2$. This was later confirmed by Jack in personal communication. %

\section{Vector model analysis} \label{sec:vectorModelAnalysis}

We reduce the general analysis to the $\gO(N_b) \times \gU(N_f)$ vector model, with $\phi_{a=1,\ldots,N_b}$, $\psi_{i=1,\ldots,N_f}$ at finite $N_b,N_f$, by keeping only maximal-trace couplings: 
\begin{equation}\begin{aligned}
V_{\text{int}}(\phi, \psi) = \frac{\lambda}{2} \phi_a \phi_a \bar{\psi}^i \psi_i + \frac{h}{6!} (\phi_a \phi_a)^3.
\end{aligned}\end{equation} 
In the case $N_b =N_f$, we could also permit the coupling $\half \lambda_{d_O} (\bar\psi_i \phi_i)(\phi_j \psi_j)$, which, if our scalar was complex, would correspond to the Popović model studied in \cref{sec:Popovic} -- we will not do so here. Substituting into \eqref{eq:generalVectorResults}:
\begin{subequations}
\begin{align}
\beta[\lambda] &= -\epsilon \lambda +\tfrac{8(T N_f +N_b+3)}{3}  \lambda^3 s^2  + \tfrac{16}{3} \tfrac{(N_b+4)(N_b+2)}{15} \tfrac{h^2}{15} \lambda \, s^4 + O(\lambda^4, \ldots)\\
\begin{split}
\beta[h] &= -2 \epsilon h + 4 (3 N_b + 22)\tfrac{h^2}{15} s^2 + 32 T N_f h \lambda^2 s^2 - 720 T N_f \lambda^4\\
& \quad \quad -\half \left( \splitfrac{\pi ^2 (((N_b+34) N_b+620) N_b+2720)}{+8 ((53 N_b+858) N_b+3304)}\right) \tfrac{h^3}{(15)^2} s^4 + O(h^4, \lambda^5,\ldots)
\end{split}\\
\gamma_\phi &=  \tfrac{(N_b+2) (N_b+4)}{15} \tfrac{h^2}{15} \tfrac{s^4}{6} +\tfrac{1}{3}  TN_f \lambda ^2 s^2 + O(h^3, \ldots) \label{eq:vectorGammaPhi}\\
\gamma_\psi &= \tfrac{1}{3} N_b  \lambda ^2s^2 + O(\lambda^3,\ldots) \label{eq:vectorGammaPsi}
 \end{align}
\end{subequations} %
For the bosonic sector, this agrees with the results of \cite{Pisarski:1982vz}, up to $\frac{h}{6!} = \frac{\pi^2 \lambda}{3}$. %

Thus for all $N_b, N_f, T$, six perturbative fixed points solve $\beta[\lambda]=\beta[h]=0$; to leading order in $\epsilon$, which manifestly match up in terms of $\epsilon$ dependence to the prismatic-type fixed points that were identified in \cref{sec:findingPerturbativeBetas}:
\begin{align*}
\text{trivial}: \; &s \lambda= 0, \; s^2 h = 0;\\
\hprismatic:\; &s\lambda = 0, \;  \frac{s^2 h}{15} = \frac{1}{2(22 + 3N_b)}\epsilon;\\
\hlamprismatic:\; &s\lambda = \pm_1 \left(\frac{3}{8(3 + N_b  + TN_f)}\right)^\half \sqrt{\epsilon}, \\
&\frac{s^2 h}{15} = \frac{ \left(1 -\frac{6 T N_f}{T N_f+N_b+3} \pm_2 \sqrt{\frac{3 T N_f \left(8 T N_f+23 N_b+186\right)}{\left(T N_f+N_b+3\right){}^2}+1}\right)}{4 (3 N_b+22)}\epsilon.
\end{align*}
Thus, these prismatic theories have not only a large-$N$ vector precursor, but also a finite-$N$ precursor. 
The presence of the $h^2$ and $\lambda^4$ terms in $\beta[h]$ means that we are missing any precursors of the melonic fixed points; these terms are of course suppressed by the melonic limit.

\subsection{Leading order conformal analysis} \label{app:conformalcalc}
\subsubsection{Free propagators}

To confirm the matching between Lorentzian and Euclidean signature, and also reproduce some of our $\beta$s, we consider the Lorentzian Lagrangian
\begin{equation}\begin{aligned}
\cL = \cL_b + \cL_f -V_\text{int}, \quad \cL_b = -\half \phi (-\partial^2 + m^2) \phi, \quad \cL_f = - \bar\psi (\slashed{\partial} + m) \psi,
\end{aligned}\end{equation}
in mostly positive signature $(-++\cdots)$, with $\bar\psi = \psi^\dagger \beta, \beta = i \gamma^0$, and $V_\text{int}$ from \eqref{eq:Neq1Lagrangian}.
The free Lorentzian position-space propagator is \cite{Zhang:2008jy} (where we set $m=0$): 
\begin{equation} \label{eq:freeBosonPropagator}
    G^\phi_0(x)=\expval{T \phi(x) \phi(0)} = \int \tfrac{d^d k}{(2\pi)^d} \tfrac{-i}{k^2 + m^2 -i \epsilon} e^{+i k \cdot x} = \tfrac{1}{(d-2) \Omega_{d-1}} \left( \tfrac{1}{x^2 + i \epsilon} \right)^{\frac{d-2}{2}}. %
\end{equation}

We define $\Omega_x$ as the surface area of an $x$-sphere, i.e. $\Omega_x = 2 \pi^{(x+1)/2} /\Gamma((x+1)/2)$.
Once we know this, the fermion propagator is trivial:
\begin{equation}\begin{aligned}%
    G^\psi_0(x) &= \expval{T \psi(x) \bar{\psi}(0)} = \expval{T \psi_\alpha(x) \bar{\psi}_\beta(0)} = \int \tfrac{d^d k}{(2\pi)^d} \tfrac{-i(-i \slashed{k}+m)}{k^2 +m^2 -i \epsilon} e^{+i k \cdot x}\\
    &= \int_k \tfrac{-i(-\not\partial +m)}{k^2 +m^2 -i \epsilon} e^{+i k \cdot x} \stackrel{m=0}{=} -\slashed{\partial} G_0^\phi(x) = \tfrac{\slashed{x}}{\Omega_{d-1}} \left( \tfrac{1}{x^2 +i \epsilon} \right)^{\frac{d}{2}}
\end{aligned}\end{equation}
\subsubsection{Anomalous dimensions}
Since we are interested in the fixed points of the Lorentzian version of \eqref{eq:Neq1Lagrangian}, we perform a simple conformal analysis in $d=3-\epsilon$. The equations of motion are
\begin{align}
    &\Box \phi = \partial^\mu \partial_\mu \phi = \lambda \phi \bar{\psi}\psi + \frac{h}{5!} \phi^5 \label{eq:bosonEOM}, \quad
    \left(\slashed{\partial} + \frac{\lambda}{2} \phi^2\right) \psi = 0, \quad \bar\psi\left(-\overleftarrow{\slashed{\partial}} + \frac{\lambda}{2} \phi^2\right) =0.
\end{align}
Considered as operator equations, these must be normal ordered. However, at the conformal fixed point, we know that the propagator looks like %
\begin{align}
    G(x) = \kappa_{\text{full}} \left(\tfrac{1}{x^2 +i\epsilon}\right)^{(d-2)/2 + \gamma_\phi} = G_0^\phi(x) + O(\gamma_\phi)
\end{align}
for some constant $\kappa_{\text{full}}$. At precisely $d=3$, we know that the theory is free, and hence near $d=3$ we can assume that $\gamma_\phi$, $\lambda$, and $h$ are small (at least $O(\sqrt\epsilon)$), as with the usual Wilson-Fisher fixed point. Squaring \eqref{eq:bosonEOM}, we can then calculate the following quantity\footnote{We use $\partial_\mu(1/\abs{x}^a) = - a x^\mu / \abs{x}^{a+2}$, which gives $\partial^\mu\partial_\mu(1/\abs{x}^a) = a (a +2-d)/\abs{x}^{a+2}$}:
\begin{align*}
    \expval{:\Box{\phi}(x)::\Box{\phi}(0):} &=\Box^2 G(x)|_{x=0} = \tfrac{4}{(d-2)\Omega_{d-1}} (d-2) d \gamma_\phi \left(\tfrac{1}{x^2 + i\epsilon}\right)^{(d+2)/2} + O(\gamma_\phi^2)\\
    &=  \expval{:\lambda \phi \bar{\psi}\psi +\frac{h}{5!} \phi^5:|_{x} :\lambda \phi \bar{\psi}\psi + \frac{h}{5!} \phi^5|_{0}:}
\end{align*}
We can assume the propagators are free to this order, and therefore we obtain
\begin{align}
   \frac{12}{\Omega_{2}} \gamma_\phi \left(\tfrac{1}{x^2 + i\epsilon}\right)^{5/2} + O(\gamma_\phi^2)= \lambda^2 \expval{:\phi_x\phi_0:}\expval{:\bar{\psi}\psi_x::\bar{\psi}\psi_0:} + \tfrac{h^2}{(5!)^2}\expval{:\phi_x^5: :: \phi_0^5:}.
\end{align}
 Precisely the same calculation can be performed for the fermionic equation of motion. From the free theory, we know $\kappa_{\text{full}} = (\Omega_2)^{-1}= 2s + O(\epsilon)$, where $s = 1/(8\pi)$ as usual. So performing the free Wick contractions, we obtain
\begin{align}
    \gamma_\phi = \frac{1}{3} T \lambda^2 s^2 + \frac{4}{3} \frac{h^2}{5!}, \quad \gamma_\psi = \frac{\lambda^2}{3} s^2.\label{eq:conformalAnalysisDims} 
\end{align}
to leading order in $\epsilon$, with $\Tr[\spinid]= T$. 
This is trivially extensible to the general vector case, which yields precisely \cref{eq:generalVecGammaPhi,eq:generalVecGammaPsi}. In the case of $N_f$ fermions and $N_b$ scalars, satisfying an $\gO(N_b)\times \gU(N_f)$ symmetry, adding the appropriate $\delta_{ij}$ and $\delta^m_n$s to the Lagrangian we find exactly \cref{eq:vectorGammaPhi,eq:vectorGammaPsi}.

\subsubsection{Beta functions}
As demonstrated by Cardy \cite{Cardy:1996:SRinSP,Gaberdiel:2008fn}, we can use conformal perturbation theory \cite{Komargodski:2016auf} (see also section 4 of \cite{Fei:2015oha}) to obtain the quadratic terms in the $\beta[h]$, in a particular scheme, using the OPE coefficients $C^I_{JK}$ of the CFT about which we are perturbing. Specifically, if perturbing a CFT with the set of dimensionless couplings $S_{\text{int}} = \sum_I \lambda^I \mu^{d-\Delta_I} \int \cO_I$, their beta functions should be
\begin{equation}\begin{aligned}\label{eq:CardyBetas}
\beta^I = (\Delta_I - d) \lambda^I + \half \Omega_{d-1} C^I_{JK} \lambda^J \lambda^K + O(\lambda^3),
\end{aligned}\end{equation}
where $C^{I}_{JK}$ is the OPE coefficient in the conformal theory: $\cO_J\times \cO_K \sim \sum_I C^I_{JK} \cO_I$. The particular scheme is similar to but not precisely the same as $\overline{\mathrm{MS}}$ scheme; however, to this order there is no difference in the beta function, because we are only considering the marginal couplings and $\overline{\mathrm{MS}}$ scheme is massless. Higher orders will be scheme dependent. 
For our case, take $\lambda^I = h$, with $\cO_I =\phi^6/6!$ as the operator in the Gaussian CFT. From the Gaussian OPE, it is easy to calculate that the OPE coefficient $C^h_{hh}$ is $\binom{6}{3} \times 6 \times 5 \times 4 = 2400$ multiplied by terms coming from the normalization of $h$ and $\phi$: %
\begin{equation}\begin{aligned}
\frac{\phi^6(x)}{6!} \times \frac{\phi^6(y)}{6!} \supset 2400 \times \frac{1}{6!} G_0(x-y)^3 \frac{\phi^6(y)}{6!}+ \cdots,
\end{aligned}\end{equation}
with $G_0^\phi(x)$ the usual free propagator, which in mostly positive or positive signature is \eqref{eq:freeBosonPropagator}.
Hence, the coefficient of $h^2$ is indeed as in \eqref{eq:generalVectorResults},
\begin{equation}\begin{aligned}
\half \Omega_{d-1} C^h_{hh} = \left[\half \Omega_{d-1} \times\frac{2400}{6!} \times \left(\tfrac{1}{(d-2)\Omega_{d-1}}\right)^3\right]_{d=3} = \frac{20 }{3} s^2.
\end{aligned}\end{equation}

\section{Analysis of the Popović model} \label{sec:Popovic}

Here we resolve some unclear points with the original model of \cite{Popovic:1977cq}, which is the large-$N$ theory of two $N$-vector fields, a Dirac fermion and a complex scalar. The corrected\footnote{The Lagrangian of \cite{Popovic:1977cq} contains a ``$\half \phi^\dagger \phi$'', but the $\phi$ field is clearly intended to be complex (for example, the symmetry factor of the one-boson-loop correction to the $\psi$ propagator is $1$). Additionally, we will leave $T$ general here for clarity, rather than taking $T=\Tr[\spinid] = 2$, which applies only in $d=3$.} $\gU(N)$-symmetric Lagrangian there corresponds to our model but with a complex $\phi$ and the only non-zero coupling being $\lambda_{d_O}$, which we give a bare value of $g_0$ to match conventions:
\begin{equation}\begin{aligned}
\cL &= Z_\phi \phi_i^\dagger (-\partial^2 + m_0^2)\phi_i + Z_\psi \bar\psi_i(i \gamma_\mu \partial^\mu - M_0) \psi_i + g_0 Z_\phi Z_\psi \sum_{i,j=1}^N (\bar\psi_i \phi_i^\dagger) (\phi_j \psi_j).
\end{aligned}\end{equation}
We shall set the masses to zero, which is possible in this DREG-like approach. We stress that we use the metric signature of \cite{Popovic:1977cq}, which is $(+--)$, and take
\begin{equation}\begin{aligned}
\Tr[\spinid] &= T, \quad \phi_{i=1,\ldots,N}, \quad \psi_{i=1,\ldots,N}.
\end{aligned}\end{equation}
They analyse this theory in the large-$N$ limit, finding that the beta function is zero in exactly three spacetime dimensions. Beginning with the four-point vertex function
\begin{equation}\begin{aligned}
i g_0 \Gamma_4(p) = \frac{i g_0}{1-ig_0\tilde\Sigma(p)} = \frac{1}{1-N g_0 G \slashed{p} (-p^2)^{(d-4)/2}},\quad G\equiv \frac{\Gamma(\frac{4-d}{2})B(\frac{d}{2}, \frac{d-2}{2})}{(4\pi)^{d/2}}
\end{aligned}\end{equation}
We define the renormalized and rescaled coupling $g$ by the renormalization condition
\begin{equation}\begin{aligned}
g_0 N &\equiv Z_g g \mu^{3-d}, \quad i g \equiv i g_0 [\Gamma_4(p^2 =\mu^2)]_{\mathbb{I}},
\end{aligned}\end{equation}
where $[\cdot]_{\mathbb{I}}$ indicates that we drop the part proportional to $\slashed{p}$. Note that this is not minimal subtraction.
The fields are not renormalized to leading order in $N$, so $Z_{\phi/\psi}=1+ O(1/N)$. Hence, we find that $Z_g = 1 + g_0^2 N^2 G^2 \mu^{2(d-3)} + O(1/N) = 1+ Z_g^2 g^2 G^2 +O(1/N)$.
Thus, almost every value of the renormalized coupling in the allowed range $g \in [-1/(2G), +1/(2G)]$ can be reached by two values of the bare coupling (which can take any real value):
\begin{equation}\begin{aligned}
Z_g &= \frac{1\pm \sqrt{1-4 g^2 G^2}}{2 g^2 G^2} +O(1/N)\\
\implies \beta[g] &= \mu \od{g}{\mu} = \pm (3-d)\sqrt{1- 4g^2 G^2} +O(1/N)
\end{aligned}\end{equation}
Note that in the limit $d=4-\epsilon$, $G \propto 1/\epsilon$, so this theory has a Wilson-Fisher-like fixed point that is perturbative around $d=4$. They then move on to calculate the anomalous dimension of $\phi$. However, in their equation (3.4) the $\frac{1}{\sqrt{-k^2}}$ should be a $\frac{1}{(-k^2)^{(d-2)/2}}$ -- they have taken the limit too early. However, in trying to resolve this, we then have to evaluate the integral %
\begin{equation}\begin{aligned}
i\tilde\Pi(p^2)&= (-1) \Tr \int \frac{d^d k}{(2\pi)^d} \frac{i}{\cancel{(k-p)}} (ig_0) i\Gamma_4(k) = i g_0\int_k \frac{1}{(k-p)^2} \Tr\left[\cancel{(k-p)} \Gamma_4(k)\right]\\
&= -i Tg_0 \int_k   \frac{(k-p)\cdot k}{[-(k-p)^2]^1} \frac{g_0 N G (-k^2 )^{\frac{d-4}{2}}}{1 +  g_0^2 N^2 G^2 (-k^2)^{d-3}}\\ 
&\propto \int \odif{q}\, q^{d-1}  \frac{q^2}{q^2}\frac{q^{d-4}}{q^{2d-6}} \sim \int_0^\infty \odif{q}\, q^{d-1} \frac{1}{q^{d-2}} \to \infty,
\end{aligned}\end{equation}
which is manifestly divergent for all values of $d$. 
This is a standard feature of the vector models; a neat way to proceed in precisely $d=3$ is to regulate by consistently shifting the conformal scaling dimension of the Hubbard-Stratonovich field \cite{Fraser-Taliente:2025udk} (see also \cite{Goykhman:2019kcj,Vasiliev:1975mq,Ciuchini:1999wy}); that is, we switch $(-k^2)^{\frac{d-4}{2}} |_{d=3} \to (-k^2)^{-\half -\eta} \mu^{2\eta}$, for some infinitesimal $\eta$, which becomes the regulator:
\begin{equation}\begin{aligned}
\tilde\Pi(p^2)|_{d=3} &= -T \frac{g_0^2 N G }{1 +  g_0^2 N^2 G^2}\int_k   \frac{(k-p)\cdot k}{[-(k-p)^2]^1(-k^2 )^{\half + \eta}} \\ 
 &= \frac{-i(-p^2)^{1-\eta}}{\eta} T \frac{\tilde{A}}{N} \times \mu^{2\eta} + O(\eta^0), \\\tilde{A}&\equiv \frac{1}{12 \pi^2} \frac{(Ng_0)^2 G}{1+(Ng_0 G)^2} =\frac{4}{3 \pi^2} \frac{Z_g -1}{Z_g} 
\end{aligned}\end{equation}
where $T\tilde{A} =A_{\text{Popović}}$.
Then we can define the field renormalization $Z_\phi \phi_R \equiv \phi$ with the renormalization condition $\Delta_R(p^2) =\frac{i}{p^2}$ at $p^2=-\mu^2$. This gives $Z_\phi = (1 + \frac{T\tilde{A}}{N} \frac{1}{\eta})^{-1}$, and so the anomalous dimension is simply $\gamma_\phi = \odv{\log \sqrt{Z_\phi}}{\log \mu} = T\frac{\tilde{A}}{N} + O(1/N^2)$.
Playing the same game for the fermion, we find an anomalous dimension $\gamma_\psi =N_b \frac{\tilde{A}}{N} + O(1/N^2)$, where we define $N_b=2$ as the number of real degrees of freedom of each entry of the scalar field. These results agree with those of \cite{Popovic:1977cq}. 
In the strong coupling limit, $g_0 N\to\infty$, $Z_g \to \infty$, $g\to 0$ these anomalous dimensions also precisely match the leading order results of a $\phi_i \bar\psi_i \lambda$-type melonic theory in $d=3$, %
where $\lambda$ is a fermionic Hubbard-Stratonovich field. 
This can be checked using the $\Ft$-extremization procedure in \cref{sec:fundamentalClaim}. 
We note that with canonical kinetic terms, the IR scaling for this melonic theory ($\gamma_\phi, \gamma_\psi>0, \gamma_\lambda >0$) is in fact only consistent for $2<d<3$; in particular, the requirement $\gamma_\lambda \ge 0$ invalidates these solutions for $d \ge 3$.

\section{SDE Euclidean integral calculations} \label{app:loopIntegrals}

For ease of use, we put here some standard Euclidean integrals. These are trivial in position space, so can be calculated straightforwardly by Fourier transforming, as described in \href{https://arxiv.org/pdf/1706.05362.pdf#appendix.B}{\color{black}{appendix B}} of \cite{Murugan:2017eto}. %

The first is symmetric in $\alpha \longleftrightarrow \beta$, which we emphasise by defining $X_2 \equiv d-\alpha-\beta$.
\begin{equation}\begin{aligned}
    I_{\alpha,\beta}(p) &= \int \frac{\mathrm{d}^d k} {(2\pi)^d} \frac{1}{(k^2)^\alpha ((k+p)^2)^\beta} = \tfrac{1}{(4\pi)^{d/2}} \tfrac{\Gamma \left(\frac{d}{2}-\alpha \right) \Gamma \left(\frac{d}{2}-\beta \right) \Gamma \left(\alpha +\beta -\frac{d}{2}\right)}{\Gamma (\alpha ) \Gamma (\beta ) \Gamma (d -\alpha -\beta)(p^2)^{\alpha+\beta-\frac{d}{2}}}\\
    &= \tfrac{1}{(4\pi)^{d/2}} \tfrac{1}{(p^2)^{\frac{d}{2}-X_2}} \tfrac{\Gamma \left(\frac{d}{2}-\alpha \right) \Gamma \left(\frac{d}{2}-\beta \right) \Gamma \left(\frac{d}{2}-X_2\right)}{\Gamma (\alpha ) \Gamma (\beta ) \Gamma (X_2)}. \label{eq:singlemelonstep} %
\end{aligned}\end{equation}
Another basic integral is
\begin{align}
    \int_{k} \frac{k \cdot p}{k^{2\alpha} (k+p)^{2\beta}}  &= \frac{1}{2} \int_k \frac{(k+p)^2 - k^2 -p^2}{k^{2\alpha} (k+p)^{2\beta}} = \frac{1}{2}\left(I_{\alpha,\beta-1}(p) - I_{\alpha-1, \beta}(p) - p^2 I_{\alpha,\beta}(p) \right).
\end{align}
Next, the quartic bosonic melon, symmetric in $\alpha,\beta,\gamma$, with $X_3\equiv \frac{3d}{2} -\alpha -\beta -\gamma$:
\begin{subequations}
\begin{align}
I_{\alpha\beta\gamma} = \int_{k,l} \frac{1}{(k+l+p)^{2\alpha} k^{2\beta} l^{2\gamma}} = \frac{1}{(4\pi)^d} \tfrac{\Gamma(\frac{d}{2} - \alpha)\Gamma(\frac{d}{2} - \beta)\Gamma(\frac{d}{2}- \gamma) \Gamma(\frac{d}{2} - X_3)}{\Gamma(\alpha)\Gamma(\beta)\Gamma(\gamma) \Gamma(X_3) (p^2)^{\frac{d}{2} - X_3}},
\end{align}
which has similar variations
\begin{align}
J_{\alpha\beta\gamma}(p) = \int_{k,l} \frac{k\cdot p}{(k+l+p)^{2\alpha} k^{2\beta} l^{2\gamma}} = \tfrac{-(d-2\beta) p^2}{3d - 2 (\alpha+ \beta+\gamma)}I_{\alpha\beta\gamma}(p),
\end{align}
\begin{align}
K_{\alpha\beta\gamma}(p) = \int_{k,l} \frac{k\cdot l}{(k+l+p)^{2\alpha} k^{2\beta} l^{2\gamma}} = \tfrac{-(d-2\beta)(d-2\gamma) p^2 \, I_{\alpha\beta\gamma}(p)}{2(\alpha+\beta + \gamma - d- 1)(3d - 2 (\alpha+ \beta+\gamma))}.
\end{align}
\end{subequations}
Hence, using that integrals containing $\slashed{k}$ or $\slashed{l}$ must be proportional to $\slashed{p}$, we find:
\begin{equation} \begin{aligned}
    \Sigma_F^{FBB}(p) &= \frac{\lambda_{t}^2}{2} FB^2 \left(I_{\alpha\beta\beta} + \frac{1}{p^2} (J_{\alpha \beta \beta} + J_{\alpha\beta\beta})\right),\\
    \Sigma_B^{BBB}(p) &= \frac{g_{t}^2}{3!} B^3 I_{\beta\beta\beta}, \quad \Sigma_B^{BFF}(p) = -\frac{\lambda_{t}^2}{1} B F^2 K_{\beta\alpha\alpha} \Tr[\spinid].
\end{aligned}\end{equation}
We also require the sextic bosonic melon, which in Euclidean signature is \cite{Benedetti:2019rja}:
\begin{equation}
    \Sigma_B^{B^5}(p) = \tfrac{h_t^2 B^5}{5!} \left(\tfrac{p^{4d-10}}{(4\pi)^{2d}} \frac{\Gamma(\tfrac{d}{2}- b)^5 \Gamma(5b- 2d)}{\Gamma(b)^5 \Gamma(5\frac{d}{2} - 5 b )} \right).
\end{equation}

 \subsection{Euclidean integrals for the \texorpdfstring{$\phi^q$}{phi\^q}-model} \label{sec:qModelIntegrals}

The general Euclidean scalar melon with $q-1$ different arbitrary conformal propagators $\sim p^{-2 \alpha_i}$ is found by either induction or Fourier transforming to be
\begin{subequations}
\begin{equation}\begin{aligned}
&I_{\alpha_1, \alpha_2, \ldots, \alpha_{q-1}} (p) = \int \prod_{i=1}^{q-2} \left(\frac{d^d k_i}{(2\pi)^d} \frac{1}{k_i^{2\alpha_i}}\right) \frac{1}{(\sum_i^{q-2} k_i + p)^{2\alpha_{q-1}}} \\
&=\frac{1}{(4\pi)^{d(q-2)/2}} \left( \prod_i^{q-1} \frac{\Gamma(\frac{d}{2} - \alpha_i)}{\Gamma(\alpha_i)}\right) \frac{\Gamma(\frac{d}{2} - \sum_i^{q-1} (\frac{d}{2} -\alpha_i))}{\Gamma( \sum_i^{q-1} (\frac{d}{2} -\alpha_i))} \frac{1}{p^{2(\frac{d}{2} - \sum_i^{q-1} (\frac{d}{2} -\alpha_i))}}.
 \ee
The induction proceeds by noting that $I_{\alpha_1, \alpha_2, \ldots, \alpha_{q-1}} (p) = \int_k k^{-2 \alpha_{q-1}} I_{\alpha_1, \alpha_2, \ldots, \alpha_{q-2}} (p+k)$; the base case and inductive step are both \eqref{eq:singlemelonstep}. The fermion integrals, with numerator $k \cdot l$, can also be found by induction from the base case, with the result:
\begin{equation}\begin{aligned}
&K_{mel}(p, \{\alpha_1, \alpha_2\}, \{\alpha_3, \ldots, \alpha_{q-1}\})\\
&= \int \frac{d^d k_1}{(2\pi)^d} \frac{d^d k_2}{(2\pi)^d} \frac{k_1 \cdot k_2}{k_1^{2\alpha_1 +1} k_2^{2\alpha_2+ 1}} \, \prod_{i=3}^{q-2} \left(\frac{d^d k_i}{(2\pi)^d} \frac{1}{k_i^{2\alpha_i}}\right) \frac{1}{(\sum_{i=1}^{q-2} k_i + p)^{2\alpha_{q-1}}} \\
&=\frac{-1}{(4\pi)^{d(q-2)/2}} \left( \prod_{i=1}^{2} \frac{\Gamma(d/2 - \alpha_i + 1/2)}{\Gamma(\alpha_i + 1/2)}\right) \left( \prod_{i=3}^{q-1} \frac{\Gamma(d/2 - \alpha_i)}{\Gamma(\alpha_i)}\right) \\ & \quad \times \frac{\Gamma(d/2 - \sum_i^{q-1} (d/2 -\alpha_i))}{\Gamma( \sum_i^{q-1} (d/2 -\alpha_i))} \frac{1}{p^{2(d/2 - \sum_i^{q-1} (d/2 -\alpha_i))}}. %
\end{aligned}\end{equation}
We can work out the propagators with $k \cdot p$ in the same way:
\begin{equation}\begin{aligned}
&J_{mel}(p, \{\alpha_1\}, \{\alpha_2, \ldots, \alpha_{q-1}\})\\
&= \int \frac{d^d k_1}{(2\pi)^d} \frac{k_1 \cdot p}{k_1^{2\alpha_1 +1}} \, \prod_{i=2}^{q-2} \left(\frac{d^d k_i}{(2\pi)^d} \frac{1}{k_i^{2\alpha_i}}\right) \frac{1}{(\sum_{i=1}^{q-2} k_i + p)^{2\alpha_{q-1}}} \\
&=\frac{-1}{(4\pi)^{d(q-2)/2}} \left(\frac{\Gamma(d/2 - \alpha_1 + 1/2)}{\Gamma(\alpha_1 + 1/2)}\right) \left( \prod_{i=2}^{q-1} \frac{\Gamma(d/2 - \alpha_i)}{\Gamma(\alpha_i)}\right) \\ & \quad \times \frac{\Gamma(d/2 +1/2 - \sum_i^{q-1} (d/2 -\alpha_i))}{\Gamma(1/2 + \sum_i^{q-1} (d/2 -\alpha_i))} \frac{1}{p^{2(d/2 -1/2 - \sum_i^{q-1} (d/2 -\alpha_i))}}. %
\end{aligned}\end{equation}
\end{subequations}

\end{subappendices}

%\chapter{The quartic Yukawa model and general features of melonic CFTs}\label{chap:3dyuk}

\chapter{Outlook}\label{chap:conclusion}

The melonic CFTs represent a large class of solvable large-$N$ conformal field theories in general dimension. 
The work of this thesis has been to understand that the fixed points found by these large-$N$ field theories can be understood by the process of $\Ft$-extremization.

We began by investigating the large-$N$ expansion and renormalization using a pair of toy models: a zero-dimensional vector model, and the flow between a pair of generalized free fields. We then studied the large-$N$ vector models, and outlined the melonic field theories that are currently known.
In \cref{chap:fextr}, we then proved that the conformal fixed points of the melonic field theories must necessarily be free conformal fields with an extremized number of degrees of freedom (measured by $\Ft$); then we discussed the practical results of this extremization procedure.

In \cref{chap:3dyuk}, we studied these theories in greater detail: we considered the RG flows and spectrum of one particular group of melonic theories that come from the quartic Yukawa QFT. This allowed us to observe generic features of the spectra, stability, and unitarity of the family of melonic CFTs.

\section{\FttextOrPDF-extremization} \label{sec:FextrOutlook}

In \cref{chap:fextr} we established our fundamental claim \eqref{eq:FmaxSummary}; \mccorrect{that is, the melonic CFTs in the strict large-$N$ limit have the same conformal scaling dimensions as a mean field theory, also called a collection of generalized free fields, with extremal sphere free energy -- subject to the linear melonic constraint on the IR scaling dimensions enforced by the interaction}.

This holds regardless of any finite symmetry groups and the interaction structure: we therefore have a complete solution and classification of melonic CFTs. 
Now, we expected mean-field behaviour in the large-$N$ limit, due to factorization. However, since $\Ft$ is also thought to count the number of degrees of freedom in a CFT, this has the pleasing interpretation that we extremize the number of IR degrees of freedom.

To establish this claim, we used the fact that the quantum solution of a field theory lies at the extremum of the two-particle-irreducible effective action $\Gamma_\mathrm{2PI}$. 
This holds even when $\Gamma_\mathrm{2PI}$ is evaluated on the IR conformal slice of solutions; thus, the quantum solution lies at an extremum with respect to the trial scaling dimensions of the fundamental fields, which leads directly to a non-perturbative definition of $\Ft(\{\Delta_\phi\})$. 
We then illustrated this procedure and its results using various example melonic CFTs, and saw that much of the IR structure can be understood by generalizing that seen in the case of the large-$N$ vector models (including the $\Box^k$ CFTs).

Future questions fall into two categories. The first concerns the nature of the IR structures that arise here: 
\begin{enumerate}
\item We often find a discrete choice of vacua in the IR, some of which are maxima, some of which are minima, and some of which are saddle points. %
Can we identify a mechanism to select which one is physically realised? Equivalently, in the SUSY mechanism, unitarity requires that $F$ and $a$ must be maximized; is the same true here?
Of course, these theories can only be unitary in integer dimensions, due to the presence of evanescent operators \cite{Hogervorst:2015akt,Ji:2018yaf}; nonetheless, is there a sense in which $\Ft$-maximization holds for QFTs that are the analytic continuations in $d$ of unitary theories?
\item Along similar lines: is $\Ft_\mathrm{IR} \le \Ft_{UV}$ always true, at least within the IR wedge? 
This would give further hints towards a generalized $\Ft$-theorem, valid in continuous dimension \cite{Giombi:2014xxa,Fei:2015oha,Giombi:2015haa}. 
The fact that the usual free field values are local maxima of $\Ft_\phi$ almost -- but not quite -- shows this, as an unconstrained maximum always gives an upper bound for a constrained maximum. %
\item The additional vacua lying outside the unitarity wedge can be understood in the case of the vector model as corresponding to the $\Box^k$ CFTs -- these also have known AdS duals \cite{Sun:2020ame,Bekaert:2013zya,Brust:2016zns}. 
It would be nice to understand the additional vacua in other melonic models to the same degree.
\item \cite{Gulotta:2011si,Taylor:2016kic} argued that the decrease of $\Ft$ between the two fixed points should be related to the increase of the size of the compact manifold \cite{Lobachevsky} in the AdS dual. Can this connection be made precise for the melonic case? %
\item In certain integer dimensions $d$, we find no solutions for the field scaling dimensions, despite the existence of perturbative solutions around that $d$.
This is just as in \cite{Biggs:2023mfn}; it was also observed in a melonic approximation to QCD in \cite{Vasiliev:1981sf}. 
What, therefore, happens in the IR of these theories \cite{Schaub:2024rnl}? 
\item We can also compute the scaling dimensions of the bilinear operators in the OPE of the fundamental fields, as well as their OPE coefficients, at least for the tensorial realisation of the SYK model \cite{Benedetti:2021wzt}; what can we discover by analysing these? Various divergences are evident in the bilinear scaling dimensions \cite{Giombi:2018qgp,Fraser-Taliente:2024rql} in the case when the scaling dimension of one of the fields hits zero, despite the fact that $d$ is non-integer. 
Does the $\Ft$-extremization perspective aid in understanding the true fate of these theories in the IR, perhaps as logarithmic CFTs?
\item The melonic and supersymmetric CFT mechanisms are identical in practice. The only difference is that for the latter, at finite $N$ the Lagrange multiplier is a more complicated function of the coupling constant, i.e. $\lambda \sim g^2 \prod_\phi \cZ_\phi^{q_\phi} + O(g^3)$. 
This coincidence arises because the form of the potential is protected in both, by supersymmetry and large-$N$ respectively; hence only field renormalization occurs. However, is there more to this than coincidence? %
In any case, we can directly import the results from $F$ and $a$-maximisation to the melonic SCFTs. This may lead to a neat explanation of their various spectral divergences: for example, in \cite{Popov:2019nja}, we obtain $\Delta =0$ in $d=1$, and therefore a missing tower of operators in (at least) the $BB$-type bilinears; likewise in \cite{Lettera:2020uay}. 
\end{enumerate}
The second category involves extending this procedure in various directions:
\begin{enumerate}
\item It is clear that the contribution of non-melonic diagrams will break the interpretation of the IR solution as extremizing the free energy of a collection of free fields (subject to a constraint). 
Of course, we must still lie at an extremum of $\Gamma_\mathrm{2PI}$. It would therefore be interesting to compute the first correction at subleading order in $N$. 
In the SYK model, the NLO 2PI vacuum graphs take the form of the periodic ladders with $n \ge 1$ rungs, with or without one twist of the rails \cite{Benedetti:2018goh}; in the tensor models, diagrams containing other couplings enter at NLO, and ladder-like diagrams appear at NNLO \cite{Benedetti:2021wzt}. 
We note that there are certain subtleties associated with the computation of the 2PI action \cite{Benedetti:2018goh}.
We performed this computation for the vector models in \cite{Fraser-Taliente:2025udk}, and observed the related fact that the standard short-range models lie at the extrema of $\Ft$ of the long-range models. This therefore extended the $\Ft$-extremization that we observed in the vector model case to the next order\footnote{This is proven nonperturbatively in \cite{Fraser-Taliente2026LRSR}.}.
\item Along similar lines: we often find complex solutions for the scaling dimension of some operator -- these are thought to indicate that in the true IR vacuum that operator would condense \cite{Benedetti:2021qyk}. Thus: what happens when we permit the possibility of symmetry breaking, as in \cite{Kim:2019upg,zhaoSymmetryBreakingMelonic2022}?
\item One obvious generalization that suggests itself is to formally modify the melonic constraint to some generic $f(\{\Delta_\phi\})=0$. This was first explored in \cite{Shen:2023srk}, where a bilocal interaction was used to set the melonic constraint of an SYK-like theory to $q \Delta_\phi = d -\alpha$ for tunable $\alpha$.
\item The so-called higher melonic theories \cite{Gubser:2018yec} fit very neatly into this framework. %
We just compute $\Ft$ for irreps of $\SO(d+1,1)$ over non-Archimedean fields, and then perform a constrained extremization as usual. Their multiple-field generalizations may then prove rich. 
This also suggests considering the properties of the melonic-type theories defined over non-compact groups $G$ other than the conformal group; these are then a solvable sector of the $G$-theories proposed in \cite{Gadde:2017sjg}. %
\end{enumerate}

\section{General features of melonic CFTs} \label{sec:3dyukOutlook}

In \cref{chap:3dyuk} we considered a set of novel interacting conformal field theories, which generalize and extend the known set of higher-dimensional tensor field theories to include fermions, and the RG flow relations between them. 

The CFTs we identified could be broadly classed within the existing categories of melonic and prismatic -- however, the QFT contained both types of fixed point, and also included fermions, which led to additional structure in the network of CFTs.

However, we observed that the melonic limit is more fundamental, as the prismatic limit is simply that of melonic diagrams where some propagators in the melons are the propagators of non-dynamical auxiliary fields. 
At the perturbative level of the $3-\epsilon$ expansion, without an auxiliary field, we can access both of these fixed points: the diagrams contributing to each are identical in terms of momentum structure, but differ in tensorial structure. However, if we want to find the prismatic fixed points using the Schwinger-Dyson equations (non-perturbatively) or indeed the $\Ft$-extremization framework of \cref{chap:fextr}, we must introduce the auxiliary fields.

While studying the perturbative RG flow between these theories, we identified a candidate line of fixed points perturbatively around $d=3$, and established the non-perturbative existence of the melonic fixed points in general dimension.
In \cref{sec:generalCharacteristicsPrismatic,sec:bilinearsCalculationResults}, by considering the simple multi-field model of just scalars (the \hprismatic fixed point), we found the various elements that enter the analysis of a melonic CFT in arbitrary $d$. 
Namely: the IR wedge; the infinite branches of UV and IR solutions in general dimension (including collision of these branches, leading to complex solutions); the apparent breakdown of the conformal analysis at exceptional values of $d$ and $\Delta_\phi$, and the associated breaks in the contours for both the two-point and four-point functions at those exceptional values. 
We expect these features to be generic for melonic CFTs, and indeed saw them reproduced in the \lammelonic CFT.

Understanding the generic features of the melonic CFTs given above at a deeper level is clearly essential. To that end, further avenues of exploration are the following:
\begin{enumerate}[noitemsep]
    \item We have focused here on the branch of \lammelonic descending from the free theory in $d=3-\epsilon$. What is described by the other branches of solutions, typically lying outside the IR wedge, of which there are an infinite number for almost every $d$? This was partially answered for the critical vector model in \cref{sec:vectorModels}; they should correspond to the theories defined with local variants of the UV kinetic term, but can this be understood systematically\footnote{See \cite{Fraser-Taliente2026LRSR} for the resolution of this question.}?
    \item Can the fate of the theory in the dimensions that exhibit the spectral divergences, discussed in \cref{sec:zeroScaling}, be understood? How does the logarithmic character of the two-point function modify the true IR physics?
    \item We have fixed points that exist only perturbatively around exceptional dimensions, $d=d_0 + \epsilon$, but not for $\epsilon=0$. These exist in the short and long-range models, and there are hints that we should regard these fixed points in $d=d_0$ as the limit of $\epsilon \to 0$, but their nature remains unclear \cite{Benedetti:2019ikb}.
    \item In \cref{sec:stabmats}, we found that in $d=3-\epsilon$, \hmelonic and \hlammelonic appear to sit on a perturbative line of fixed points in the direction of $-\hp + 3 h_5 - 3 h_7 + h_8$; it is similar to the perturbative line of fixed points that was found in $d=3-\epsilon$ in the $\gO(N)^4$ $\phi^6$ model in \cite[\S 3.4]{Jepsen:2023pzm}.
    \mccorrect{It is known for the vector $\gO(N)$ $\tfrac{\tilde{h}}{N^2}\phi^6$ model in exactly $d=3$ that in the strict $N\to\infty$ limit there exists another apparent line of fixed points that terminates at some finite $\tilde{h}=\tilde{h}_\star$ due to the non-perturbative Bardeen-Moshe-Bander (BMB) phenomenon} \cite{Kvedaraite:2025lgi}. 
    However, said line of fixed points is now understood as an artifact of an incorrect approach to the large-$N$ limit; it also disappears for any $d\neq 3$\footnote{Essentially, this is due to an incorrect (non-optimal) scaling of the coupling $\tilde{h}$ that makes the entire beta function subleading: $\beta_{\tilde{h}} = -2 \epsilon \tilde{h} + O(1/N)$. 
    When the coupling is scaled correctly, $\tfrac{h}{N^{3/2}} \phi^6$, no such line of fixed points is seen (see the discussion around III.B of \cite{Kvedaraite:2025lgi}).}. 
    \mccorrect{This suggests that the fictitious $\phi^6$ line is likely unrelated to the apparent marginal operator observed in the $h(\lambda)_{\mathrm{melonic}}$ cases, since our marginal operator exists for $\epsilon\neq 0$. 
    It would be interesting to consider the next orders in $N$ and $h_i$, to see what happens to this line of fixed points; however, can this line be understood or eliminated without resorting to perturbative six-loop computations?}
    \item The $4$d Yukawa model is not accessible with rank-$3$ tensors; is there a tractable tensorial $\gO(N)^r$ realisation of a melonic $4$d Yukawa model? Such a model would enable comparison of the $4$d Yukawa model with the $3$D result presented here. The non-perturbative singlet spectrum has been studied already via a disorder realisation in \cite{Prakash:2022gvb}.
    \item We have tacitly assumed that there is no symmetry breaking in the IR, whether of the $\gO(N)^3$, $\gU(1)$, or conformal symmetry. This is despite the clear indication in \cref{sec:windowsOfStability} of dimensional windows in which the theory is unstable \cite{Benedetti:2021qyk}. By loosening these assumptions, can we understand the true IR phase? One particular symmetry breaking pattern has been studied in \cite{Benedetti:2019sop}, but this does not apply, for example, to melonic theories realised via disorder.
    \item A number of generalizations of standard melonic theories that exist in the literature could be pursued with this model, made more interesting by the presence of fermions in a higher-dimensional CFT under good analytic control: for example, consideration of the CFT induced on a defect in this theory, along the lines of \cite{Popov:2022nfq,Gimenez-Grau:2022ebb}; or consideration of the long-range version of this model \cite{Gross:2017vhb,Benedetti:2019rja,Harribey:2021xgh,Shen:2023srk}.
    \item A number of general field-theoretical questions are suggested by this model: The short-range quartic tensor model at its fixed point -- with imaginary tetrahedral coupling -- was found to be asymptotically free in the large-$N$ limit \cite{Berges:2023rqa}, giving access to a strongly coupled theory where analytic progress can be made. The long-range bosonic model was used to verify the F-theorem, even in the absence of unitarity, in \cite{Benedetti:2021wzt}. To what extent can these behaviours be extended to the other theories in this class, such as this one? The presence of fermions complicates the computations, due to the nontrivial $\SO(d)$ representations.
    \item Finally: the bulk dual of these melonic theories remains unknown \cite{Harribey:2022esw}. However, because of the existence of a vast number of gauge-invariant operators involving higher powers of the tensor field, the gravity dual is expected to be complicated \cite{Klebanov:2016xxf,Rosenhaus:2018dtp}.
\end{enumerate}

%\end{comment}
%\chapter{Discussion}
%Discuss.
%\include{text/ch2-litreview}
%% APPENDICES %% 
% Starts lettered appendices, adds a heading in table of contents, and adds a
%    page that just says "Appendices" to signal the end of your main text.
% Add or remove any appendices you'd like here:
%\startappendices

%The authors are grateful to M\'ark Mezei and John March-Russell for discussions. LFT is supported by a Dalitz Scholarship from the University of Oxford and Wadham College.

%For the purpose of open access, the authors have applied a CC BY public copyright licence to any Author Accepted Manuscript (AAM) version arising from this submission.

%%%%% REFERENCES

% JEM: Quote for the top of references (just like a chapter quote if you're using them).  Comment to skip.
%\begin{savequote}[8cm]
%The first kind of intellectual and artistic personality belongs to the hedgehogs, the second to the foxes \dots
%  \qauthor{--- Sir Isaiah Berlin \cite{Kravchuk:2016qvl}}
%\end{savequote}

\setlength{\baselineskip}{0pt} % JEM: Single-space References
{
\cleardoublepage % modify margins on the new page. 
%\newgeometry{left=2cm, right=2cm, top=2cm, bottom=2cm}
\newgeometry{left=2.5cm,right=2.5cm,bottom=3cm}
\renewcommand*\MakeUppercase[1]{#1}%
\printbibliography[heading=bibintoc,title={\bibtitle}]
}

\end{document}